\documentclass[aps,twocolumn,footinbib, notitlepage,superscriptaddress, longbibliography]{revtex4-1}
\usepackage{graphicx}
\usepackage{amssymb}
\usepackage{amsmath}
\usepackage{amsfonts}
\usepackage{overpic}
\usepackage[normalsize]{subfigure}
\usepackage{color}
\usepackage{xcolor}
\usepackage{xspace}
\usepackage{hyperref}
\usepackage{bm}
\usepackage[english]{babel}

\DeclareFontFamily{OT1}{pzc}{}
\DeclareFontShape{OT1}{pzc}{m}{it}{<-> s * [1.10] pzcmi7t}{}
\DeclareMathAlphabet{\mathpzc}{OT1}{pzc}{m}{it}

\providecommand{\st}[1]{_{\text{#1}}}

\providecommand{\sfrac}[2]{#1/#2}
\providecommand{\ut}[1]{^{\text{#1}}}

\def\onehalf{\frac{1}{2}}
\def\onequarter{\frac{1}{4}}

\def\bra{\ensuremath{\langle}}
\def\ket{\ensuremath{\rangle}}
\def\eq{\st{eq}}

\def\const{\mathrm{const}}

\def\pd{\partial}

\def\uv{\bv{u}}

\def\rv{\bv{r}}

\def\b0{\bv{0}}

\def\eff{\st{eff}}

\def\Fcal{\mathcal{F}}

\def\Hcal{\mathcal{H}}
\def\Hc2{\Hcal^{(2)}}

\def\Kcal{\mathcal{K}}
\def\Lcal{\mathcal{L}}
\def\Mcal{\mathcal{M}}

\def\hyp13{{_1 F_3}}
\def\sqtau{\sqrt{\tau}}
\def\sqtscal{\sqrt{\tscal}}
\def\mtscal{\hat\tscal}
\def\mtau{\hat\tau}

\def\cap{\text{cap}}

\def\res{\st{res}}
\def\cas{\st{cas}}
\def\gc{\ut{(gc)}}
\def\can{\ut{(c)}}
\def\cgc{\ut{(c,gc)}}
\def\br0{{(0)}}

\def\sgn{\mathrm{sgn}}
\def\tscal{x}
\def\amplPhit{\phi_t^\br0}
\def\amplPhimu{\phi_\mu^\br0}
\def\amplXipm{\xi_\pm^\br0}
\def\amplXip{\xi_+^\br0}
\def\amplXim{\xi_-^\br0}
\def\amplXimu{\xi_{\mu}^\br0}
\def\lenPhi0{l_\mden^\br0}
\def\lenH1{l_{h_1}^\br0}
\def\lenSpH1{l_{h_1,\mathrm{sp}}^\br0}
\def\lenOrdH1{l_{h_1,\mathrm{ord}}^\br0}
\def\CCF{CCF\xspace}
\def\CCFs{CCFs\xspace}
\def\mass{\Phi}
\def\Mass{\Mcal}
\def\mden{\varphi}

\newcommand{\bitem}{\begin{itemize}}
\newcommand{\eitem}{\end{itemize}}
\newcommand{\benum}{\begin{enumerate}}
\newcommand{\eenum}{\end{enumerate}}
\newcommand{\bblock}[1]{\begin{block}{#1}}
\newcommand{\eblock}{\end{block}}
\newcommand{\bmini}[1]{\begin{minipage}{#1}}
\newcommand{\emini}{\end{minipage}}
\newcommand{\btab}[1]{\begin{tabular}{#1}}
\newcommand{\etab}{\end{tabular}}
\newcommand{\btabn}[1]{\begin{tabular}{#1}}
\newcommand{\etabn}{\end{tabular}}
\newcommand{\beq}{\begin{equation}}
\newcommand{\eeq}{\end{equation}}
\newcommand{\bal}{\begin{align}}
\newcommand{\eal}{\end{align}}
\newcommand{\baln}{\begin{align*}}
\newcommand{\ealn}{\end{align*}}
\newcommand{\beqn}{\begin{equation*}}
\newcommand{\eeqn}{\end{equation*}}
\newcommand{\bmult}{\begin{multline}}
\newcommand{\emult}{\end{multline}}
\newcommand{\bsplit}{\begin{split}}
\newcommand{\esplit}{\end{split}}
\newcommand{\bmat}{\begin{pmatrix}}
\newcommand{\emat}{\end{pmatrix}}

\newcommand{\bv}[1]{\mathbf{#1}}

\begin{document}

\title{Critical adsorption and critical Casimir forces in the canonical ensemble}
\author{Markus Gross}
\email{gross@is.mpg.de}
\affiliation{Max-Planck-Institut f\"{u}r Intelligente Systeme, Heisenbergstra{\ss}e 3, 70569 Stuttgart, Germany}
\affiliation{IV.\ Institut f\"{u}r Theoretische Physik, Universit\"{a}t Stuttgart, Pfaffenwaldring 57, 70569 Stuttgart, Germany}
\author{Oleg Vasilyev}
\affiliation{Max-Planck-Institut f\"{u}r Intelligente Systeme, Heisenbergstra{\ss}e 3, 70569 Stuttgart, Germany}
\affiliation{IV.\ Institut f\"{u}r Theoretische Physik, Universit\"{a}t Stuttgart, Pfaffenwaldring 57, 70569 Stuttgart, Germany}
\author{Andrea Gambassi}
\affiliation{SISSA -- International School for Advanced Studies and INFN, via Bonomea 265, 34136 Trieste, Italy}
\author{S. Dietrich}
\affiliation{Max-Planck-Institut f\"{u}r Intelligente Systeme, Heisenbergstra{\ss}e 3, 70569 Stuttgart, Germany}
\affiliation{IV.\ Institut f\"{u}r Theoretische Physik, Universit\"{a}t Stuttgart, Pfaffenwaldring 57, 70569 Stuttgart, Germany}
\date{\today}

\begin{abstract}
Critical properties of a liquid film between two planar walls are investigated in the canonical ensemble, within which the total number of fluid particles, rather than their chemical potential, is kept constant. The effect of this constraint is analyzed within mean field theory (MFT) based on a Ginzburg-Landau free energy functional as well as via Monte Carlo simulations of the three-dimensional Ising model with fixed total magnetization. Within MFT and for finite adsorption strengths at the walls, the thermodynamic properties of the film in the canonical ensemble can be mapped exactly onto a grand canonical ensemble in which the corresponding chemical potential plays the role of the Lagrange multiplier associated with the constraint. However, due to a non-integrable divergence of the mean field order parameter (OP) profile near a wall, the limit of infinitely strong adsorption turns out to be not well-defined within MFT, because it would necessarily violate the constraint. The critical Casimir force (CCF) acting on the two planar walls of the film is generally found to behave differently in the canonical and grand canonical ensembles. For instance, the canonical CCF in the presence of equal preferential adsorption at the two walls is found to have the opposite sign and a slower decay behavior as a function of the film thickness compared to its grand canonical counterpart. We derive the stress tensor in the canonical ensemble and find that it has the same expression as in the grand canonical case, but with the chemical potential playing the role of the Lagrange multiplier associated with the constraint. The different behavior of the CCF in the two ensembles is rationalized within MFT by showing that, for a prescribed value of the thermodynamic control parameter of the film, i.e., density or chemical potential, the film pressures are identical in the two ensembles, while the corresponding bulk pressures are not.  

\end{abstract}

\maketitle

\section{Introduction}

The thermodynamic equivalence of statistical ensembles is known to break down in systems of finite extent \cite{hill_thermodynamics_1964,lebowitz_ensemble_1967}. 
Fixing the particle number in a finite volume but still allowing heat exchange with a bath leads to the canonical ensemble as the proper description.
The vast majority of theoretical studies on critical phenomena in confinement have been performed in the grand canonical ensemble \cite{zinn-justin_quantum_2002,brankov_theory_2000}, whereas there are relatively few studies concerning the canonical ensemble (see, e.g., Refs.\ \cite{eisenriegler_helmholtz_1987,blote_three-dimensional_2000, caracciolo_finite-size_2001,pleimling_crossing_2001, gulminelli_transient_2003, deng_constrained_2005}).
On the other hand, in many circumstances the experimental setup is naturally realizing the canonical ensemble \cite{ravikovitch_capillary_1995, christenson_confinement_2001, williams_effect_2014}, which easily justifies corresponding theoretical studies.
For instance, important consequences of the conservation of the particle number arise concerning the structure and the phase transitions of (off-critical) fluids confined in nanoscopic pores or capillaries \cite{rao_computer_1978, furukawa_two-phase_1982, gonzalez_density_1997, gonzalez_how_1998,neimark_inside_2002,binder_theory_2003, neimark_bridging_2003,macdowell_evaporation/condensation_2004,neimark_phase_2006, berim_fluid_2006}, which motivated the development of canonical density functional methods \cite{white_density-functional_2000,de_las_heras_full_2014}.
Furthermore, there is a strong, intrinsically theoretical, interest in such analyses, in particular stemming from numerical methods.
Molecular dynamics simulations \cite{frenkel_understanding_2001} as well as lattice gas \cite{rothman_lattice-gas_2004} or lattice Boltzmann simulations \cite{succi_lattice_2001} naturally operate in the canonical ensemble and have been applied for studying static and dynamic critical phenomena \cite{puri_dynamics_1993, das_critical_2006, das_static_2006, roy_transport_2011,das_finite-size_2012, gross_simulation_2012, gross_critical_2012, yabunaka_phase_2013}. 
In order to properly extract physical properties of bulk systems from such simulations, a detailed understanding of finite-size effects in the canonical ensemble is required \cite{lebowitz_long-range_1961, salacuse_finite-size_1996, roman_fluctuations_1997,rovere_block_1988,rovere_gas-liquid_1990,rovere_simulation_1993, blote_three-dimensional_2000, caracciolo_finite-size_2001, deng_constrained_2005}.
Constraining \emph{non-ordering} degrees-of-freedom results in the so-called Fisher renormalization of critical exponents \cite{fisher_renormalization_1968, imry_theory_1973, achiam_phase_1975,anisimov_general_1995, krech_critical_1999,mryglod_corrections_2001}. 
Recently, it has been shown that the choice of the ensemble may also affect the phase behavior of and the \CCFs on colloids in a critical solvent \cite{hobrecht_many-body_2015}.

In the present study, we consider films of one-component or binary fluids close to their bulk critical point within the canonical ensemble, i.e., with a fixed  total number of particles (of each species).
Simple one-component liquids undergoing liquid-vapor transitions as well as binary liquid mixtures undergoing liquid-vapor or liquid-liquid segregation transitions belong to the Ising universality class.
Their static critical behavior is properly captured by a statistical field theory based on a Ginzburg-Landau free energy functional \cite{zinn-justin_quantum_2002}.
Our analysis employs the mean field approximation of this field theory, which provides the leading contribution to the canonical and grand canonical partition functions. 
The conclusions drawn on this basis will be supported by Monte Carlo simulations of the three-dimensional Ising model.
Fluctuation corrections will be analyzed in detail in a forthcoming study, in which a statistical field theory in the canonical ensemble will be developed systematically.
For concreteness, here we use the vocabulary appropriate for a simple fluid which may separate into phases of different densities, noting that the notion of \emph{density}---which represents the OP of the transition---can be replaced by that of \emph{concentration} or \emph{magnetization} and the notions of \emph{chemical and substrate potential} by those of \emph{bulk and surface (magnetic) field}, respectively. (For more details, see Sec.~\ref{sec_scal_CA}.)
The spatial integral of the OP will henceforth be called the \emph{``mass''} $\mass$ of the film.
The film is bounded in one spatial direction by two planar and parallel walls, which are assumed to be of macroscopic lateral extent. 
In this case, thermodynamically extensive quantities such the mass $\mass$ or the free energy have to be defined as \emph{per transverse area} of the film.
In the Monte Carlo simulations discussed here, periodic boundary conditions are employed along the lateral directions.
We focus on the one-phase region of this confined system, i.e., on temperatures above the capillary critical point \cite{fisher_scaling_1981, nakanishi_critical_1983}.
The walls have an adsorption preference, modeled by appropriate surface chemical potentials, for one of the two phases of the film, leading to \emph{critical adsorption} on the walls. 
We consider the cases of either symmetric or antisymmetric boundary conditions for finite values of the substrate potential, referred to, in the case of equal strengths of the surface fields, by $(++)$ and $(+-)$ boundary conditions, respectively.
This setup covers the so-called normal surface universality class \cite{binder_critical_1983, diehl_field-theoretical_1986} and describes the typical adsorption behavior of a near-critical binary liquid mixture \cite{desai_critical_1995, cho_critical_2001, fukuto_critical_2005, rafai_repulsive_2007, hertlein_direct_2008,nellen_tunability_2009, gambassi_critical_2009}.

First, we shall address the basic phenomenology of critical adsorption in a film bounded by two parallel walls in the canonical ensemble.
Previous studies of critical adsorption have been performed in the grand canonical ensemble, i.e., assuming that the film can exchange particles with its surroundings at fixed bulk chemical potential and allowing the spatial integral $\mass$ of the OP profile to fluctuate \cite{fisher_wall_1978, au-yang_wall_1980, fisher_scaling_1981, nakanishi_critical_1983, binder_critical_1983, burkhardt_universal_1985, evans_fluids_1986, diehl_field-theoretical_1986, marconi_critical_1988,evans_fluids_1990, parry_influence_1990, parry_novel_1992, borjan_order-parameter_1998, maciolek_critical_1998,  drzewinski_effect_2000, borjan_off-critical_2008,  mohry_crossover_2010, okamoto_casimir_2012, dantchev_exact_2015}.
We investigate the mapping between a canonical and a grand canonical system with the chemical potential $\mu$ chosen such that the imposed value of $\mass$ is recovered. 
Due to the absence of closed analytical solutions of the Ginzburg-Landau model in the presence of arbitrary bulk and surface fields, we address this problem numerically as well as via a suitable perturbation theory and by a short-distance expansion.
As a crucial consequence of the mass constraint, $\mu$ acquires dependences on the physical parameters of the system, such as temperature, surface field, total mass, and film thickness. 
As expected, for all boundary conditions studied here, the dependence of $\mu$ on the total mass $\mass$ differs from that of a system with periodic boundary conditions in all spatial directions. 
This finding will have important repercussions on the behavior of the \CCF in the two ensembles.
In the case of $(++)$ boundary conditions we point out that, as a consequence of the mass constraint, one may not simply set the strength of the surface fields to infinity in order to obtain the universal OP profile corresponding to the case of strong adsorption. The reason is the non-integrable short-distance divergence $\sim 1/z$ of the mean field OP profiles near both walls, with $z$ denoting the distance from them. This is a mean field specific effect which is expected to be eliminated by critical fluctuations, as they give rise to a weaker, integrable divergence $\sim z^{-\beta/\nu}$ in the strong adsorption regime, with $\beta/\nu\simeq 0.52$ for the three-dimensional Ising universality class \cite{pelissetto_critical_2002}, where $\beta$ and $\nu$ are standard bulk critical exponents.

Building upon the analysis of the OP profiles of the critically adsorbed film, we proceed to a study of \CCFs in the canonical ensemble.
Critical Casimir forces generally arise from confining a near-critical medium (see, e.g., Refs.\ \cite{krech_casimir_1994, brankov_theory_2000, gambassi_casimir_2009} for reviews).
As a consequence, the fluctuation spectrum is modified and the mean OP acquires a spatial dependence; the latter effect lends itself to a description within MFT.
Similarly to the case of critical adsorption, \CCFs seem to have been investigated so far only in the grand canonical ensemble (see, e.g., Refs.\ \cite{krech_casimir_1994, brankov_theory_2000, gambassi_casimir_2009} and references therein).
We study the effect of a mass constraint on the \CCF by computing numerical solutions of the Ginzburg-Landau model in the mean field approximation and by performing Monte Carlo simulations of the three-dimensional Ising model.
The salient features of our numerical results are rationalized within linear MFT (i.e., upon neglecting the quartic coupling in the Ginzburg-Landau free energy functional), within which analytical calculations for arbitrary bulk and surface fields can be carried out and the \CCF can be easily extracted from the residual finite-size part of the free energy.
In the grand canonical ensemble, it is well-known that the \CCF is attractive for $(++)$ boundary conditions and that, up to prefactors, its scaling function decays exponentially as a function of the scaling variable $L/\xi$, where $\xi$ is the bulk correlation length and $L$ is the film thickness \cite{krech_casimir_1997, mohry_crossover_2010}.
Notably, for $(++)$ boundary conditions in the canonical ensemble, we find that, upon varying the total mass, the \CCF may change sign and thus becomes repulsive. 
Furthermore, we find that its scaling function decays rather slowly, i.e., as a power law of the scaling variable $L/\xi$, and may attain significantly larger values compared to the grand canonical one.

As an alternative to computing the residual finite-size contributions to the free energy, the \CCF can also be determined via the stress tensor as the difference between the wall stresses of the film and the stresses of the surrounding fluid. 
We prove, within MFT, that the stress tensor in the canonical ensemble assumes the same analytic expression as in the grand canonical one, with the Lagrange multiplier being equal to the bulk field.
Accordingly, within MFT and for the same thermodynamic conditions (i.e., total mass or chemical potential), the canonical and grand canonical film pressures are identical.
Consequently, the difference in the behavior of the \CCF between the two ensembles is due to a difference in the bulk pressures which are subtracted in order to obtain the \CCF.
While in the grand canonical ensemble the bulk fluid and the film are thermodynamically coupled via the overall, spatially constant chemical potential, in the canonical ensemble the number density of the bulk fluid---and hence its pressure---is in principle arbitrary and depends on the actual experimental setup.
In this respect it is quite natural to assume that the bulk surrounding the film and the film itself are governed by the same thermodynamic control parameter, corresponding to the chemical potential in the grand canonical ensemble and the mean density in the canonical one.
We will show that this assumption indeed leads to different values of the bulk pressures for the two ensembles and thus explains the difference between the \CCFs in the two ensembles, being the force obtained by subtracting from the same film pressure two different bulk pressures.
In the canonical case we shall furthermore demonstrate that defining the \CCF as the difference between the film and the bulk pressure does not necessarily yield the same result as extracting it from the residual finite size contribution to the free energy. The reason is that certain terms in the free energy which, as a consequence of the canonical constraint, depend on the film thickness $L$ can, based on finite-size scaling arguments, still be identified as ``surface-like'', i.e., as contributions to the surface free energy. 

The outline of the paper is as follows: 
In Sec.~\ref{sec_critads} we focus on the critical adsorption in a film in the canonical ensemble. In particular, we first discuss (Sec.~\ref{sec_scal_CA}) the expected general scaling properties and then introduce the mean field Ginzburg-Landau model (Sec.~\ref{sec_ads_model}). We then proceed to the first central part of this study, which concerns the investigation of the OP profiles and the mapping between the canonical and the grand canonical ensembles within MFT. These results are obtained from a perturbation theory about the solution of the linearized Euler-Lagrange equations (Sec.~\ref{sec_ads_pert}), as well as from a short-distance expansion (Sec.~\ref{sec_sde}). These two approaches are already sufficient to illustrate the essential effects emerging from the mass constraint, as the comparison with the numerical solutions of the full, nonlinear mean field model shows (Sec.~\ref{sec_ads_num}).
We also briefly discuss the OP profiles obtained from Monte Carlo simulations of the Ising model in three spatial dimensions and discuss the influence of the lateral system size (Sec.~\ref{sec_MC_prof}). 
The second central aspect of the present study is the investigation of the \CCF in the canonical ensemble, presented in Sec.~\ref{sec_Casimir}. 
After outlining the general scaling behavior of the \CCF (Sec.~\ref{sec_scal_Casimir}), we analytically study the \CCF within linear MFT and compare it with the full numerical solutions of the nonlinear mean field model (Sec.~\ref{sec_Casimir_mft}), as well as with the results of Monte Carlo simulations of the three-dimensional Ising model (Sec.~\ref{sec_MC_Casimir}).
Appendix \ref{app_mapping} discusses general scaling properties of a film within MFT and presents a useful mapping relation.
Appendix \ref{app_pert} contains a generalized perturbative treatment of the MFT considered in Sec.~\ref{sec_ads_model}, while Appendix \ref{app_stressten} presents a derivation of the mean field stress tensor in the canonical case, which is an essential tool for determining \CCFs.
A glossary of the most frequently used quantities is provided in Table \ref{tab_glossary}.

\begin{table*}[t]
	\begin{center}
		\begin{tabular}{c c c}
			\hline\hline
			quantity & description & definition in \\
			\hline
			$L$ & film thickness  &  Sec.~\ref{sec_scal_CA} \\
			$z$ & coordinate in the direction perpendicular to the walls  & Sec.~\ref{sec_scal_CA}\\
			$\hat z$ & distance from a wall & Sec.~\ref{sec_sde}\\
			$t$ & reduced temperature & Eq.~\eqref{eq_gen_t} \\
			$\phi$ & order parameter (OP) profile & Sec.~\ref{sec_scal_CA} \\
			$\mass$ & total mass of the film$^\dag$ & Eq.~\eqref{eq_Mass0} \\
			$\mden$ & mean mass density of the film & Eq.~\eqref{eq_mden} \\
			$\mu$ & chemical potential / bulk field & Sec.~\ref{sec_scal_CA}\\
			$h_1$ & substrate potential / surface field & Sec.~\ref{sec_scal_CA}\\
			$\amplXip$ & correlation length amplitude associated with $t$ & Eq.~\eqref{eq_gen_correlt} \\
			$\amplXimu$ & correlation length amplitude associated with $\mu$ & Eqs.~\eqref{eq_gen_correlmu}, \eqref{eq_gen_amplXimu} \\
			$\lenH1$ & amplitude associated with $h_1$ & Eqs.~\eqref{eq_gen_h1len_sp}, \eqref{eq_gen_h1len_ord} \\
			$\amplPhit$ & amplitude associated with $\phi$ and $t$ & Eq.~\eqref{eq_gen_phit} \\
			$\lenPhi0$ & amplitude associated with $\phi$ & Eq.~\eqref{eq_lenPhi0} \\
			$\zeta$ & reduced coordinate ($\zeta=z/L$) & Eq.~\eqref{eq_zeta} \\
			$\hat\zeta$ & reduced distance from a wall & Sec.~\ref{sec_sde}\\
			$m$ & scaled OP & Eq.~\eqref{eq_gen_prof_scal} \\ 
			$\Mass$ & finite-size scaling variable associated with $\mden$ & Eq.~\eqref{eq_Mass}\\ 
			$x$ & finite-size scaling variable associated with $t$ & Eq.~\eqref{eq_tscal}\\
			$B$ & finite-size scaling variable associated with $\mu$ & Eq.~\eqref{eq_Bscal}\\
			$H_1$ & finite-size scaling variable associated with $h_1$ & Eq.~\eqref{eq_H1scal}\\
			$\Fcal_f$ & total film free energy$^\dag$ & Eqs.~\eqref{eq_Landau_func_c}, \eqref{eq_Landau_func_gc}\\
			$\Delta_0$ & amplitude in mean field free energy functional & Eq.~\eqref{eq_Delta0}\\
			$\phi_0$, $m_0$ & equilibrium OP profile within linear MFT & Eqs.~\eqref{eq_m0_sol}, \eqref{eq_phi0_sol} \\
			$\tilde \phi_0$, $\tilde m_0$ & constrained equilibrium OP profile within linear MFT & Eqs.~\eqref{eq_m0_constr}, \eqref{eq_phi0_constr}\\
			$\tilde \mu_0$, $\tilde B_0$ & constraint-induced bulk field & Eqs.~\eqref{eq_B0}, \eqref{eq_h0}\\
			$x_{c}$, $B_{c}$ & bulk critical point ($x_c=B_c=0$) & Sec.~\ref{sec_scal_CA}\\
			$x_{c,\text{cap}}$, $B_{c,\text{cap}}$ & capillary critical point & Secs.~\ref{sec_scal_CA}, \ref{sec_ads_num} \\
			$\Fcal\res$ & residual finite-size free energy$^\dag$ & Eq.~\eqref{eq_Ffilm_split}\\
			$\Kcal$ & critical Casimir force (CCF) & Eqs.~\eqref{eq_pCas_dFdL}, \eqref{eq_pCas_pdiff} \\
			$p_b$ & bulk pressure & Eq.~\eqref{eq_Ffilm_split}\\
			$p_f$ & film pressure & Eq.~\eqref{eq_Ffilm_split}\\
			$T_{ij}$, $\bar T_{ij}$ & stress tensor & Eq.~\eqref{eq_stressten_dFdL}, Eq.~\eqref{eq_stressten_gc}\\
			$\Theta$ & scaling function of the residual free energy & Eqs.~\eqref{eq_gen_Fres_gc}, \eqref{eq_gen_Fres_c}\\
			$\Xi$ & scaling function of the \CCF & Eqs.~\eqref{eq_Casi_force_gc_scalf}, \eqref{eq_Casi_force_c_scalf}\\
			$\Delta_{++}$, $\Delta_{++,*}$, $\tilde\Delta_{++}$ & amplitude of the \CCF [for $(++)$ boundary conditions]$^\ddagger$ & Eqs.~\eqref{eq_cas_ampl}, \eqref{eq_cas_ampl_pp_gc_lim}\\
			\hline\hline
		\end{tabular}
	\end{center}
	\caption{Glossary of quantities frequently used in the present study. The notions \emph{nonlinear} and \emph{linear} MFT refer to the Ginzburg-Landau model [see Eqs.~\eqref{eq_Landau_func_c} and \eqref{eq_Landau_func_gc}] \emph{with} and \emph{without} the quartic nonlinearity in $\phi$. The notion ``total mass'' refers to the total OP and can be understood as ``total number of particles'' in the case of a single-component fluid. $^\dag$Unless otherwise indicated, thermodynamic extensive quantities such as $\mass$ and $\Fcal_f$ are considered as \emph{per transverse area}. $^\ddagger$Distinct from the usual notation, we define $\Delta_{++}$ and related quantities as amplitudes of the \CCF $\Kcal$ rather than of the residual finite size free energy $\Fcal\res$.}
	\label{tab_glossary}
\end{table*}

\section{Critical adsorption}
\label{sec_critads}

\subsection{General scaling considerations}
\label{sec_scal_CA}
Before turning to MFT, here we describe the general setup and the properties which can be expected from general finite-size scaling arguments \cite{binder_critical_1983, diehl_field-theoretical_1986, privman_finite-size_1990, brankov_theory_2000}. 
Consider a $d$-dimensional film bounded by two $(d-1)$-dimensional planar and parallel walls a distance $L$ apart.
The walls are located at positions $z=\pm L/2$ and carry the two corresponding surface fields $h_1^{\pm}$ which lead to preferential adsorption of the OP at the walls.
Here, we exclusively consider the case in which $h_1^\pm$ are of equal strength and have equal or opposite signs. 
For notational convenience, we thus shall occasionally drop the superscript of $h_1$.
Furthermore, near surfaces, fluids typically have a reduced tendency to order, which can be modeled by an effective ``surface temperature'' or ``surface enhancement'' field $c$, which acts like an inverse extrapolation length \cite{binder_critical_1983, diehl_field-theoretical_1986}.
In the present study, we shall focus on the dependences on $h_1$ for a fixed value of $c$. 
As discussed below, the presence of a non-vanishing surface enhancement $c$ affects, however, the scaling laws involving $h_1$.
In the case of a one-component fluid, the OP $\phi$ is proportional to the deviation of the density $n$ from its critical value $n_c$, i.e., $\phi\propto n-n_c$, whereas for a binary fluid $\phi$ is proportional to the difference of the concentration $C_A$ of one of the species from its bulk critical value $C_{A,c}$, i.e., $\phi\propto C_A-C_{A,c}$.
The system is described by a reduced temperature
\beq t= \frac{T-T_c}{T_c},
\label{eq_gen_t}\eeq 
where $T_c$ is the \emph{bulk} critical temperature of the fluid medium.
For the present scaling considerations we assume the film in the directions parallel to the walls to be of macroscopic extent and, in particular, much larger than any correlation length in the fluid. 
Thus we effectively take the aspect ratio of the system to be $\rho\equiv L/A^{1/(d-1)}\to 0$, where $A$ is the area of the walls, so that $\rho$ does not appear in the scaling relations presented below. 
The effects associated with a nonzero value of $\rho$ will, however, be briefly addressed in the context of analyzing our Monte Carlo simulation data (Secs.~\ref{sec_MC_prof} and \ref{sec_MC_Casimir}).

We first discuss the scaling properties of a film in the \emph{grand canonical} ensemble, in which the film can exchange particles with an external reservoir with a prescribed chemical potential $\mu$.
For a one-component fluid the quantity $\mu$ is actually the deviation of the chemical potential from its critical value in the bulk, while for a binary fluid, it is the deviation of the difference in the chemical potentials of the two species A and B from its bulk critical value: $\mu\equiv (\mu_A-\mu_B) - (\mu_{A,c}-\mu_{B,c})$.
For $(++)$ boundary conditions, it is known that the critical point of the film is shifted to a lower value $T_{c,\cap}<T_c$ of the temperature and a negative chemical potential difference $\mu_{c,\cap}<\mu_c=0$ \cite{nakanishi_critical_1983}.
Below $T_{c,\cap}$, phase separation in a film can occur. While we shall occasionally comment on these aspects in the course of this study, a detailed investigation is beyond the scope of the present analysis and therefore we focus here on temperatures above the capillary critical point.
In this case, the translational symmetry along the directions parallel to the walls is not broken and the OP field $\phi(z)$ depends on $z$ only.
Therefore, unless stated otherwise, we shall henceforth consider all thermodynamically extensive quantities, such as the total number of particles or the free energy, as quantities \emph{per transverse area} $A$. 
The integrated OP per transverse area, i.e., the so-called total mass $\mass$ (which for a simple fluid essentially corresponds to the total number of particles and should not be confused with the actual mass of the fluid), is thus given by
\beq  \mass \equiv  \int_{-L/2}^{L/2} dz\,\phi(z).
\label{eq_Mass0}\eeq 

The asymptotic critical behavior of thermodynamic quantities is governed by the re\-norm\-alizat\-ion-group fixed points in the phase diagram spanned, \emph{inter alia}, by the variables $t$, $\mu$, $h_1$, and $c$.
For the present study, the relevant fixed points are: $(t=0,\mu=0,h_1=0,c=0)$ corresponding to the so-called special phase transition \footnote{We assume that $c$ is shifted by a constant such that its fixed-point value is $c=0$ \cite{diehl_field-theoretical_1986}.}, $(t=0,\mu=0,h_1=0,c=\infty)$ corresponding to the so-called ordinary phase transition, and $(t=0,\mu=0,h_1=\infty,|c|<\infty)$ corresponding to the so-called normal phase transition \cite{diehl_field-theoretical_1986}. 
We remark that, depending on the boundary conditions, the presence of a mass constraint requires to keep the value of the surface fields $h_1$ finite within MFT; this will be discussed in detail below.
In order to avoid a clumsy notation and for the purpose of discussing general scaling relations, in this subsection we consider thermodynamic control parameters such as $\mu$, $h_1$, or $c$ to be renormalized quantities which are dimensionless due to splitting off suitable dimensional factors carrying the proper units.

\begin{table}[t]
	\begin{center}
		\begin{tabular}{c c c}
			\hline\hline
			exponent & \hspace{0.3cm}$d=4$\hspace{0.3cm} & \hspace{0.3cm}$d=3$\hspace{0.3cm} \\
			\hline
			$\nu$ & $1/2$  & $0.630$ \\
			$\eta$ & $0$ & $0.0336$ \\
			$\beta$ & $1/2$ & $0.326$ \\
			$\gamma$ & $1$ & $1.24$ \\
			$\delta$ & $3$ & $4.80$ \\
			$\Delta$ & $3/2$ & $1.56$ \\
			$\Delta_1\ut{sp}$ & $1$ & $1.05$ \\
			$\Delta_1\ut{ord}$ & $1/2$ & $0.46$ \\
			\hline\hline
		\end{tabular}
	\end{center}
	\caption{Values of bulk and surface critical exponents for the Ising universality class in spatial dimensions $d=4$ (MFT) and $d=3$ (rounded to three significant digits) \cite{diehl_field-theoretical_1986, pelissetto_critical_2002}. Within the context of the present MFT study, one has $\Delta_1\equiv \Delta_1\ut{sp}$, whereas for the MC simulation data of the Ising model the appropriate surface critical exponent is $\Delta_1\equiv \Delta_1\ut{ord}$ [see the discussion in Sec.\ \ref{sec_scal_CA} and, in particular, Eq.\ \eqref{eq_H1scal}].}
	\label{tab_crit_exp}
\end{table}

All surface phase transitions share the same bulk critical behavior, which we discuss first.
Based on the exponential decay of the two-point correlation function of the OP in the bulk, the correlation length $\xi_t$ at zero bulk field ($\mu=0$) and $\xi_\mu$ at zero reduced temperature ($t=0$) can be defined as \cite{pelissetto_critical_2002} \footnote{Note that we consider here the so-called true bulk correlation length $\xi_{t,\mu}$ defined by the exponential decay of the two-point correlation function of the order parameter, whereas in Ref.\ \cite{pelissetto_critical_2002}, the so-called second-moment correlation length $\xi_{t,\mu}\ut{(2nd)}$ is used. The corresponding amplitude ratios are given by $\amplXip/\xi_{+}^{(0),\text{2nd}}\simeq 1$ and $\amplXimu/\xi_{\mu}^{(0),\text{2nd}}\simeq 1$ for the three-dimensional Ising model.}
\begin{subequations}\begin{alignat}{3} 
 \xi_t &= \amplXipm |t|^{-\nu},\qquad &&\text{for }\mu=0\text{ and }t\to 0^\pm,\label{eq_gen_correlt}\\
 \xi_\mu &= \amplXimu |\mu|^{-\nu/\Delta},\qquad &&\text{for }t=0\text{ and }\mu\to 0. \label{eq_gen_correlmu}
\end{alignat}\label{eq_gen_correl}\end{subequations}
Here, $\nu$ and $\Delta$ are the standard universal bulk critical exponents (see Table \ref{tab_crit_exp}), while $\amplXipm$ and $\amplXimu$ are the corresponding non-universal amplitudes. 
However, the amplitude ratio $U_\xi= \amplXip / \amplXim$ forms a universal number, with $U_\xi\simeq 1.9$ in $d=3$ and $U_\xi=\sqrt{2}$ in $d=4$ spatial dimensions \cite{pelissetto_critical_2002}.
A further relevant quantity is the value $\phi_b$ of the OP parameter in the bulk, which, near criticality, behaves as
\begin{subequations}\begin{alignat}{3}
 \phi_{b,t} &= \theta(-t)\amplPhit |t|^\beta, \qquad &&\text{for }\mu=0 \text{ and }t\to 0,\label{eq_gen_phit}\\
 \phi_{b,\mu} &= \sgn(\mu) \amplPhimu |\mu|^{1/\delta}, \qquad &&\text{for }t=0\text{ and }\mu\to 0, \label{eq_gen_phimu}
\end{alignat}\label{eq_gen_phi}\end{subequations}
where $\theta(t)$ is the step function [$\theta(t>0)=1$, $\theta(t<0)=0$], $\amplPhit$ and $\amplPhimu$ are non-universal amplitudes, and $\delta=\Delta/\beta$ is a universal critical exponent. 
For later use, we note that the amplitudes $\amplXimu$ and $\amplPhimu$ in Eqs.~\eqref{eq_gen_correlmu} and \eqref{eq_gen_phimu} can be expressed in terms of $\amplXip$ and $\amplPhit$ in Eqs.~\eqref{eq_gen_correlt} and \eqref{eq_gen_phit} as \cite{pelissetto_critical_2002}
\begin{subequations}\bal
\amplPhimu &= \left(\frac{C^+}{R_\chi}\right)^{1/\delta} \left(\amplPhit\right)^{1-1/\delta},\\
\amplXimu &= \amplXip\left(\frac{Q_2}{ C^+}\right)^{\sfrac{1}{(2-\eta)}} \left(\amplPhimu\right)^{\sfrac{1}{(2-\eta)}} \nonumber \\
&= \amplXip \left(\frac{Q_2}{ R_\chi^{1/\delta}}\right)^{1/(2-\eta)} \left( \frac{\amplPhit}{C^+}\right)^{\nu/\Delta},\label{eq_gen_amplXimu}
\end{align}\label{eq_gen_amplMu_t}\end{subequations}
where $C^+$ is a non-universal amplitude entering the definition of the susceptibility $\chi$ via $\chi=C^+t^{-\gamma}$ for $t>0$, $\eta$ and $\gamma$ are further standard bulk critical exponents, and $R_\chi$ and $Q_2$ are universal amplitude ratios, with $R_\chi\simeq 1.6$, $Q_2\simeq 1.2$ in $d=3$ and $R_\chi=1$, $Q_2=1$ in $d=4$ spatial dimensions for the Ising universality class \cite{pelissetto_critical_2002}.
Note that in our definition in Eq.~\eqref{eq_gen_amplXimu} the amplitude $\amplXimu$ is by a (universal) factor of $\delta^{1/(2-\eta)}$ larger than the one considered in Ref.~\cite{pelissetto_critical_2002}. 
This permissible rescaling, which could be alternatively understood as a change of the definition of the field $\mu$, is performed here in order to cast MFT (see Sec.~\ref{sec_ads_model} below) into its mathematically most simple scaling form.

We now briefly recall the critical behavior induced by the scaling variables $h_1$ and $c$ associated with the presence of surfaces.
For $c \ll |t|^\Psi$ with $\mu$ and $h_1$ small, a scaling behavior characteristic of the special transition is expected \cite{diehl_field-theoretical_1986}. Here, $\Psi$ is a surface critical exponent having the value $\Psi\simeq 0.68$ in $d=3$ and $\Psi=1/2$ in $d=4$ spatial dimensions \footnote{In the literature, $\Psi$ is commonly denoted as $\Phi$, see, e.g., Ref.\ \cite{diehl_field-theoretical_1986}}.
Analogously to the bulk field $\mu$, one can associate a length scale $l_{h_1}\ut{sp}$ with the surface field $h_1$ \cite{leibler_magnetisation_1982, brezin_critical_1983}:
\beq l_{h_1}\ut{sp} \equiv \lenSpH1 | h_1|^{-\nu/\Delta_1\ut{sp}},
\label{eq_gen_h1len_sp}\eeq 
where $\lenSpH1$ denotes the corresponding non-universal amplitude and $\Delta_1\ut{sp}$ is another surface critical exponent (see Table \ref{tab_crit_exp}). 
At bulk criticality and for distances $\hat z \ll l_{h_1}\ut{sp}$, the OP behaves as $\phi(\hat z)\sim \hat z^{(\beta_1\ut{sp}-\beta)/\nu}$, with $(\beta_1\ut{sp}-\beta)/\nu$ having the value $-0.15$ in $d=3$, while this exponent is zero in $d=4$ \cite{brezin_critical_1983, ciach_critical_1997}.
For $\hat z\gg l_{h_1}\ut{sp}$, instead, a crossover to the behavior characteristic of the normal universality class occurs, for which $\phi(\hat z)\sim \hat z^{-\beta/\nu}$ in the limit $h_1\to\infty$ \cite{binder_critical_1983}.
Off criticality, the OP decays exponentially for distances $\hat z\gg \xi_{t,\mu}$, independently of the value of $l_{h_1}\ut{sp}$.
Heuristically, the length $l_{h_1}$ can be interpreted as an extrapolation length $l_{\mathrm{ex}}\propto l_{h_1}$, such that the OP profile behaves as $|\phi(\hat z\to 0)|\sim (\hat z+l_\mathrm{ex})^{-\beta/\nu}$ near a wall \cite{binder_critical_1983}.
Although the very concept of an extrapolation length strictly applies only to MFT, it is useful for the interpretation of experimental or simulation data \cite{schlossman_order-parameter_1985,smock_universal_1994,vasilyev_critical_2011, vasilyev_critical_2013} and provides an effective means to take into account scaling corrections to the leading critical behavior.

For $c\gg |t|^\Psi$ and sufficiently small $\mu$ and $h_1$, scaling properties are governed by the so-called ordinary fixed point, for which the relevant scaling field is a combination of $h_1$ and $c$ \cite{diehl_field-theoretical_1986, diehl_theory_1997}:
\beq \mathrm{h_1} \equiv h_1/c^Y
\label{eq_h1ord_scalv}\eeq 
where $Y\equiv (\Delta_1\ut{sp}-\Delta_1\ut{ord})/\Psi$ and $\Delta_1\ut{ord}$ is a further surface critical exponent (see Table \ref{tab_crit_exp}).
The corresponding length scale $l_{h_1}\ut{ord}$, analogous to the one in Eq.\ \eqref{eq_gen_h1len_sp}, is defined as
\beq l_{h_1}\ut{ord} \equiv \tilde{l}_{h_1,\mathrm{ord}}^\br0 | \mathrm{h_1}|^{-\nu/\Delta_1\ut{ord}} = \lenOrdH1 | h_1|^{-\nu/\Delta_1\ut{ord}}.
\label{eq_gen_h1len_ord}\eeq
In the second equality, we have absorbed the factor $c^{-Y}$, which here we consider to be a constant, into the amplitude $\lenOrdH1$.
At bulk criticality, the OP profile behaves for distances $l_c\ll \hat z\ll l_{h_1}\ut{ord}$ as $\phi(\hat z)\sim \hat z^{(\Delta_1\ut{ord}-\beta)/\nu}$, with a length scale $l_c\sim c^{-\nu/\Psi}$ \cite{ciach_critical_1997, diehl_theory_1997}.
The decay $\phi(\hat z)\sim \hat z^{-\beta/\nu}$ characteristic for the normal surface universality class occurs for $\hat z\gg  l_{h_1}\ut{ord}$, while for $\hat z \ll l_c$, one recovers the special universal behavior $\phi(\hat z) \sim \hat z^{(\beta_1\ut{sp}-\beta)/\nu}$.
We remark that, in three dimensions, the exponent $(\Delta_1\ut{ord}-\beta)/\nu$ is positive, giving rise to a non-monotonic behavior of the OP profile \cite{czerner_near-surface_1997, ciach_critical_1998}.
Generically, fluids exhibit a nonzero surface enhancement $c$ and are strongly adsorbed at the surfaces of the container walls \cite{liu_universal_1989, floter_universal_1995}. 
Accordingly, one expects critical behavior to occur which corresponds to the normal ($h_1\to \infty$) or the ordinary ($h_1\to 0$) surface universality class, including cross-over phenomena.
For $h_1\to 0$ and sufficiently small values of $c$, however, a large portion of the scaling region falls into the domain of the special surface universality class.
In order to keep the focus of the discussion on the role of the ensemble, we shall confine our mean field investigation below (Sec.\ \ref{sec_ads_model}), as far as $c$ is concerned, to the case $c=0$.

In a film of thickness $L$, the finite-size scaling behavior is described by universal scaling functions which depend on the following set of scaling variables \cite{binder_critical_1983, schlesener_critical_2003, gambassi_critical_2006}:
\begin{subequations}\begin{alignat}{2} 
	\zeta &\equiv z/L, \label{eq_zeta}\\
	\tscal &\equiv  \left(\frac{L}{\amplXip}\right)^{1/\nu}t,\label{eq_tscal}\\
	B  &\equiv  \left(\frac{L}{\amplXimu}\right)^{\Delta/\nu} \mu, \label{eq_Bscal}\\
	H_1 &\equiv  \left(\frac{L}{\lenH1}\right)^{\Delta_1/\nu}h_1,\label{eq_H1scal}\\
	\Mass &\equiv \left(\frac{L}{\amplXip}\right)^{\beta/\nu} \frac{\mden}{\amplPhit} = \left(\frac{L}{\lenPhi0}\right)^{\beta/\nu} \mden,
	\label{eq_Mass}
	\end{alignat}\label{eq_scalvar}\end{subequations} 
where, in Eq.~\eqref{eq_Mass},
\beq \mden \equiv \frac{\Phi}{L}
\label{eq_mden}\eeq 
is the mean mass density of the film \footnote{We mention that for actual fluids, the proper scaling fields are linear combinations of $t$ and $\mu$ \cite{onuki_phase_2002}. Such field mixing effects are, however, beyond the scope of the present study and will be neglected.}. 
The scaling variable $H_1$ in Eq.\ \eqref{eq_H1scal} is written in a form which applies, upon inserting the corresponding exponent and length scale defined in Eqs.\ \eqref{eq_gen_h1len_sp} and \eqref{eq_gen_h1len_ord}, to both the crossover from the normal to the special as well as the crossover from the normal to the ordinary phase transition. We shall keep this unified description.
Accordingly, the finite-size scaling relation for the OP profile reads
\begin{widetext}
\beq\begin{split}
 \phi(z, t, \mu, h_1, L) 
    &= \left(\frac{L}{\lenPhi0}\right)^{-\beta/\nu} m\left(\frac{z}{L}, \left(\frac{L}{\amplXip}\right)^{1/\nu}t, \left(\frac{L}{\amplXimu}\right)^{\Delta/\nu} \mu , \left(\frac{L}{\lenH1}\right)^{\Delta_1/\nu}h_1 \right)
    = \amplPhit \left(\frac{L}{\amplXip}\right)^{-\beta/\nu} m,
\end{split}\label{eq_gen_prof_scal}\eeq 
\end{widetext}
where we have introduced the quantity
\beq \lenPhi0 \equiv \amplXip \left(\amplPhit\right)^{\nu/\beta}
\label{eq_lenPhi0}\eeq 
for convenience.
The scaling variable $\Mass$ in Eq.\ \eqref{eq_Mass} is related to the universal scaling function $m$ via
\beq \Mass =\int_{-1/2}^{1/2} d\zeta\, m(\zeta).
\eeq 
Equation \eqref{eq_gen_prof_scal} follows from the homogeneity relation \cite{diehl_field-theoretical_1986}
\beq 
\phi(z, t, \mu, h_1, L) = b^{-\beta/\nu} \phi\left(z/b, t b^{1/\nu}, \mu b^{\Delta/\nu}, h_1 b^{\Delta_1/\nu}, L/b\right),
\label{eq_gen_hom_phi}\eeq 
upon choosing as rescaling factor $b=L$ and upon introducing the appropriate length scales and amplitudes according to Eq.\ \eqref{eq_scalvar}.
Equation ~\eqref{eq_Mass0} may be inverted in order to obtain the bulk field $\mu$ as function of $\mass$, which obeys the scaling relation 
\begin{widetext}
\beq \begin{split} \mu(t,\mden,h_1,L) 
    &= \left(\frac{L}{\amplXimu}\right)^{-\Delta/\nu} \mathcal B\left( \left(\frac{L}{\amplXip}\right)^{1/\nu}t, \left(\frac{L}{\lenPhi0}\right)^{\beta/\nu} \mden , \left(\frac{L}{\lenH1}\right)^{\Delta_1/\nu}h_1 \right),
\end{split}\label{eq_gen_mu_scaling}\eeq
where $\mathcal B$ is the corresponding universal scaling function.
In writing Eq.~\eqref{eq_gen_mu_scaling} we have taken into account that, as implied by Eq.~\eqref{eq_Mass}, the density $\mden$ rather than the total mass $\mass$ is the appropriate quantity entering into the finite-size scaling relations (see, e.g., Ref.\ \cite{eisenriegler_helmholtz_1987}).

In the \emph{canonical} ensemble, instead of the chemical potential, the total mass $\mass$ [Eq.~\eqref{eq_Mass0}] is fixed.
Therefore, in this ensemble the natural counterpart of Eq.~\eqref{eq_gen_prof_scal} is
\beq\begin{split}
 \phi(z, t, \mden, h_1, L) 
    &= \left(\frac{L}{\lenPhi0}\right)^{-\beta/\nu} m\left( \frac{z}{L}, \left(\frac{L}{\amplXip}\right)^{1/\nu}t, \left(\frac{L}{\lenPhi0}\right)^{\beta/\nu} \mden, \left(\frac{L}{\lenH1}\right)^{\Delta_1/\nu} h_1 \right).
\end{split}\label{eq_gen_prof_scal_c}\eeq 
\end{widetext}
For notational convenience, we use the same symbol $\phi$ for the profile in the canonical and in the grand canonical ensemble.
In the case of a binary fluid, we remark that, since the OP is given by $\phi\propto C_A-C_{A,c}$ (with the concentration of species A defined as $C_A=n_A/(n_A+n_B)$ in terms of the individual number densities $n_{A,B}$ of the two species A and B), fixing $\mass$, i.e., the number of particles of species A, does \emph{a priori} not impose a constraint on the other component (B) or on the total density $\int_V d^3r\,(n_A+n_B)$ of the mixture. However, from an experimental point of view it appears to be natural to require that, within the canonical setup, the particle number of \emph{each} species is conserved individually.
The ensuing constraint of a non-ordering parameter such as the total density may, depending on the location of the phase transition in the phase diagram, lead to Fisher renormalization of the critical exponents (see Refs.\ \cite{fisher_renormalization_1968, imry_theory_1973, achiam_phase_1975, krech_critical_1999,mryglod_corrections_2001} and, in particular, Refs.\ \cite{vause_coexistence_1982, anisimov_general_1995}). A detailed analysis of such effects is, however, beyond the scope of the present study. 
Scaling relations analogous to those in Eqs.~\eqref{eq_gen_prof_scal} and \eqref{eq_gen_prof_scal_c} can be formulated for any observable in the grand canonical and canonical ensemble, respectively.
The scaling behaviors of the (residual) free energy and of the \CCF will be discussed separately in Sec.~\ref{sec_scal_Casimir}.

\subsection{Model}
\label{sec_ads_model}
We study MFT based on the Ginzburg-Landau free-energy functional in the film geometry, which is the standard model to describe universal quantities of systems undergoing second-order phase transitions. 
The setup here is the same as the one described in Sec.~\ref{sec_scal_CA}.
However, in order to focus on the effect of the ensemble, we consider in the present context only the simplest possible model, which amounts to setting the surface enhancement $c=0$ and to keeping only a surface field $h_1$. Accordingly, within our mean field model, exponents and amplitudes appropriate for the crossover from the normal to the special surface universality class [see Eq.\ \eqref{eq_gen_h1len_sp}] are to be used in the definition of the scaling variable $H_1$ in Eq.\ \eqref{eq_H1scal}.
We assume that the translational symmetry in the directions parallel to the walls is not broken, so that the OP field $\phi$ depends on $z$ only and we can consider all extensive quantities as quantities per transverse area.
Note that, while in Sec.\ \ref{sec_scal_CA} quantities like $\mu$, $h_1$, and $c$ were considered to be dimensionless in order to keep the notation simple, in the following we use the same symbols to denote their bare (dimensional) counterparts entering the Ginzburg-Landau model.

In the canonical ensemble (c), the free-energy functional of the \emph{f}ilm (per transverse area and $k_B T$), including bulk and surface contributions, is given by \footnote{$\Fcal_f\can$ may additionally depend on an external bulk field such as a magnetic field in the case of a magnet. Such a field can be included in $\Fcal_f\can$ in Eq.~\eqref{eq_Landau_func_c} via a term $-h\phi$ in the integrand. This merely amounts to a shift of the bulk field $\mu$ in $\Fcal_f\gc$ in Eq.~\eqref{eq_Landau_func_gc}.}
\begin{multline} \Fcal_f\can[\phi] \equiv \int_{-L/2}^{L/2} dz \left[\onehalf (\pd_z \phi)^2 + \onehalf \tau \phi^2 +\frac{1}{4!} g \phi^4 \right] \\
-  \left[ h_1^-\phi(z=-L/2)+ h_1^+\phi(z=L/2) \right],
\label{eq_Landau_func_c}\end{multline}
which is to be minimized under the constraint of a prescribed, fixed total mass $\mass$, given by Eq.~\eqref{eq_Mass0}.
Since the statistical weight of an OP configuration $\phi$ is $\exp(-\Fcal_f\can)$, $\Fcal_f\can[\phi]$ is dimensionless.
In Eq.~\eqref{eq_Landau_func_c}, the coupling constant $\tau$ is proportional to the reduced temperature $t= (T-T_c)/T_c$, where $T_c$ is the bulk critical temperature. 
Within MFT, $\tau = (\amplXip)^{-2}t$, which follows from the expression of the correlation length in $d>4$, while the non-universal amplitudes [see Eq.~\eqref{eq_gen_correlt}] $\amplXip$ and $\amplXim$ form the universal ratio $\amplXim/\amplXip=1/\sqrt{2}$.
Within MFT, the coupling constant $g>0$ is a free parameter the dimensionless counterpart of which attains a fixed-point value only under renormalization-group flow, which accounts for the effect of fluctuations.
Within MFT, some of the universal amplitude ratios turn out to be related to the parameter $g$, which is dimensionless in $d=4$; e.g., from Eq.~\eqref{eq_gen_phit} one finds:
\beq \Delta_0\equiv \left(\amplXip \amplPhit \right)^2=\frac{6}{g} ,
\label{eq_Delta0}\eeq 
where we denote this product of amplitudes, which will appear frequently in expressions related to the mean field free energy and \CCF, as $\Delta_0$. Since $\nu/\beta=1$ in MFT, in this case we have also $\Delta_0^{1/2} =\lenPhi0$ [see Eq.~\eqref{eq_lenPhi0}].
Analogously, the non-universal amplitudes in Eqs.~\eqref{eq_gen_correlmu} and \eqref{eq_gen_phimu} can be obtained from Eq.~\eqref{eq_gen_amplMu_t} by noting that, within MFT, $\eta=0$ and that one has for the susceptibility amplitude $C^+=(\amplXip)^2$, yielding
$\amplPhimu =(\amplXip)^{2/\delta} (\amplPhit)^{1-1/\delta}= (6/g)^{1/3}$  and $\amplXimu = (\amplXip \amplPhimu)^{1/2} =  (\amplXip \amplPhit)^{1/3} = (6/g)^{1/6}$.
The amplitude $\lenH1=\lenSpH1$ appropriate to the special surface phase transition [see Eq.~\eqref{eq_gen_h1len_sp}] can be extracted from the MFT result presented in Eq.~\eqref{eq_sde_gc_weak_Tc_0} below, based on the concept of an extrapolation length [see the discussion following Eq.~\eqref{eq_gen_h1len_sp}], yielding $\lenH1=(6/g)^{1/4}$ \footnote{Similarly to the amplitude $\amplXimu$ we have included an additional factor of $2^{-1/4}$ in $\lenH1$ in order to end up with a simple form of the MFT expressions [see Eqs.~\eqref{eq_ELE} and \eqref{eq_ELE_BC}]}.

In order to obtain the equilibrium states from Eqs.~\eqref{eq_Landau_func_c} and \eqref{eq_Mass0}, we minimize the extended, unconstrained  functional (per transverse area and $k_B T$)
\begin{widetext}
\begin{equation} \Fcal_f\gc([\phi];\mu) \equiv \int_{-L/2}^{L/2} dz \left[\onehalf (\pd_z \phi)^2 + \onehalf \tau \phi^2 +\frac{1}{4!} g \phi^4 - \mu\phi\right] - \left[h_1^-\phi(-L/2)+ h_1^+\phi(L/2)\right] 
\label{eq_Landau_func_gc}\end{equation}
with respect to $\phi$ and determine the Lagrange multiplier $\mu$ such that the constraint in Eq.~\eqref{eq_Mass0} is fulfilled.
As indicated by the notation, $\Fcal_f\gc$ represents the free-energy functional of a film in the grand canonical (gc) ensemble, in which $\mu$ plays the role of a chemical potential (or a bulk field).
Minimization of the functional in Eq.~\eqref{eq_Landau_func_gc} leads to the Euler-Lagrange equation (ELE)
\beq \pd_z^2 \phi - \tau \phi - \frac{g}{6}\phi^3 + \mu = 0,
\label{eq_ELE0}\eeq
together with the boundary conditions
\beq \begin{split}
	\pd_z \phi\big|_{z=-L/2} = -h_1^-, \qquad
	\pd_z\phi\big|_{z=L/2} = h_1^+ .
\end{split}\label{eq_ELE0_bc}\eeq 

In order to highlight the scaling behavior it is instructive and convenient to introduce the dimensionless finite-size scaling variables in Eqs.~\eqref{eq_gen_prof_scal} and \eqref{eq_scalvar}, which, within MFT, take the form
\beq 
\tscal = L^{2} \tau,\quad 
B = \sqrt{\frac{g}{6}} L^{3} \mu,\quad
H_1 = \sqrt{\frac{g}{6}} L^{2} h_1 ,\quad
m(\zeta) = \sqrt{\frac{g}{6}} L \phi(\zeta L),\quad\text{and}\quad
\Mass = \sqrt{\frac{g}{6}} L \mden.
\label{eq_scalvar_mft}\eeq  
Accordingly, the functional $\Fcal_f\gc$ in Eq.~\eqref{eq_Landau_func_gc} can be expressed as
\beq \Fcal_f\gc([m];B) =\frac{\Delta_0}{L^3} \left\{\int_{-1/2}^{1/2} d\zeta \left[\onehalf ( m')^2 + \onehalf \tscal  m^2 +\frac{1}{4}  m^4 - B  m\right] - [H_1^- m(-1/2)+H_1^+ m(1/2)] \right\},
\label{eq_Landau_func_ndim}\eeq 
\end{widetext}
where $\Delta_0$ is defined in Eq.~\eqref{eq_Delta0}.
Within MFT, the film thickness $L$ as well as the unknown coupling constant $g$ can be scaled out and enter into Eq.~\eqref{eq_Landau_func_ndim} only as prefactors.
As a consequence they neither appear in the dimensionless ELE
\beq  m''(\zeta) - \tscal  m(\zeta) -  m^3(\zeta) + B = 0,
\label{eq_ELE}\eeq
nor in the boundary conditions
\beq \begin{split}
 m'\big|_{\zeta=-1/2} = -H_1^-,\qquad 
 m'\big|_{\zeta=1/2} = H_1^+.
\end{split}\label{eq_ELE_BC}\eeq 

The critical properties emerging from Eq.~\eqref{eq_ELE} have been analyzed in Refs.~\cite{fisher_scaling_1981, nakanishi_critical_1983} and, for $|H_1|=\infty$, analytical solutions of Eq.~\eqref{eq_ELE} in terms of elliptic functions are given in Refs.~\cite{krech_casimir_1997, gambassi_critical_2006, dantchev_exact_2015, dantchev_exact_2016}.
Solutions of the linearized Eq.~\eqref{eq_ELE} have been discussed, for instance, in Ref.~\cite{binder_critical_1983}, while the behavior of the OP near the boundaries has been investigated in Refs.\ \cite{brezin_critical_1983, peliti_strong_1983}.
In order to provide a self-contained presentation, we recall some of these results as they are relevant for the present purpose.
The case of finite $H_1$ and arbitrary values of $\tscal$ and $B$ is studied in the following via perturbation theory and numerical methods, with a particular focus on the effect of introducing the mass constraint.
In Appendix \ref{app_mapping} the mean field scaling properties are utilized in order to derive a mapping between the OP profile in a film with finite surface fields and the profile in a film in which they are infinite. 
While this relationship is not needed in the remaining part of this work, it may be used in conjunction with the known analytical solutions for $|H_1|=\infty$ as an alternative to the perturbative expansion discussed below.

\subsection{Perturbative solution}
\label{sec_ads_pert}
In order to proceed analytically, we solve the ELE in Eq.\ \eqref{eq_ELE} for arbitrary bulk ($B$) and surface ($H_1$) fields employing a perturbative expansion in terms of the nonlinear term. 
To this end, we introduce a book-keeping parameter $\epsilon$ (eventually set to 1) into Eq.~\eqref{eq_ELE},
\beq  m'' = \tscal  m + \epsilon  m^3 - B,
\label{eq_ELE_eps}\eeq 
and expand the OP profile $m(\zeta)$ and the bulk field $B$ accordingly:
\begin{subequations}
\begin{align}
 m &=  m_0 + \epsilon  m_1 + \epsilon^2  m_2 + \ldots , \label{eq_m_eps}\\
B &= B_0 + \epsilon B_1 + \epsilon^2 B_2 + \ldots\,. \label{eq_Hb_eps}
\end{align}\label{eq_eps_expansion}
\end{subequations}
We consider the total mass $\Mass$ to be a quantity of $O(\epsilon^0)$ and therefore we enforce the mass constraint in Eq.~\eqref{eq_Mass} completely at this order, i.e.,
\beq \Mass_0=\Mass,\quad \Mass_{i\geq 1}=0\quad \text{with } \Mass_{i\geq 0}\equiv \int_{-1/2}^{1/2} d\zeta\,  m_i(\zeta). \label{eq_mass_eps}
\eeq
As a consequence, the total mass $\Mass$ will affect the higher-order corrections $m_i$ only implicitly via their dependence on $m_0$.
In the following, we consider a system with equal surface fields, $H_1\equiv H_1^-=H_1^+$, i.e., with symmetric boundary conditions. Results for the case of opposite surface fields will be summarized briefly in Sec.~\ref{sec_ads_pert_pm}.
The surface field is considered to be a quantity of $O(\epsilon^0)$. Hence the boundary conditions in Eq.~\eqref{eq_ELE_BC} turn into
\beq \begin{split}
 m_0'(-1/2) &= - m_0'(1/2) = -H_1,\\
 m_i'(-1/2) &=  m_i'(1/2) = 0\quad \text{for $i\geq 1$}.
\end{split}\label{eq_ELE_bc_eps}\eeq 
Equation \eqref{eq_ELE_eps} can be considered also in the grand canonical ensemble, i.e., without enforcing a mass constraint. In this case $B$ is a certain assigned external field of $O(\epsilon^0)$ and it is not expanded in terms of $\epsilon$ (i.e., $B_{i}=0$ for $i\geq 1$).

\subsubsection{Solution at $O(\epsilon^0)$}

In the absence of the nonlinear term, i.e., for $\epsilon=0$, Eq.~\eqref{eq_ELE_eps} reduces to
\beq  m_0'' = \tscal  m_0 - B_0,
\label{eq_ELE_lin}\eeq 
with the solution
\beq  m_0(\zeta) = \frac{B_0}{\tscal} + \frac{H_1}{\sqrt{\tscal}} \frac{\cosh(\zeta \sqrt{\tscal})}{\sinh(\sqrt{\tscal}/2)}.
\label{eq_m0_sol}\eeq
Occasionally, we shall refer to the above two equations as linear MFT. 
By using elementary properties of hyperbolic functions, Eq.~\eqref{eq_m0_sol} can be cast into the equivalent form
\beq  m_0(\zeta) = \frac{B_0}{\tscal} - \frac{H_1}{\sqrt{-\tscal}} \frac{\cos(\zeta \sqrt{-\tscal})}{\sin(\sqrt{-\tscal}/2)}\,,
\label{eq_m0_sol_negtau}\eeq 
which is particularly suited for the case $\tscal<0$ corresponding to temperatures below the bulk critical point.
For completeness, we report also the profile in terms of the unscaled quantities, as this form will be useful further below in Sec.~\ref{sec_Casimir} for studying \CCFs:
\beq \phi_0(z) = \frac{\mu_0}{\tau} + \frac{h_1}{\sqrt{\tau}} \frac{\cosh(z\sqrt{\tau})}{\sinh(L\sqrt{\tau}/2)}.
\label{eq_phi0_sol}\eeq 
These profiles formally diverge for $\tscal = -4\pi^2 n^2$ with $n=0,1,2,\ldots$, which can be considered to be an artifact of linear MFT. We remark that, while for $\tscal<0$ linear MFT does not allow the occurrence of a stable ordered bulk phase, the critical point in a film is actually shifted to a temperature $T_{c,\text{cap}}$ below the bulk critical temperature $T_c$ \cite{nakanishi_critical_1983}. 
Accordingly, the confined system is still in a stable disordered phase even within a certain range of negative values of $\tscal$.
In particular, it will be shown below that, once the mass constraint is imposed, the divergence of the profiles in Eqs.~\eqref{eq_m0_sol} and \eqref{eq_m0_constr_negtau} for $\tscal=0$ is eliminated and one obtains a well-defined profile for all $\tscal>-4\pi^2$.

Upon inserting Eq.~\eqref{eq_m0_sol} into Eq.~\eqref{eq_Mass} and by imposing the mass constraint according to Eq.~\eqref{eq_mass_eps}, one obtains the dependence of the bulk field $B$ on $\Mass$ at the zeroth order:
\beq \tilde B_0 = \Mass \tscal -2 H_1, 
\label{eq_B0}\eeq
yielding
\beq \tilde m_0(\zeta) = \Mass - \frac{2H_1}{\tscal} + \frac{H_1}{\sqrt{\tscal}} \frac{\cosh(\zeta \sqrt{\tscal})}{\sinh(\sqrt{\tscal}/2)},
\label{eq_m0_constr}\eeq
or, in terms of unscaled quantities [see Eq.~\eqref{eq_scalvar}],
\beq \tilde \mu_0 = \frac{\mass \tau -2 h_1}{L}
\label{eq_h0}\eeq 
and
\beq \tilde \phi_0(z) = \frac{\mass}{L} - \frac{2 h_1}{L \tau} + \frac{h_1}{\sqrt{\tau}} \frac{\cosh(z\sqrt{\tau})}{\sinh(L\sqrt{\tau}/2)}\,.
\label{eq_phi0_constr}\eeq 
Here and in the following, a tilde is used to indicate a quantity evaluated under the mass constraint. 
Up to this order in perturbation theory, enforcing the mass constraint results in a certain, spatially constant shift of the corresponding grand canonical profile obtained for $B=0$.
We note that for fixed $B$, the profile $m_0$ in Eq.~\eqref{eq_m0_sol} diverges for $\tscal\to 0$, whereas in that limit the constrained profile $\tilde m_0$ in Eq.~\eqref{eq_m0_constr} remains finite: 
\beq \tilde  m_0(\zeta,x\to 0) = \Mass + H_1\left(\zeta^2-\frac{1}{12}\right).
\label{eq_m0_tau0_lim}\eeq
Similarly to Eq.~\eqref{eq_m0_sol_negtau}, the constrained profile in Eq.~\eqref{eq_m0_constr} can be expressed in the equivalent form
\beq \tilde m_0(\zeta) = \Mass - \frac{2H_1}{\tscal} - \frac{H_1}{\sqrt{-\tscal}} \frac{\cos(\zeta \sqrt{-\tscal})}{\sin(\sqrt{-\tscal}/2)},
\label{eq_m0_constr_negtau}\eeq
which is convenient for $\tscal<0$.
The constrained profile diverges for $\tscal = -4 \pi^2 n^2$ with $n=1,2,\ldots$, but not for $n=0$.
For thick films in the supercritical region, i.e., $x\to +\infty$, and independently of $H_1$, one has asymptotically 
\beq \tilde m_0(\zeta,x\to +\infty) \simeq 
\begin{cases}
 \Mass + \frac{H_1}{\sqtscal},\qquad \zeta=\pm \onehalf,\\
 \Mass -2\frac{H_1}{\tscal},\qquad \zeta=0.
\end{cases}
\label{eq_m0_constr_largeTau}\eeq

\subsubsection{Solution at $O(\epsilon)$}

At first order in $\epsilon$, the ELE in Eq.\ \eqref{eq_ELE_eps} turns into
\beq  m_1''(\zeta) = \tscal  m_1(\zeta) +  m_0^3(\zeta) - B_1,
\label{eq_ELE_eps1}\eeq
with the boundary conditions
\beq  m_1'(-1/2) =  m_1'(1/2) = 0.
\label{eq_ELE_eps1_BC}\eeq 
The complete analytic expression for $ m_1$ is rather lengthy and therefore we do not report it here. 
For the special case $B=0$ (and thus $B_1=0$), one has
\begin{multline} m_1(\zeta, B=0) =  m_0(\zeta,B=0) \frac{H_1^2}{18\tscal^2 \sinh^2(\sqrt{\tscal}/2)}\\
 \Big\{\cosh(2\zeta\sqrt{\tscal}) - 3\cosh\sqrt{\tscal}-8 - 3\sqrt{\tscal}\big[\coth(\sqtscal/2)\\ -2\zeta\tanh(\zeta\sqtscal)\big]\Big\},
\label{eq_m1_Hb0}\end{multline} 
which holds both for positive and for negative values of $\tscal$.
As expected, $ m_1$ vanishes for $H_1=0$, i.e., in the absence of an ordering field.
By comparing (in the supercritical region, i.e., for $\tscal>0$) the relative magnitude of the various terms in Eq.~\eqref{eq_m1_Hb0} one finds, that for $\tscal\lesssim |H_1|$ the perturbative correction $ m_1$ becomes larger in magnitude than the profile $ m_0$. 
This leads us to introduce a coarse smallness parameter 
\beq \sigma \equiv \frac{H_1^2}{\tscal^2} = \frac{g}{6} \frac{h_1^2}{\tau^2}.
\label{eq_smallness_param}\eeq 
For a fixed bulk field $B$, the results obtained perturbatively are reliable only as long as $\sigma \lesssim 1$.
Note that $\sigma$ does not depend on the thickness $L$ of the film.

Upon inserting the full solution of Eqs.~\eqref{eq_ELE_eps1} and \eqref{eq_ELE_eps1_BC} into Eq.~\eqref{eq_Mass} and enforcing the mass constraint according to Eq.~\eqref{eq_mass_eps}, i.e., $\Mass_1=0$, one finds
\begin{multline} \tilde B_1 = \frac{H_1^3}{\tscal^2}\left[ \frac{2}{3} + \frac{16}{\tscal} - \frac{16 \coth(\sqtscal/2)}{\sqtscal} - \frac{1}{\sinh^2(\sqtscal/2)}\right]\\ + 3\Mass \frac{ H_1^2}{\tscal^2 } \frac{ 4+\tscal-4\cosh\sqtscal + \sqtscal\sinh\sqtscal}{\cosh\sqtscal -1} + \Mass^3.
\label{eq_B1}\end{multline}
Due to its perturbative nature, this expression holds only for $\tscal\gtrsim |H_1|$. 
In contrast to the zeroth-order constraint field $\tilde B_0$ [Eq.~\eqref{eq_B0}], $\tilde B_1$ as well as the higher-order corrections depend on the scaling variable $\tscal$ even for $\Mass=0$.
For large $\tscal$, $\tilde B_1$ behaves asymptotically as
\beq \tilde B_1 (\tscal\to\infty)\simeq \frac{2}{3}\frac{H_1^3}{\tscal^2} + 3 \Mass \frac{H_1^2}{\tscal^{3/2}} + \Mass^3.
\label{eq_B1_largetau}\eeq
Under the constraint, the profile behaves asymptotically for large $\tscal$ as 
\beq \begin{split}
\tilde  m_1(\zeta=0, \tscal\to\infty)&\simeq -\tilde  m_0(0)\left[\frac{H_1^2}{3\tscal^2} + \frac{6H_1\Mass}{\tscal^2}\right],\\
\tilde  m_1(\zeta=\pm 1/2, \tscal\to\infty) &\simeq -\tilde  m_0(\pm 1/2)\left[\frac{H_1^2}{4\tscal^2}+ \frac{3H_1\Mass}{2\tscal^{3/2}}\right].
\end{split}\label{eq_m1_tauLarge} \eeq
Interestingly, both $\tilde B_1$ [Eq.~\eqref{eq_B1}] and the constrained profile $\tilde  m_1$ [not reported here---and in contrast to the unconstrained $m_1$ in Eq.~\eqref{eq_m1_Hb0}], attain a \emph{finite} value for $\tscal\to 0$. 
We shall show below that, in fact, for sufficiently small $H_1$ and $\Mass$, the solution of the linear MFT provides an accurate approximation to the one of the full nonlinear theory.

\subsubsection{Solution at $O(\epsilon^2)$}
While the perturbative solution can be extended to higher orders in $\epsilon$ without basic problems, the resulting expressions become increasingly lengthy. Since no novel qualitative features emerge by accounting for the higher-order contributions, in the following we only report certain limiting behaviors. 

At second order in $\epsilon$, one has
\beq  m_2''(\zeta) = \tscal  m_2(\zeta) +  m_0^2(\zeta)  m_1(\zeta) - B_2,
\eeq
with the boundary conditions
\beq  m_1'(-1/2) =  m_1'(1/2) = 0.
\eeq 
Enforcing the constraint $\Mass_2 = 0$ yields the bulk field $\tilde B_2$, which behaves asymptotically far from the bulk critical point as
\beq \tilde B_2 (\tscal\to\infty) \simeq -\frac{9 H_1^2}{64 \tscal^{5/2}} \Mass^3 - \frac{H_1^5}{160 \tscal^4 }.
\label{eq_B2_largetau}\eeq 
In contrast to $\tilde B_1$ [Eq.~\eqref{eq_B1_largetau}], $\tilde B_2$ vanishes for large $\tscal$ even for $\Mass\neq 0$.
Similarly to $\tilde  m_1$, the perturbative correction $\tilde m_2$ has a finite value for $\tscal\to 0$.

\subsubsection{Summary}
\label{sec_ads_pert_summary}

Due to the preferential adsorption at the walls, the OP profile increases upon approaching them and generically takes values of the same sign as that of the closest surface field. Accordingly, for $(++)$ boundary conditions the contribution to $\Mass$ stemming from the region close to the walls is positive and, consequently, a negative bulk field $\tilde B$ is expected to be necessary in order to yield $\Mass=0$, in agreement with the above perturbative results [see Eq.~\eqref{eq_B0}].
Concerning the special case $\Mass=0$, it is interesting to note that the bulk field behaves asymptotically as a polynomial in the smallness parameter $\sigma$ [Eq.~\eqref{eq_smallness_param}]: 
\begin{multline} \tilde B(\Mass=0,\tscal\gg H_1) = \tilde B_0 + \epsilon \tilde B_1 + \epsilon^2 \tilde B_2 + O(\epsilon^3) \\
= -2H_1 \left(1 - \epsilon\frac{H_1^2}{\tscal^2}\left(\frac{1}{3} - \epsilon \frac{H_1^2}{80\tscal^2}\right)\right)+O(\epsilon^3).
\end{multline} 
On the other hand, in the opposite limit, i.e., at bulk criticality $\tscal=0$, the constrained bulk field $\tilde B$ is also nonzero, but, for $\Mass=0$, it is a polynomial in $H_1$:
\begin{multline} \tilde B(\Mass=0, \tscal=0) = -2H_1 \\ \left(1 - \epsilon H_1^2 \left(\frac{1}{7560} - \epsilon \frac{149}{174\:356\:583\:400} H_1^2\right)\right) + O(\epsilon^3).
\end{multline}
We emphasize that, for $H_1=0$, the OP profile is flat and the constraint-induced field $\tilde B$ reduces to the one of a homogeneous bulk system, $\tilde B = \tilde B_0 + \tilde B_1 = \Mass \tscal + \Mass^3$ [Eqs.~\eqref{eq_B0} and \eqref{eq_B1}], while $\tilde B_{i\geq 2}=0$.

\begin{figure}[t]\centering
	\includegraphics[width=0.8\linewidth]{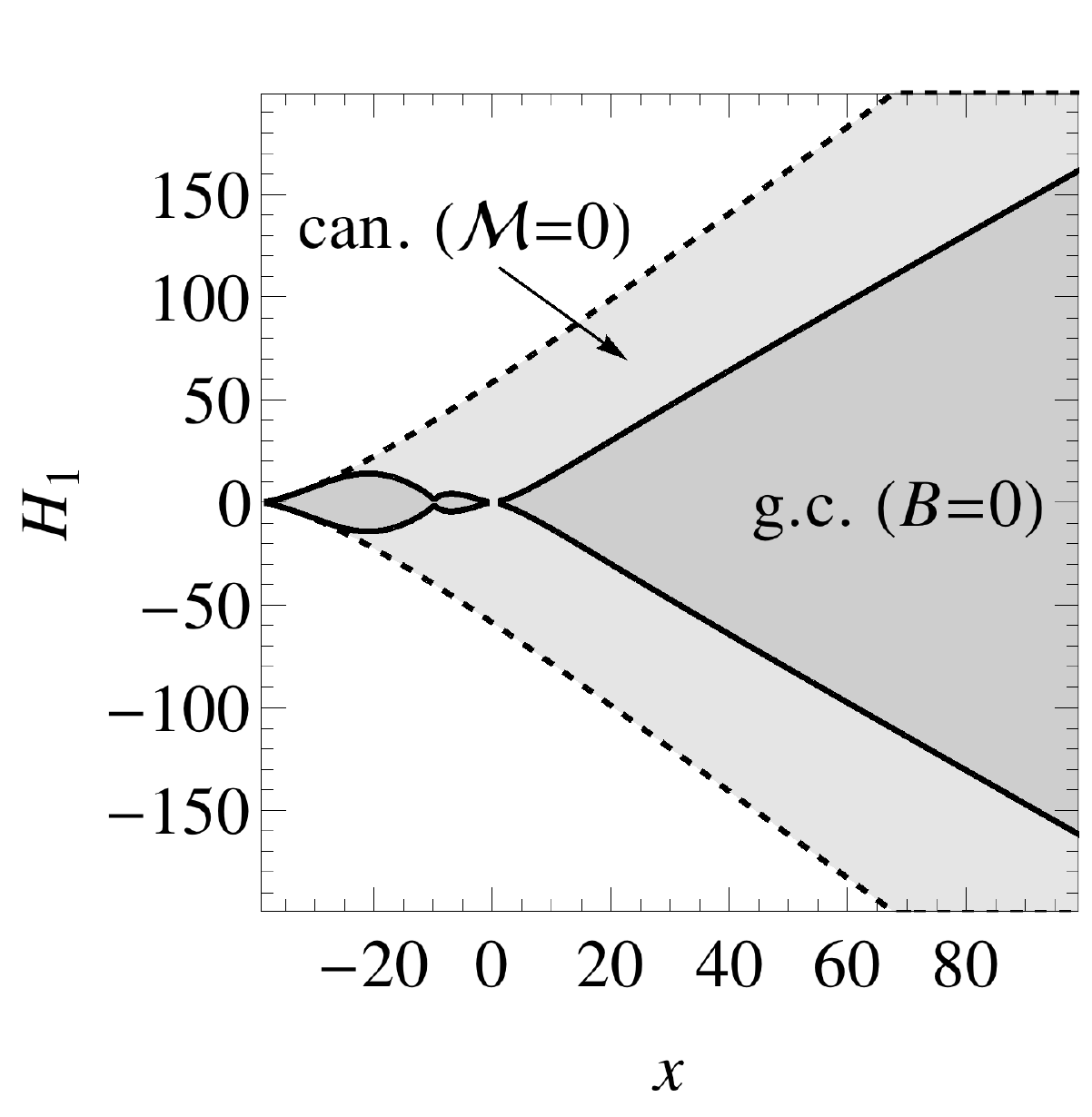}
	\caption{Ranges of values of the parameters $\tscal$ and $H_1$ within which the grand canonical [$m_0$, Eq.~\eqref{eq_m0_sol}] and the canonical [$\tilde m_0$, Eq.~\eqref{eq_m0_constr}] mean field solutions are reliable for a film with symmetric surface fields $H_1$, obtained by requiring $|m_1(\zeta)|\leq |m_0(\zeta)|$ and $|\tilde m_1(\zeta)|\leq |\tilde m_0(\zeta)|$, respectively. The shaded areas represent the regions in which these conditions are fulfilled at $\zeta=\pm 1/2$, $\zeta=0$, and for a vanishing bulk field $B$ or mass $\Mass$, as indicated. For nonzero $B$ or $\Mass$, the ranges of parameters are qualitatively similar. Note that the lightly shaded region encompasses also the darkly shaded one.}
	\label{fig_linMFT_valid}
\end{figure}

Since in the remaining part of the present study we shall focus on the solution of the linearized ELE, it is important to determine the parameter region for which it provides an accurate description.
A simple estimate can be obtained by requiring that the first-order perturbative correction $m_1$ is small compared to the zeroth-order solution $m_0$.
In the case of the grand canonical ensemble, we have derived from this requirement a smallness parameter $\sigma=H_1^2/\tscal^2$ [Eq.~\eqref{eq_smallness_param}] which signals the onset of a strongly nonlinear regime for $\sigma\gtrsim 1$.
The solution [Eq.~\eqref{eq_m0_sol}] of the unconstrained linear ELE may thus be expected to provide an accurate approximation to the full theory only if $\tscal\gg |H_1|$.
In the canonical case, instead, Eq.~\eqref{eq_m1_tauLarge} indicates that the constrained solution $\tilde m_0$ in Eq.~\eqref{eq_m0_constr} ceases to be accurate for large mass $|\Mass|$ because the subsequent terms in the perturbative expansion are dominant for $|\Mass|\gg 1$.
This is expected, because neglecting the nonlinear term $m^3$ in the ELE implicitly assumes that the mean OP, hence $\Mass$ [Eq.~\eqref{eq_Mass}], are small as well. Thus, effectively, the present perturbation theory is constructed around $\Mass=0$. 
Alternatively, one could develop a perturbation scheme around the proper mean $\Mass$ of the OP, taking into account already at leading order the dominant terms proportional to powers of $\Mass$ arising from an expansion of the nonlinearity in Eq.~\eqref{eq_ELE}. 
Such an approach is outlined in Appendix \ref{app_pert}, where, \textit{inter alia}, expressions for the (grand-)canonical OP profiles are derived which are applicable for $|\Mass| \gg 1$.
Those results will be used further below in order to rationalize the asymptotic behavior of the \CCF for large values of $\Mass$.
The results derived in the present section so far (which in fact follow from the generalized perturbation theory of Appendix \ref{app_pert} in the limit of small $|\Mass|$) will, however, be sufficient for most parts of the subsequent discussion and therefore we continue with their analysis.
For $\tscal=0$ and $\Mass\ll 1$, one finds from Eqs.~\eqref{eq_m0_constr} and \eqref{eq_B1} and from the corresponding full expression for $\tilde m_1$ (not reported)
\beq \begin{split}
	\frac{\tilde  m_1}{\tilde  m_0}\Big|_{\zeta=\pm 1/2,\tscal\to 0} &= -\frac{H_1^2}{5040}-\frac{H_1\Mass}{840}-\frac{3\Mass^2}{70} + O(\Mass^3),\\
	\frac{\tilde B_1}{\tilde B_0} \Big|_{\tscal\to 0} &= -\frac{H_1^2}{7560} - \frac{H_1\Mass}{120} - \frac{\Mass^3}{2 H_1} ,
\end{split} \label{eq_m1_tau0}\eeq
indicating that the constrained solution $\tilde m_0$ together with $\tilde B_0$ [Eqs.~\eqref{eq_m0_constr} and \eqref{eq_B0}] remains larger than $\tilde m_1$ and $\tilde B_1$ even around the bulk critical point ($\tscal=0$), provided $H_1$ and $\Mass$ are sufficiently small in magnitude \footnote{The divergence of $\tilde B_1/\tilde B_0$ as $H_1\to 0$ for $\Mass\neq 0$ in Eq.~\eqref{eq_m1_tau0} implies that $\Mass$ must in fact vanish faster than $H_1^{1/3}$ as $H_1\to 0$ in order for the constrained solution to be valid for $\tscal=0$.
}.

In order to complete this picture, in Fig.~\ref{fig_linMFT_valid} we visualize the range of values of $\tscal$ and $H_1$ within which the condition $|m_1/m_0|\leq 1$ in the grand canonical and $|\tilde m_1/\tilde m_0|\leq 1$ in the canonical ensemble, respectively, are fulfilled.
For simplicity, we evaluate these conditions at the positions $\zeta=0$ and $\zeta=\pm 1/2$ in the film and take the most stringent one.
We find that the range of allowed values of $H_1$ at fixed $\tscal$ widens essentially linearly upon increasing $\tscal$ both for the constrained and the unconstrained solution, consistently with Eqs.~\eqref{eq_smallness_param} and \eqref{eq_m1_tauLarge}.
The allowed domain of $H_1$ shrinks to zero for those values of $\tscal$ for which the solutions of the linear MFT diverge [see Eqs.~\eqref{eq_m0_sol_negtau} and \eqref{eq_m0_constr_negtau}].
In agreement with Eq.~\eqref{eq_m1_tau0}, for $x=0$ in the constrained case, the crossover between the domain of validity of linear MFT and the nonlinear regime occurs at $|H_1|\simeq 50$.

\begin{figure*}[t]\centering
    \subfigure[]{\includegraphics[width=0.45\linewidth]{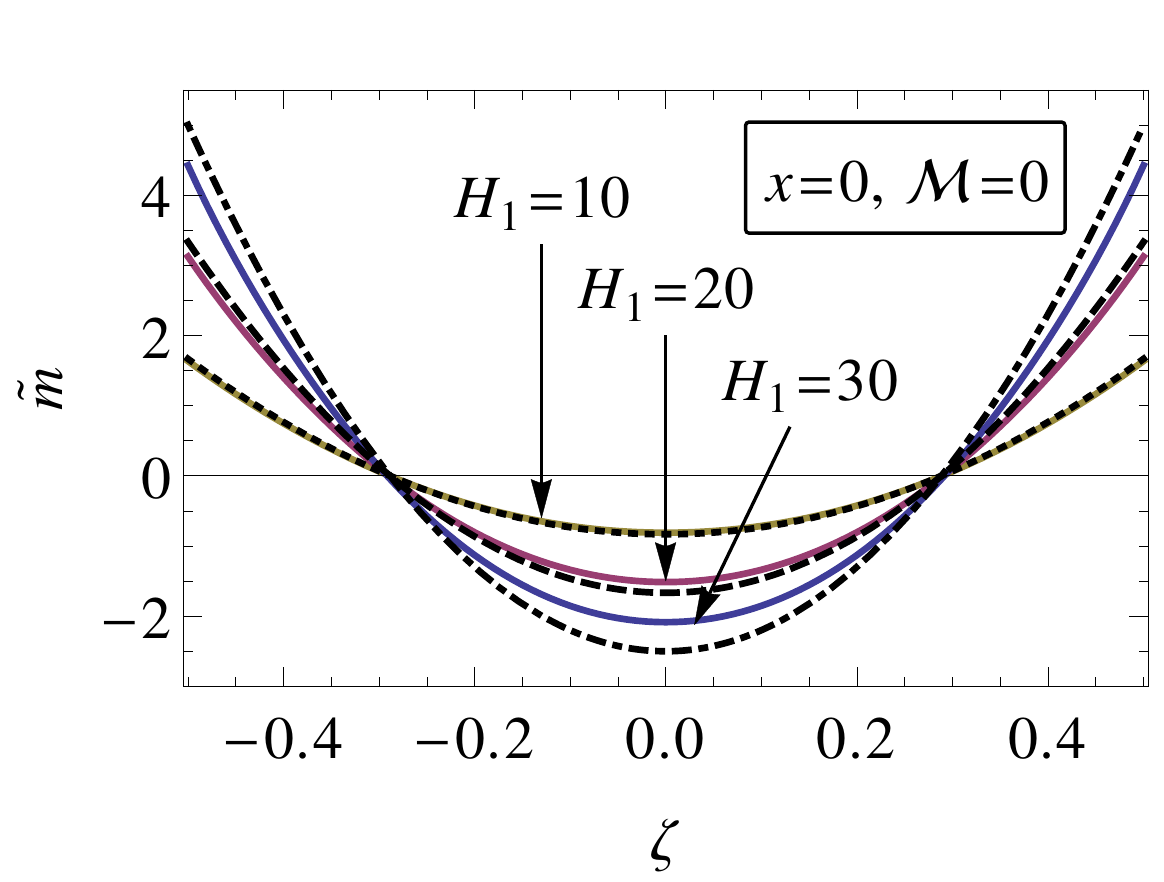}}\qquad
    \subfigure[]{\includegraphics[width=0.458\linewidth]{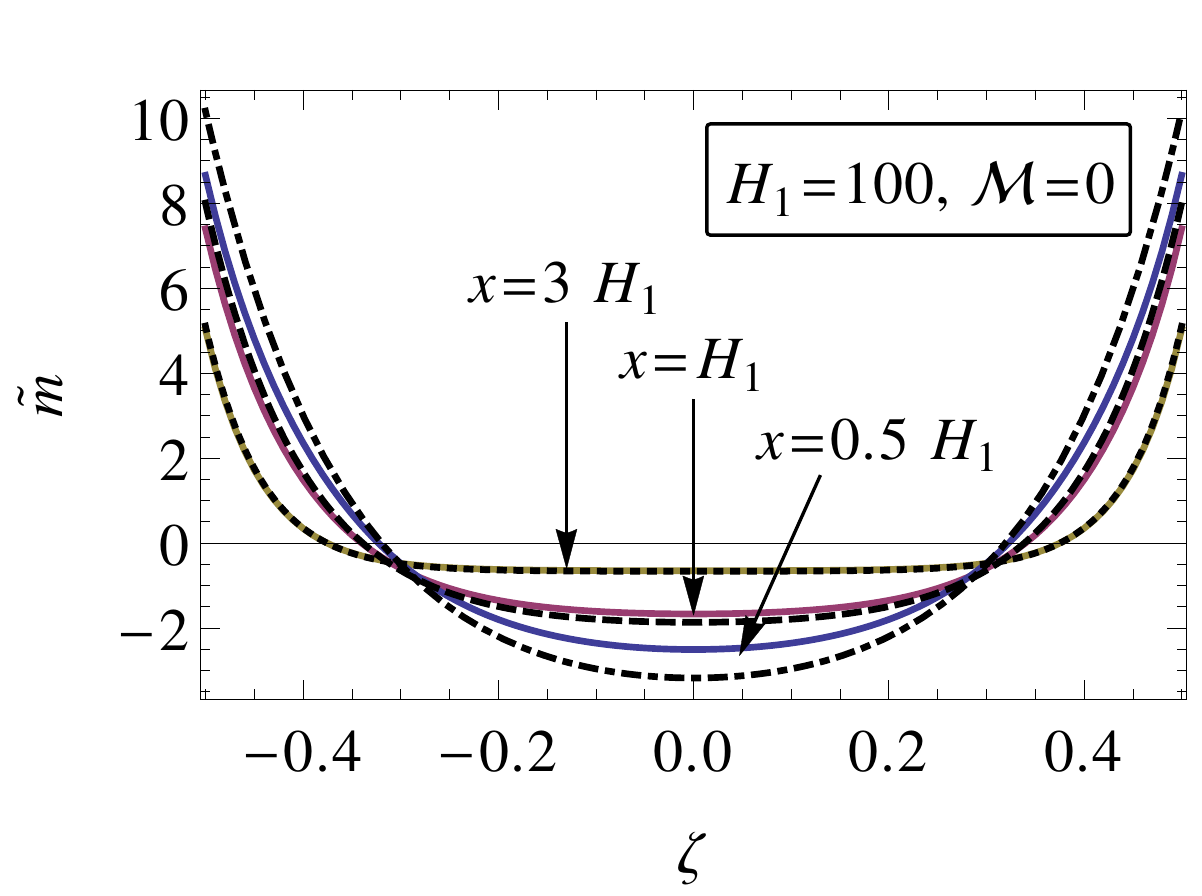}}
    \caption{Comparison of the constrained solution $\tilde m_0$ of the linear ELE (broken lines) with the numerical solution (solid lines) of the full ELE in Eq.\ \eqref{eq_ELE} with the mass constraint $\Mass=0$ and for various scaled surface fields $H_1$ [(a), for $x=0$] and scaled temperatures $x$ [(b), for $H_1=100$]. The walls are located at $\zeta=\pm 1/2$ and impose $(++)$ boundary conditions in accordance with Eq.~\eqref{eq_ELE_BC}. These numerical results explicitly confirm that linear MFT is reliable for sufficiently large $x$ and, provided $H_1$ is sufficiently small, even for $x\simeq 0$  (see the discussion in the main text).}
    \label{fig_prof_test}
\end{figure*}

In Fig.\ \ref{fig_prof_test}, we compare the analytical solution $\tilde m_0$ [Eq.~\eqref{eq_m0_constr}] of the linearized MFT with the numerical solution of the full ELE in Eq.\ \eqref{eq_ELE} for a selected set of parameters. 
For $\Mass=0$ and $x=0$, we expect, according to Eq.~\eqref{eq_m1_tau0} and Fig.~\ref{fig_linMFT_valid}, that the linear solution of MFT is accurate for sufficiently small $H_1$.
This is confirmed in Fig.~\ref{fig_prof_test}(a), where we observe good agreement between the numerical and analytical profiles for $H_1=10$, but increasing deviations for larger $H_1$.
In panel (b), we fix the surface field to the rather large value $H_1=100$, in which case we expect linear MFT to remain valid only for $x\gg H_1$. Indeed, we observe good agreement between $\tilde m_0$ and the full numerical solution for $x\gg H_1$, whereas deviations become noticeable for $x\lesssim H_1$.

\subsubsection{Antisymmetric boundary conditions}
\label{sec_ads_pert_pm}
Here, we briefly summarize the relevant features of the solution of the linearized mean field ELE with the mass constraint in the case that the surface fields $H_1^-$ and $H_1^+$ have equal strength but opposite signs, i.e., $H_1\equiv H_1^-=-H_1^+$. In this case, the boundary conditions for the perturbative solutions are
\beq \begin{split}
 m_0'(-1/2) &=  m_0'(1/2) = -H_1,\\
 m_i'(-1/2) &=  m_i'(1/2) = 0\quad \text{for $i\geq 1$}.
\end{split}\eeq 
Proceeding as above for symmetric boundary conditions, we obtain the solution of the linear ELE in Eq.\ \eqref{eq_ELE_lin}:
\beq m_0(\zeta) = \frac{B_0}{\tscal} - \frac{H_1}{\sqtscal} \frac{\sinh(\zeta \sqtscal)}{\cosh(\sqtscal/2)}.
\label{eq_pm_m0}\eeq 
In order to fulfill the mass constraint, the bulk field has to take the value
\beq \tilde B_0 = \Mass\tscal.
\label{eq_pm_B0}\eeq 
In terms of unscaled variables, these results are given by
\begin{subequations}\bal \phi_0(z) &= \frac{\mu_0}{\tau} - \frac{h_1}{\sqtau} \frac{\sinh(z\sqtau)}{\cosh(L\sqtau/2)}\label{eq_pm_sol0_bare}
\intertext{and}
\tilde\mu_0 &=\frac{\mass \tau}{L} \label{eq_pm_h0}.
\end{align}\label{eq_pm_sol0_bare_all}\end{subequations}
While, as an artifact of the linearized MFT, $m_0$ diverges upon approaching the bulk critical point $\tscal= 0$, the constrained profile $\tilde m_0$ remains finite in that limit:
\beq \lim_{\tscal\to 0} \tilde m_0(\zeta) = \Mass - H_1 \zeta,
\label{eq_pm_m0_constr_Tc}\eeq 
as it was the case for symmetric boundary conditions [see Eq.~\eqref{eq_m0_tau0_lim}].
Still as an artifact, both the constrained and the unconstrained profiles $\tilde m_0$ and $m_0$, respectively, diverge for $\tscal=-\pi^2(2n+1)$ with $n=0,1,\ldots$, similar to the case of symmetric boundary conditions [see Eq.~\eqref{eq_m0_constr} and the related discussion]. 
In contrast to the symmetric case [Eq.~\eqref{eq_B0}], the adsorption strength $H_1$ does not affect the lowest-order constraint field $\tilde B_0$ [Eq.~\eqref{eq_pm_B0}] but enters only via perturbative corrections.
In particular, at $O(\epsilon)$ in the nonlinear term, we find
\beq \tilde B_1 = \frac{3 H_1 \Mass }{\tscal} \left[ \frac{\tanh(\sqtscal/2)}{\sqtscal} - \frac{1}{1+\cosh\sqtscal }\right] +  \Mass^3.
\label{eq_pm_B1}\eeq 
The first-order correction $m_1$ to the OP profile is also given by a rather lengthy expression which we do not report here.
The parameter range within which the profiles $m_0$ and $\tilde m_0$ obtained from linear MFT provide an accurate approximation of the solution of the full ELE [Eq.~\eqref{eq_ELE}] can be estimated as in the previous subsection. It turns out that this occurs for sufficiently small magnitudes of the mass $\Mass$ and for $\tscal\gg |H_1|$.
An exception here is the constrained case, in which, for sufficiently small $|H_1|$, even a small region around $\tscal\simeq 0$ is still allowed in the above sense, similar to Fig.~\ref{fig_linMFT_valid}.

\subsection{Short-distance expansion}
\label{sec_sde}
Outside the domain within which linear MFT and its successive perturbative corrections [see Sec.~\ref{sec_ads_pert}] are accurate one can determine the OP profile \emph{close to} one of the confining walls by applying a short-distance expansion (SDE) \cite{brezin_critical_1983, peliti_strong_1983, diehl_field-theoretical_1986, schlesener_critical_2003}. 
We first consider the case of the grand canonical ensemble, i.e., with fixed external field $\mu$, which can be later mapped onto the canonical ensemble with fixed mass $\mass$.
Within the grand canonical ensemble, the SDE in semi-infinite geometry has been studied previously \cite{brezin_critical_1983, peliti_strong_1983, diehl_field-theoretical_1986, schlesener_critical_2003}.
In the corresponding parallel-plate geometry, the presence of a second wall typically induces a so-called ``distant wall correction'' to the semi-infinite OP profile $\phi_{\infty/2}$ such that the critical OP profile $\phi$ in a film behaves, in the strong adsorption regime ($|H_1|=\infty$), as \cite{fisher_wall_1978, au-yang_wall_1980, fisher_critical_1980, rudnick_critical_1982, cardy_universal_1990} 
\beq \phi(\hat z\ll L) \simeq \phi_{\infty/2}(\hat z)\left[1+\mathcal{C}\left(\frac{\hat z}{L}\right)^{d^*}+\ldots\right]
\label{eq_SDE_distwallcorr}\eeq 
for small distances $\hat z$ from the nearest wall.
Here, $\mathcal{C}$ is a universal constant which depends on the boundary conditions at the two walls \cite{eisenriegler_absence_1993} and $d^*$ is a universal exponent which, in the case of critical adsorption, is equal to the spatial (bulk) dimensionality of the film, $d^*=d$.
In the following, we derive, within MFT, the SDE for the OP profile in the film geometry by locally solving the corresponding ELE near one wall via a (partly re-summed) power-series ansatz in terms of $\hat z$.  
The solution is constructed by inserting such an ansatz into the ELE [Eq.~\eqref{eq_ELE}] and by successively requiring the lowest order terms of the expansion to satisfy the ELE and the boundary conditions.
In general, as long as one is interested in a power-series solution with a sufficiently small number of expansion terms, it turns out that the single nearby boundary condition is in fact already sufficient to fix the required unknown coefficients.
Specifically, if $H_1$ is infinite, the coefficients appearing up to and including $O(\hat z^2)$ in the SDE can be determined uniquely by the near wall boundary condition and, as a consequence of Eq.~\eqref{eq_SDE_distwallcorr}, the distant wall affects the SDE only at $O(\hat z^3)$ or higher.
If, in contrast, $H_1$ is finite, it turns out that the near wall boundary condition allows one to unambiguously determine only the constant term $\propto \hat z^0$ in the SDE. 
However, this term necessarily carries a non-vanishing error because in this case the effect of the distant wall enters already at $O(\hat z^0)$.

\subsubsection{Grand canonical ensemble} 
\label{sec_sde_gc}

We first consider the case of finite $H_1>0$. (The case of $H_1=\infty$ is discussed further below.)
The expression of the OP profile corresponding to $H_1<0$ is obtained as the negative of the one for $H_1>0$ but for a bulk field of reversed sign.
This property follows from the invariance of the ELE in Eq.\ \eqref{eq_ELE} and of the boundary conditions in Eq.~\eqref{eq_ELE_BC} at the two walls under a change of sign of the symmetry-breaking bulk and surface fields and of the OP profile. 
Since we seek a local solution of the OP profile in a film, but \emph{near} one wall, it is appropriate to continue using the quantities in Eq.~\eqref{eq_scalvar} which turn into dimensionless scaling variables upon rescaling them by appropriate powers of the film thickness $L$.
We furthermore do not consider here the effects of capillary condensation transitions associated with the emergence of the equilibrium of two metastable OP profiles (cf.\ Sec.~\ref{sec_ads_num}) and therefore focus only on the supercritical region $\tscal\geq 0$.

Owing to the enhancement of $m$ near a wall, we assume that $|\tscal m| \ll |m^3|$ and $|B|\ll |m^3|$ hold close to the bulk critical point, i.e., for $\tscal$ and $B$ sufficiently small. 
We therefore determine the leading contribution to the desired SDE from the solution of the ELE [Eq.\ \eqref{eq_ELE}] at criticality, i.e., from $m''= m^3$, and the boundary condition in Eq.~\eqref{eq_ELE_BC} for $\zeta=-1/2$.
The corresponding solution is identical to the one obtained in the semi-infinite geometry \cite{brezin_critical_1983, peliti_strong_1983} and is given by 
\beq  m(\zeta=-1/2+\hat\zeta) = \frac{\sqrt{2}}{\hat\zeta+2^{1/4}/\sqrt{H_1}},
\label{eq_sde_gc_weak_Tc_0}\eeq 
for $\hat\zeta\ll 1$, where $\hat\zeta=\hat z/L$ denotes the rescaled distance from the near wall.
This expression does not carry a dependence on $\tscal$ or $B$. Instead, in order to account for the effect of nonzero $\tscal$ and $B$, polynomial terms can be added to the r.h.s.\ of Eq.~\eqref{eq_sde_gc_weak_Tc_0}. 
The simplest ansatz involves only an additional constant term $a_0$, i.e.,
\beq  m(-1/2+\hat\zeta) = \frac{\sqrt{2}}{\hat\zeta + 2^{1/4}/\sqrt{H_1}} + a_0.
\label{eq_sde_gc_weak_Tc}\eeq 
While this ansatz does not render an exact solution of the full ELE in Eq.\ \eqref{eq_ELE}, we can nevertheless determine $a_0$ such that the ansatz approximately fulfills the ELE for small $\hat\zeta$. 
To this end, we insert Eq.~\eqref{eq_sde_gc_weak_Tc} into Eq.\ \eqref{eq_ELE}, expand in terms of powers of $\hat \zeta$, and require that the ELE is satisfied at $O(\hat\zeta^0)$.
Consequently, the coefficient of the term $\hat\zeta^0$, which constitutes the lowest order in the expansion of Eq.~\eqref{eq_ELE} [for the ansatz in Eq.~\eqref{eq_sde_gc_weak_Tc}], must vanish.
This yields
\beq \begin{split}
a_0 &= k - \frac{\tscal}{3k} - 2^{1/4}\sqrt{H_1},\\
k &= \frac{1}{2^{1/3}}\Bigg[ B + 2^{3/4} H_1^{3/2} \\ &\qquad\qquad + \sqrt{\left(B+2^{3/4} H_1^{3/2}\right)^2 + \frac{4}{27}\tscal^3}\Bigg]^{1/3}.
\end{split}
\label{eq_sde_gc_weak_Tc_coeff}\eeq 
Upon approaching $\tscal=B=0$, as well as independently of the values of $\tscal$ and $B$ for $H_1\to\infty$, one has $a_0\to  0$; accordingly, the characteristic behavior $m(-1/2+\hat\zeta)\propto  1/\hat\zeta$ of the profile in MFT is recovered from Eq.~\eqref{eq_sde_gc_weak_Tc} for $H_1\gg 1$.
The strong adsorption regime is approached asymptotically for large $H_1$ (such that $\tscal/\sqrt{H_1}\ll 1$ and $B/H_1\ll 1$) according to
\beq a_0\simeq  - \frac{\tscal}{3 \times 2^{1/4} \sqrt{ H_1}} + \frac{B}{3\sqrt{2} H_1} +O\left(H_1^{-2}\right).
\label{eq_sde_gc_weak_Tc_coeff_asymp}\eeq 
In the same limit of large $H_1$, at the wall $\hat\zeta=0$, the approximate expression in Eq.~\eqref{eq_sde_gc_weak_Tc} turns into
\beq  m(\zeta =-1/2) \simeq 2^{1/4} \sqrt{H_1} - \frac{\tscal}{3\times 2^{1/4} \sqrt{H_1}} + \frac{B}{3\sqrt{2} H_1}. 
\label{eq_sde_gc_weak_OPwall}\eeq 
We remark that, if terms of $O(\hat \zeta)$ or higher are included in the ansatz in Eq.~\eqref{eq_sde_gc_weak_Tc}, these, in contrast to $a_0$, do not vanish in the limit $H_1\to\infty$, but constitute corrections to the leading  behavior. The SDE which emerges in this limit is constructed below.
Furthermore, although the ansatz in Eq.~\eqref{eq_sde_gc_weak_Tc} solves the ELE up to an error of $O((B\sqrt{H_1}+\tscal H_1)\hat\zeta)$, this does not imply that the value of $ m$ at the wall ($\hat\zeta=0$) is predicted \emph{exactly} by Eq.~\eqref{eq_sde_gc_weak_Tc}. 
The reason is that including higher-order terms in $\hat\zeta$ in the ansatz in Eq.~\eqref{eq_sde_gc_weak_Tc} (not only in the case $\tscal=B=0$ for which $a_0=0$) leads to a coupling between their coefficients, affecting in particular also $a_0$ (and thus the dependence on $\tscal$ and $B$). 
We thus conclude that, for \emph{finite} $H_1$, the distant wall affects the SDE in general already at $O(\hat\zeta^0)$. 
This is also expected from the fact \cite{ascher_numerical_1995} that the boundary value problem described by Eqs.~\eqref{eq_ELE} and \eqref{eq_ELE_BC} can be equivalently represented by an initial value problem, in which $m'(\zeta=-1/2)=-H_1^-$ is given and the value $m(\zeta=-1/2)$ is a free parameter which must be determined such that the imposed boundary condition for $m'$ at the distant wall ($\zeta=1/2$) is obtained at the end of the integration. This implies a dependence of $m(\zeta=-1/2)$ on $m'(\zeta=1/2)$, i.e., on the properties of the distant wall.

\begin{figure}[t]\centering
	\includegraphics[width=0.8\linewidth]{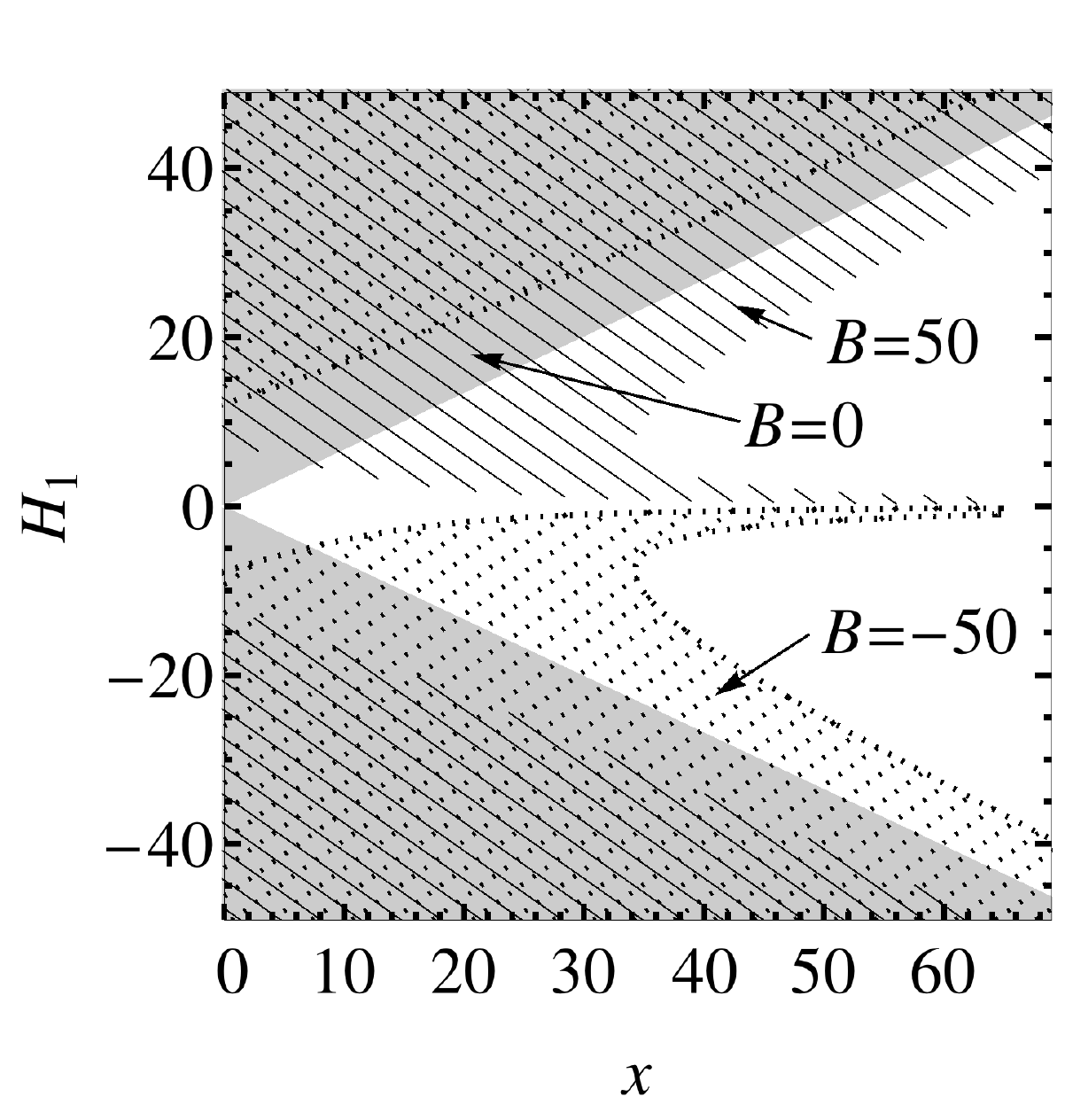}
	\caption{Ranges of the parameters $H_1$ and $\tscal$ within which the SDE in Eq.~\eqref{eq_sde_gc_weak_Tc} provides an accurate solution of the ELE in Eq.\ \eqref{eq_ELE} close to the walls. The shaded and hatched areas represent, for three different values of the bulk field $B$ (indicated by the corresponding labels), the regions where $a_0$ [Eq.~\eqref{eq_sde_gc_weak_Tc_coeff}] is less than an arbitrarily chosen factor of 1/3 of the leading term $m(-1/2)$ of the OP at the wall given by Eq.~\eqref{eq_sde_gc_weak_Tc_0} for $\hat\zeta=0$, i.e., $|a_0|< 2^{1/4} \sqrt{H_1}/3$. The shaded areas correspond to $B=0$, the hatched areas with full lines correspond to $B=50$, and the hatched areas with broken lines correspond to $B=-50$.}
	\label{fig_sde_valid}
\end{figure}

The regime in which the SDE in Eq.~\eqref{eq_sde_gc_weak_Tc} provides a reliable approximation of the solution of the full ELE [Eq.~\eqref{eq_ELE}] close to the walls can be self-consistently estimated by requiring the perturbative correction $a_0$ to remain smaller than the dominant term given by Eq.~\eqref{eq_sde_gc_weak_Tc_0}. 
In Fig.~\ref{fig_sde_valid}, this condition is graphically evaluated for $\hat\zeta=0$.
A detailed calculation reveals furthermore that, asymptotically for large $H_1$, the condition reduces to
\beq \left| \frac{B}{2^{1/4}H_1^{3/2}}-\frac{\tscal}{H_1} \right| \ll 1.
\label{eq_sde_gc_weak_valid}\eeq 
Consistently with Fig.~\ref{fig_sde_valid}, this implies that, for fixed $B$ and sufficiently large $H_1$, the SDE is only valid for $\tscal\ll H_1$.
The comparison with Fig.~\ref{fig_linMFT_valid} shows, in particular, that the SDE is not reliable deep in the domain of validity of linear MFT. This is expected, because, in contrast to the solution of linear MFT, the SDE [Eq.~\eqref{eq_sde_gc_weak_Tc_0}] is constructed as a solution of the nonlinear ELE which becomes accurate at criticality and sufficiently close to one wall, accounting for the effect of nonzero $\tscal$ or $B$ via small corrections.

For $H_1=\infty$, Eq.~\eqref{eq_sde_gc_weak_Tc_coeff_asymp} implies $a_0=0$ and the leading dependencies of $ m_*\equiv  m|_{H_1\to\infty}$ on $\tscal$ and $B$ can be incorporated by higher order polynomial terms of the dimensionless distance $\hat\zeta$ from the near wall:
\beq  m_*(-1/2+\hat\zeta) = \frac{\sqrt{2}}{\hat\zeta} + \sum_{i=1}^\infty a_i \hat\zeta^i. 
\label{eq_sde_gc_strong_ansatz}
\eeq  
The coefficients $a_i$ are fixed by inserting this ansatz into the ELE in Eq.\ \eqref{eq_ELE}, expanding the result in terms of $\hat\zeta$ and requiring, by setting the corresponding coefficients to zero, that the ELE is fulfilled up to a certain order in $\hat\zeta$.
In general, a term $\propto \hat\zeta^i$, $i\geq 1$, in the ansatz in Eq.~\eqref{eq_sde_gc_strong_ansatz} produces, at leading order, a contribution $\propto \hat\zeta^{i-2}$ in the ELE. The term $a_3 \hat\zeta^3$ is exceptional because, when inserted into the ELE together with the term $\sqrt{2}/\hat\zeta$, it appears at $O(\hat\zeta^3)$ instead at $O(\hat\zeta)$.
We find that, by this procedure, the coefficients $a_1$ and $a_2$ of the ansatz in Eq.~\eqref{eq_sde_gc_strong_ansatz} are determined uniquely and the ELE is satisfied up to an error of $O(\hat\zeta^2)$ or higher.
We remark that, in this way, one would also obtain $a_0=0$ in the case that a term $a_0 \hat\zeta^0$ is added to the right hand side of Eq.~\eqref{eq_sde_gc_strong_ansatz} instead of invoking the limit $H_1\to \infty$ in Eq.~\eqref{eq_sde_gc_weak_Tc_coeff_asymp} beforehand.
In summary, in the case $H_1=\infty$, we obtain the SDE
\beq  m_*\left(-1/2+\hat\zeta,\tscal,B\right) = \sqrt{2}\left[ \frac{1}{\hat\zeta} - \frac{1}{6} \tscal \hat\zeta + \frac{1}{4\sqrt{2}} B\hat\zeta^2 \right],
\label{eq_sde_gc_strong}\eeq
which, in terms of the original dimensional quantities, reads
\beq \phi_*\left(-L/2+\hat z,\tau,\mu\right) = \sqrt{\frac{12}{g}}\left[\frac{1}{\hat z} - \frac{1}{6} \tau \hat z  + \frac{\sqrt{g}}{8\sqrt{3}} \mu \hat z^2 \right].
\eeq
In order to uniquely determine the coefficients $a_i$ for $i\geq 3$, the boundary condition at the distant wall would have to be considered as well. 
However, here we do not aim at the full power-series solution of the OP profile. Instead, we are content with the result in Eq.~\eqref{eq_sde_gc_strong}, which represents the dominant contribution to the OP profile in the film near one wall in the limit $H_1\to\infty$.
Note that the expression in Eq.~\eqref{eq_sde_gc_strong} coincides with the corresponding SDE for the semi-infinite geometry up to $O(\hat\zeta^2)$ \cite{schlesener_critical_2003}.
This is consistent with Eq.~\eqref{eq_SDE_distwallcorr}, which predicts [since $d^*=4$ in MFT and $m_{\infty/2}(\hat\zeta\to 0)\simeq m_*(\zeta\to 0)\sim 1/\hat\zeta$] the distant wall correction to affect the SDE in the strong adsorption regime only at $O(\hat \zeta^3)$ or higher.
We thus conclude that, while the solution of Eq.~\eqref{eq_ELE} requires the boundary conditions [Eq.~\eqref{eq_ELE_BC}] at both walls, up to $O(\hat\zeta^2)$ the SDE for $|H_1|=\infty$ reflects only the boundary condition at the near wall. The second boundary condition enters into the solution at $O(\hat\zeta^3)$.

\subsubsection{Canonical ensemble}
\label{sec_sde_canonical}
For $(++)$ conditions with $|H_1|=\infty$, the divergence $\propto 1/\hat \zeta$ of the mean field OP upon approaching the wall, as implied by the SDE [Eq.~\eqref{eq_sde_gc_strong}], is not integrable and therefore violates the constraint of a fixed and finite total amount of mass (per area) in the system. 
In fact, as will be demonstrated below, the constraint-induced bulk field $\tilde B$ diverges logarithmically with $|H_1|\to\infty$.
Accordingly, within MFT of the canonical ensemble, $H_1$ must necessarily be kept finite in this case and the SDE must formally start with a constant term, as given by Eqs.~\eqref{eq_sde_gc_weak_Tc} and \eqref{eq_sde_gc_weak_Tc_coeff}.
The SDE can be understood as a local approximation of the actual OP profile obtained for fixed $\tilde B=\tilde B(\Mass)$, which is asymptotically accurate upon approaching each single wall. 
The field $\tilde B$, however, is determined by imposing the mass constraint, which therefore requires the knowledge of the whole profile. For instance, this can be obtained numerically, as discussed below.
In the canonical case, the value of the OP at the wall thus depends on the total mass $\Mass$, which is a global property. 
Despite these restrictions, the result in Eq.~\eqref{eq_sde_gc_weak_Tc} together with Eq.~\eqref{eq_sde_gc_weak_Tc_coeff_asymp} demonstrates that also in the canonical case the mean field OP profile approaches the characteristic behavior $\propto 1/\hat\zeta$ near the wall for sufficiently large $H_1$. 
For $(+-)$ boundary conditions the limit $|H_1|\to \infty$ is well defined since the diverging contributions to the total mass from the profile at the two walls asymptotically cancel [see Eq.~\eqref{eq_sde_gc_strong}], at least as long the surface fields are taken to be of equal strength.

\subsection{Numerical results: MFT}
\label{sec_ads_num}

Numerical results for the nonlinear MFT are obtained by directly solving the associated ELE in Eqs.\ \eqref{eq_ELE} and \eqref{eq_ELE_BC}, as well as by explicit minimization of the free-energy functional in Eq.~\eqref{eq_Landau_func_ndim} via the conjugate-gradient method. We find the latter approach to be slightly more robust if bulk or surface fields are strong. We have checked in a number of cases that the results provided by both methods agree. 
In the film geometry, the critical point is generally shifted from its bulk value $(\tscal_{c},B_{c})=(0,0)$ to $(\tscal_{c,\text{cap}},B_{c,\text{cap}})$ with $\tscal_{c,\text{cap}}<0$ and $B_{c,\text{cap}}<0$ [for $(++)$ boundary conditions] or $B_{c,\text{cap}}=0$ [for $(+-)$ boundary conditions] \cite{nakanishi_critical_1983,parry_novel_1992}. 
For $(++)$ boundary conditions and temperatures below the capillary critical point, the film undergoes a first-order ``capillary'' phase transition between two OP profiles corresponding to two competing free energy minima \cite{nakanishi_critical_1983}.
Similarly to two-phase coexistence in the bulk, the transition occurs upon crossing the capillary condensation line $B\st{cap}(\tscal)$ such that one of the two possible profiles is stable for $B$ infinitesimally above or below $B\st{cap}$.
For $(+-)$ boundary conditions and $T<T_c$ the $+/-$ interface undergoes a transition between a configuration localized near one of the two walls and a delocalized configuration positioned in the middle of the film \cite{parry_influence_1990}.
However, we focus in the following mostly on the region above the capillary critical point, where the film necessarily remains homogeneous in the lateral directions.

\begin{figure*}[t]
	\subfigure[]{\includegraphics[width=0.456\linewidth]{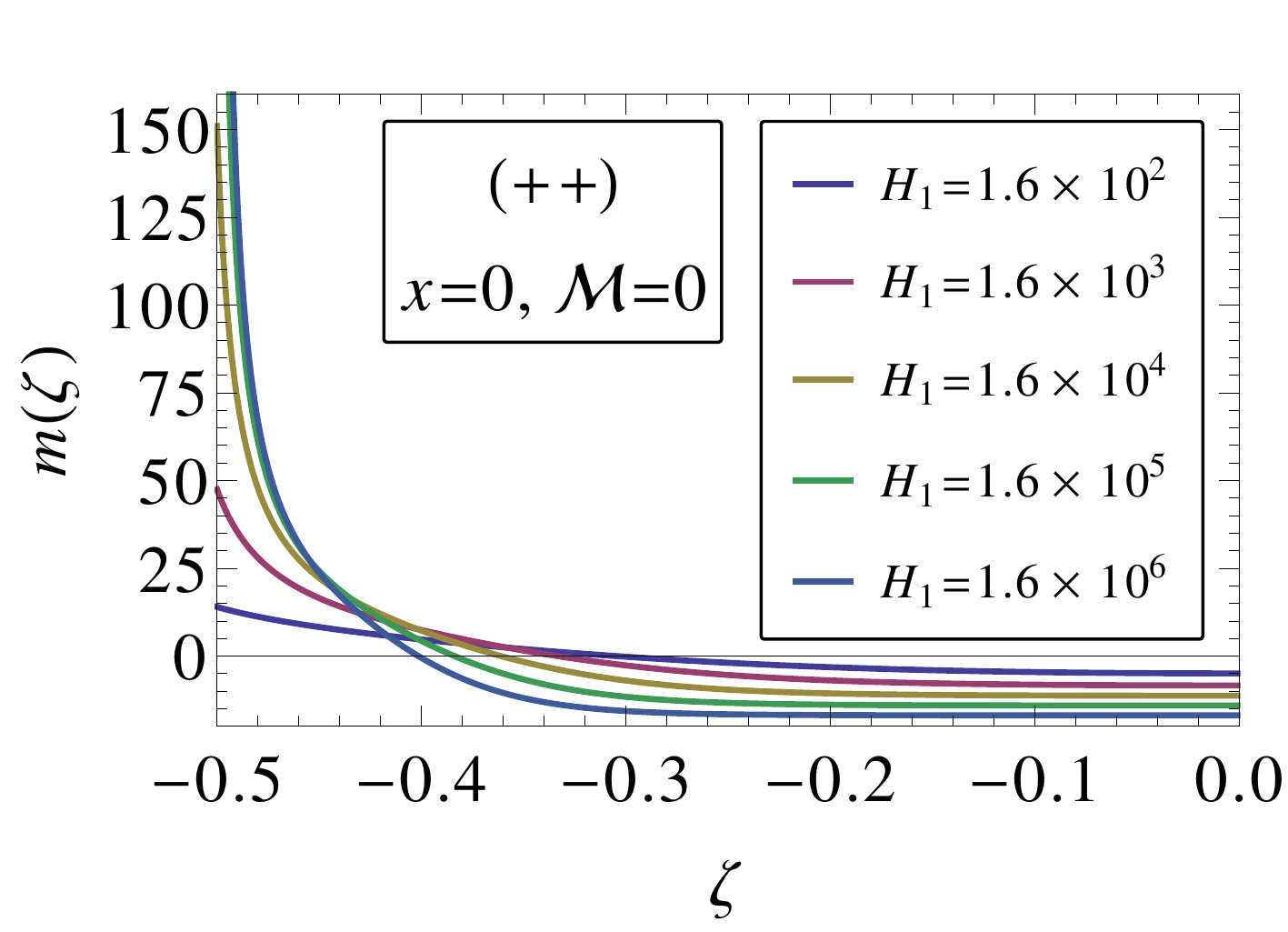}}\qquad
   	\subfigure[]{\includegraphics[width=0.44\linewidth]{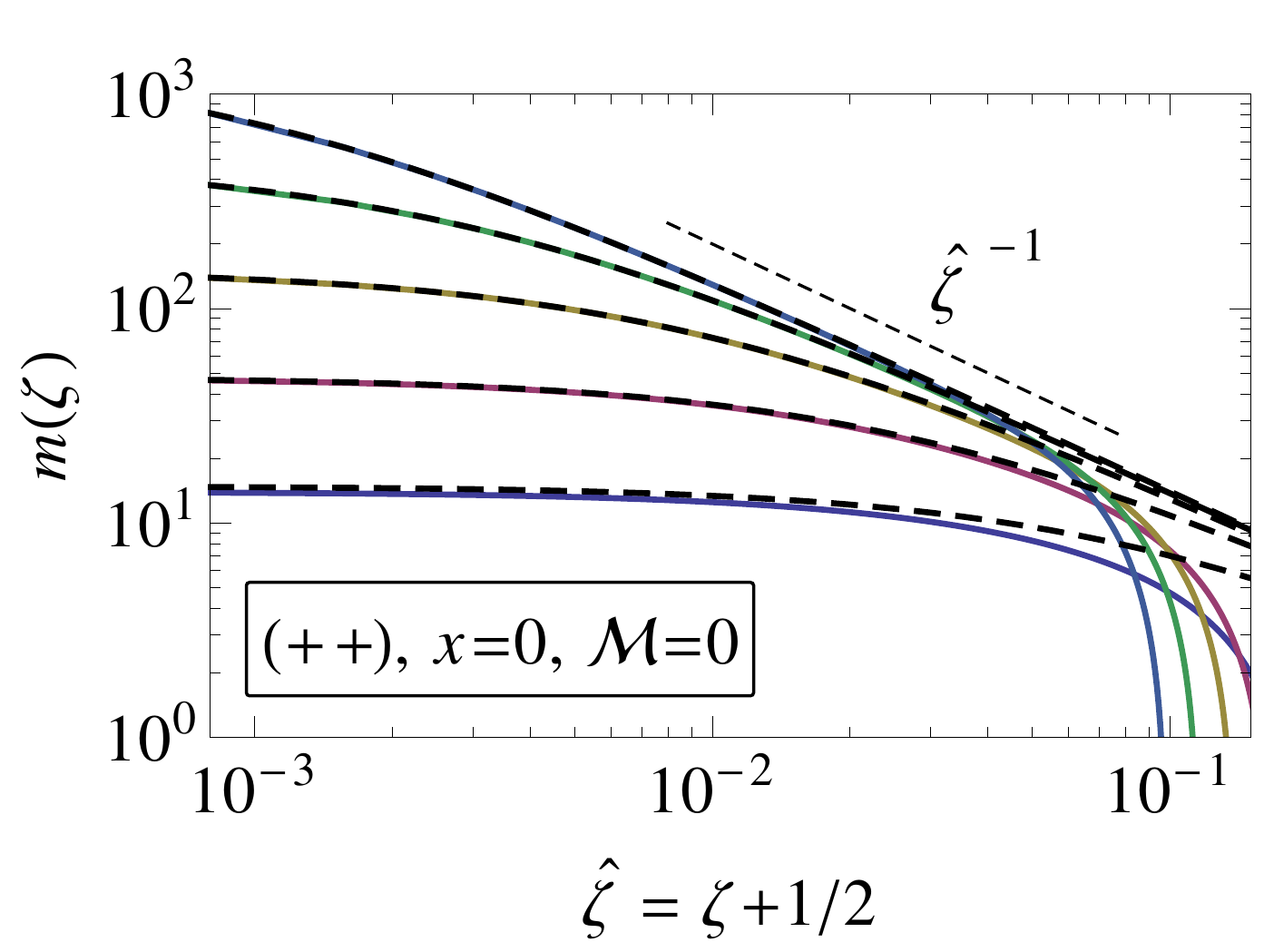}}
	\subfigure[]{\includegraphics[width=0.45\linewidth]{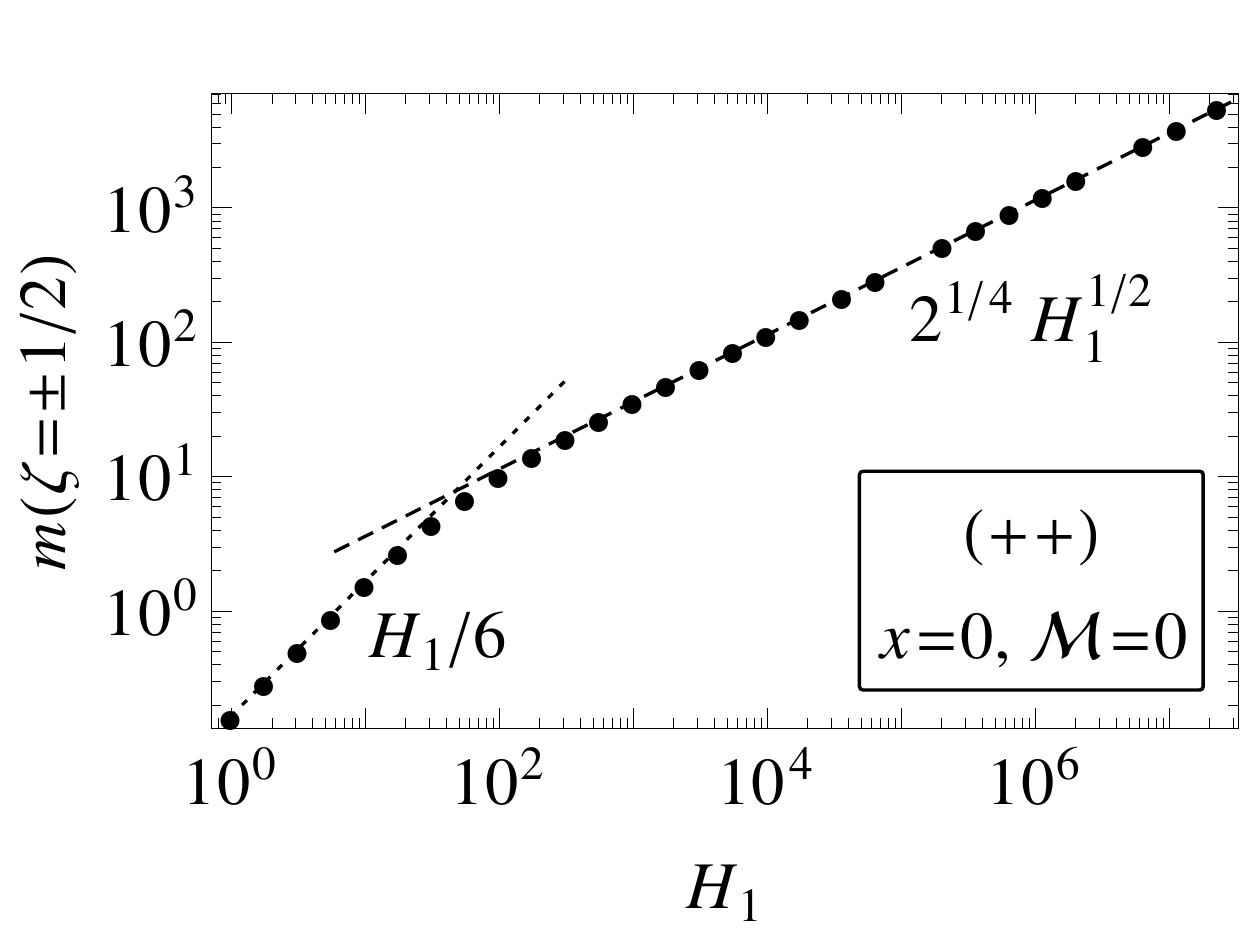}}\qquad
   	\subfigure[]{\includegraphics[width=0.455\linewidth]{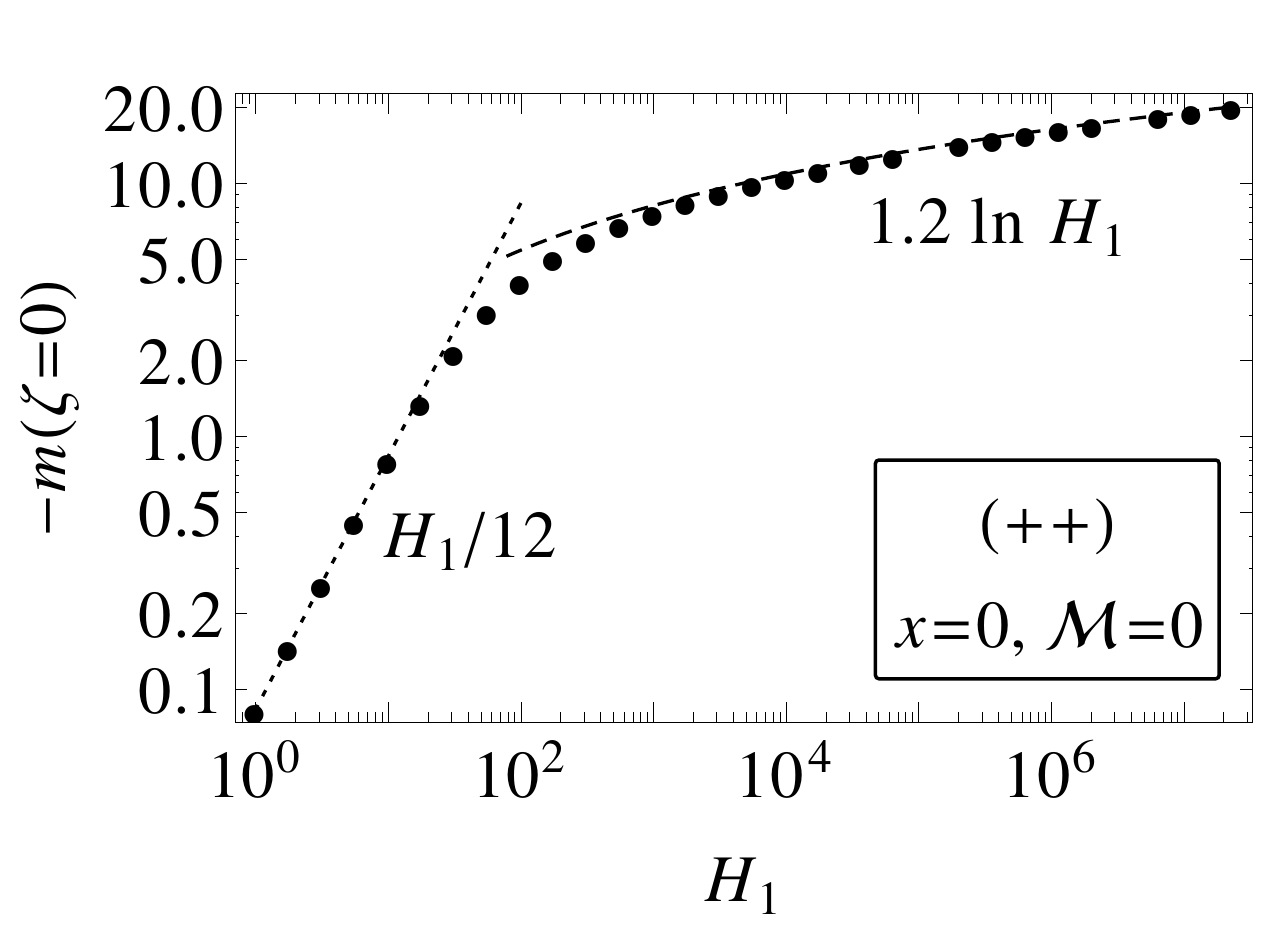}}
    \caption{(a) OP profiles across a film with $(++)$ boundary conditions fulfilling $\Mass=0$, as obtained from the numerical solution of the ELE in Eq.\ \eqref{eq_ELE} for $\tscal=0$ and various values of $H_1$ ranging from $H_1=1.6\times 10^2$ (center top) to $H_1=1.6\times 10^6$ (center bottom). In order to have $\Mass=0$, the bulk field $B$ is fixed, correspondingly, at $B=2.0\times 10^2, 6.4\times 10^2, 1.5\times 10^3, 2.9\times 10^3, 4.9\times 10^3$. The confining walls are located at $\zeta=\pm  1/2$. Due to the spatial symmetry of the problem, the profile is plotted only across the left half of the film. (b) Short-distance behavior of the profiles in (a) (solid lines), compared with the theoretical prediction of the SDE [Eqs.~\eqref{eq_sde_gc_weak_Tc} and \eqref{eq_sde_gc_weak_Tc_coeff}, dashed lines, same color code as in (a)]. (c) Value of the OP at the boundary ($\zeta=\pm 1/2$) as a function of $H_1$. The numerical results [$\bullet$] agree well with the theoretical predictions of the SDE [Eq.~\eqref{eq_sde_gc_weak_OPwall}, dashed line]. (d) Value of the OP profile at the center of the film ($\zeta=0$) as a function of $H_1$. The logarithmic dependence on $H_1$ is predicted theoretically for large $H_1$ by Eq.~\eqref{eq_OPscale_largeH1} (dashed line, with the proportionality constant $c\approx 1.2$). The dotted lines in (c) and (d) represent the prediction of Eq.~\eqref{eq_OPscale_smallH1}, which is accurately recovered for small $H_1$.}
    \label{fig_prof_H1range}
\end{figure*}

\begin{figure*}[t]\centering
    \subfigure[]{\includegraphics[width=0.48\linewidth]{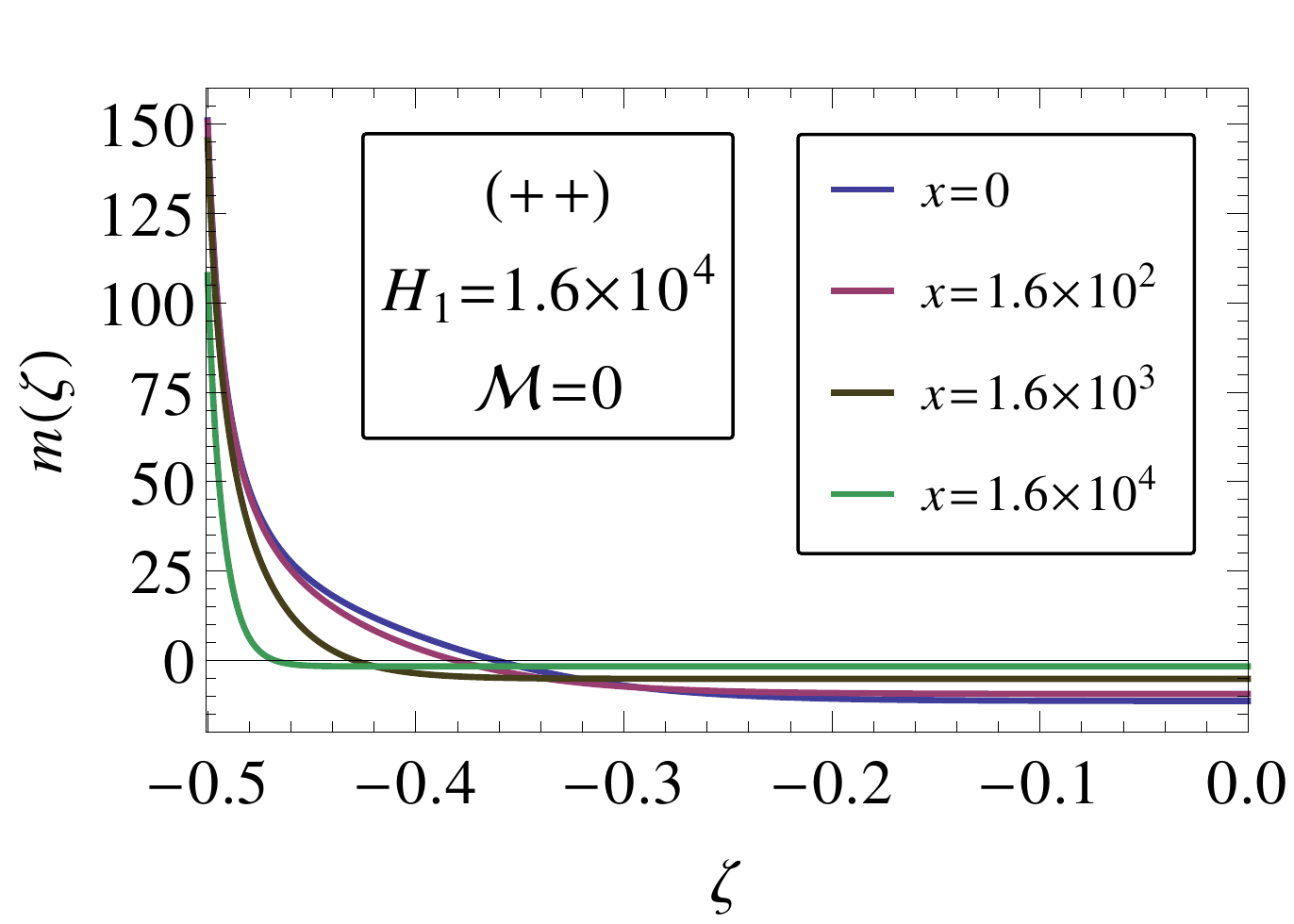}}\qquad
    \subfigure[]{\includegraphics[width=0.48\linewidth]{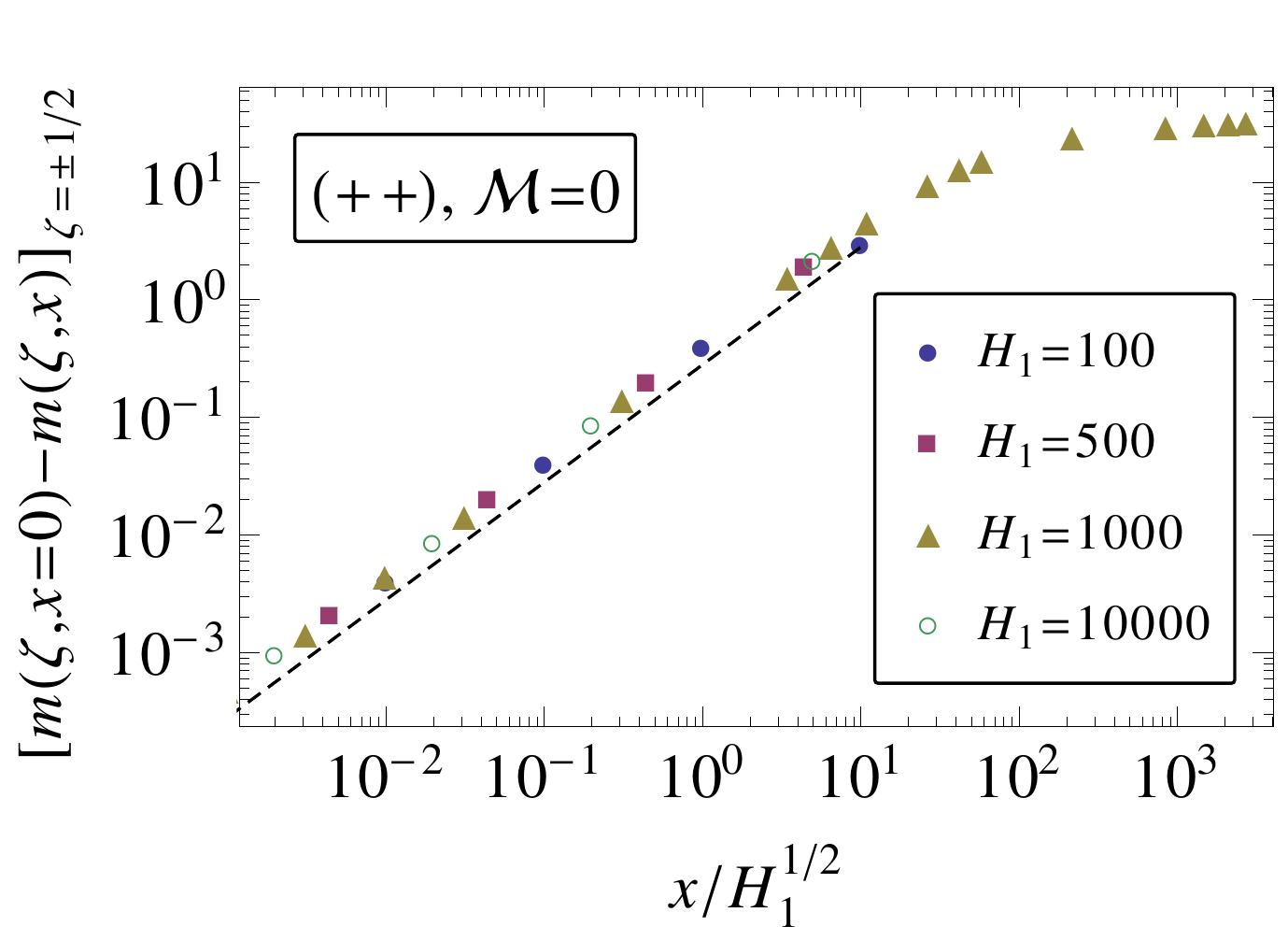}}
    \caption{(a) OP profiles across a film with $(++)$ boundary conditions and fulfilling $\Mass=0$, as obtained from the numerical solution of the ELE in Eqs.\ \eqref{eq_ELE} and \eqref{eq_ELE_BC} for $H_1=1.6\times 10^4$ and various values of $\tscal$ ranging from $\tscal=1.6\times 10^4$ (center top) to $\tscal=0$ (center bottom). In order to have $\Mass=0$, the bulk field $B$ is fixed, correspondingly, at $B=2.6\times 10^4, 8.4\times 10^3, 2.3\times 10^3, 1.5\times 10^3$. Due to the symmetry of the problem, the profile is plotted only for the left half of the film. (b) Approach of the OP at the wall ($\zeta=\pm 1/2$) to its value at bulk criticality ($\tscal=0$) in the supercritical regime. The figure shows the difference between $m(\zeta,\tscal=0)$ and $m(\zeta,\tscal)$ at $\zeta=\pm 1/2$ as a function of $\tscal/\sqrt{H_1}$ for various values of $\tscal$ and $H_1$ (symbols). The dashed line represents the expression $\tscal/(3\times 2^{1/4}\sqrt{H_1})$, which is a prediction for $\left[m(\zeta,\tscal=0)-m(\zeta,\tscal)\right]_{\zeta=\pm 1/2}\geq 0$ as obtained from the SDE in the limit of large $H_1$ [see Eq.~\eqref{eq_sde_gc_weak_OPwall} and the discussion in the main text].}
    \label{fig_prof_Trange}
\end{figure*}

\begin{figure}[t]\centering
    \includegraphics[width=0.96\linewidth]{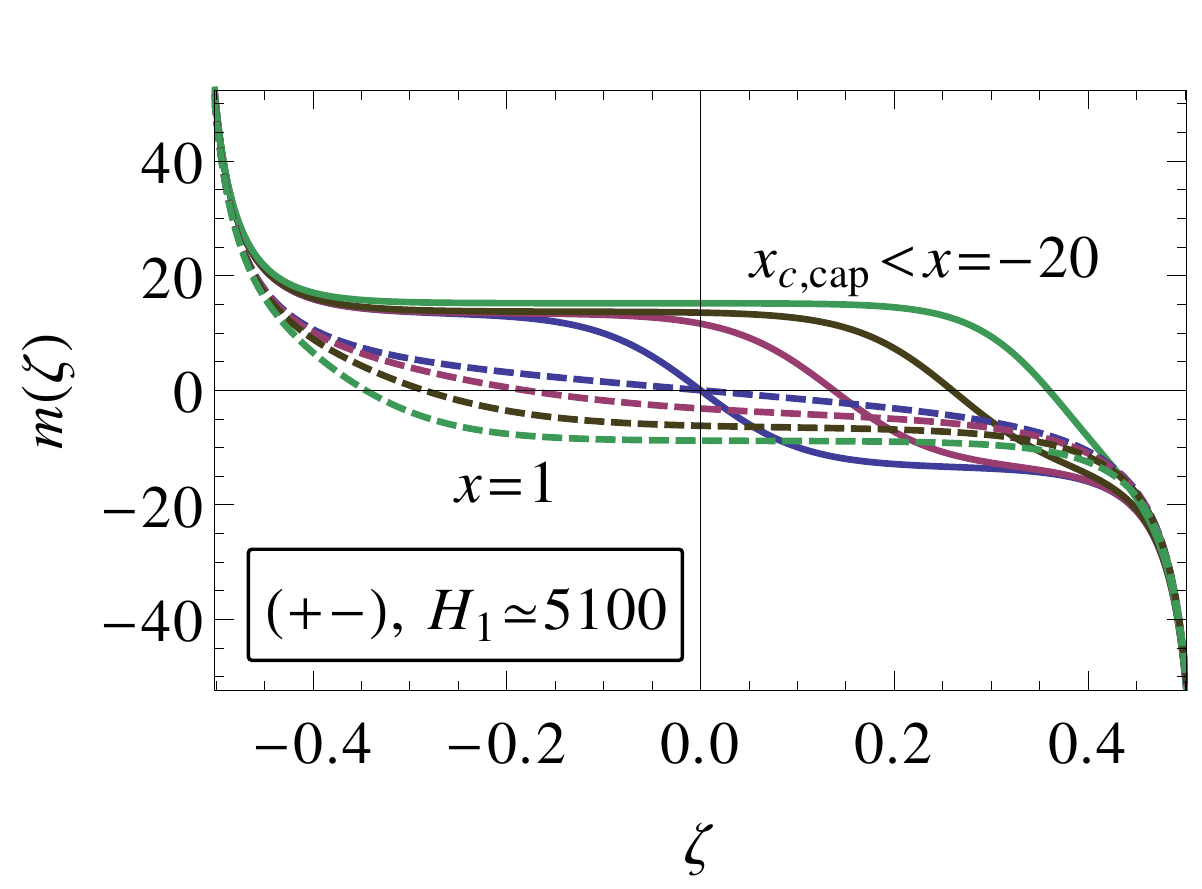}
    \caption{Typical behavior of the OP profiles across a film with $(+-)$ boundary conditions. Profiles shown by solid lines are obtained for $\tscal=-20$ and four bulk fields $B=0, 11, 110, 770$ (from bottom to top), while the dashed lines are obtained for $\tscal=1$ and bulk fields $B=0,-110,-330,-770$ (from top to bottom). The values of the mass corresponding to these bulk fields are $\Mass=0,3.7, 7.1, 11.0$ for $\tscal=-20$ and $\Mass=0, -1.7, -3.8, -6.2$ for $\tscal=1$, respectively. The walls are located at $\zeta=\pm 1/2$ and exert opposing surface fields of magnitude $H_1\simeq 5100$.}
    \label{fig_prof_pm_sample}
\end{figure}

\subsubsection{OP profiles}
\label{sec_OPprof_num}

We first discuss the OP profiles for $(++)$ boundary conditions and $\Mass=0$.
Figure \ref{fig_prof_H1range}(a) shows typical OP profiles across a film at bulk criticality ($\tscal=\Mass=0$) for various strengths of the surface field $H_1$, as obtained numerically from the ELE in Eqs.~\eqref{eq_ELE} and \eqref{eq_ELE_BC}.
We observe that, for large $H_1$, the profile varies most strongly near the boundaries ($\zeta=\pm 1/2$), while it is practically constant in the center of the film ($\zeta= 0$).
As seen in Fig.~\ref{fig_prof_H1range}(b), the SDE in Eq.~\eqref{eq_sde_gc_weak_Tc} accurately captures the profile of the OP near the boundary, including the characteristic $1/\hat\zeta$-behavior expected for large but finite $H_1$ at intermediate distances from the wall.
For finite $H_1$, the effective power law $\propto 1/\hat \zeta$ always crosses over towards a constant upon approaching $\hat\zeta=0$ [see Eq.~\eqref{eq_sde_gc_weak_Tc}].

The OP at the boundary and at the center of the film exhibits characteristic scaling behaviors both in the weak and strong surface adsorption regime.
For $\Mass=0$ and sufficiently \emph{weak} surface fields one can approximate the OP profile by the constrained solution in Eq.~\eqref{eq_m0_constr} of the linear ELE [Sec.~\ref{sec_ads_pert_summary}].
In particular, for $\tscal=0$, Eq.~\eqref{eq_m1_tau0} implies that linear MFT is valid for $|H_1|\ll 100$.
From Eq.~\eqref{eq_m0_tau0_lim}, we then infer the behavior of the OP as 
\begin{equation}
 m(\pm 1/2)\simeq H_1/6,\qquad m(0)\simeq -H_1/12
\label{eq_OPscale_smallH1}\end{equation}
$\text{for }|H_1|\ll 100\text{ and } \Mass=0$ at the wall and in the center of the film, respectively, in agreement with Figs.~\ref{fig_prof_H1range}(c) and (d).
For \emph{strong} surface fields $H_1>0$, the behavior of the OP at the wall follows from the SDE [Sec.~\ref{sec_sde}], which predicts [see Eq.~\eqref{eq_sde_gc_weak_OPwall}] for $\tscal=0$ and asymptotically for $H_1\gg B$ (which is fulfilled in the present case, see below): $m(\pm 1/2)\simeq 2^{1/4} \sqrt{H_1}$, in agreement with Fig.~\ref{fig_prof_H1range}(c).
In order to rationalize the behavior for large $H_1$ observed at the center of the film, we recall that the profile varies most strongly close to the boundaries [see Fig.~\ref{fig_prof_H1range}(a)], while the central part is approximately spatially constant so that its contribution to $\Mass$ is proportional to $m(0)$. 
This allows one to estimate the dependence of the total amount of mass $\Mass$ on $H_1$ by integrating the profile over a small interval from $\zeta=-1/2$ to a certain position $\zeta=-1/2+\Delta\zeta$ (with $0<\Delta\zeta < 1/2$) which, upon making use of Eq.~\eqref{eq_sde_gc_weak_Tc} (with $a_0\simeq 0$ due to $H_1\gg 1$), yields
\beq \Delta \Mass \sim \ln (\Delta\zeta\times H_1).
\label{eq_sde_int_largeH1}\eeq 
In order to keep $\Mass=0$, the profile in the film center must thus behave as
\beq  m(0)\sim -\Delta \Mass \simeq -s \ln H_1,
\label{eq_OPscale_largeH1}\eeq
in agreement with the numerical results shown in Fig.~\ref{fig_prof_H1range}(d) for $H_1\gg 100$, with a numerical prefactor $s\simeq 1.2$ as determined from a fit.

\begin{figure*}[t]\centering
	\subfigure[]{\includegraphics[width=0.45\linewidth]{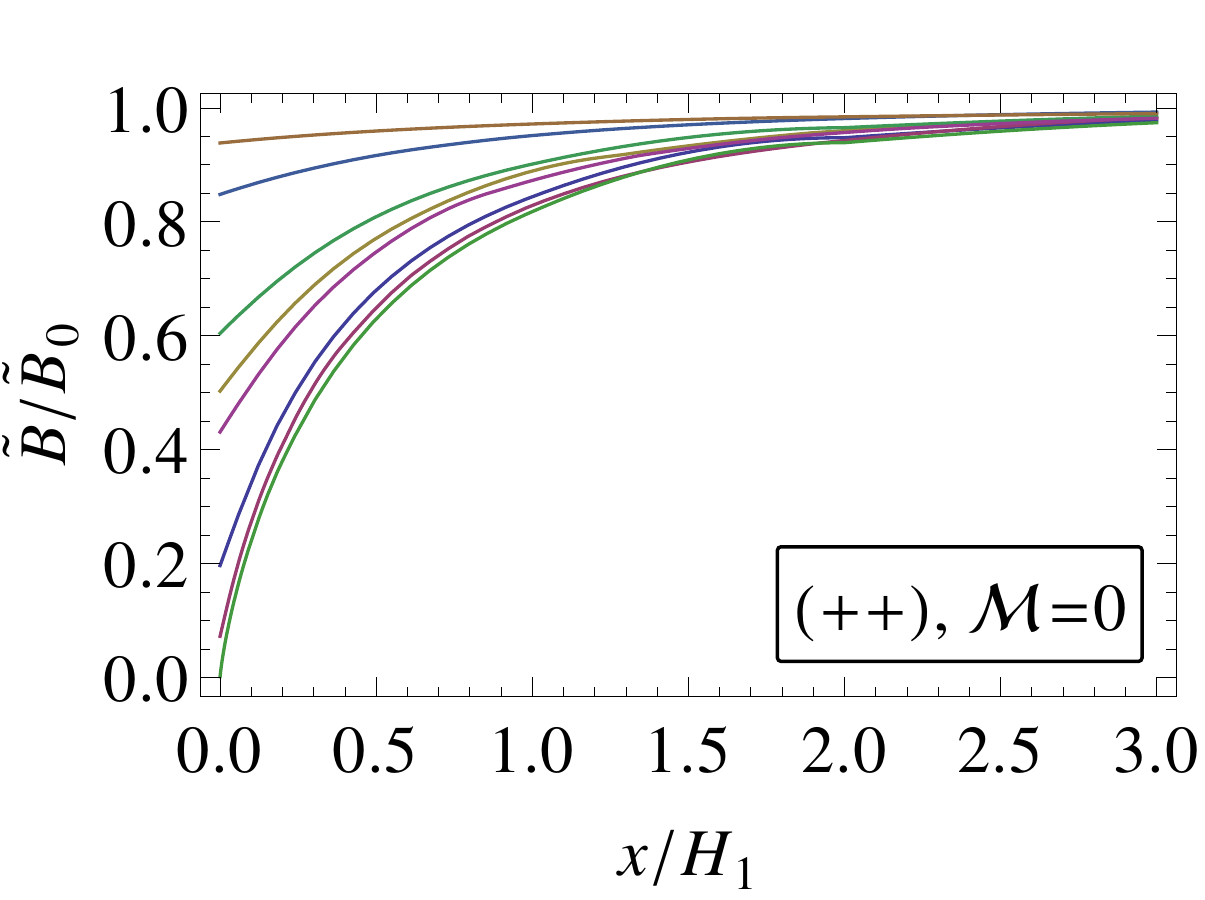}}\qquad
	\subfigure[]{\includegraphics[width=0.45\linewidth]{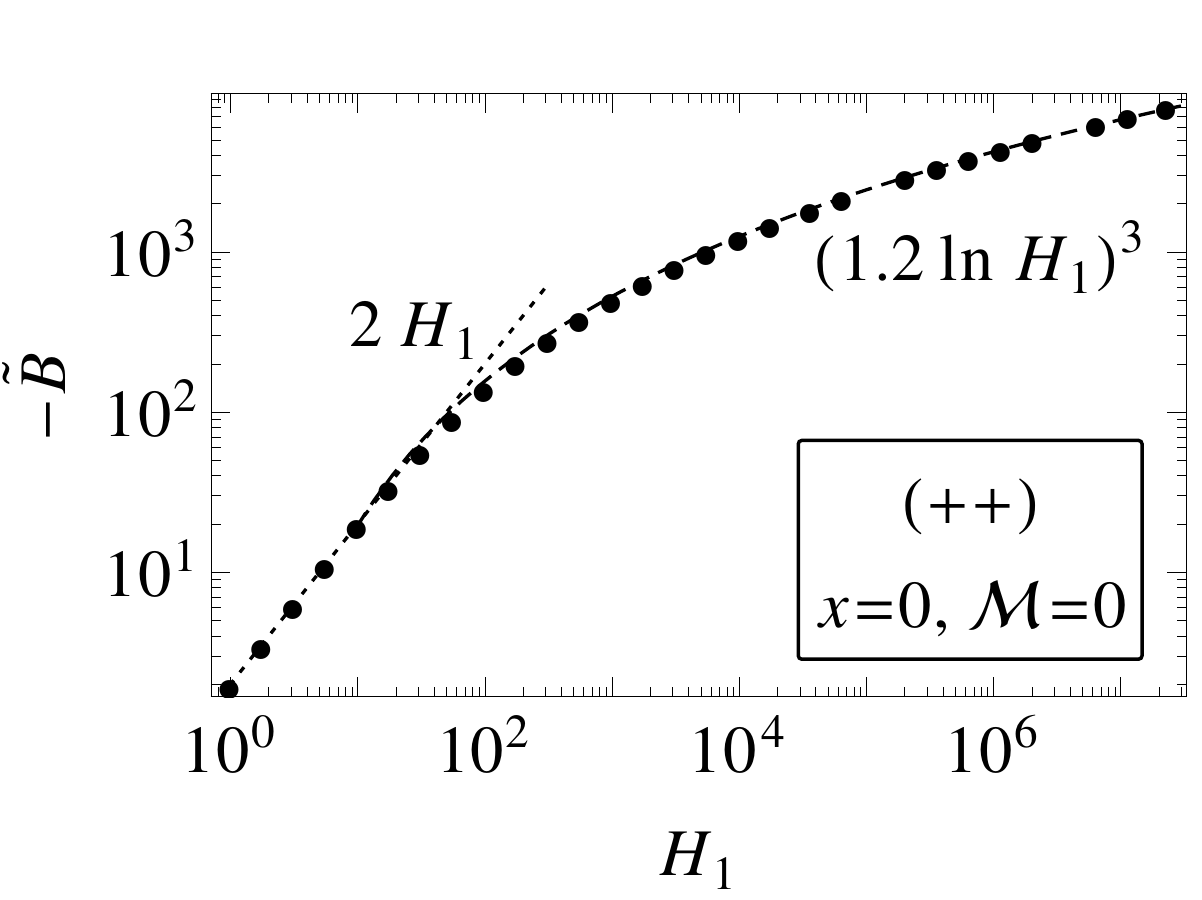}}
	\caption{Bulk chemical potential $\tilde B$ numerically determined in such a way that the constraint $\Mass=0$ of vanishing mass is fulfilled in a film with $(++)$ boundary conditions. (a) Dependence of $\tilde B$ on the ratio between the scaled temperature and the scaled surface field $\tscal/H_1$ [$=1/\sqrt{\sigma}$, see Eq.~\eqref{eq_smallness_param}], normalized by its limiting value $\tilde B_0$ obtained for $\tscal\to\infty$ [Eq.~\eqref{eq_B0}], for surface fields $H_1=25, 52, 160, 260, 370, 1600, 8200, 1.6\times 10^6$ (from top to bottom). (b) Dependence of $\tilde B$ on the surface field $H_1$ for $\tscal=0$. For small $H_1$, $\tilde B$ calculated numerically (symbols) approaches the prediction given by Eq.~\eqref{eq_Bconstr_smallH1} (dotted line). For large $H_1$, instead, $\tilde B\simeq -(s\ln H_1)^3$ [dashed line, Eq.~\eqref{eq_Bconstr_largeH1}], with a constant $s\simeq 1.2$ [see Eq.~\eqref{eq_OPscale_largeH1}].} 
	\label{fig_bulkfield_constr}
\end{figure*}

Figure \ref{fig_prof_Trange}(a) displays OP profiles for various values of the rescaled temperature $\tscal$ in the supercritical regime $\tscal>0$. As in the case $\tscal=0$ illustrated in Fig.~\ref{fig_prof_H1range}(a), these profiles are obtained via a numerical solution of the ELE in Eqs.~\eqref{eq_ELE} and \eqref{eq_ELE_BC} with $\Mass=0$.
Upon decreasing $\tscal$ from large values, the OP value at the wall (at the center of the film) gradually increases (decreases) and the spatial variation of the profiles becomes more pronounced. 
This behavior is expected because, correspondingly, the bulk correlation length increases and so does the distance from the walls at which the effect of the boundaries is present.
For large $\tscal$, the behavior of the OP at the boundaries and at the center follows the linear MFT predictions given in Eq.~\eqref{eq_m0_constr_largeTau}, independently of $H_1$ (not shown).
We find from the present numerical solution that, for $\tscal \ll \sqrt{H_1}$ and $\tilde B\ll H_1$, the bulk field $\tilde B$ required to keep $\Mass=0$ does not vary anymore significantly with $\tscal$ upon approaching bulk criticality $\tscal=0$. Under these circumstances, the SDE predicts, according to Eq.~\eqref{eq_sde_gc_weak_OPwall}, the OP at the wall to approach its value for $\tscal=0$ as $\left[m(\zeta,\tscal=0)-m(\zeta,\tscal)\right]_{\zeta=\pm 1/2}\simeq \tscal/(3\times 2^{1/4}\sqrt{H_1})$. 
As shown in Fig.~\ref{fig_prof_Trange}(b), the numerical solution of the ELE follows this scaling behavior for sufficiently small values of $\tscal/\sqrt{H_1}$. The slight difference in the overall magnitude between the data and the analytical prediction reflects the influence of the distant wall, which is not accounted for in the SDE [see the discussion in Sec.~\ref{sec_sde}; we emphasize that the actual values $m(\zeta=\pm 1/2,\tscal\simeq 0)$ of the numerically obtained profile differ by less than one percent from the corresponding prediction of the SDE in Eq.~\eqref{eq_sde_gc_weak_OPwall}; see also Fig.~\ref{fig_prof_H1range}(b)].
While the above results pertain to $\Mass=0$, we expect similar scaling behaviors to hold, at least asymptotically for $|H_1|\to\infty$, for any nonzero value of $\Mass$.
This is so because the scaling properties of the profile for $H_1\gg 1$ are controlled by the SDE in Eq.~\eqref{eq_sde_gc_weak_Tc}, which, in this limit, is dominated by its first term. This term, however, is independent of $\Mass$ (or, correspondingly, the associated bulk field $B$).

For $(+-)$ boundary conditions, the constraint $\Mass=0$ is realized for $B=0$ for all temperatures above the capillary critical temperature $\tscal_{c,\text{cap}}$, while $B\neq 0$ yields $\Mass\neq 0$. 
As a rather large value of $H_1$ is used to obtain the numerical solution in the present case, $\tscal_{c,\text{cap}}$ is located well below the bulk critical point \cite{parry_influence_1990,parry_novel_1992} and is not covered by the present results.
Figure~\ref{fig_prof_pm_sample} shows the OP profiles obtained numerically, which are representative of systems with temperatures above ($\tscal=1$) or below ($\tscal=-20$), respectively, the bulk critical point with various values of the bulk chemical potential and thus of $\Mass$.
The observed qualitative behavior of the profiles is well known \cite{parry_novel_1992} and therefore we do not discuss it further.

\subsubsection{Bulk chemical potential}

\begin{figure*}[t]\centering
	\subfigure[]{\includegraphics[width=0.473\linewidth]{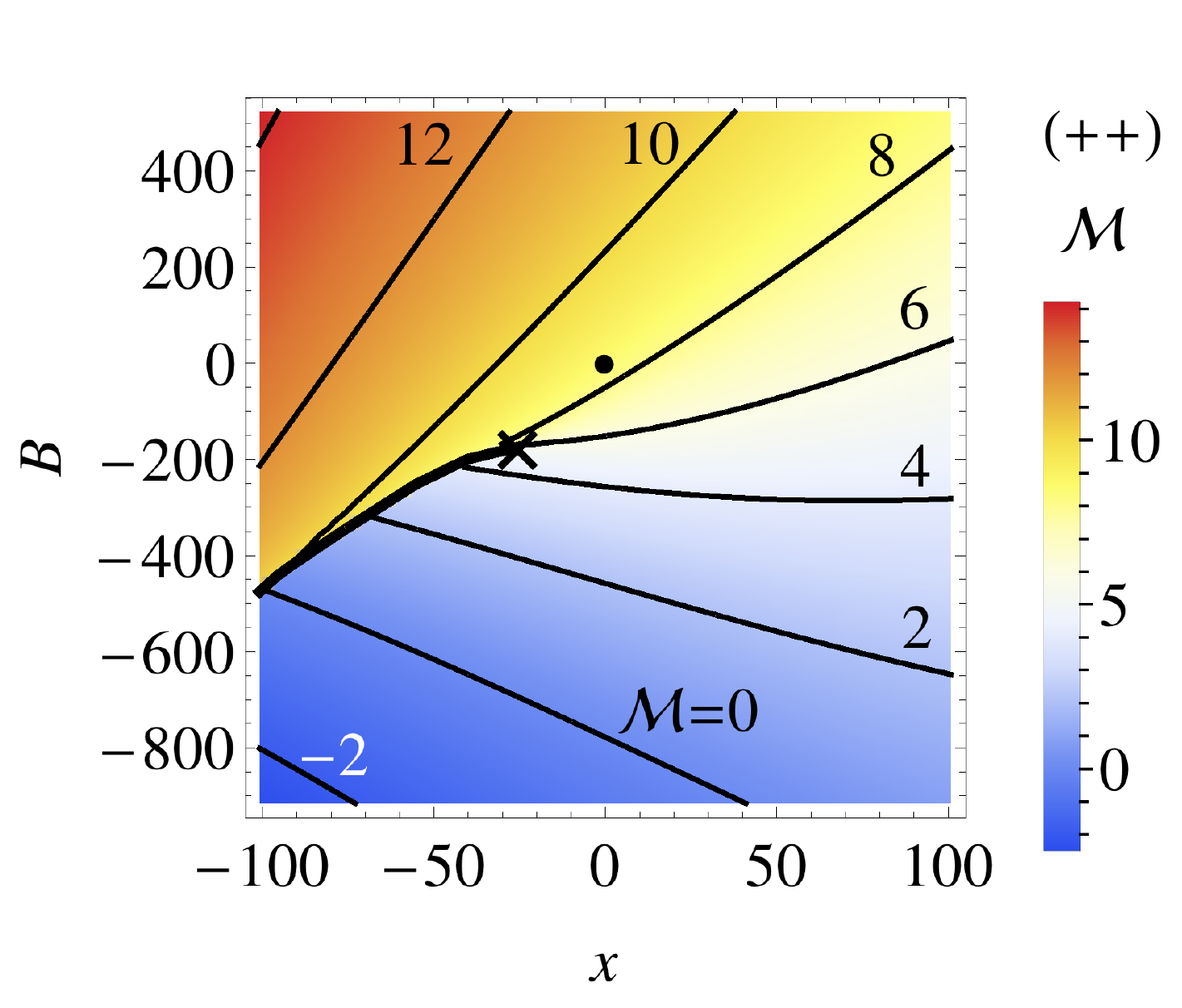}}\qquad
	\subfigure[]{\includegraphics[width=0.47\linewidth]{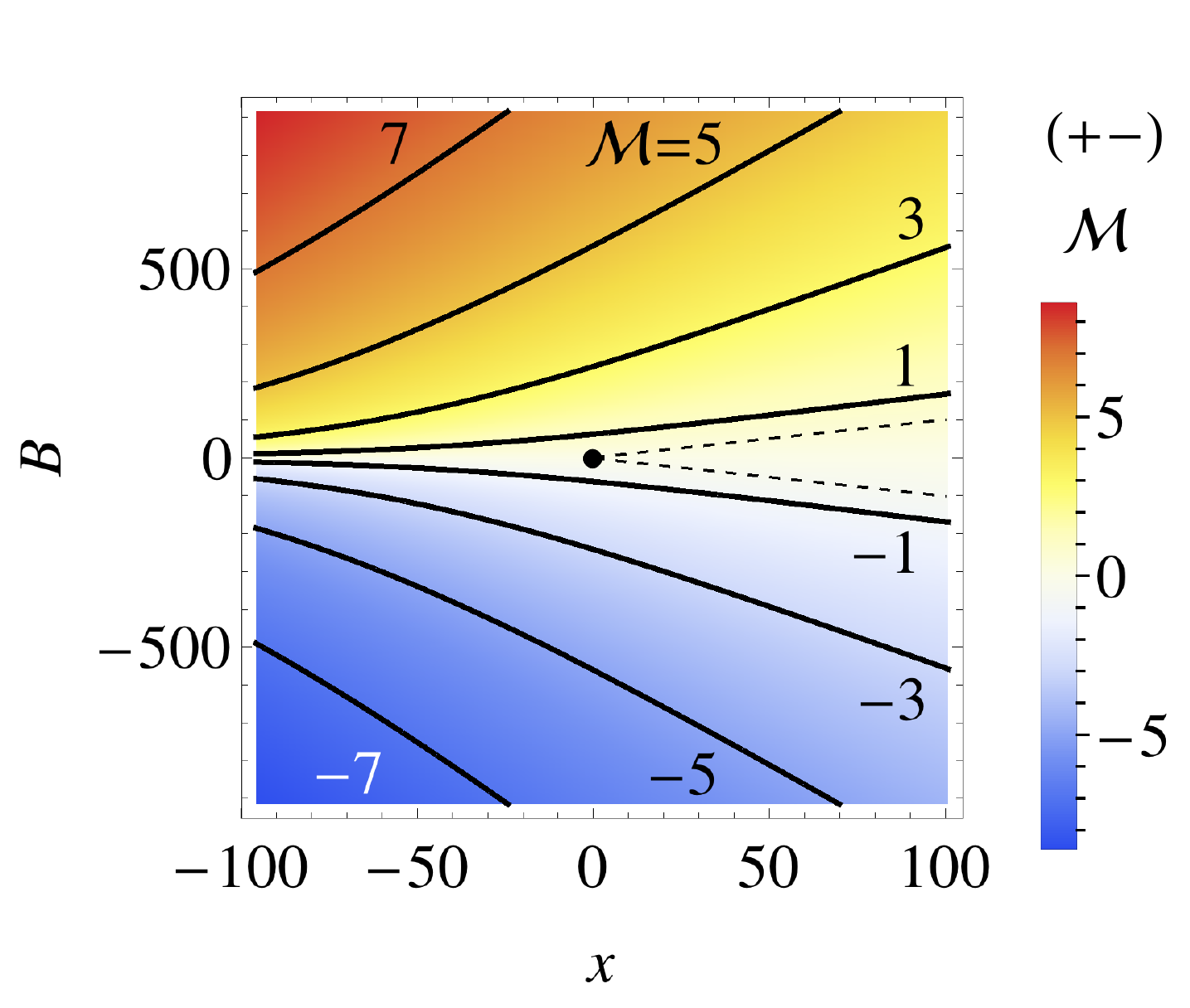}}
	\caption{Total mass $\Mass$ (corresponding to a scaled mean mass per volume, see Eq.~\eqref{eq_Mass}; color coding) as a function of the scaled temperature $\tscal$ and the bulk field $B$ for the Ginzburg-Landau model [Eq.~\eqref{eq_Landau_func_ndim}] with $H_1\simeq 5100$ and (a) $(++)$ boundary conditions and (b) $(+-)$ boundary conditions. The solid lines are curves of constant mass, with the values of $\Mass$ indicated by the labels. The bulk critical point $(\tscal_{c,b},B_{c,b})=(0,0)$ is marked by a dot ($\bullet$) and the capillary critical point $(\tscal_{c,\text{cap}},B_{c,\text{cap}})\simeq (-25,-166)$ by a cross ($\times$). In (a) the thick line ending at the cross is the line of first-order capillary condensation transitions. The dashed line in (b) represents, for comparison, the constraint-induced bulk field $\tilde B\st{hom}$ of a homogeneous bulk system with a density corresponding to $\Mass=1$. In (b) the lines are symmetric with respect to $B=0$. }
	\label{fig_phasediag_MFT}
\end{figure*}

Figure~\ref{fig_bulkfield_constr} shows the bulk chemical potential (bulk field) $\tilde B$ which has to act in the film in order to enforce the constraint $\Mass=0$ of vanishing total mass in the presence of $(++)$ boundary conditions. 
The corresponding data have been obtained via a numerical solution of the ELE  \eqref{eq_ELE}, including the nonlinear term.
In Fig.~\ref{fig_bulkfield_constr}(a), we display the constraint-induced bulk field, normalized by its analytical value $\tilde B_0$ [Eq.~\eqref{eq_B0}] for $\tscal\to\infty$, as a function of the ratio between the scaled temperature and surface field, $\tscal/H_1=1/\sqrt{\sigma}$, which was identified in Eq.~\eqref{eq_smallness_param} as an inverse smallness parameter controlling the onset of the strongly nonlinear regime for $\sigma\gg 1$.
In agreement with the perturbative study [Sec.~\ref{sec_ads_pert}], the bulk field $\tilde B$ approaches the limit $\tilde B_0$ for $\tscal\gg |H_1|$, whereas nonlinear effects dominate for $\tscal\lesssim |H_1|$ and increase in magnitude upon increasing $H_1$.
The dependence of $\tilde B$ on $H_1$ for $\tscal=0$ is shown in Fig.~\ref{fig_bulkfield_constr}(b).
For $\tscal=0$ and for \emph{weak} surface fields [more precisely, for $|H_1|\ll 100$ as implied by Eq.~\eqref{eq_m1_tau0}], the behavior of $\tilde B$  can be rationalized from Eq.~\eqref{eq_B0}, which predicts
\beq \tilde B \simeq -2 H_1,
\label{eq_Bconstr_smallH1}\eeq
consistent with the numerical data in Fig.~\ref{fig_bulkfield_constr}(b).
In the opposite limit of \emph{strong} surface fields, we recall that the OP profile in the central region of the film is almost constant [see Fig.~\ref{fig_prof_H1range}(a)]. 
Accordingly, as a direct consequence of the ELE in Eq.\ \eqref{eq_ELE}, we may approximately relate the value of the OP at the center to the chemical potential via the \emph{bulk} equation of state, $m(0) \sim \mathrm{sgn}(B) |B|^{1/\delta}$ with $\delta=3$ and, in the present case ($H_1>0$), $\mathrm{sgn}(B)=-1$. 
Together with Eq.~\eqref{eq_OPscale_largeH1} this yields the scaling of the constraint-induced bulk field (for $\tscal=0$ and $H_1\gg 100$) as
\beq \tilde B \sim -(s\ln H_1)^3,
\label{eq_Bconstr_largeH1}\eeq 
where $s$ is a numerical factor determined previously.
Figure~\ref{fig_bulkfield_constr}(b) shows that this prediction is in good agreement with the numerical mean field solution.

Figure~\ref{fig_phasediag_MFT} shows the total mass $\Mass$ (color coding) of a film near criticality as a function of the scaled temperature $\tscal$ and bulk field $B$ for $(++)$ and $(+-)$ boundary conditions. 
The data in Fig.~\ref{fig_phasediag_MFT} have been obtained via a numerical conjugate-gradient minimization of the Ginzburg-Landau functional in Eq.~\eqref{eq_Landau_func_ndim}.
We recall that the mass $\Mass$ is the appropriate scaling variable corresponding to the mass density [see Eq.~\eqref{eq_Mass}]. 
The solid curves in the plot serve to illustrate the temperature dependence of the constrained bulk field $\tilde B(\tscal)|_{\Mass}$ for selected values of $\Mass$.
Owing to the large value of $H_1\simeq 5100$ used in the present case, the data pertain to the strongly nonlinear mean field regime according to Eq.~\eqref{eq_smallness_param}.
The overall behavior of the bulk field resulting from imposing the constraint is consistent with the trends revealed by the perturbative solution (see Sec.~\ref{sec_ads_pert}; note that the absolute values are different, as they depend, in particular, on $H_1$).
For comparison, the constraint-induced field for a homogeneous bulk system would be simply given by $\tilde B\st{hom}(\tscal) = \tscal  \Mass +  \Mass^3$, corresponding, as a function of $\tscal$ with fixed $\Mass$, to straight lines of slope $\Mass$. 
As seen in Fig.~\ref{fig_phasediag_MFT}, the presence of the surface field $H_1$ leads to significant changes compared to the homogeneous case, which are most evident for $(++)$ boundary conditions. 
In fact, for $(+-)$ boundary conditions, the bulk field asymptotically approaches such a straight line already for rather small positive values of $\tscal$.

The thick solid line in Fig.~\ref{fig_phasediag_MFT}(a) indicates the capillary condensation line, ending at the capillary critical point $(\tscal_{c,\text{cap}},B_{c,\text{cap}})\simeq (-25,-166)$. We have determined the location of this point from a study of the OP value at the center of the film \cite{nakanishi_critical_1983} and its location is in good agreement with the one previously reported for the strong adsorption regime \cite{schlesener_critical_2003}.
Due to the large value of $H_1$ chosen in our analysis, the capillary critical point in the case of $(+-)$ boundary conditions is actually outside the range of values of $\tscal$ considered in Fig.~\ref{fig_phasediag_MFT}(b) \cite{parry_novel_1992}.
For $\tscal<\tscal_{c,\text{cap}}$, we infer from Fig.~\ref{fig_phasediag_MFT}(a) that there ceases to exist a corresponding value of $B$ for certain values of the mass $\Mass$ (which is visible in the plot by a gap in the color coding upon crossing the capillary condensation line). In the canonical ensemble, preparing a spatially extended but finite system with a value of $\Mass$ within this gap results in phase-separation into domains with densities corresponding to those just above and below the capillary condensation line.

\subsection{Numerical results: 3D Ising model}
\label{sec_MC_prof}
\begin{figure*}[t]\centering
	\subfigure[]{\includegraphics[width=0.48\linewidth]{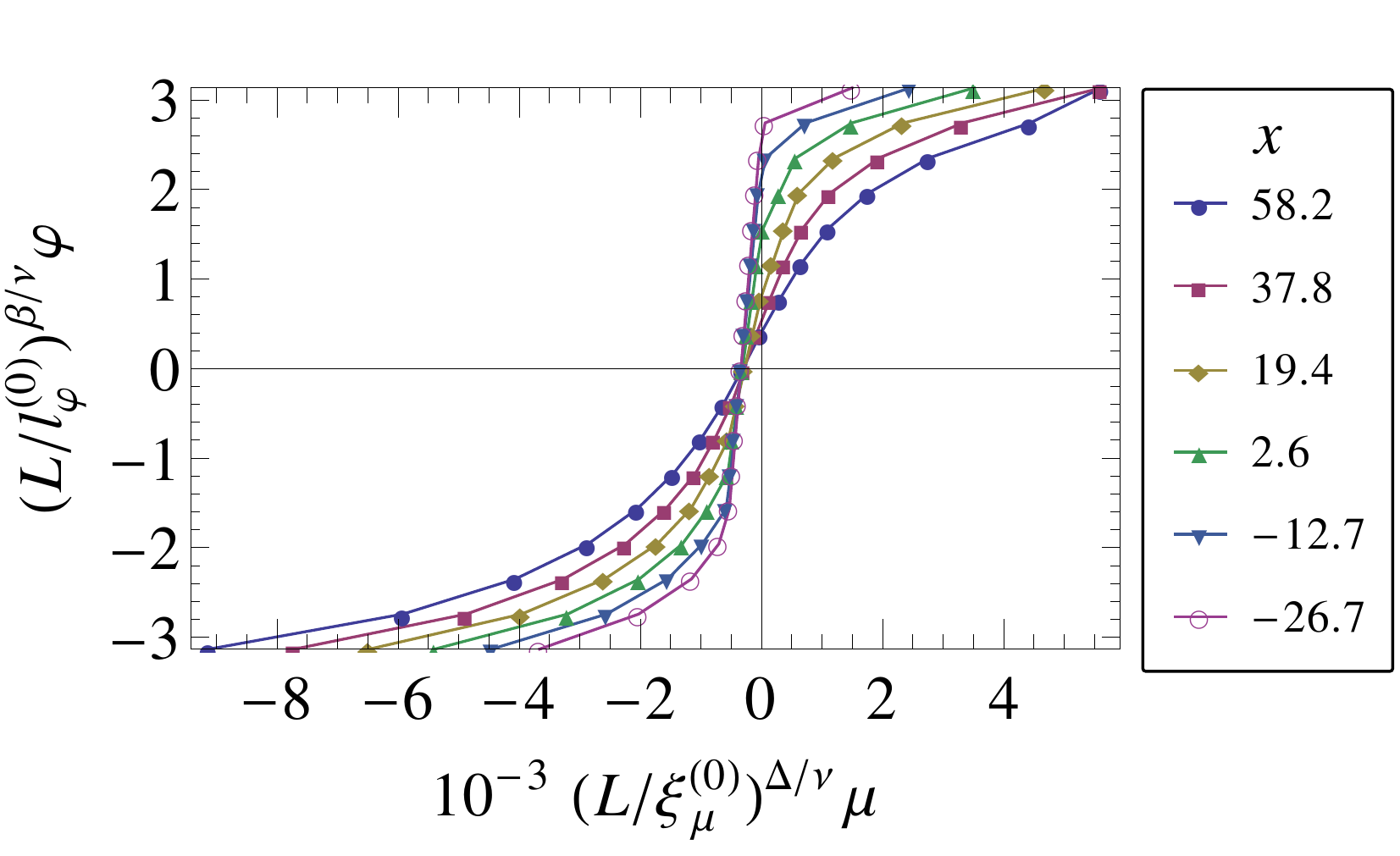}}\qquad
	\subfigure[]{\includegraphics[width=0.48\linewidth]{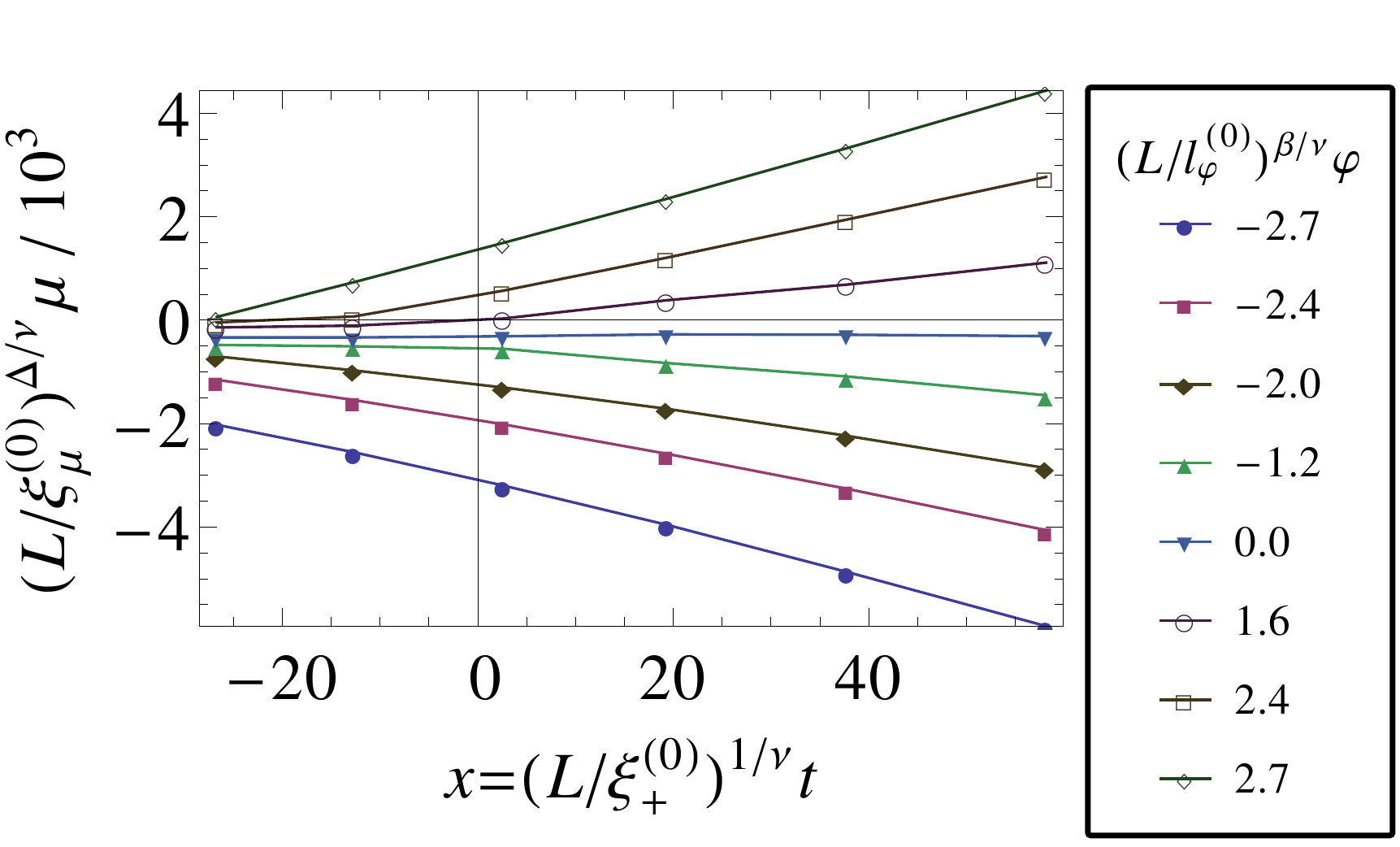}}
	\caption{MC simulation data of the Ising model in a film of thickness $L\equiv L_z$ with $(++)$ boundary conditions in the grand canonical ensemble. (a) Dependence of the scaled mean magnetization $(L/\lenPhi0)^{\beta/\nu}\mden$ [see Eq.~\eqref{eq_Mass}] on the bulk magnetic field $\mu$ for various scaled temperatures $\tscal=(L/\amplXip)^{1/\nu}(T/T_c-1)$. (b) Dependence of the scaled bulk magnetic field $\mu$ on $\tscal$ for various values of the scaled mean magnetization. Solid lines are drawn as a guide to the eye. A system of size $L_x\times L_y\times L_z=100\times 100\times 20$ is used. }
	\label{fig_bulkfield_MC}
\end{figure*}

\begin{figure*}[t]\centering
	\subfigure[]{\includegraphics[width=0.45\linewidth]{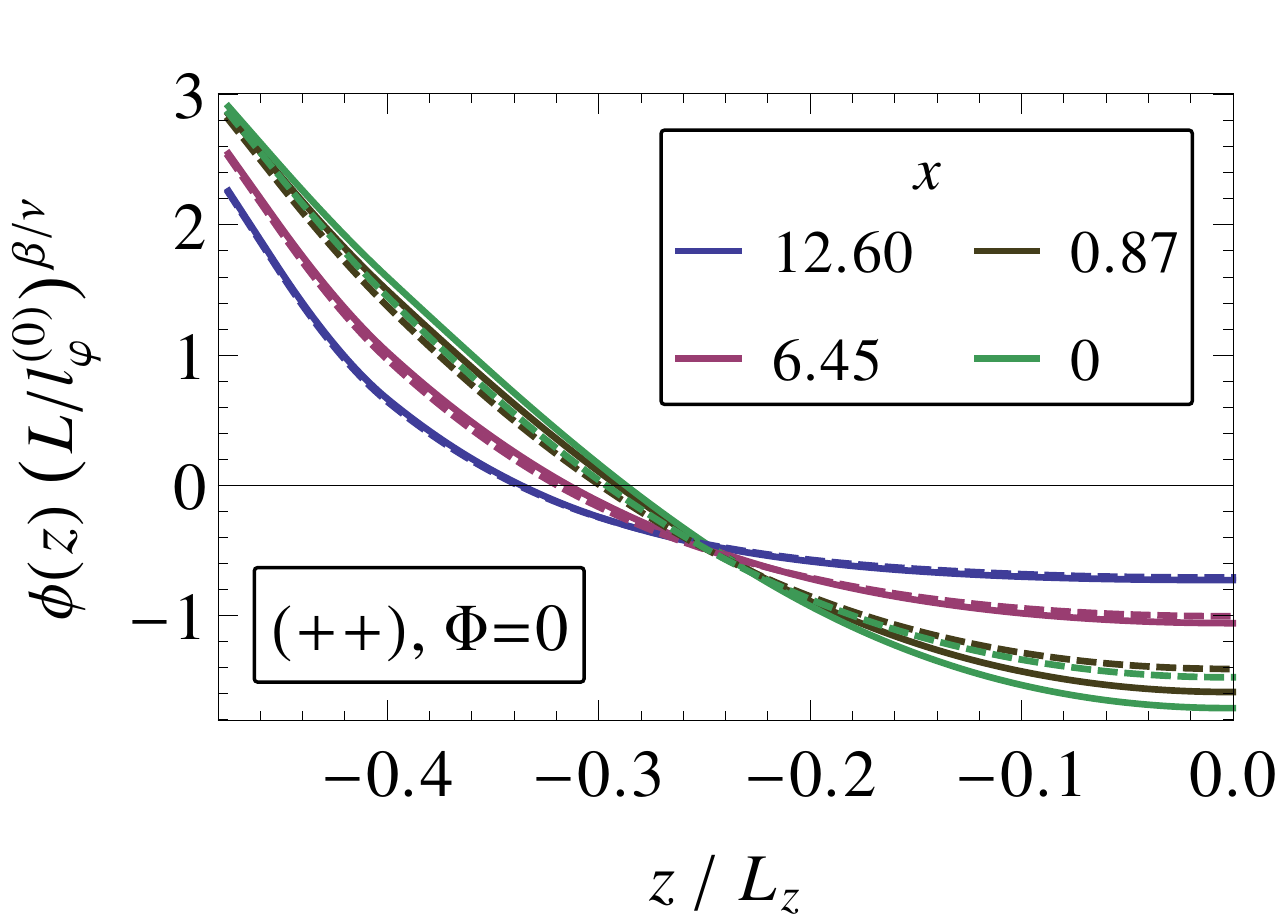}}\qquad
	\subfigure[]{\includegraphics[width=0.45\linewidth]{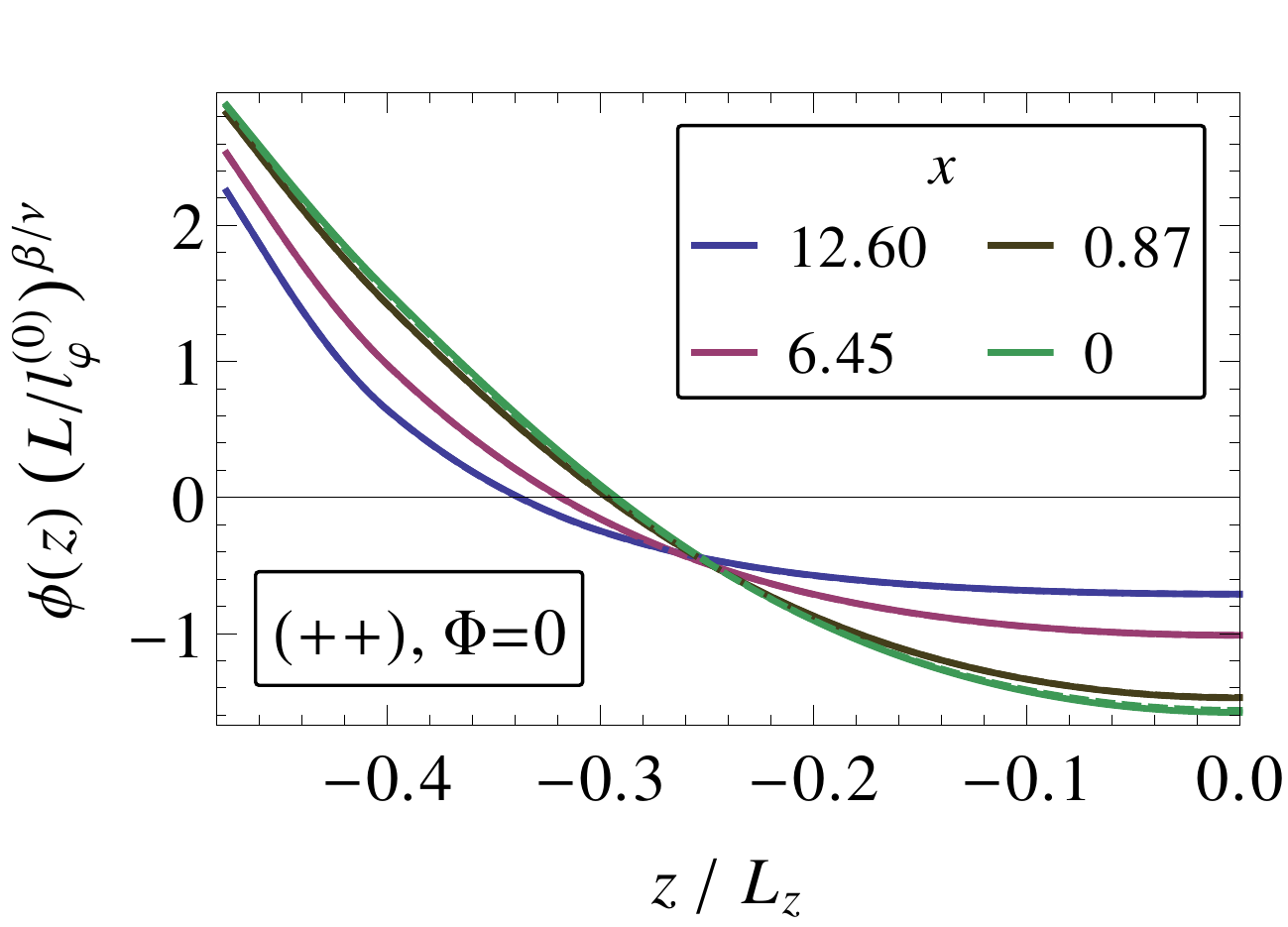}}
	\caption{OP profiles [interpolated and scaled according to Eq.~\eqref{eq_gen_prof_scal}] resulting from MC simulations of the Ising model for total magnetization $\mass=0$ and $(++)$ boundary conditions. Profiles in the canonical ensemble (solid lines) are compared with those in the grand canonical ensemble [with the bulk field $\mu$ inferred from Fig.~\ref{fig_bulkfield_MC}(b); dashed lines] for various scaled temperatures $\tscal$ (increasing from bottom to top at $z=0$) and system sizes $L_x\times L_y\times L_z $ of (a) $10\times 10\times 20$ and (b) $100\times 100\times 20$. In panel (b), the corresponding canonical and grand canonical results are practically indistinguishable. In order to enhance visibility, only the left half of the spatially symmetric profiles is shown. Contrary to the visual appearance, the profiles do not all intersect at the same point.}
	\label{fig_prof_MC_Trange}
\end{figure*}

In the grand canonical ensemble it is well known (see, e.g., Ref.~\cite{diehl_field-theoretical_1986}) that, in the limit of strong adsorption, the MFT divergence $\propto 1/\hat z$ of the critical OP profile near the wall at a distance $\hat z$ is changed by non-Gaussian fluctuations, leading to a weaker singularity 
\beq \phi(z)\sim \hat z^{-\beta/\nu}\,,
\label{eq_OPwall_decay}\eeq
where $\beta/\nu\simeq 0.52$ is the value of the exponent for the three-dimensional Ising universality class (see Tab.~\ref{tab_crit_exp}).
This behavior has been investigated via field-theoretical methods in the semi-infinite geometry \cite{bray_critical_1977, fisher_wall_1978, diehl_field-theoretical_1981, leibler_magnetisation_1982,brezin_critical_1983, ohno_critical_1989, diehl_critical_1993}, and by Monte Carlo simulations of the Ising model in a film geometry \cite{smock_universal_1994, czerner_near-surface_1997}. The prediction in Eq.~\eqref{eq_OPwall_decay} has been confirmed also experimentally (see, e.g., Refs. \cite{floter_universal_1995, law_wetting_2001} as well as references therein).

An explicit numerical test of the scaling behavior given by Eq.~\eqref{eq_OPwall_decay} within the \emph{canonical} ensemble is postponed to future studies, as it requires rather large wall-to-wall distances of the necessarily finite simulation cell and is therefore computationally demanding.
Here, instead, we consider smaller systems and focus on the dependence of the profiles on the transverse system size $L_z$ in the canonical and grand canonical ensembles. 
To this end, we present Monte Carlo (MC) simulation data of the three-dimensional Ising model on a cubic lattice of volume $L_{x} L_{y} L_{z} = A L_z$ with unit lattice spacing $a=1$ and $L_z$ even. 
A spin  $s_{i}=\pm 1$ is located at each site $i=(1 \le x \le L_{x}, 1 \le y \le L_{y}, -L_z/2+1 \le z \le L_{z}/2)$  of the lattice.
Along the $x$ and $y$ directions periodic boundary conditions are applied.
The Hamiltonian of the Ising model is given by
\begin{equation}
{\mathcal H} = -  J\left[\sum_{\bra i,j\ket}  s_{i}  s_{j}
+h_{1}^{-} \sum_{(\mathrm{ bot.} )}s_{j}+
h_{1}^{+}\sum_{(\mathrm{ top} )}s_{j} +\mu\sum_{ k }s_{k}\right],
\label{eq_Ising}\end{equation}
where $J$ is an interaction energy (which rescales the thermal energy $k_B T$) and $\mu$ is a bulk field (chemical potential); $h_1^-$ and $h_1^+$ are surface fields acting on the bottom ($z=-L_z/2 +1$) and the top ($z=L_{z}/2$) layer, respectively.
The first sum in Eq.~\eqref{eq_Ising} runs over nearest neighbor sites $\bra i,j\ket$ on the lattice, while the last one runs over all lattice spins. The sum with the subscript $(\mathrm{ bot.})$ is taken over the bottom layer $z=-L_z/2+1$ and the one with $( \mathrm{ top})$ is taken over the top layer $z=L_{z}/2$.
Note that, for simplicity, we denote the Ising model parameters by the same symbols as their counterparts in the Ginzburg-Landau free energy functional [Eq.~\eqref{eq_Landau_func_gc}]. However, the former carry no engineering dimensions and we therefore just report their numerical values as used in our simulations.
We generally use finite values of $h_1^\mp\in \{+ 1,-1\}$ for the surface fields in order to realize $(++)$ and $(+-)$ boundary conditions and henceforth, for convenience, suppress the superscript $\mp$ of $h_1$. 
(Here we choose a rather small value of $h_1$ in order to facilitate the simulation via the multi-spin technique in the canonical ensemble, see below.)
The fact that, in Eq.\ \eqref{eq_Ising}, the interaction constant $J$ in the bulk is the same as at the surface gives rise, within MFT, to a nonzero surface enhancement $c=1/a$ in the coarse-grained continuum counterpart of the Ising model \cite{binder_critical_1983,diehl_field-theoretical_1986}.
Accordingly, the asymptotic critical behavior is governed by the ordinary surface universality class (see Sec.\ \ref{sec_scal_CA}) and the scaling variable $H_1$ in Eq.\ \eqref{eq_H1scal} is given by $H_1=h_1 (L_z/\lenOrdH1)^{\Delta_1\ut{ord}/\nu}$, where $\Delta_1\ut{ord}\simeq 0.46$ (see Table \ref{tab_crit_exp}) and the length scale $\lenOrdH1\simeq 0.21$ \cite{hasenbusch_thermodynamic_2011}; this differs from the special surface universality class studied within the above continuum MFT.
The total magnetization $\mass$ is given by the thermal average $\mass= \big\bra \sum \limits_{ x,y,z }  s_{x,y,z} \big\ket $.
Note that, differently from before, here we do not consider thermodynamically extensive quantities to be implicitly normalized by the transverse area $A$.

Simulations in the grand canonical ensemble are performed via a hybrid MC algorithm~\cite{landau_guide_2009}.
Each MC step consists of a flip of a Wolff cluster followed by $L_{x} L_{y} L_{z}$ attempts to flip a randomly selected spin in accordance with the Metropolis criterion. The mean magnetization per spin $\mden=\mass / (L_{x}L_{y}L_{z})$ as well as the magnetization profile (per transverse area) $\phi(z)= \big\bra \sum \limits_{ x ,y } s_{x,y,z} \big\ket / (L_{x}L_{y})$ are computed as a thermal average $\bra\cdots \ket$, based on the statistical weight $\exp(-\beta \Hcal)$ and $\beta\equiv 1/(k_B T)$, over $10^{6}$ MC steps which are split into 10 series in order to determine the statistical accuracy.

As a preliminary step, we first determine the relationship between the bulk field $\mu$ and the magnetization $\mass$, which will be needed also later for the computation of the critical Casimir force.
In order to compute the value of $\mu$ which yields a certain assigned $\mass $ we proceed as follows: 
For a given value of the reduced temperature $t$ and of the system size $L$ we compute the mean magnetization $\mden$ as a function of the bulk field $\mu$, which is reported in Fig.~\ref{fig_bulkfield_MC}(a) in terms of the rescaled quantities $(L/\lenPhi0)^{\beta/\nu}\mden $ vs.\ $(L/\amplXimu)^{\Delta /\nu}\mu $, with $L=L_{z}$.
Here $\lenPhi0 =\amplXip \left(\amplPhit\right)^{\nu/\beta}\simeq 1.36$ [Eq.\ \eqref{eq_lenPhi0}], where we have used $\amplXip\simeq 0.50$ \cite{ruge_correlation_1994} and $\amplPhit\simeq 1.69$ \cite{caselle_universal_1997}. Furthermore we have $\amplXimu\simeq 0.617$ \cite{vasilyev_monte_2015}, which, in contrast to the standard definition \cite{pelissetto_critical_2002}, includes a factor $\delta^{1/(2-\eta)}\simeq 2.22$ according to our convention [see Eq.~\eqref{eq_gen_amplXimu}].
The values of the critical exponents can be found in Table~\ref{tab_crit_exp}.
In a second step, for a given magnetization $\mass$, the equation $\mden(\mu,T)=\mass/(L_x L_y L_z)$ is solved numerically for $\mu$, resulting in the plot of Fig.~\ref{fig_bulkfield_MC}(b) in terms of the scaled temperature $\tscal=(L/\amplXip)^{1/\nu}t$, with $\beta_c J\simeq 0.22165452(8)$ \cite{deng_simultaneous_2003}.

Simulations in the canonical ensemble have been performed by using Kawasaki dynamics~\cite{kawasaki_diffusion_1966} and the multi-spin technique~\cite{van_gemmert_phase_2005} which allows 64 independent systems to be simultaneously simulated by taking advantage of bitwise operations.  
Briefly, each site in the lattice is represented by a 64-bit integer variable, where the $k$th bit corresponds to the $k$th system. The average is performed over $2 \times 10^{5}$ MC steps, one MC step consisting of $10L_{x} L_{y} L_{z}$ attempts of pair Kawasaki exchanges. 
  
Since fluctuations are restricted in the canonical ensemble, one may expect the canonical OP profiles to increasingly deviate from the grand canonical ones upon decreasing the lateral system size. 
This is indeed corroborated by Fig.~\ref{fig_prof_MC_Trange}, where OP profiles for various temperatures and lateral system sizes $L_{x,y}$ (keeping $L_z$ fixed) are compared.
While for $L_{x,y}=100$ [Fig.~\ref{fig_prof_MC_Trange}(b)] the corresponding profiles in the two ensembles are practically indistinguishable, visible deviations appear for $L_{x,y}=20$ [Fig.~\ref{fig_prof_MC_Trange}(a)] and their magnitude increases upon approaching $T_c$.
Accordingly, for sufficiently large lateral system size, a film in the canonical ensemble can, at least as far as the behavior of the OP profiles is concerned, be fully described by a film in the grand canonical ensemble once the relation $\mu(t, \mass,h_1,L)$ is known.

\section{Critical Casimir force}
\label{sec_Casimir}

In this section we study the critical Casimir force (\CCF) in the canonical and in the grand canonical ensemble, taking advantage of our analysis of critical adsorption in the previous section.
The \CCF $\Kcal$ is defined in terms of the singular contribution to the residual finite-size free energy (per transverse area $A$ and in the limit $A\to \infty$) $\Fcal\res$ of a film of thickness $L$ as
\beq \Kcal = -\frac{d\Fcal\res}{dL},
\label{eq_pCas_dFdL}\eeq 
where $\Fcal\res$ is obtained by subtracting the bulk and surface contributions from the total film free energy $\Fcal_f$ (per area). 
In an expansion in decreasing powers of the system size $L$ one has, for films of sufficiently large lateral extent,
\beq \Fcal_f = L (-p_b) + f_s + \Fcal\res,
\label{eq_Ffilm_split}\eeq
where $p_b$ is the bulk pressure and $f_s$ is the surface free energy per area associated with the presence of the two walls confining the system. 
In the grand canonical ensemble $f_s$ does not contribute to $\Kcal$.
The residual finite-size part of the free energy per area \emph{at} bulk criticality is known to vary (in units of $k_B T_c$) asymptotically for large $L$ as $\Fcal\res = \hat \Delta/L^{d-1}$, where $d$ is the spatial dimensionality of the bulk system and $\hat\Delta$ is a universal critical Casimir amplitude which depends on the bulk universality class of the system and on the surface universality classes of the two confining walls \cite{privman_finite-size_1990, brankov_theory_2000}. (Note that, below, we shall introduce a critical Casimir amplitude $\Delta$ in terms of the scaling function of the \CCF rather than the residual free energy.)
Intriguingly, in the canonical ensemble it will turn out that, if the decomposition of $\Fcal_f$ in Eq.~\eqref{eq_Ffilm_split} follows the usual finite-size scaling arguments, one may obtain a surface free energy for which $d f_s/ d L\neq 0$, yielding a nonzero ``surface pressure'' contribution to the \CCF.

Alternatively, the \CCF may be determined directly as the difference between the pressure of the confined fluid film, $p_{f} = -d\Fcal_{f}/dL$, and the pressure $p_b$ of the surrounding bulk fluid phase:
\beq \Kcal = p_f - p_b.
\label{eq_pCas_pdiff}\eeq 
This definition is natural if one assumes that the confining surfaces of the film are exposed to a bulk fluid surrounding the film;
therefore it directly relates to typical experiments.
In the grand canonical ensemble, the bulk fluid can exchange mass with the film and, being in thermodynamic equilibrium, both are governed by the same chemical potential $\mu$. 
In contrast, in the canonical case, the film is isolated with respect to particle exchange from its surroundings. Hence the pressure $p_b$ of the bulk fluid---and thus also the \CCF $\Kcal$---depends on the experimental setup and thus is required to be fixed by a definition. 
In our investigation we will generally follow the convention of defining $p_b$ as the limit of $p_f$ as $L\to\infty$:
\beq p_b = \lim_{L\to\infty} p_f.
\label{eq_pB_lim_pF}\eeq 
The limit is taken such that, besides temperature, the relevant thermodynamic control parameter of the bulk fluid is the same as the one of the film.
In the grand canonical ensemble, this control parameter is the chemical potential $\mu$, whereas in the canonical ensemble it is the mean mass density $\mden = \mass/L$. 
It will turn out that only this convention leads to a force which can be interpreted as a \CCF.
Furthermore we shall see that it is precisely the different nature of the respective control parameters in the two ensembles which, within MFT, is ultimately responsible for the difference between the \CCFs in the two ensembles.

Given a mean field free energy functional such as the one in Eq.~\eqref{eq_Landau_func_gc}, a stress tensor $T_{ij}$ can be constructed \cite{krech_casimir_1994, onuki_phase_2002} which expresses the change of the total free energy $\Fcal_f$ of the film upon a change of the configuration of the boundaries.
Specifically, for a film of thickness $L$, which is homogeneous in the lateral, i.e., $x$ and $y$, directions, the film pressure $p_f=T_{zz}$ is equal to the change of the free energy of the film upon varying $L$:
\beq p_f = T_{zz}[\phi\eq] = -\frac{d }{dL}\Fcal_f[\phi\eq].
\label{eq_stressten_dFdL}\eeq 
Here, $\phi\eq$ is the equilibrium OP profile which minimizes the free energy functional $\Fcal_f$.
Furthermore, in equilibrium, the stress tensor is constant across the system; in particular, $T_{zz}$ does not depend on the distance from the confining surfaces. 
Analogously to Eq.~\eqref{eq_stressten_dFdL}, the pressure $p_b$ of the bulk system surrounding the film can be obtained as $p_b=T_{zz}(\phi_{b})$, with $\phi_{b}$ being the corresponding equilibrium value of the bulk OP.
The stress tensor thus allows one to circumvent the calculation of the free energy and its derivative, rendering the definition in Eq.~\eqref{eq_pCas_pdiff} convenient whenever these quantities are difficult to determine explicitly. This applies, for instance, to the analysis of the nonlinear Ginzburg-Landau model considered here. 
The stress tensor and, in particular, Eq.~\eqref{eq_stressten_dFdL} are typically considered in the grand canonical ensemble, in which $T_{ij}$ and $\Fcal_f$ depend on the externally imposed bulk field $\mu$. In the present study we shall show that an analogous equation [see Eqs.~\eqref{eq_stressten_can} and \eqref{eq_pf_equal} below] applies in the canonical ensemble as well.
We shall see further that, even if Eq.~\eqref{eq_pB_lim_pF} is used to define $p_b$, the definitions in Eqs.~\eqref{eq_pCas_dFdL} and \eqref{eq_pCas_pdiff} are not fully equivalent in the canonical case, because in Eq.~\eqref{eq_pCas_pdiff} only bulk and no surface contributions are subtracted from the film pressure.

In the following, we shall study the \CCFs in the canonical and grand canonical ensembles for symmetric [$(++)$] and antisymmetric [$(+-)$] boundary fields. In order to simplify the notation, we explicitly indicate the boundary conditions only when confusion might occur.
While some results in the grand canonical ensemble are known from previous studies \cite{marconi_critical_1988, krech_casimir_1997, borjan_order-parameter_1998, schlesener_critical_2003, borjan_off-critical_2008, mohry_crossover_2010,okamoto_casimir_2012, mohry_critical_2014, dantchev_exact_2016}, they are briefly re-derived here along with the canonical ones in order to provide a self-contained and easily accessible presentation.

\begin{widetext}\subsection{General scaling considerations}
\label{sec_scal_Casimir}
Before turning to the analysis of a specific model, we first consider the general scaling behavior expected for the \CCF, building upon the discussion in Sec.~\ref{sec_scal_CA}.
In the \emph{grand canonical} ensemble, the scaling form of the residual finite-size free energy [Eq.~\eqref{eq_Ffilm_split}] (per transverse area and per $k_B T$) can be written as \cite{diehl_field-theoretical_1986, privman_finite-size_1990} 
\beq \begin{split}
\Fcal\res\gc(t,\mu,h_1,L) 
&= L^{-d+1} \Theta\gc \left(  \left(\frac{L}{\amplXip}\right)^{1/\nu}t, \left(\frac{L}{\amplXimu}\right)^{\Delta/\nu}\mu, \left(\frac{L}{\lenH1}\right)^{\Delta_1/\nu} h_1\right),
\end{split}\label{eq_gen_Fres_gc}\eeq 
with $\Theta\gc(x,B,H_1)$ as the corresponding scaling function.
The scaling form of the \CCF (per transverse area and $k_B T$) follows from Eq.~\eqref{eq_pCas_dFdL} as (note that here $\Delta$ is a critical exponent and not a Casimir amplitude)
\beq \Kcal\gc(t,\mu,h_1,L) = L^{-d} \Xi\gc\left(  \left(\frac{L}{\amplXip}\right)^{1/\nu}t, \left(\frac{L}{\amplXimu}\right)^{\Delta/\nu}\mu, \left(\frac{L}{\lenH1}\right)^{\Delta_1/\nu} h_1\right), 
\label{eq_Casi_force_gc}\eeq 
with the scaling function 
\begin{multline}
\Xi\gc(\tscal, B, H_1) =  (d-1) \Theta\gc(\tscal, B, H_1) - \frac{1}{\nu} \tscal \pd_{\tscal}\Theta\gc(\tscal, B, H_1) - \frac{\Delta}{\nu} B \pd_{B}\Theta\gc (\tscal, B, H_1)- \frac{\Delta_1}{\nu} H_1 \pd_{H_1}\Theta\gc(\tscal, B, H_1).
\label{eq_Casi_force_gc_scalf}\end{multline} 

In the \emph{canonical} ensemble, instead of the chemical potential $\mu$, the total mass $\mass$ is fixed. In Sec.~\ref{sec_scal_CA} we have identified the mean mass density $\mden$ and its scaled counterpart $\Mass$ as the proper scaling variables. Accordingly, the scaling form of the canonical residual finite-size free energy is proposed as
\beq \begin{split}
\Fcal\res\can(t,\mden,h_1,L) 
&= L^{-d+1} \Theta\can \left( \left(\frac{L}{\amplXip}\right)^{1/\nu}t , \left(\frac{L}{\lenPhi0}\right)^{\beta/\nu}\mden , \left(\frac{L}{\lenH1}\right)^{\Delta_1/\nu} h_1\right),
\label{eq_gen_Fres_c}\end{split}\eeq 
where $\Theta\can(\tscal, \Mass,H_1)$ is the corresponding scaling function.
Crucially, in order to compute the \CCF [based on Eq.~\eqref{eq_pCas_dFdL}] in the canonical ensemble, we have to take into account the constraint of having a fixed total mass $\mass$. This implies a dependence of the mean density $\mden=\mass/L$ [Eq.~\eqref{eq_mden}] on $L$:
\beq \frac{\pd\mden}{\pd L} = -\frac{\mden}{L}.
\label{eq_dPhi_dL}\eeq
The canonical \CCF follows from Eqs.~\eqref{eq_pCas_dFdL} and \eqref{eq_gen_Fres_c} as
\beq \Kcal\can(t,\mden,h_1,L) = L^{-d} \Xi\can\left(  \left(\frac{L}{\amplXip}\right)^{1/\nu}t, \left(\frac{L}{\lenPhi0}\right)^{\beta/\nu}\mden, \left(\frac{L}{\lenH1}\right)^{\Delta_1/\nu} h_1\right), 
\label{eq_Casi_force_c}\eeq 
with the scaling function
\begin{multline} 
\Xi\can(\tscal, \Mass, H_1) =  (d-1) \Theta\can(\tscal, \Mass, H_1) - \frac{1}{\nu} \tscal \pd_{\tscal}\Theta\can(\tscal, \Mass, H_1)\\  - \left(\frac{\beta}{\nu}-1\right) \Mass \pd_{\Mass}\Theta\can(\tscal, \Mass, H_1) - \frac{\Delta_1}{\nu} H_1 \pd_{H_1}\Theta\can(\tscal, \Mass, H_1).
\label{eq_Casi_force_c_scalf}\end{multline}
\end{widetext}
Note that, within MFT, one has $\beta/\nu=1$ so that in Eq.~\eqref{eq_Casi_force_c_scalf} the term involving $\pd_\Mass\Theta\can$ vanishes.
In Eq.~\eqref{eq_Casi_force_c_scalf} the presence of the term $\Mass\pd_\Mass\Theta\can$ [in addition to $-(\beta/\nu)\Mass\pd_\Mass\Theta\can$] is a genuine consequence of the mass constraint. 
Since $\Xi\gc$ and $\Xi\can$ are functions of different variables ($B$ and $\Mass$, respectively), in order to asses their difference based on the general scaling relations above, an equation of state relating $B$ and $\Mass$ must be specified. 
Instead of following this route further, below we shall explicitly compute the \CCF analytically within linear MFT and numerically within full MFT.

The values of the scaling functions $\Theta$ and $\Xi$ at the fixed-points of the renormalization group flow define universal critical Casimir amplitudes $\hat \Delta$ and $\Delta$, respectively \cite{krech_casimir_1994, brankov_theory_2000, gambassi_casimir_2009}. 
The fixed point of the normal surface universality class \cite{diehl_field-theoretical_1981} corresponds to $\tscal=B=0$ and $|H_1|=\infty$. Under these conditions one obtains from Eqs.~\eqref{eq_Casi_force_gc_scalf} or \eqref{eq_Casi_force_c_scalf} the simple relation $\Delta=(d-1)\hat\Delta$ between the amplitude $\Delta$ of the \CCF and the amplitude $\hat\Delta$ of the residual free energy scaling function.
However, as discussed in Sec.~\ref{sec_critads} (and, in particular, in Sec.~\ref{sec_sde_canonical}), within MFT and for $(++)$ boundary conditions, the limit $|H_1|\to\infty$ violates the constraint of a fixed total number of particles. 
Consequently, in this case, in Eq.~\eqref{eq_Casi_force_c_scalf} one cannot simply set $|H_1|=\infty$ but, instead, one must take into account that the value of the scaling function at criticality still depends on $H_1 \neq \infty$.

\subsection{Mean field theory}
\label{sec_Casimir_mft}
Within MFT it can be shown [see Appendix~\ref{app_stressten} and, in particular, Eq.~\eqref{eq2_stress_ten} therein that the stress tensor $T_{ij}\can$ of a system in the canonical ensemble is given by the grand canonical stress tensor $T_{ij}\gc$ with the bulk field $\mu$ taking the value $\tilde \mu(\mass)$ of the Lagrange multiplier required to satisfy the OP constraint in Eq.~\eqref{eq_Mass0}:
\beq T_{ij}\can[\phi\eq] = T_{ij}\gc([\phi\eq];\mu=\tilde \mu).
\label{eq_stressten_can}\eeq  
Here, $\phi\eq$ is the solution of the ELE which minimizes the corresponding free energy functional and satisfies the constraint in the case of the canonical ensemble.
We recall that within MFT (see Sec.~\ref{sec_ads_model}) and for a given value of the total mass $\mass$, the equilibrium profile $\phi\eq$ is, by construction, exactly the same in the canonical and the grand canonical ensemble.
As shown in Appendix \ref{app_stressten}, Eq.\ \eqref{eq_stressten_can} holds for any free energy functional and boundary conditions, as long as the system is in a unique thermodynamic equilibrium state. 
For the pressure $p_b$ of a bulk system, a relation corresponding to the one in Eq.~\eqref{eq_stressten_can} with $T_{zz}(\phi_{\text{eq},b})=p_b$ is in fact well known and can be easily derived from thermodynamics [see Eq.~\eqref{eq2_pbulk_thermo} in Appendix \ref{app_stressten}].

In this study, we shall consider only films which are translationally invariant along the lateral directions (i.e., perpendicular to the $z$ coordinate), excluding the case of lateral phase separation.
(In the presence of two-phase coexistence, stresses in the system are not necessarily anymore homogeneous and the analysis of this case requires an extension of the present model.) 
For a laterally homogeneous film, Eq.~\eqref{eq_stressten_can} implies the equality of the canonical and grand canonical film pressures $p_f^{\text{(c,gc)}} \equiv T_{zz}^{\text{(c,gc)}}$ in MFT, i.e.,
\beq\begin{split} -\frac{d}{dL} \Fcal_f\can[\phi\eq] &= p_f\can[\phi\eq] = p_f\gc([\phi\eq];\tilde \mu) \\ &= -\frac{d}{dL} \Fcal_f\gc([\phi\eq];\tilde \mu) ,\end{split}\label{eq_pf_equal}
\eeq 
provided $p_f$ is evaluated for \emph{both} ensembles under the same thermodynamic conditions, i.e., using, in the grand canonical case, the bulk field $\tilde\mu$ corresponding to the imposed total mass $\mass$ [Eq.~\eqref{eq_Mass0} with $\phi\equiv \phi\eq$].
Here and in the following pressures are given per $k_B T$ and therefore have the unit of an inverse volume.

For the grand canonical Ginzburg-Landau free energy functional in Eq.~\eqref{eq_Landau_func_gc}, the mean-field stress tensor is given by (see Appendix \ref{app_stressten}):
\begin{multline} T_{ij}\gc([\phi\eq];\mu) = (\pd_i\phi\eq) (\pd_j\phi\eq) \\ -\left[\onehalf (\pd_k\phi\eq)(\pd_k\phi\eq) +  \onehalf \tau \phi\eq^2 + \frac{1}{4!} g\phi\eq^4 - \mu \phi\eq \right]\delta_{ij},
\label{eq_stressten_gc}\end{multline}
(where summation over repeated indices is implied) and the corresponding film pressure for a laterally homogeneous film is
\beq \begin{split}
	p_f^{\text{(c,gc)}} &= T_{zz}^{\text{(c,gc)}} =  \onehalf (\pd_z \phi\eq)^2  - \onehalf \tau \phi\eq^2 - \frac{1}{4!} g \phi\eq^4 + \tilde \mu \phi\eq  \\
	&=\frac{\Delta_0}{L^{4}} \left[ \onehalf \left(\pd_\zeta m\eq\right)^2  - \onehalf \tscal m\eq^2 - \frac{1}{4} m\eq^4 + \tilde B m\eq  \right].
\end{split}\label{eq_pf_landau}\eeq
The second line above follows from introducing the scaling variables defined in Eqs.~\eqref{eq_scalvar} and \eqref{eq_scalvar_mft}, where $\Delta_0$, defined in Eq.~\eqref{eq_Delta0}, is a non-universal mean-field amplitude, which eventually will be absorbed in the Casimir amplitudes studied further below.

Having discussed the pressure within a film, we now turn to the description of the corresponding bulk systems.
In the grand canonical ensemble the bulk free energy functional analogous to Eq.~\eqref{eq_Landau_func_gc} is
\beq \Fcal_b\gc(\phi_b,\mu_b\gc) \equiv \int d^dr \left[\onehalf \tau \phi_b^2 +\frac{1}{4!} g \phi_b^4 - \mu_b\gc \phi_b \right].
\label{eq_Landau_func_gc_blk}\eeq 
The bulk pressure $p_b\gc$ of a homogeneous grand canonical system is identical to the negative of the equilibrium bulk free energy density and follows from the expression of $T_{zz}\gc(\phi_b,\mu_b\gc)$ [Eq.~\eqref{eq_stressten_gc}] as
\beq 
p_b\gc(\mu_b\gc) =  \onehalf \tau\phi^2_b + \frac{1}{8}g\phi^4_b,
\label{eq_pB_gc}\eeq 
where $\phi_b$ is the spatially constant equilibrium solution of the bulk equation of state: 
\beq \tau\phi_b + \frac{1}{6}g\phi_b^3 = \mu_b\gc =\mu\,.
\label{eq_bulk_EOS_gc}\eeq 
Accordingly, $p_b\gc$ is a function of the bulk chemical potential $\mu_b\gc$, which in the grand canonical ensemble, is identical to the one of the film ($\mu$) due to the thermodynamic coupling between the bulk and the film.
For $\mu_b\gc=0$ and $\tau<0$, $\Fcal_b\gc$ [Eq.~\eqref{eq_Landau_func_gc_blk}] yields the two coexisting, symmetric equilibrium states $\pm \phi_{b,\text{eq}}$ with $\phi_{b,\text{eq}}\equiv \sqrt{-6\tau /g}$, which give rise to identical bulk pressures in Eq.~\eqref{eq_pB_gc}.
We anticipate here that for a film with, e.g., $(++)$ boundary conditions, we shall find that the two equilibrium densities at the capillary condensation line lead to different film pressures [see Fig.~\ref{fig_pCas_MFT}(d) further below and the related discussion].

For a film in the canonical ensemble, particle exchange between film and bulk is prohibited and the mass densities in the bulk and in the film are therefore \emph{a priori} not related. 
It turns out, however, that, for the purpose of isolating the canonical \CCF, it is crucial that the bulk system has the \emph{same} mass density $\mden=\mass/L$ as the film.
This way the equivalence between a film in the limit $L\to\infty$ and a bulk system is ensured [see Eq.~\eqref{eq_pB_lim_pF}].
In contrast, the choice of the ensemble used to describe the bulk is immaterial, because in the thermodynamic limit all ensembles are equivalent.
Furthermore, Eq.~\eqref{eq_stressten_can} implies that, within MFT, the stress tensor is identical in both ensembles under the same thermodynamic conditions, i.e., for the same mass density $\mden$. 
Thus, when referring to a bulk system, we shall use henceforth the notion ``canonical''  in order to indicate that the bulk is coupled to the film by imposing the \emph{same mass density} in both.
In the case $\tau>0$, the canonical bulk pressure coincides, according to Eq.~\eqref{eq_stressten_can}, with the expression in Eq.~\eqref{eq_pB_gc} with $\phi_b=\mden$.
In the case $\tau<0$, the possibility of phase separation precludes a direct application of Eq.\ \eqref{eq_stressten_can}. Now $\phi_b=\mden$ minimizes the free energy $\Fcal_b\gc$ in Eq.~\eqref{eq_Landau_func_gc_blk} with $\mu_b\gc$ given by Eq.~\eqref{eq_bulk_EOS_gc} only if $\mden$ is outside the binodal region, i.e., if $|\mden|\geq \phi_{b,\text{eq}}=\sqrt{-6\tau /g}$.
The case $\tau<0$ and $|\mden|<\phi_{b,\text{eq}}$ does not admit such a spatially uniform minimum $\phi_b$ and can, instead, be only realized via phase-separation into domains of local density $\pm \phi_{b,\text{eq}}$ corresponding to the symmetric equilibrium states of $\Fcal_b\gc$ for $\mu_b\gc=0$. As noted previously, the associated bulk pressure $p_b\gc$ is, however, insensitive to this phenomenon.
In summary, the pressure of a bulk system which is ``canonically'' coupled to the film, i.e., has the same mass density $\mden$ as the latter, is given by
\begin{widetext}\beq p_b\can(\phi_b) = 
 \onehalf \tau \phi_{b}^2 + \frac{1}{8}g\phi_{b}^4 ,\qquad \text{with}\quad  
 \begin{cases}
   \phi_b = \pm \phi_{b,\text{eq}},\qquad &\tau<0\quad \text{and}\quad -\phi_{b,\text{eq}} \leq \varphi \leq \phi_{b,\text{eq}},\\
   \phi_b = \varphi,\qquad &\text{otherwise.}
  \end{cases}
\label{eq_pB_can}\eeq
Furthermore, the chemical potential associated with a canonical bulk system of mass density $\mden$ is given by
\beq \mu_b\can = \begin{cases}
 0,\qquad &\tau<0\quad \text{and}\quad -\phi_{b,\text{eq}} \leq \varphi \leq \phi_{b,\text{eq}},\\
 \tau\varphi + \frac{1}{6}g\varphi^3,\qquad &\text{otherwise.}
 \end{cases}
\label{eq_bulk_EOS_can}\eeq 
\end{widetext}

The pressures in Eqs.~\eqref{eq_pB_gc} and \eqref{eq_pB_can} are identical in bulk systems with the same mean density $\mden$, as expected on general grounds due to the equivalence of ensembles in the thermodynamic limit. 
For the present purposes, however, we have to compare bulk pressures which emerge by coupling the bulk either canonically or grand canonically to a film of a given total mass $\mass$.
In the case of a grand canonical coupling, the corresponding bulk pressure $p_b\gc$ is a function of the chemical potential $\mu=\tilde\mu(\mass)$ required to satisfy the constraint of fixed mass $\mass$ in the \emph{film} \footnote{For notational simplicity, we shall use the same symbol $p_b\gc$ both for the bulk pressure, which is a function of the prescribed bulk chemical potential $\mu$, and for that pressure, which follows from evaluating $p_b\gc$ at $\mu=\tilde\mu$ where $\tilde \mu$ is the constraint-induced chemical potential of the film. In the latter case, $p_b\gc$ therefore turns into an implicit function of the thermodynamic variables and parameters of the film.}.
In contrast, in the case of a canonical coupling, the pressure $p_b\can$ is not a function of $\tilde \mu$, but it is determined from the condition that the mean mass densities $\mden$ in the film and in the bulk must be the same.
Since, in the presence of surface fields, the chemical potential of a film generally assumes a value different from the one for a bulk system with the same mean density (see Secs.~\ref{sec_ads_pert} and \ref{sec_ads_num} and, in particular, Fig.~\ref{fig_phasediag_MFT}), we have $p_b\gc\neq p_b\can$. 
However, because generally $p_f\gc=p_f\can$ [Eq.~\eqref{eq_pf_equal}], the difference between the bulk pressures directly implies [via Eq.~\eqref{eq_pCas_pdiff}] a difference in the \CCFs between the canonical and the grand canonical ensembles.
This crucial insight will be confirmed in the following by analytical calculations within linear MFT and numerically for the nonlinear MFT.

\subsubsection{Linear MFT: Critical Casimir force deduced from the free energy}
Linear MFT of the Ginzburg-Landau model [i.e., Eq.~\eqref{eq_Landau_func_gc}, neglecting the quartic interaction term involving the coupling constant $g$] offers the advantage that analytical results, which already capture essential features of the nonlinear case, can be obtained easily. 
The linear theory considered in the following is based on the results of Sec.~\ref{sec_ads_pert} and thus it is generally expected to provide an accurate approximation of the nonlinear model for large values of the scaled temperature $\tscal$ and sufficiently small values of the scaled mass $\Mass$ and of the surface fields $H_1$ (see Sec.~\ref{sec_ads_pert_summary} and the related discussion).

Neglecting the quartic coupling in Eq.~\eqref{eq_Landau_func_gc}, the grand canonical free energy (per transverse area) of a near-critical film with symmetric [$(++)$] boundary conditions is given by
\begin{widetext}
\begin{equation} \Fcal_f\gc([\phi];\tau,\mu,h_1,L) = \int_{-L/2}^{L/2} dz \left[\onehalf (\pd_z\phi)^2 + \onehalf \tau \phi^2 - \mu\phi\right] - 2 h_1 \phi_w,
\label{eq_lin_Landau_func_gc}\end{equation}
where $\phi_w\equiv \phi(-L/2)=\phi(L/2)$ is the value of the OP at the walls.
Evaluated for the equilibrium solution $\phi=\phi_0$ obtained within linear MFT [Eq.~\eqref{eq_phi0_sol}], we find
\begin{equation}
 \Fcal_f\gc(\tau, \mu,h_1,L)  = \underbrace{-L \frac{\mu^2}{2\tau} \vphantom{\frac{h_1^2}{\sqrt{\tau}}} }_\text{bulk} \underbrace{-\frac{2 h_1 \mu}{\tau} - \frac{h_1^2}{\sqrt{\tau}}}_\text{surface}  \underbrace{-\frac{2h_1^2}{\sqrt{\tau}}\frac{1}{\exp(L\sqtau)-1}}_\text{residual}.
\label{eq_filmF_gc}\end{equation}
Since, in the grand canonical ensemble, $\mu$ is independent of $L$, this expression displays the expected decomposition of the total free energy (per area) $\Fcal_f\gc$ into a bulk ($\propto L$), a surface ($\propto L^0$), and a residual finite size part [see Eq.~\eqref{eq_Ffilm_split}], which are highlighted in Eq.~\eqref{eq_filmF_gc} by the braces.
We remark that the bulk part in Eq.\ \eqref{eq_filmF_gc} coincides, within the linear mean field approximation, with the result obtained by evaluating Eq.\ \eqref{eq_Landau_func_gc_blk} for the equilibrium solution $\tau\phi_b=\mu$ as determined by Eq.\ \eqref{eq_bulk_EOS_gc}.

In the canonical ensemble, instead, the film free energy $\Fcal_f\can$ for $(++)$ boundary conditions follows, after neglecting the quartic coupling in Eq.~\eqref{eq_Landau_func_c}, as
\beq \Fcal_f\can([\phi];\tau, h_1,L) = \int_{-L/2}^{L/2} dz \left[\onehalf (\pd_z\phi)^2 + \onehalf \tau \phi^2 \right] - 2 h_1 \phi_w.
\label{eq_lin_Landau_func_c}\eeq 
Evaluated for the constrained mean field solution $\phi=\tilde\phi_0$ reported in Eq.~\eqref{eq_phi0_constr}, $\Fcal_f\can$ becomes
\begin{equation} \Fcal_f\can(\tau, \mass,h_1,L) =  \underbrace{\onehalf L \tau \left(\frac{\mass}{L}\right)^2}_\text{bulk}  \underbrace{-2 h_1 \frac{\mass}{L} - \frac{h_1^2}{\sqrt{\tau}}}_\text{surface} + \underbrace{\frac{2 h_1^2}{L\tau} - \frac{2h_1^2}{\sqrt{\tau}}\frac{1}{\exp(L\sqtau)-1}.}_\text{residual}
\label{eq_filmF_can}\end{equation}
Alternatively, $\Fcal_f\can$ can also be obtained directly from $\Fcal_f\gc$ via a Legendre transform [noting that, in accordance with Eqs.~\eqref{eq_Mass0} and \eqref{eq_lin_Landau_func_gc}, $\pd \Fcal_f\gc / \pd\mu = -\mass$]: 
\begin{equation} \Fcal_f\can(\tau,\mass,h_1,L) = \Bigg[\Fcal_f\gc([\phi];\tau,\mu,h_1,L)  + \mu\int_{-L/2}^{L/2} dz\, \phi(z) \Bigg]_{\phi=\phi_0,\, \mu=\tilde \mu_0(\mass)}, 
\label{eq_Legendre}\end{equation}\end{widetext}
with the bulk field taking (within linear MFT) the value $\tilde \mu_0=\mden \tau -2h_1/L$ [Eq.~\eqref{eq_h0}] in order to satisfy the constraint of constant mass $\mass$. 
Note that also the bulk contributions in Eqs.\ \eqref{eq_filmF_gc} and \eqref{eq_filmF_can} are related via a Legendre transform, which explains, in particular, the different signs of these terms.

For a meaningful comparison of a film and a bulk system in the canonical ensemble, both must have the same mass density $\mden=\mass/L$.
Indeed, it has turned out in Sec.\ \ref{sec_scal_CA} that in this case $\mden$ is the natural finite-size scaling variable.
Accordingly, in Eq.\ \eqref{eq_filmF_can}, the various finite-size contributions to the canonical free energy have been identified based on their scaling behavior with $L$, assuming, for this purpose, $\mass/L$ to be fixed
\footnote{If one would keep $\mass$ fixed rather than $\mden$ while performing the limit $L\to \infty$, one would obtain vanishing bulk and surface contributions to $\Fcal_f\can$. As a consequence, independently of the thermodynamic conditions of the film, the limit $L\to\infty$ would always yield a bulk system with vanishing density $\mden=0$.}. 
We emphasize, however, that the quantity which, by definition, is actually constant for a system in the canonical ensemble is the total mass $\mass$.
This implies, in particular, that computing a derivative with respect to $L$ in the canonical ensemble must take into account Eq.\ \eqref{eq_dPhi_dL}.

Note that, if one were interested merely in the free energies $\Fcal_f\cgc$ of the film in the two ensembles, the value $\tilde \mu_0$ for the constraint field would have to be used instead of $\mu$ in Eq.~\eqref{eq_filmF_gc} in order to determine the grand canonical free energy. This would, however, not be appropriate for deriving the finite-size scaling behavior, which is based on the idea of comparing systems of different $L$ while keeping all other thermodynamic parameters fixed. In particular, in the limit $L\to\infty$, the canonical film is supposed to match a homogeneous system with the specified mean density $\mden$, whereas the grand canonical film acquires the mean density $\mden(\mu)$ which is determined by the value of the external field $\mu$.

\begin{figure*}[t]\centering
    \subfigure[]{\includegraphics[width=0.46\linewidth]{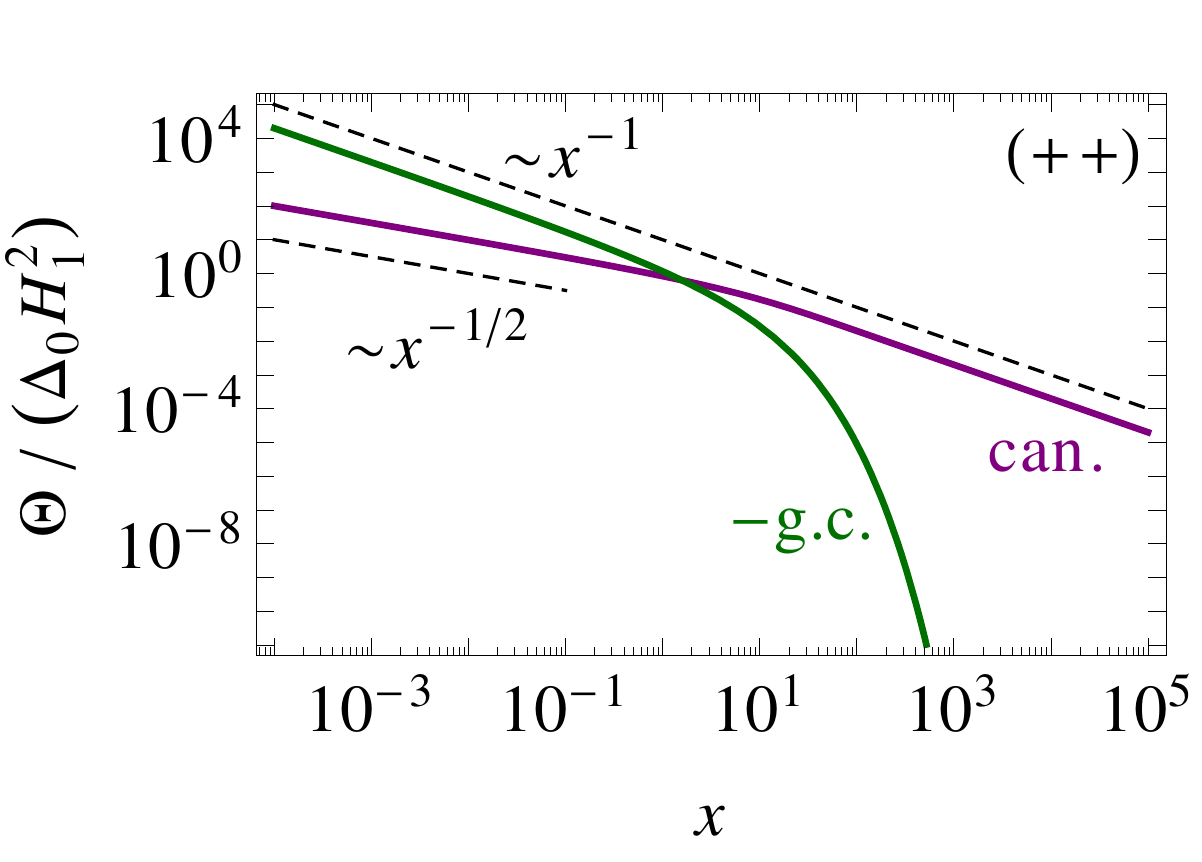}}\qquad
    \subfigure[]{\includegraphics[width=0.46\linewidth]{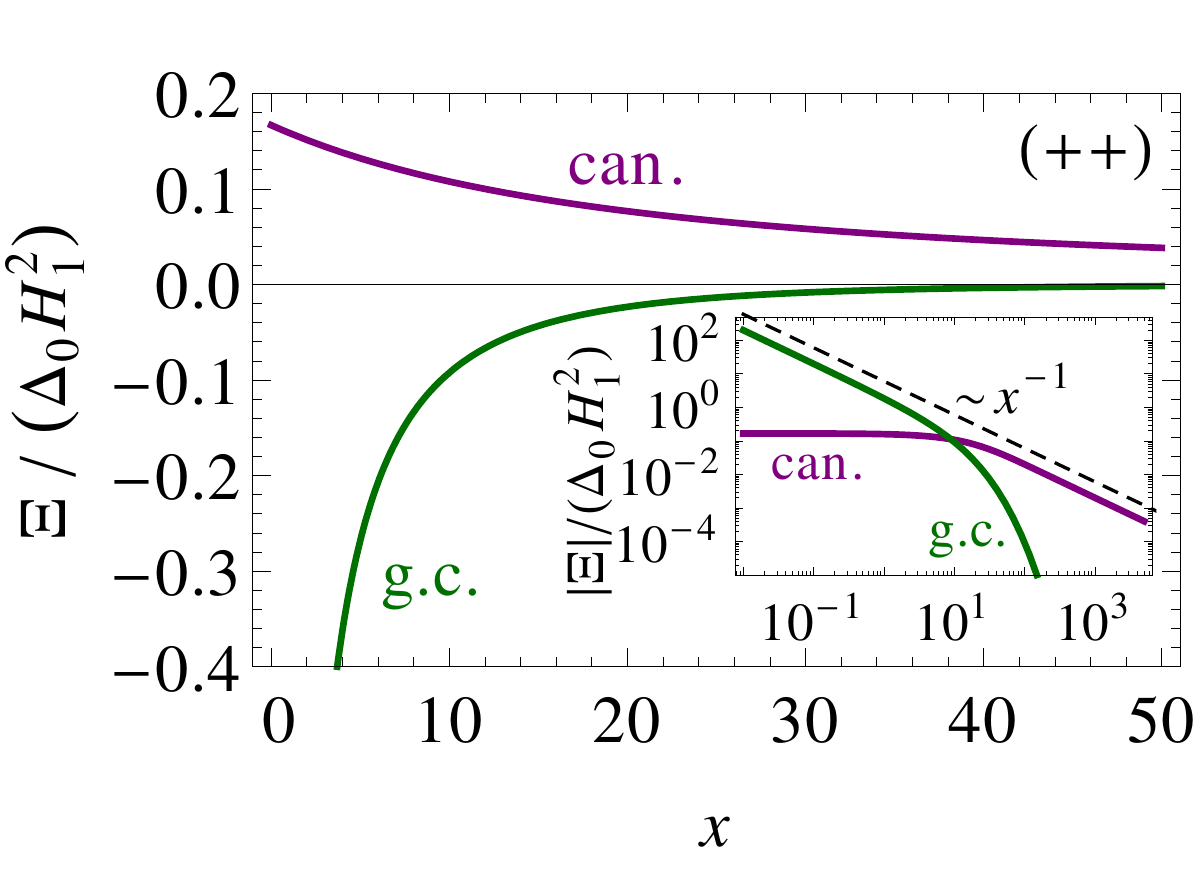}}
    \caption{Scaling functions $\Theta$ and $\Xi$ of (a) the residual finite-size free energy [Eq.~\eqref{eq_Theta_linMF}] and of (b) the \CCF [Eq.~\eqref{eq_Xicas_linMF}], respectively, within linear MFT for the canonical and the grand canonical ensemble of a film with $(++)$ boundary conditions as a function of the temperature scaling variable $x=L^2\tau=(L/\amplXip)^{1/\nu}t$. The dashed lines indicate the characteristic power laws for small and large values of $\tscal$ as given in the main text. In (a) the negative of $\Theta\gc$ is plotted. The inset in (b) shows the scaling functions of the main panel in a double-logarithmic scale in order to highlight their different asymptotic behaviors. $\Theta$ and $\Xi$ are normalized by $\Delta_0 H_1^2$, which, within MFT, appears as a common overall prefactor. For large $\tscal$ one has $\Theta\gc(\tscal\gg 1)/(\Delta_0 H_1^2)\simeq 2\exp(-\sqtscal)/\sqtscal$ and $\Xi\gc(\tscal \gg 1)/(\Delta_0 H_1^2)\simeq -2\exp(-\sqtscal)$. As an artifact of linear MFT, here the residual finite-size free energies as well as the grand canonical \CCF do not depend on the imposed mass $\Mass$ or the bulk field $B$ and diverge for $\tscal\to 0$. However, linear MFT correctly predicts the sign as well as the decay behavior for large $\tscal$ of the \CCF (see Sec.~\ref{sec_Casimir_full_mft} further below).}
    \label{fig_cas_limMF_tau}
\end{figure*}

As an artifact of linear MFT, the residual free energies in Eqs.~\eqref{eq_filmF_gc} and \eqref{eq_filmF_can} turn out to be independent of the chemical potential $\mu$ and of the imposed total mass $\mass$.
In the nonlinear mean field model (see Sec.~\ref{sec_Casimir_full_mft} further below), they do  acquire a dependence on $\mu$ and $\mass$, respectively.
Apart from this deficiency, $\Fcal\res\ut{(gc,c)}$ in Eqs.~\eqref{eq_filmF_gc} and \eqref{eq_filmF_can} can, after introducing the scaling variables defined in Eqs.~\eqref{eq_scalvar} and \eqref{eq_scalvar_mft}, be cast into the scaling forms given in Eqs.~\eqref{eq_gen_Fres_gc} and \eqref{eq_gen_Fres_c} (for $d=4$)
with the scaling functions
\begin{subequations}\bal
\Theta\gc(\tscal, H_1)/\Delta_0 &= -\frac{2 H_1^2}{\sqrt{\tscal}}\frac{1}{\exp \sqrt{\tscal}-1} \label{eq_Theta_gc}\qquad \text{and}\\
\Theta\can(\tscal,H_1)/\Delta_0 &= \frac{2 H_1^2}{\sqrt{\tscal}}\left[\frac{1}{\sqtscal}-\frac{1}{\exp\sqtscal -1}\right], \label{eq_Theta_can}
\end{align}\label{eq_Theta_linMF}\end{subequations}
respectively.
The non-universal amplitude $\Delta_0=6/g$ [Eq.~\eqref{eq_Delta0}] contains the coupling $g$, which is unknown within MFT \footnote{In order to be in line with the general scaling theory we keep the amplitude factor $\Delta_0$ even within linear MFT, in which, in fact, $g$ does not appear in the free energy functional. Indeed, in Eq.\ \eqref{eq_Theta_linMF} $\Delta_0 H_1^2$ is independent of $g$.}.
The scaling functions in Eq.~\eqref{eq_Theta_linMF} are displayed in Fig.~\ref{fig_cas_limMF_tau}(a).
Remarkably, they have opposite signs and decay differently for large $\tscal$: while the grand canonical scaling function $\Theta\gc$ decays exponentially modified by a power law, i.e., $\Theta\gc(\tscal\gg 1)/\Delta_0 \simeq -2 H_1^2 \exp(-\sqtscal)/\sqtscal$, the canonical scaling function $\Theta\can$ decays algebraically, i.e., $\Theta\can(\tscal\gg 1)/\Delta_0  \simeq 2H_1^2/\tscal$.
As an artifact of linear MFT, $\Theta\gc$ and $\Theta\can$ diverge for $\tscal\to 0$, i.e., upon approaching the bulk critical point. Specifically, we have $\Theta\gc(\tscal \ll 1)/\Delta_0 \simeq -2H_1^2/\tscal$ and $\Theta\can(\tscal \ll 1)/\Delta_0 \simeq H_1^2/\sqrt{\tscal}$ for small positive values of $\tscal$.
Accordingly, one cannot infer a proper critical Casimir amplitude from these expressions. This will be achieved in Sec.~\ref{sec_Casimir_full_mft} below when discussing the nonlinear model, which renders finite critical Casimir free energies and forces. There, it will turn out that, despite the incorrect behavior of linear MFT for $\tscal\to 0$, the essential features of the \CCFs in nonlinear MFT---in particular their sign and decay for $\tscal\gg 1$---are captured correctly by the linear approximation.

The \CCFs $\Kcal\ut{(gc,c)}(\tau,h_1,L) = -d \Fcal\res\ut{(gc,c)}(\tau,h_1,L)/dL$ following from the residual free energies defined in Eqs.~\eqref{eq_filmF_gc} and \eqref{eq_filmF_can} obey the scaling form given in Eqs.~\eqref{eq_Casi_force_gc} and \eqref{eq_Casi_force_c} 
with $d=4$, where the scaling functions $\Xi\ut{(gc,c)}$ are obtained from Eq.~\eqref{eq_Theta_linMF} as
\begin{subequations}\bal
\Xi\gc(\tscal, H_1)/\Delta_0 &= \frac{H_1^2}{1-\cosh \sqtscal}\qquad \text{and} \label{eq_Xicas_gc}\\
\Xi\can(\tscal, H_1)/\Delta_0 &= \frac{2H_1^2}{\tscal} + \frac{H_1^2}{1-\cosh \sqtscal}. \label{eq_Xicas_can}
\end{align}\label{eq_Xicas_linMF}\end{subequations}
Analogously to the residual free energies [Eq.~\eqref{eq_Theta_linMF}], these \CCFs do not depend on the chemical potential $B$ or on the imposed total mass $\Mass$.
This artifact of the linear mean field approximation disappears in the nonlinear theory (see Sec.~\ref{sec_Casimir_full_mft} below). 
The leading dependence of the \CCF on large $\Mass$ can be derived from the generalized perturbation theory discussed in Appendix \ref{app_pert}, which amounts to replacing the temperature scaling variable $\tscal$ by $\tscal+3\Mass^2$ in Eq.~\eqref{eq_Xicas_linMF}.
The scaling functions in Eq.~\eqref{eq_Xicas_linMF} are plotted in Fig.~\ref{fig_cas_limMF_tau}(b).
Similarly to the free energies, also the \CCF behaves differently in the two ensembles: most strikingly, $\Kcal$ is repulsive ($\Kcal\can>0$) in the canonical case, in contrast to being attractive ($\Kcal\gc<0$) in the grand canonical case. 
For large $\tscal$, the scaling functions of the canonical \CCF decay within linear MFT as
\begin{subequations}\bal
\Xi\gc(\tscal\gg 1)/\Delta_0 &\simeq -2 H_1^2\exp(-\sqtscal)\qquad \text{and}\label{eq_Xicas_gc_asympt}\\
\Xi\can(\tscal\gg 1)/\Delta_0 &\simeq \frac{2 H_1^2}{\tscal}. \label{eq_Xicas_can_asympt}
\end{align}\label{eq_Xicas_linMF_asympt}\end{subequations}
For small positive values of $x$ the scaling function of the grand canonical \CCF diverges as
$\Xi\gc(\tscal \to 0)/\Delta_0 \simeq -2 H_1^2/\tscal,$ 
which, however, is an artifact on linear MFT.
In contrast, the scaling function of the canonical \CCF attains a finite limit within linear MFT: 
\beq \Xi\can(\tscal \to 0)/\Delta_0  =\frac{1}{6} H_1^2.
\label{eq_Xicas_can_Tc}\eeq 
As discussed in Sec.~\ref{sec_ads_pert_summary}, in the canonical ensemble linear MFT for critical adsorption renders an accurate approximation of the nonlinear theory even for $\tscal\to 0$, provided $H_1$ is sufficiently small. We expect this to be the case also for the \CCF discussed here.

We now briefly summarize the case of a film with antisymmetric [$(+-)$] boundary conditions. Using the results for the profile given in Sec.~\ref{sec_ads_pert_pm} and proceeding as above, we find within linear MFT
\beq 
\Fcal_{f,+-}\gc(\tau, \mu,h_1,L) = \underbrace{-L \frac{\mu^2}{2\tau} \vphantom{\frac{h_1^2}{\sqrt{\tau}}} }_\text{bulk} \underbrace{-\frac{h_1^2}{\sqrt{\tau}} }_\text{surface} + \underbrace{\frac{h_1^2}{\sqrt{\tau}}\frac{2}{1+\exp(L\sqrt{\tau})}}_\text{residual}
\label{eq_filmF_pm_gc}\eeq
in the grand canonical and
\beq 
\Fcal_{f,+-}\can(\tau, \mden,h_1,L) = \underbrace{\onehalf L \tau \mden^2 \vphantom{\frac{h_1^2}{\sqrt{\tau}}} }_\text{bulk} \underbrace{-\frac{h_1^2}{\sqrt{\tau}} }_\text{surface} + \underbrace{\frac{h_1^2}{\sqrt{\tau}}\frac{2}{1+\exp(L\sqrt{\tau})}}_\text{residual}
\label{eq_filmF_pm_can}\eeq
in the canonical ensemble.
The bulk part in Eq.\ \eqref{eq_filmF_pm_gc} coincides, within linear MFT, with the one obtained by evaluating Eq.\ \eqref{eq_Landau_func_gc_blk} for the equilibrium solution determined by Eq.\ \eqref{eq_bulk_EOS_gc}. 
The expression for $\Fcal_{f,+-}\can$ in Eq.\ \eqref{eq_filmF_pm_can} can be obtained from a Legendre transform of $\Fcal_{f,+-}\gc$ according to Eq.\ \eqref{eq_Legendre} with $\phi_0$ given by Eq.\ \eqref{eq_pm_sol0_bare} and $\tilde\mu(\mass)=\mass \tau/L$ [Eq.\ \eqref{eq_pm_h0}].
The two indicated residual finite-size free energies [see Eq.~\eqref{eq_Ffilm_split}] are identical and yield the scaling functions for the \CCF,
\beq \Xi\gc_{+-}(\tscal,H_1)/\Delta_0 = \Xi\can_{+-}(\tscal,H_1)/\Delta_0 =  \frac{H_1^2}{1+\cosh \sqtscal},
\label{eq_pm_Xicas_linMF}\eeq 
which feature an exponential decay to zero for large $\tscal$ and a finite value at $\tscal=0$:
\begin{subequations}\bal
\Xi\ut{(gc,c)}_{+-}(\tscal\gg 1)/\Delta_0 &\simeq  H_1^2 \exp{(-\sqtscal)}\qquad \text{and} \label{eq_pm_Xicas_asympt_largeTau}\\
\Xi\ut{(gc,c)}_{+-}(\tscal=0)/\Delta_0 &= \onehalf H_1^2. \label{eq_pm_Xicas_Tc}
\end{align}\label{eq_pm_Xicas_asympt}\end{subequations}
As in the case of symmetric boundary conditions and as an artifact of linear MFT, these results are independent of the chemical potential $B$ and of the imposed total mass $\Mass$, but none of the two diverges upon approaching bulk criticality.

The asymptotic results in Eqs.\ \eqref{eq_Xicas_gc_asympt} and \eqref{eq_pm_Xicas_asympt_largeTau} pertaining to linear MFT can be compared with the generally expected behavior within the Ising universality class in the grand canonical ensemble for $B=0$ and $H_1=\infty$. In this case, local-functional approaches \cite{borjan_off-critical_2008, upton_off-critical_2013}, exact results for the Ising  strip \cite{evans_solvation_1994}, and estimates based on the transfer-matrix method \cite{hasenbusch_thermodynamic_2010} indicate that
\begin{subequations} \bal
\Xi\gc_{++}(\tscal\gg 1,B=0,H_1=\infty)&\propto -\tscal^{2-\alpha} \exp(-\tscal^\nu)\text{ and} \label{eq_Xicas_nlin_asympt_pp}\\ 
\Xi\gc_{+-}(\tscal\gg 1,B=0,H_1=\infty)&\propto \tscal^{2-\alpha} \exp(-\tscal^\nu). \label{eq_Xicas_nlin_asympt_pm}
\end{align}\label{eq_Xicas_nlin_asympt}\end{subequations} 
The exact results for the asymptotic behavior of the nonlinear MFT considered in Ref.\ \cite{krech_casimir_1997} are recovered by Eq.\ \eqref{eq_Xicas_nlin_asympt} upon inserting mean-field values for the critical exponents (see Tab.\ \ref{tab_crit_exp}) \footnote{Within nonlinear MFT, we expect, for any finite $H_1$ and sufficiently large $\tscal$, $\Xi\gc$ to ultimately follow the linear mean field prediction (see Fig.~\ref{fig_linMFT_valid} and the related discussion).}.
Differently from linear MFT, $\Xi\cgc(\tscal\to 0,B,H_1=\infty)$ is finite within nonlinear MFT and, generally, within the Ising universality class, for both $(++)$ and $(+-)$ boundary conditions [see Eqs.\ \eqref{eq_cas_ampl_limits} and \eqref{eq_MC_casAmpl_c} below].

\subsubsection{Linear MFT: Critical Casimir force deduced from the stress tensor}
\label{sec_cas_stress}
In this subsection we illustrate the calculation of the \CCF based on Eq.~\eqref{eq_pCas_pdiff} and on the explicit expression in Eq.\ \eqref{eq_pf_landau} for the stress tensor.
This will not only provide an independent check of the results obtained in the previous subsection and therefore, of the fundamental relation in Eq.\ \eqref{eq_stressten_can}, but it also allows us to highlight a subtlety associated with the canonical \CCF, which will lead to an expression different from the one in Eq.~\eqref{eq_Xicas_can}.
Focusing first on a film with symmetric $(++)$ boundary conditions, the film pressure $p_f=T_{zz}$ within linear MFT follows from Eqs.~\eqref{eq_phi0_sol} (writing $\mu$ instead of $\mu_0$) and \eqref{eq_stressten_gc} in the grand canonical ensemble as
\beq p_f\ut{(gc)}(\tau,\mu,h_1,L) = \underbrace{\frac{\mu^2}{2\tau} \vphantom{\frac{L h_1^2}{\cosh(\sqtau)}}}_\text{from bulk} + \underbrace{\frac{h_1^2 }{1-\cosh(L\sqrt{\tau})}}_\text{from residual},
\label{eq_linMF_pf_gc}\eeq
which equals $-d \Fcal_f\gc/dL$ [Eq.~\eqref{eq_stressten_dFdL}] with the grand canonical film free energy $\Fcal_f\gc$ in Eq.~\eqref{eq_filmF_gc}.
The braces indicate the terms in $\Fcal_f\gc$ from which the respective contributions in $p_f\gc$ originate.
In a spatially homogeneous bulk system, within linear MFT the OP value $\phi_b$ induced by the external field $\mu$ is $\phi_b = \mu/\tau$ [Eq.~\eqref{eq_bulk_EOS_gc}], resulting in a grand canonical bulk pressure [see Eqs.~\eqref{eq_pB_gc} and \eqref{eq_bulk_EOS_gc}]
\beq p_b\ut{(gc)}(\tau,\mu) = \frac{\mu^2}{2\tau},
\label{eq_linMF_pb_gc}\eeq 
consistent with Eqs.\ \eqref{eq_pB_lim_pF} and \eqref{eq_linMF_pf_gc}.
Accordingly, the \CCF is given by
\beq \Kcal\ut{(gc)}(\tau,\mu,h_1,L) = p_f\ut{(gc)} - p_b\ut{(gc)} = \frac{h_1^2 }{1-\cosh(L\sqrt{\tau})}.
\label{eq_linMF_pCasGC_stressroute}\eeq 
After introducing the scaling variables defined in Eq.~\eqref{eq_scalvar}, this expression turns out to be identical to the one reported in Eq.~\eqref{eq_Xicas_gc}, which was obtained by taking the derivative of the grand canonical residual free energy according to Eq.~\eqref{eq_pCas_dFdL}.

In the canonical ensemble, instead, Eq.~\eqref{eq_stressten_can} implies $p_f\can(\tau,\mass,h_1,L) = p_f\gc(\tau,\tilde \mu(\mass),h_1,L)$, with the constraint-induced bulk field $\tilde\mu(\mass)= \sfrac{(\mass \tau -2 h_1)}{L}$ as given by Eq.~\eqref{eq_h0}, and $p_f\gc$ given by Eq.~\eqref{eq_linMF_pf_gc} within linear MFT.
Accordingly, the canonical film pressure for $(++)$ boundary conditions results in
\beq \begin{split} 
p_f\ut{(c)}(\tau,\mass,h_1,L) &= \frac{1}{2 \tau}\left(\mden \tau -\frac{2h_1}{L}\right)^2 + \frac{h_1^2}{1-\cosh(L\sqrt{\tau})}\\
= \underbrace{\onehalf \tau \mden^2 \vphantom{\frac{2 h_1^2}{L^2 \sqtau}}}_\text{from bulk} & \underbrace{-\frac{2 h_1\mden}{L} \vphantom{\frac{2 h_1^2}{L^2 \sqtau}}}_\text{\quad from surface} + \underbrace{\frac{2 h_1^2}{L^2 \tau} + \frac{ h_1^2}{1-\cosh(L\sqrt{\tau})}}_\text{from residual} ,
\end{split}\label{eq_linMF_pf_c}\eeq 
where $\mden = \mass/L$ is the mean mass density within the film.
This result equals $-d\Fcal_f\can/dL$ [Eq.~\eqref{eq_stressten_dFdL}, evaluated at fixed $\mass$] with the canonical film free energy $\Fcal_f\can$ in Eq.~\eqref{eq_filmF_can}.
As discussed above, in the canonical case, we generally consider a bulk system which has the same mean mass density $\mden$ as the film. 
The corresponding bulk pressure $p_b\can$ follows from Eq.~\eqref{eq_pB_can} in linear MFT immediately as 
\beq p_b\ut{(c)}(\tau,\mden) = \frac{1}{2} \tau \mden^2,
\label{eq_linMF_pb_c}\eeq 
which can be equivalently obtained from Eqs.\ \eqref{eq_pB_lim_pF} and \eqref{eq_linMF_pf_c} or, alternatively, from Eq.~\eqref{eq_linMF_pb_gc} by inserting in the latter the constraint-induced bulk field $\tilde \mu_b = \mden \tau$ [see Eq.~\eqref{eq_bulk_EOS_can} for $\tau>0$].
The canonical \CCF for $(++)$ boundary conditions resulting from Eqs.\ \eqref{eq_linMF_pf_c} and \eqref{eq_linMF_pb_c} is therefore given by 
\beq \begin{split} \Kcal\ut{(c)}(\tau,\mass,h_1,L) &= p_f\ut{(c)} - p_b\ut{(c)} \\
&= -\frac{2 h_1\mden}{L} + \frac{2 h_1^2}{L^2 \tau} + \frac{ h_1^2}{1-\cosh(L\sqrt{\tau})},
\end{split}\label{eq_linMF_pCasCan_stressroute}\eeq 
with $\mden=\mass/L$ and can, after introducing the scaling variables in Eqs.~\eqref{eq_scalvar}, be cast into the scaling form given in Eq.~\eqref{eq_Casi_force_c} with $d=4$ and
\beq \Xi\can(\tscal, \Mass, H_1)/\Delta_0 = -2 H_1\Mass + \frac{2 H_1^2}{\tscal} + \frac{H_1^2}{1-\cosh\sqtscal}.
\label{eq_pCasCan_linMF_scalform}\eeq 
Due to the first term, $-2 H_1 \Mass$ [which corresponds the second term in Eq.\ \eqref{eq_linMF_pf_c}], this result differs from Eq.~\eqref{eq_Xicas_can}; the latter equation was obtained based on identifying the residual finite-size free energy in Eq.~\eqref{eq_filmF_can}.
By calculating the derivative with respect to $L$ of the canonical film free energy $\Fcal_f\can$ in Eq.~\eqref{eq_filmF_can}, we find that this additional term originates from the contribution $-2h_1 \mass/L$, which was identified in Eq.~\eqref{eq_filmF_can} as a surface free energy term.
The appearance of such a contribution is a direct consequence of the fact that, in the canonical ensemble, the total mass $\mass$, rather than the density $\mden=\mass/L$, is kept constant.
Indeed, as indicated in Eq.~\eqref{eq_linMF_pf_c}, the first term on the right hand side of Eq.~\eqref{eq_linMF_pCasCan_stressroute} emerges precisely from inserting the constraint-induced bulk field $\tilde \mu$ [Eq.~\eqref{eq_h0}] for $\mu$ into the grand canonical film pressure [Eq.~\eqref{eq_linMF_pf_gc}].
We thus conclude that, in the canonical ensemble, the calculations of the \CCF via the residual finite-size free energy [Eq.~\eqref{eq_pCas_dFdL}] and via the stress tensor approach [Eq.~\eqref{eq_pCas_pdiff}] are not necessarily equivalent, because the subtraction procedure employed in the latter method removes only the bulk contribution to the pressure. 

We finally comment on the fact that, in the canonical ensemble, film and bulk cannot exchange particles.  
As a consequence, and in contrast to the situation in the grand canonical ensemble, the precise value of the bulk pressure $p_b\can$ which is subtracted in Eq.~\eqref{eq_linMF_pCasCan_stressroute} depends on the actual experiment performed.
However, as evidenced by Eq.~\eqref{eq_linMF_pf_c}, for any choice of $p_b\can$ other than the one in Eq.~\eqref{eq_linMF_pb_c}, the critical Casimir contribution may be concealed by a possible residual difference in the bulk pressures between film and environment. 
The present analysis of linear MFT thus provides further support to our choice [see Eqs.~\eqref{eq_pB_lim_pF} and \eqref{eq_pB_can} and the related discussion] of defining the appropriate $p_b\can$ based on the condition that the bulk has the same mass density $\mden$ as the film.

In the case of antisymmetric [$(+-)$] boundary fields, the film pressures resulting within linear MFT from Eqs.~\eqref{eq_pm_sol0_bare_all} and \eqref{eq_pf_landau} are
\begin{subequations}\bal
p_{f,+-}\gc(\tau,\mu,h_1,L) &= \frac{\mu^2}{2\tau} + \frac{H_1^2 }{1+\cosh (L\sqrt {\tau})} \quad \text{and}\\
p_{f,+-}\can(\tau,\mden, h_1,L) &= \onehalf \mden^2 \tau + \frac{H_1^2 }{1+\cosh (L\sqrt {\tau})},
\end{align}\label{eq_pm_linMF_pf_gc}\end{subequations}
which turn out to be identical if the expression $\tilde\mu = \mden \tau$ [Eq.~\eqref{eq_pm_h0}] for the constraint-induced bulk field is inserted.
Upon subtracting the corresponding bulk pressures [Eqs.~\eqref{eq_linMF_pb_gc} and \eqref{eq_linMF_pb_c}] we find the same scaling functions for the \CCF as those reported in Eq.~\eqref{eq_pm_Xicas_linMF}, which have been inferred from taking the derivative of the residual free energy.

The mean-field expressions for the \CCFs given in Eqs.\ \eqref{eq_linMF_pCasGC_stressroute}, \eqref{eq_linMF_pCasCan_stressroute}, and \eqref{eq_pm_Xicas_linMF} have been obtained from a perturbation theory constructed around an OP profile with vanishing mean value ($\mden=0$) and are therefore applicable only to rather small values of $\mden$.
Expressions valid for large $|\mden|$ are derived in Appendix \ref{app_pert} and indicate that the dominant mass dependence for large $|\mden|$ can be accounted for by replacing in the expressions in Eqs.\ \eqref{eq_linMF_pCasGC_stressroute}, \eqref{eq_linMF_pCasCan_stressroute}, and \eqref{eq_pm_Xicas_linMF} the reduced temperature $\tau$ by an effective one:
\beq \mtau = \tau+\onehalf g\mden^2.
\label{eq_mod_tau}\eeq 
In the grand canonical ensemble, the mass density $\mden$ in Eq.~\eqref{eq_mod_tau} is to be understood as a function of the given bulk field $\mu$ and can be determined by evaluating the mass constraint [see Eq.~\eqref{eq1_B_rel_mass}].

\subsubsection{Nonlinear MFT}
\label{sec_Casimir_full_mft}
We now turn to the \CCF which arises from the Ginzburg-Landau model [Eq.~\eqref{eq_Landau_func_gc}] including the quartic interaction term proportional to the coupling constant $g$.
The \CCF [Eq.~\eqref{eq_pCas_pdiff}] in the grand canonical and the canonical ensemble exhibits the general scaling forms given in Eqs.~\eqref{eq_Casi_force_gc} and \eqref{eq_Casi_force_c}, respectively.
The mean field results discussed below exhibit these scaling properties (with $d=4$) but carry an undetermined prefactor $g^{-1}$ which, as before, will be accounted for by an appropriate normalization.
Within nonlinear MFT, closed analytical expressions for the scaling functions $\Xi\ut{(gc,c)}$ for finite $H_1$ and arbitrary bulk field $B$ are not available. 
We thus solve the nonlinear Ginzburg-Landau model for the OP profile numerically by integrating the ELE in Eqs.\ \eqref{eq_ELE} and \eqref{eq_ELE_BC} as well as via conjugate-gradient minimization of the free energy functional in Eq.~\eqref{eq_Landau_func_ndim}. 
As before, the mass constraint is imposed by determining, for each value of the surface field $H_1$, the bulk field $B$ necessary to recover the prescribed total mass $\Mass$ in the film.
The \CCF is obtained from the stress tensor [Eq.~\eqref{eq_stressten_can}] as the difference between the film and the bulk pressures [Eq.~\eqref{eq_pCas_pdiff}], where the bulk pressure is computed for a homogeneous bulk system according to Eqs.~\eqref{eq_pB_gc} and \eqref{eq_pB_can}.
Equivalently, the bulk pressure may also be obtained simply as the pressure of a film in the limit of a macroscopically large thickness $L$. 
We mention that close to the boundaries, where the OP profiles rapidly increase upon approaching the surfaces, the numerical accuracy of the solution (and, in particular, its derivative) typically deteriorates.
As a consequence, and contrary to the expectation, the film pressure computed numerically from Eq.~\eqref{eq_pf_landau} is in general not fully constant across the film, but only sufficiently far from the boundaries.

Here, we mention two possibilities to overcome this problem.
One option is to solve the ELE via a so-called symplectic integration method \cite{ruth_canonical_1983, hairer_geometric_2010}, which, by construction, yields a constant pressure.
To this end one introduces the ``momentum'' $p\equiv \pd\Lcal/\pd m' = m'$ conjugate to the OP $m$, where the prime denotes a derivative with respect to $\zeta=z/L$ and $\Lcal= (m')^2/2 + \tscal m^2/2 + m^4/4 - B m$ is the bulk free energy density, i.e., the integrand in the Ginzburg-Landau functional in Eq.~\eqref{eq_Landau_func_ndim}.
In this formulation the one-dimensional ELE in Eq.~\eqref{eq_ELE} is equivalent to the Hamiltonian ``equations of motion'' $m' = \pd \Hcal/\pd p$, $p' = -\pd \Hcal/\pd m$, where the bulk Hamiltonian density $\Hcal(m,p) = p m' - \Lcal(m, m') = p^2/2 - \tscal m^2/2 - m^4/4 + B m$ is the Legendre transform of $\Lcal$.
The Hamiltonian $\Hcal$ coincides with the dimensionless, rescaled film pressure $(L^4/\Delta_0)p_f$ [Eq.~\eqref{eq_pf_landau}] and is, as the latter, conserved for a solution of the Hamiltonian equations of motion [the conservation of the ``energy'' $\Hcal$ also follows from Noether's theorem due to the absence of an explicit dependence on $\zeta$ of the bulk free energy density in Eq.~\eqref{eq_Landau_func_ndim}]. 
Numerically solving the Hamiltonian equations of motion via a symplectic integration method guarantees the conservation of $\Hcal$ and allows us to directly obtain the film pressure $p_f$ as $p_f=(\Delta_0/L^4)\Hcal(m,p)$. This approach avoids the numerically inaccurate calculation of the derivative $m'$. 
If, instead, the OP profile is obtained via a direct numerical integration of the ELE in Eq.~\eqref{eq_ELE} or by the minimization of the functional in Eq.~\eqref{eq_Landau_func_gc}, one can still obtain reliable results for the film pressure provided the latter is computed as an average over only a small interval near the center of the film where the OP profile is approximately constant and $m'$ is small.
We have found that both approaches yield similar results, except for $(+-)$ boundary conditions and large values of the bulk field, for which a symplectic integration method has to be used in order to accurately capture the film pressure.

\begin{figure*}[t]\centering
    \subfigure[]{\includegraphics[width=0.45\linewidth]{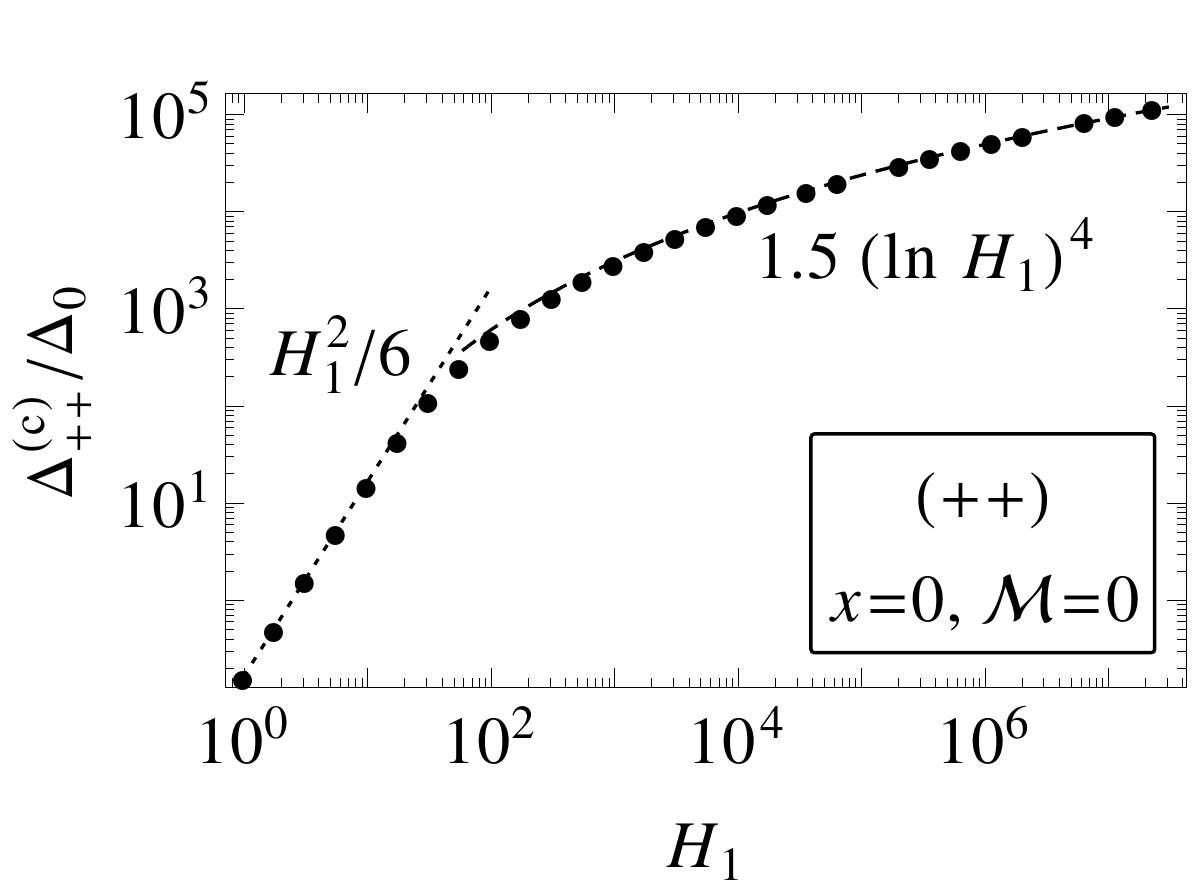}}\qquad
    \subfigure[]{\includegraphics[width=0.47\linewidth]{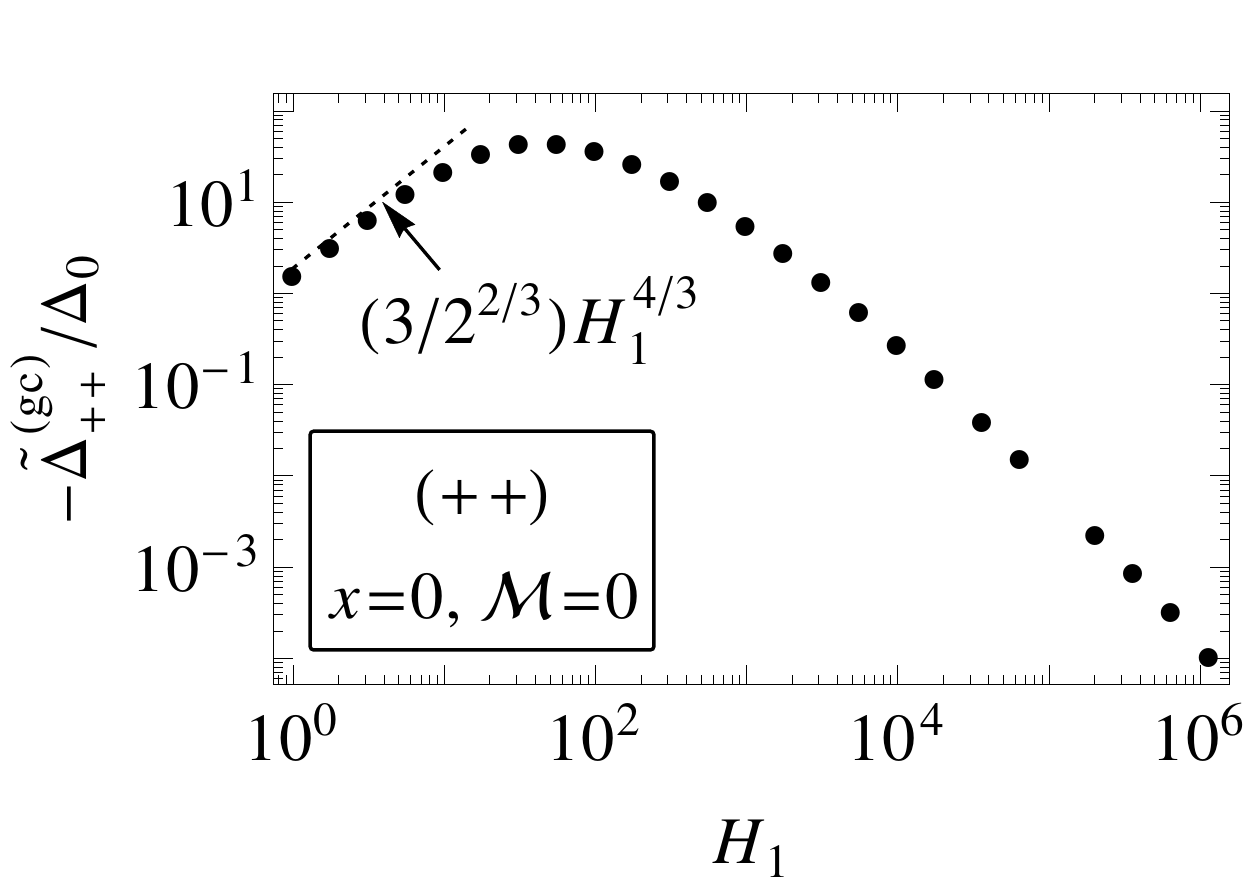}}
    \subfigure[]{\includegraphics[width=0.475\linewidth]{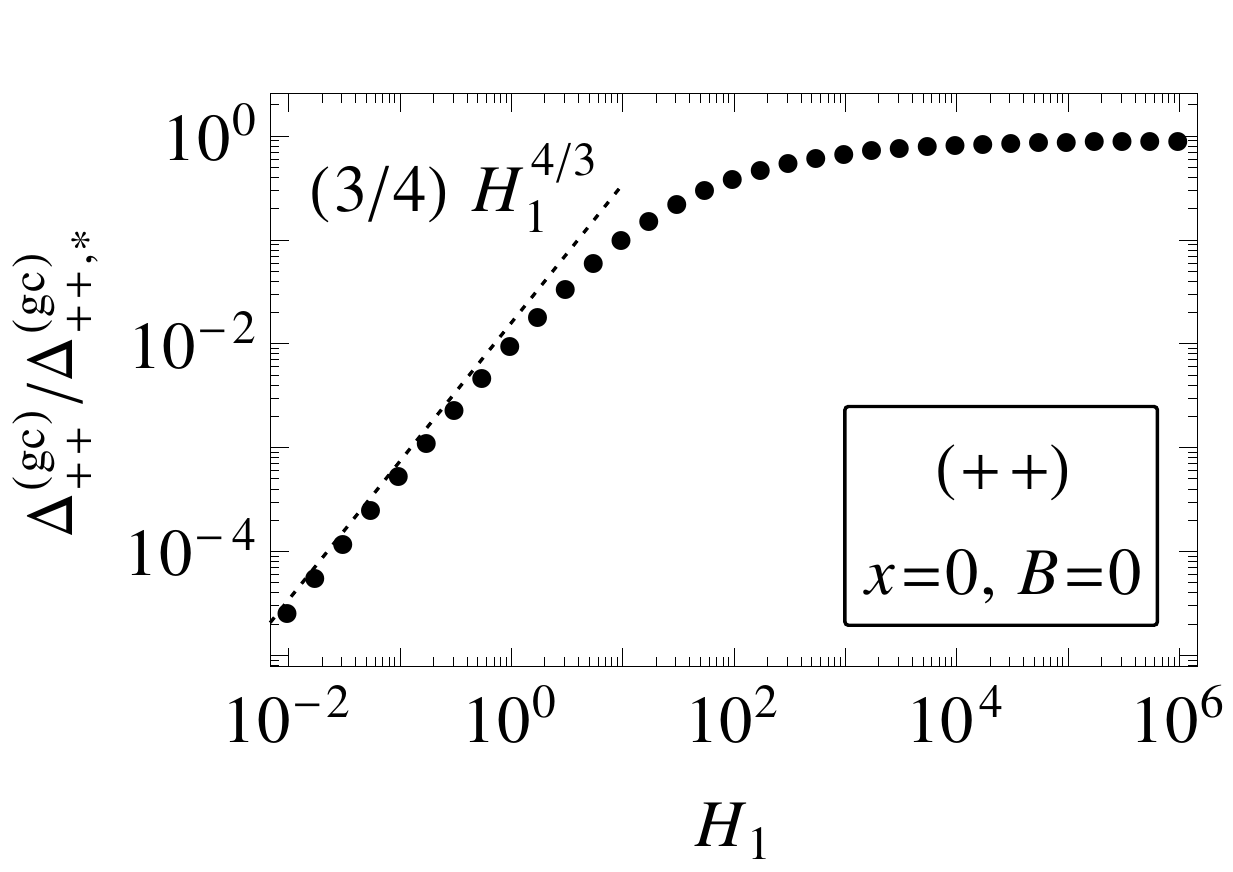}}\qquad
    \subfigure[]{\includegraphics[width=0.47\linewidth]{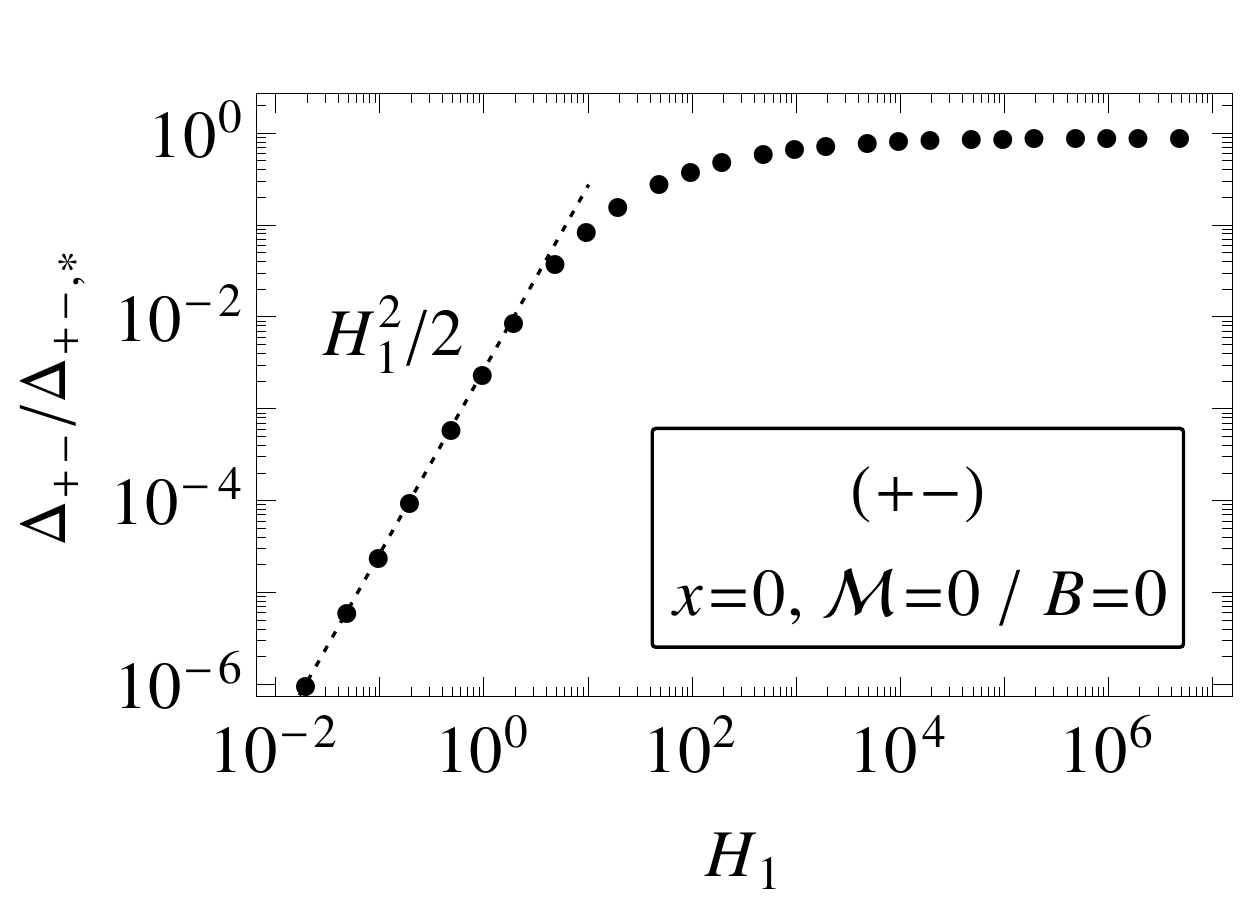}}
    \caption{Dependence of the critical Casimir amplitudes $\Delta_{++}\can$, $\tilde \Delta_{++}\gc$, $\Delta_{++}\gc$, and $\Delta_{+-}=\Delta_{+-}\cgc=\tilde \Delta_{+-}\cgc$ [Eqs.~\eqref{eq_cas_ampl} and \eqref{eq_cas_ampl_pp_gc}] on the adsorption strength $H_1$ for a film at bulk criticality $\tscal=0$ and for vanishing imposed total mass $\Mass=0$ [except (c)] within nonlinear MFT. The amplitudes in panels (a) and (b) are normalized by the mean-field amplitude $\Delta_0$ defined in Eq.~\eqref{eq_Delta0}, while the amplitudes in panels (c) and (d) are normalized by their limits $\Delta_{++,*}$ and $\Delta_{+-,*}$ for $H_1\to\infty$ given in Eqs.~\eqref{eq_cas_ampl_pp_gc_lim} and \eqref{eq_cas_ampl_pm_lim}, respectively. The symbols represent the data obtained by numerically solving the ELE in Eq.\ \eqref{eq_ELE}. In panel (b), the grand canonical amplitude $\tilde \Delta_{++}\gc$ is computed with a bulk field $B(H_1)$ chosen such as to satisfy the constraint $\Mass=0$, and has to be distinguished from the amplitude $\Delta_{++}\gc$ [Eq.~\eqref{eq_cas_ampl_pp_gc}] defined at a fixed bulk field $B=0$, which is shown in (c). The dashed and dotted lines represent the analytical predictions of Eqs.~\eqref{eq_pCasCan_H1scale}, \eqref{eq_pCasGC_M0_H1scale}, \eqref{eq_pCasGC_B0_H1scale}, and \eqref{eq_pm_Xicas_Tc}, as indicated by the corresponding labels. Panels (a) and (b) show that, for $(++)$ boundary conditions, the limit $H_1\to \infty$ is ill-defined in the presence of a mass constraint $\Mass=0$, in contrast to the situation with a fixed bulk field $B=0$ [panel (c)], for which the critical Casimir amplitude attains a finite limit for $H_1\to\infty$ \cite{krech_casimir_1997}. For $(+-)$ boundary conditions, as illustrated in panel (d), the limit $H_1\to\infty$ is well defined, too, both in the canonical and in the grand canonical case.} 
    \label{fig_casAmpl}
\end{figure*}

We first study the \emph{amplitudes} of the \CCFs, defined by [compare Eqs.\ \eqref{eq_Casi_force_gc}, \eqref{eq_Casi_force_gc_scalf}, \eqref{eq_Casi_force_c}, and \eqref{eq_Casi_force_c_scalf}]
\begin{subequations}\bal 
	\Delta_{++}\can(H_1) &\equiv \Xi\can_{++}(\tscal=0,\Mass=0, H_1), \label{eq_cas_ampl_can}\\
	\tilde \Delta_{++}\gc(H_1) &\equiv \Xi\gc_{++}(\tscal=0,B=\tilde B(H_1), H_1),\label{eq_cas_ampl_gc_constr}\\
	\Delta_{+-}\can(H_1) &\equiv \Xi\can_{+-}(\tscal=0,\Mass=0, H_1),\label{eq_cas_ampl_pm}\\
	\tilde \Delta\gc_{+-}(H_1) &\equiv \Xi\gc_{+-}(\tscal=0,B=\tilde B(H_1), H_1)=\Delta_{+-}\can(H_1).\label{eq_cas_ampl_pm_gc_constr}	
\end{align}\label{eq_cas_ampl}\end{subequations}
In Eqs.\ \eqref{eq_cas_ampl_gc_constr} and \eqref{eq_cas_ampl_pm_gc_constr}, $B$ is chosen as a function $\tilde B(H_1)$ such that, at $\tscal=0$, one has $\Mass=0$.
For $(+-)$ boundary conditions, this is realized for $B=0$ independently of $H_1$, so that $\Delta_{+-}\can = \tilde \Delta_{+-}\gc$.
Here, the distinction of being canonical or grand canonical refers to the nature of the coupling between film and bulk: in the canonical case, film and bulk are taken to have the same mean mass density $\mden$, while in the grand canonical case they have the same chemical potential $\mu$ [see also the discussion related to Eq.~\eqref{eq_pB_can}]. 
We therefore use a tilde in Eq.~\eqref{eq_cas_ampl_gc_constr} in order to distinguish $\tilde \Delta_{++}\gc$ from the more common grand canonical critical Casimir amplitude 
\beq \Delta_{++}\gc(H_1)\equiv \Xi\gc_{++}(\tscal=0,B=0,H_1)
\label{eq_cas_ampl_pp_gc}\eeq 
defined at a fixed bulk field $B=0$.
Note that the amplitudes defined in Eq.~\eqref{eq_cas_ampl} carry a dependence on $H_1$. 
As discussed below, for $(++)$ boundary conditions within MFT, the constraint $\Mass=0$ causes the limit $H_1\to\infty$ to be ill-defined (see also Sec.~\ref{sec_ads_num}).
For $(+-)$ boundary conditions, the limit $H_1\to \infty$ is well-defined within MFT even in the presence of a mass constraint $\Mass=0$.
In the grand canonical ensemble, the critical Casimir amplitudes $\Delta_{++}\gc$ and $\Delta_{+-}\gc$ are finite in the limit $H_1\to\infty$ and are known analytically \cite{krech_casimir_1997}:
\begin{widetext}
\begin{subequations}\bal 
\Delta_{++,*}\gc &\equiv \Delta_{++}\gc(H_1\to\infty) =  \Xi_{++}\gc(\tscal=0,B=0, H_1\to\infty)=-4\Delta_0  K^4(1/\sqrt{2})\simeq -47.3 \Delta_0,
\label{eq_cas_ampl_pp_gc_lim}\\
\Delta_{+-,*}\gc &\equiv \Delta_{+-}\gc(H_1\to\infty) = \Xi_{+-}\gc(\tscal=0,B=0, H_1\to\infty)=-4 \Delta_{++,*}\gc \simeq 189 \Delta_0,
\label{eq_cas_ampl_pm_gc_lim}\\ 
\Delta_{+-,*}\can &\equiv \Delta_{+-}\can(H_1\to\infty) = \Xi_{+-}\can(\tscal=0,\Mass=0,H_1\to\infty) = \Delta_{+-,*}\gc \equiv \Delta_{+-,*}\,,\label{eq_cas_ampl_pm_lim}
\end{align}\label{eq_cas_ampl_limits}\end{subequations}
\end{widetext}
where $K$ is the complete elliptic integral of the first kind \cite{gradshteyn_table_2014} and $\Delta_0$ is defined in Eq.~\eqref{eq_Delta0}.
Equation \eqref{eq_cas_ampl_pp_gc_lim} allows one to express the amplitude $\Delta_0$ used in the present study in terms of $|\Delta_{++}|\equiv|\Delta_{++,*}\gc|$ which has been used previously and which, beyond MFT, is a universal number.
As a check, we verified that in terms of $g$ these values of $\Delta_{++,*}\gc$ and $\Delta_{+-,*}$ are accurately recovered by our numerical solutions.
Within linear MFT discussed in the previous subsection, one has $\Delta_{++}\can = \Delta_0 H_1^2/6$ [Eq.~\eqref{eq_Xicas_can_Tc}] and $\Delta_{+-}\can=\Delta_0 H_1^2/2$ [Eq.~\eqref{eq_pm_Xicas_Tc}], whereas the grand canonical amplitude $\Delta_{++}\gc$ diverges for $\tscal\to 0$ [see Eq.~\eqref{eq_Xicas_gc}].
Importantly, within nonlinear MFT, the \CCF remains finite for $\tscal\to 0$ in both ensembles, justifying the definitions in Eq.~\eqref{eq_cas_ampl}.

Figure \ref{fig_casAmpl} shows the amplitudes of the \CCF defined in Eq.~\eqref{eq_cas_ampl} as a function of the scaled surface field $H_1$ within nonlinear MFT, as obtained from the numerical solution.
Considering first $(++)$ boundary conditions [panels (a) and (b)], we observe that, consistently with linear MFT [see Fig.~\ref{fig_cas_limMF_tau}(b)], the canonical \CCF is \emph{repulsive}, whereas the grand canonical \CCF, under the same condition of $\Mass=0$, is \emph{attractive}.
Furthermore, we have seen previously [see Eq.~\eqref{eq_OPscale_largeH1}] that, for $(++)$ boundary conditions and within MFT, the value of the OP at the center of the film diverges logarithmically for $H_1\to\infty$, as a consequence of imposing a certain value of the mass $\Mass$. 
As shown in Fig.~\ref{fig_casAmpl}(a) for the case $\Mass=0$, also the amplitude $\Delta_{++}\can$ of the canonical \CCF diverges logarithmically upon increasing $H_1$, implying that also for this quantity the limit $H_1\to \infty$ is ill-defined. 
The amplitude $\tilde \Delta_{++}\gc$ [Fig.~\ref{fig_casAmpl}(b)], obtained by imposing the same chemical potential both in the film and the bulk under the additional constraint $\Mass=0$ in the film, depends non-monotonically on $H_1$ and vanishes in the limit $H_1\to \infty$.
The observed limiting behaviors can be quantitatively understood from the known behavior of the OP at the center of the film [Eqs.~\eqref{eq_OPscale_smallH1} and \eqref{eq_OPscale_largeH1}] and of the constraint-induced field $\tilde B$ [Eqs.~\eqref{eq_Bconstr_smallH1} and \eqref{eq_Bconstr_largeH1}]:
since the stress is constant across the film (and therefore can be conveniently evaluated at its center, where $m'=0$) and $p_b\can=0$ in the present case [$\phi_b=0$ in Eq.~\eqref{eq_pB_can} for $\tau=\mden=0$], from Eq.~\eqref{eq_pf_landau} one asymptotically finds in the canonical ensemble for $\tscal=\Mass=0$
\beq\begin{split} \Delta_{++}\can/\Delta_0 &= \frac{L^4}{\Delta_0} \left(p_f\can-p_b\can\right) =  -\onequarter m^4(0) + \tilde B m(0) \\ &\simeq
\begin{cases}
\frac{1}{6}H_1^2,      & \quad \text{for } H_1 \to 0, \\
\frac{3}{4}s^4\:(\ln H_1)^4,   & \quad \text{for }  H_1\to\infty,
\end{cases}\end{split}
\label{eq_pCasCan_H1scale}\eeq 
in agreement with the numerical data shown in Fig.~\ref{fig_casAmpl}(a). 
The numerical constant $s\simeq 1.2$ is defined in Eq.~\eqref{eq_OPscale_largeH1}, where it has been estimated from a fit. The above prediction for $H_1\to 0$ is consistent with Eq.~\eqref{eq_Xicas_can_Tc}, confirming that linear MFT provides a reliable approximation of the full mean field behavior of the \CCF in the canonical ensemble for sufficiently small values of $\tscal$ and $\Mass$.
In the grand canonical case under the condition $\Mass=0$ [Fig.~\ref{fig_casAmpl}(b)], the chemical potential $\tilde\mu = (\Delta_0^{1/2}/L^3)\tilde B(\Mass=0)$ [Eq.~\eqref{eq_scalvar_mft}] of the film determines the pressure $p_b\gc$ [Eq.~\eqref{eq_pB_gc}] via the bulk equation of state [Eq.~\eqref{eq_bulk_EOS_gc}]. Asymptotically, one finds
\beq\begin{split} \frac{L^4}{\Delta_0} p_b\gc\Big|_{\Mass=0} &\simeq \frac{3}{4} \tilde B^{4/3} \\ &\simeq 
\begin{cases}
	(3/2^{2/3})H_1^{4/3},      & \quad \text{for } H_1 \to 0, \\
	(\sfrac{3}{4})s^{4}\:(\ln H_1)^4,   & \quad \text{for } H_1\to\infty,
\end{cases}\end{split}
\label{eq_pBGC_H1scale}\eeq 
where Eqs.~\eqref{eq_Bconstr_smallH1} and \eqref{eq_Bconstr_largeH1} have been used, respectively.
Since, under the same thermodynamic conditions, one generally has $p_f\can=p_f\gc$  [Eq.~\eqref{eq_pf_equal}], in the case $\Mass=0$ the limiting behaviors obtained in Eq.~\eqref{eq_pCasCan_H1scale} apply to $p_f\gc$ as well.
This leads to the following asymptotic behaviors of the grand canonical \CCF for $\tscal=\Mass=0$:
\beq\begin{split} \tilde \Delta_{++}\gc/\Delta_0 &= \frac{L^{4}}{\Delta_0}  \left(p_f\gc-p_b\gc\right) \\ &\simeq
\begin{cases}
-(3/2^{2/3})H_1^{4/3},      & \quad \text{for } H_1 \to 0, \\
0,   & \quad \text{for }  H_1\to\infty,
\end{cases}\end{split}
\label{eq_pCasGC_M0_H1scale}\eeq
which is confirmed by the numerical data in Fig.~\ref{fig_casAmpl}(b).
We mention that the analysis of the numerical data in Fig.~\ref{fig_casAmpl}(b) further reveals that $\tilde \Delta_{++}\gc(H_1\to\infty)$ vanishes algebraically as $\tilde \Delta_{++}\gc\propto -H_1^{-2}$.
In the grand canonical case, for a fixed bulk field $B=0$ in Fig.\ \ref{fig_casAmpl}(c) the critical Casimir amplitude $\Delta_{++}\gc$ is shown as function of $H_1$. As expected, the analytically predicted asymptotic mean-field value $\Delta_{++,*}\gc$ is approached for $H_1\to\infty$. 
In contrast to the constrained case, linear MFT is not applicable in the unconstrained situation for $(++)$ boundary conditions because the corresponding OP profile [Eq.\ \eqref{eq_m0_sol}] diverges at $\tscal=0$.  
Therefore the behavior exhibited in Fig.\ \ref{fig_casAmpl}(c) corresponds to a nonlinear effect even for small $H_1$.
In order to rationalize the scaling behavior in this limit, we note that [compare Fig.\ \ref{fig_prof_H1range}(a)] for small $H_1$ the OP profile is approximately constant across the film. Hence, one may argue \cite{mohry_crossover_2010} that the surface field acts similarly to a bulk field, such that the value of the OP at the center of the film is given by $m^3(0)\sim H_1$ in accordance with Eq.\ \eqref{eq_ELE}. 
Since $p_b\gc=0$ for $\tau=\mu=0$ [see Eqs.\ \eqref{eq_pB_gc} and \eqref{eq_bulk_EOS_gc}], Eqs.\ \eqref{eq_pCas_pdiff} and \eqref{eq_pf_landau} yield the scaling behavior of the grand canonical critical Casimir amplitude for $\tscal=B=0$ and $H_1\to 0$ \cite{mohry_crossover_2010}:
\beq \Delta_{++}\gc/\Delta_0 \simeq \frac{3}{4} H_1^{4/3},
\label{eq_pCasGC_B0_H1scale}\eeq 
in agreement with the data in Fig.\ \ref{fig_casAmpl}(c).
In the case of $(+-)$ boundary conditions, the canonical critical Casimir amplitude $\Delta_{+-}\can$, which coincides with its grand canonical counterpart studied in Ref.~\cite{mohry_crossover_2010}, is shown in Fig.~\ref{fig_casAmpl}(d) as a function of $H_1$ for $\Mass=0$. For $H_1\lesssim 10$, $\Delta_{+-}\can$ follows the prediction of linear MFT in Eq.~\eqref{eq_pm_Xicas_Tc}, while for $H_1\to \infty$ it approaches the limit $\Delta_{+-,*}\can$ given in Eq.~\eqref{eq_cas_ampl_pm_lim}.

Since, as discussed above, the scaling of the \CCF, under the mass constraint $\Mass=\const$ for $(++)$ boundary conditions and large $H_1$, is essentially a consequence of the characteristic shape of the OP profile in that limit, it is expected that the scaling behavior of the amplitudes exhibited in Figs.~\ref{fig_casAmpl}(a) and (b) will be obtained asymptotically in the limit $H_1\to \infty$ for any nonzero and finite value of $\Mass$.
We thus conclude that, for $(++)$ boundary conditions and within MFT, the presence of a mass constraint $\Mass=\const$\ introduces a nontrivial dependence of the amplitude of the \CCF on the surface adsorption strength $H_1$, rendering the limit $H_1\to \infty$ to be ill-defined in this case.

\begin{figure*}[t]\centering
	\subfigure[]{\includegraphics[width=0.48\linewidth]{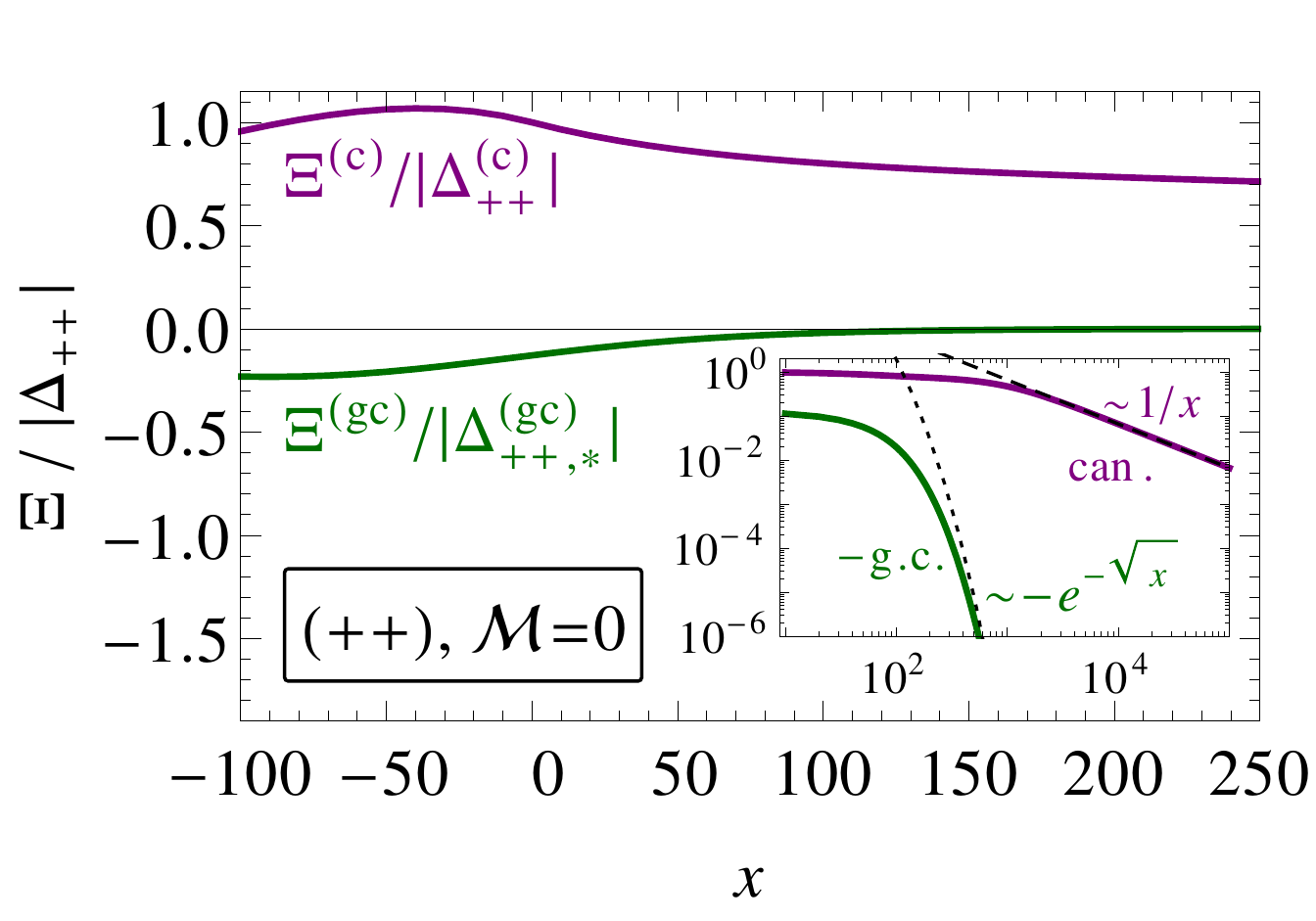}}\qquad
	\subfigure[]{\includegraphics[width=0.46\linewidth]{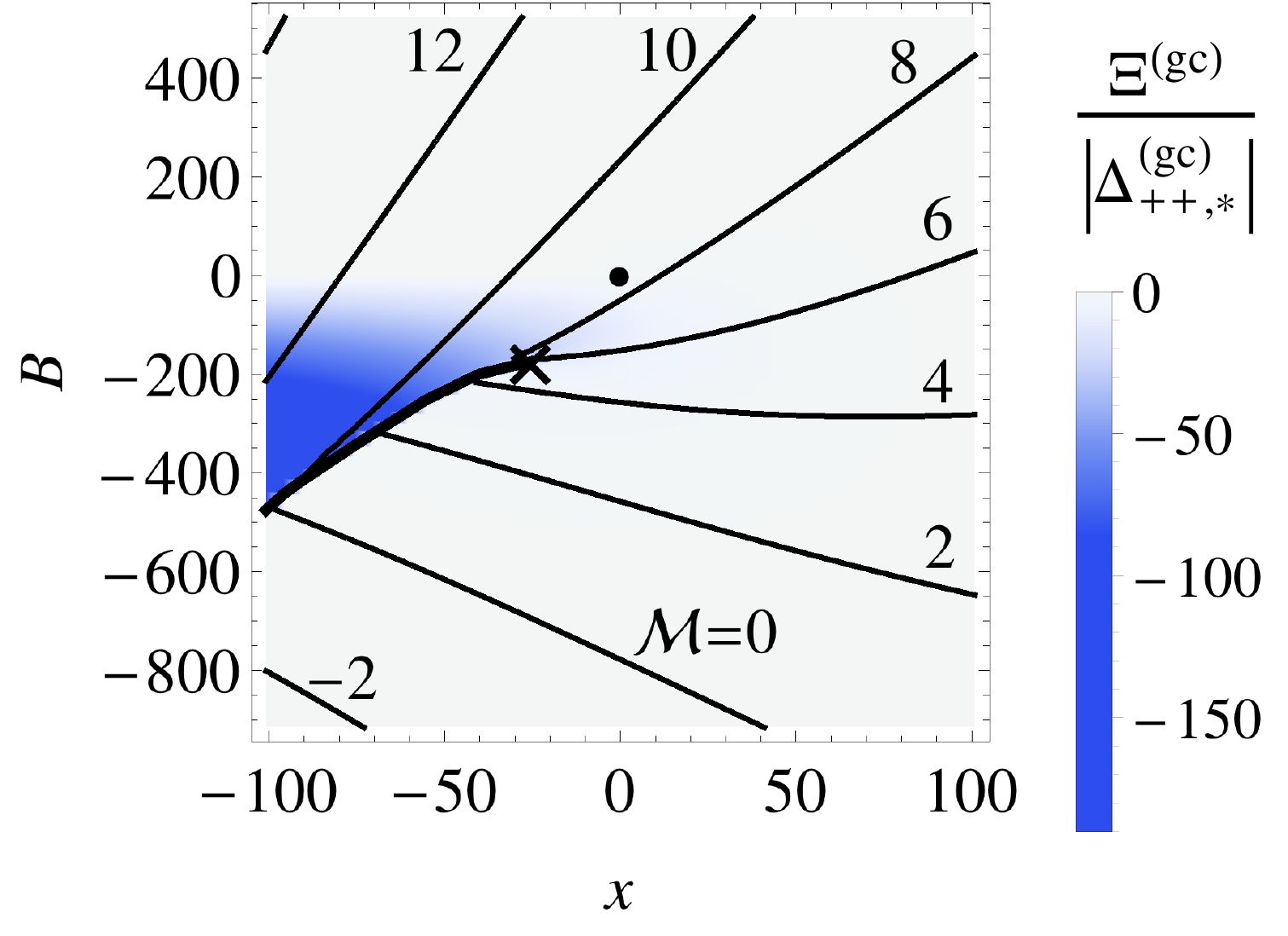}}
	\subfigure[]{\includegraphics[width=0.5\linewidth]{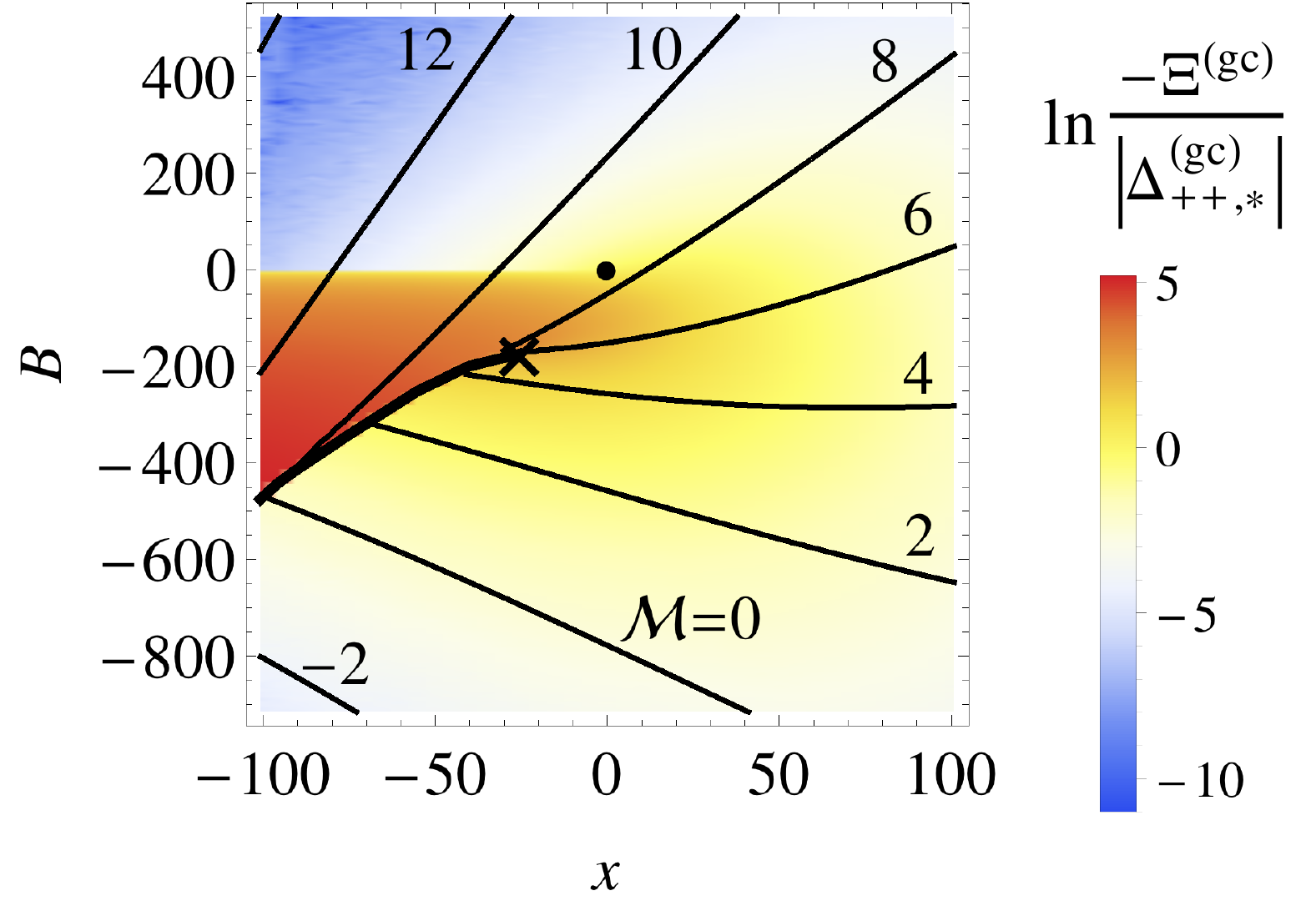}}\qquad
	\subfigure[]{\includegraphics[width=0.46\linewidth]{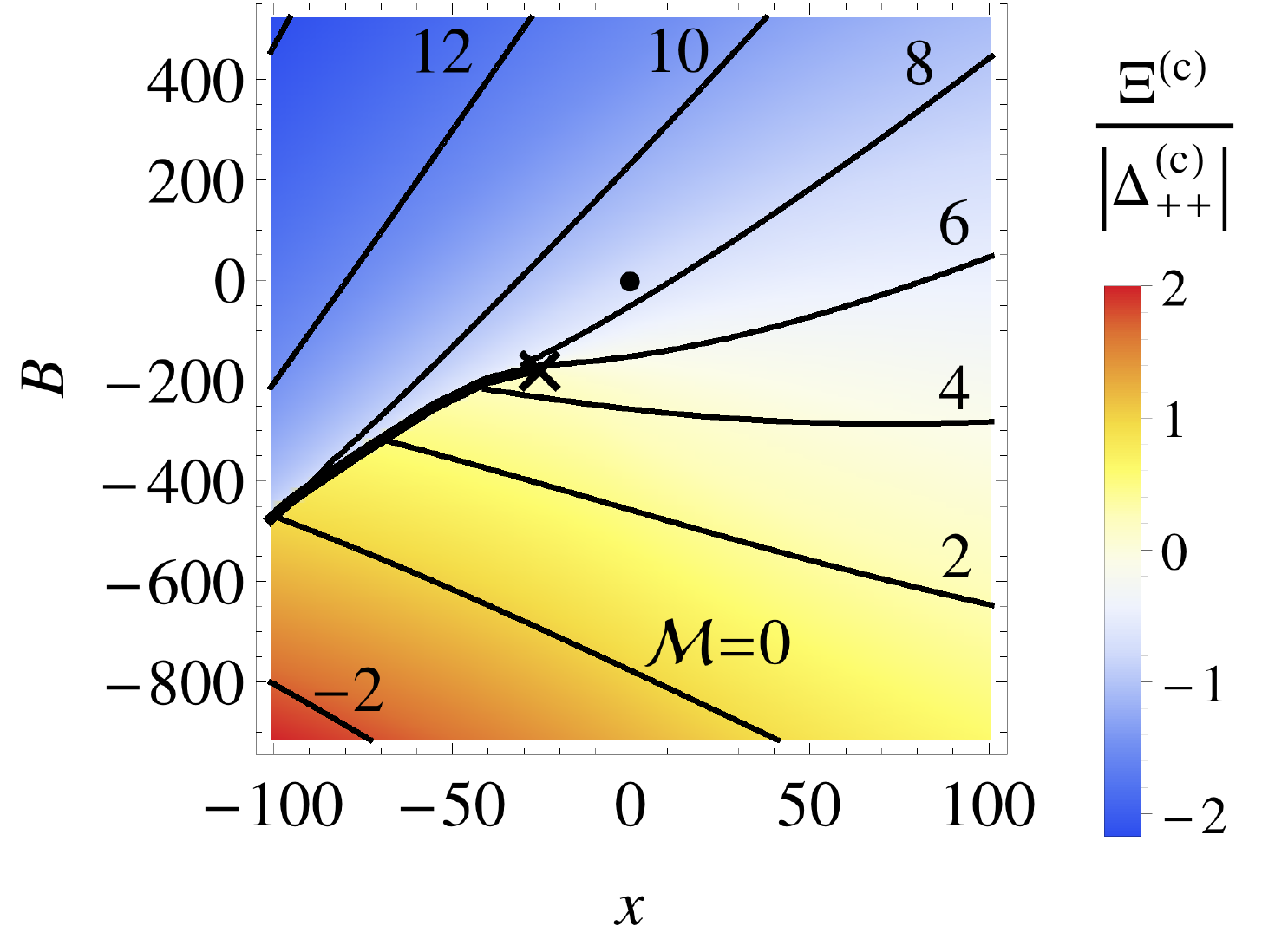}}
	\caption{Critical Casimir force in a film with $(++)$ boundary conditions in the grand canonical and the canonical ensemble within nonlinear MFT, computed according to Eq.~\eqref{eq_pCas_pdiff} via the stress tensor and the numerically determined OP profiles. (a) Dependence of the (differently) normalized scaling functions $\Xi\ut{(c,gc)}$ of the \CCF [see Eqs.~\eqref{eq_Casi_force_gc} and \eqref{eq_Casi_force_c}] on the scaled temperature $\tscal=(L/\amplXip)^{1/\nu}t$ for vanishing mass $\Mass=0$. The inset shows the normalized scaling functions (with a minus sign for $\Xi\gc$, i.e., $-\Xi\gc$) on a double-logarithmic scale together with the predictions of linear MFT [Eqs.~\eqref{eq_Xicas_gc} and \eqref{eq_Xicas_can}, corresponding to the dotted and the dashed line, respectively]; their asymptotic decay is given by Eq.~\eqref{eq_Xicas_linMF_asympt}. (b), (c), and (d) Dependence of the \CCF scaling functions on the scaled temperature $\tscal$ and on the scaled bulk field $B$. Curves of constant total mass $\Mass$ (compare Fig.~\ref{fig_phasediag_MFT}) are indicated by the black isolines labeled by the corresponding value of $\Mass$. Note that the grand canonical \CCF is negative over the whole domain plotted in panel (b), but the linear color scale is insufficient to resolve its strong variation. Therefore, panel (c) shows the same data on a logarithmic scale. All scaling functions are normalized by the absolute value of the appropriate amplitude, i.e., $\Delta_{++}\can$ [Eq.~\eqref{eq_cas_ampl_can}] and $\Delta_{++,*}\gc$ [Eq.~\eqref{eq_cas_ampl_pp_gc_lim}], respectively. The locations of the bulk and the capillary critical point in panels (b)-(d) are indicated by a dot ($\bullet$) and a cross ($\times$), respectively, while the thick line ending at the cross is the line of first-order capillary condensation transitions [see also Fig.~\ref{fig_phasediag_MFT}(a)]. In (a), $H_1=1000$ has been used and $H_1\simeq 5100$ in (b)-(d), representing the strong adsorption regime.}
	\label{fig_pCas_MFT}
\end{figure*}

Figure \ref{fig_pCas_MFT}(a) shows the scaled temperature dependence of the scaling functions of the \CCF computed from Eqs.~\eqref{eq_pCas_pdiff}, \eqref{eq_Casi_force_gc} and \eqref{eq_Casi_force_c} within nonlinear MFT in a film with $(++)$ boundary conditions, $\Mass=0$, and a fixed value of the surface field $H_1$. 
As before, in the canonical ensemble, film and bulk are assumed to have the same mean mass density (here $\mden=0$), while in the grand canonical ensemble, they have the same chemical potential.
In the case of the grand canonical ensemble, the constraint $\Mass=0$ is achieved for a certain field $B=\tilde B(\tscal,H_1)$ [see Figs.\ \ref{fig_pCas_MFT}(b)-(d)].
We find that the repulsive character of the canonical \CCF and the attractive one of the grand canonical \CCF persists for all values of $\tscal$ considered here.
In agreement with linear MFT [see Fig.~\ref{fig_cas_limMF_tau}(b) and Eq.~\eqref{eq_Xicas_can_asympt}], the canonical \CCF decays as $\Xi\can(\tscal\gg 1) \simeq 2\Delta_0 H_1^2/\tscal$ for large $\tscal$ (see the inset), while the grand canonical \CCF vanishes asymptotically as $\Xi\gc(\tscal\gg 1)\simeq -2\Delta_0 H_1^2 \exp(-\sqtscal)$ [Eq.~\eqref{eq_Xicas_gc_asympt}].
We remark here that for $B=0$, large $H_1$, and intermediate values of $\tscal$, our numerical solutions follow, as a function of $\tscal$, the asymptotic prediction given in Eq.\ \eqref{eq_Xicas_nlin_asympt_pp} which in turn has been derived for $H_1\to\infty$ within nonlinear MFT in Ref.~\cite{krech_casimir_1997}. For any finite $H_1$, however, the ultimate decay of $\Xi\gc$ for large $\tscal$ is governed by the solution of the linear MFT.
In contrast to the linear case, the grand canonical \CCF remains finite at criticality.
At $\tscal\simeq -100$, the capillary condensation line is reached for the film setup considered in Fig.~\ref{fig_pCas_MFT}(a) [compare panels (b) and (c) therein].
Note that $\Xi\gc$ is normalized by $|\Delta_{++,*}\gc|$ but $\Xi\can$ is normalized by $|\Delta_{++}\can|$ because $\Delta\can_{++}(H_1\to\infty)$ diverges [Fig.\ \ref{fig_casAmpl}(a)] so that $\Delta_{++,*}\can=\infty$.

In Figs.~\ref{fig_pCas_MFT}(b)-(d), the \CCF within nonlinear MFT is displayed for $(++)$ boundary conditions and for various values of the total mass $\Mass$. 
In order to better visualize the strong variation of the grand canonical \CCF, panel (c) shows the same data as panel (b) but on a logarithmic color scale. 
While the grand canonical \CCF $\Kcal\gc$ is sizable only within a small region below the bulk critical point [see panels (b) and (c) and compare Refs.~\cite{vasilyev_critical_2013, mohry_critical_2014, dantchev_exact_2016}], the canonical \CCF $\Kcal\can$ has a significant strength across the whole phase diagram, decaying only rather slowly for large scaled temperatures $\tscal$ [see panel (d)].
In addition, Figs.~\ref{fig_pCas_MFT}(b)-(d) explicitly demonstrate that the film pressure varies discontinuously upon crossing the capillary phase coexistence curve (thick line ending at a cross $\times$).
Within nonlinear MFT, $\Kcal\gc$ acquires a dependence on the bulk field, but remains attractive over the entire phase diagram [panel (b)]. 
By contrast, as illustrated in panel (d), the canonical \CCF $\Kcal\can$, as defined by Eq.~\eqref{eq_pCas_pdiff}, changes its sign from being repulsive ($\Mass\lesssim 5$) to attractive ($\Mass\gtrsim 5$), depending on the total mass $\Mass$ (indicated by the isolines). 
Remarkably, the decrease of the singular canonical \CCF upon increasing the mean density $\Mass$ in the film is opposite to the commonly experienced behavior of a fluid, the pressure of which increases upon increasing the density. (One should keep in mind, however, that the total pressure is the sum of an analytic background contribution and of the singular contribution considered here.)

\begin{figure*}[t]\centering
	\subfigure[]{\includegraphics[width=0.45\linewidth]{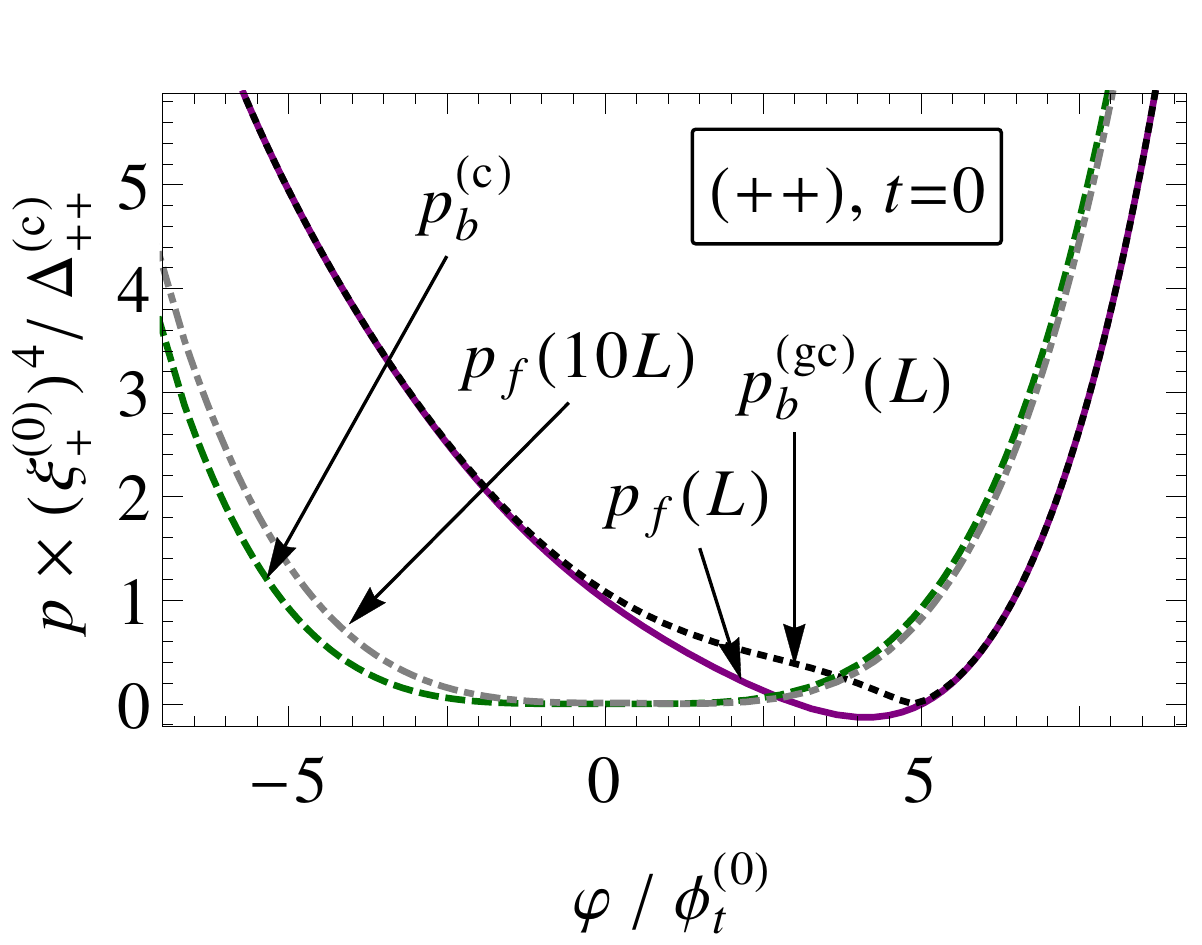}}\qquad
	\subfigure[]{\includegraphics[width=0.465\linewidth]{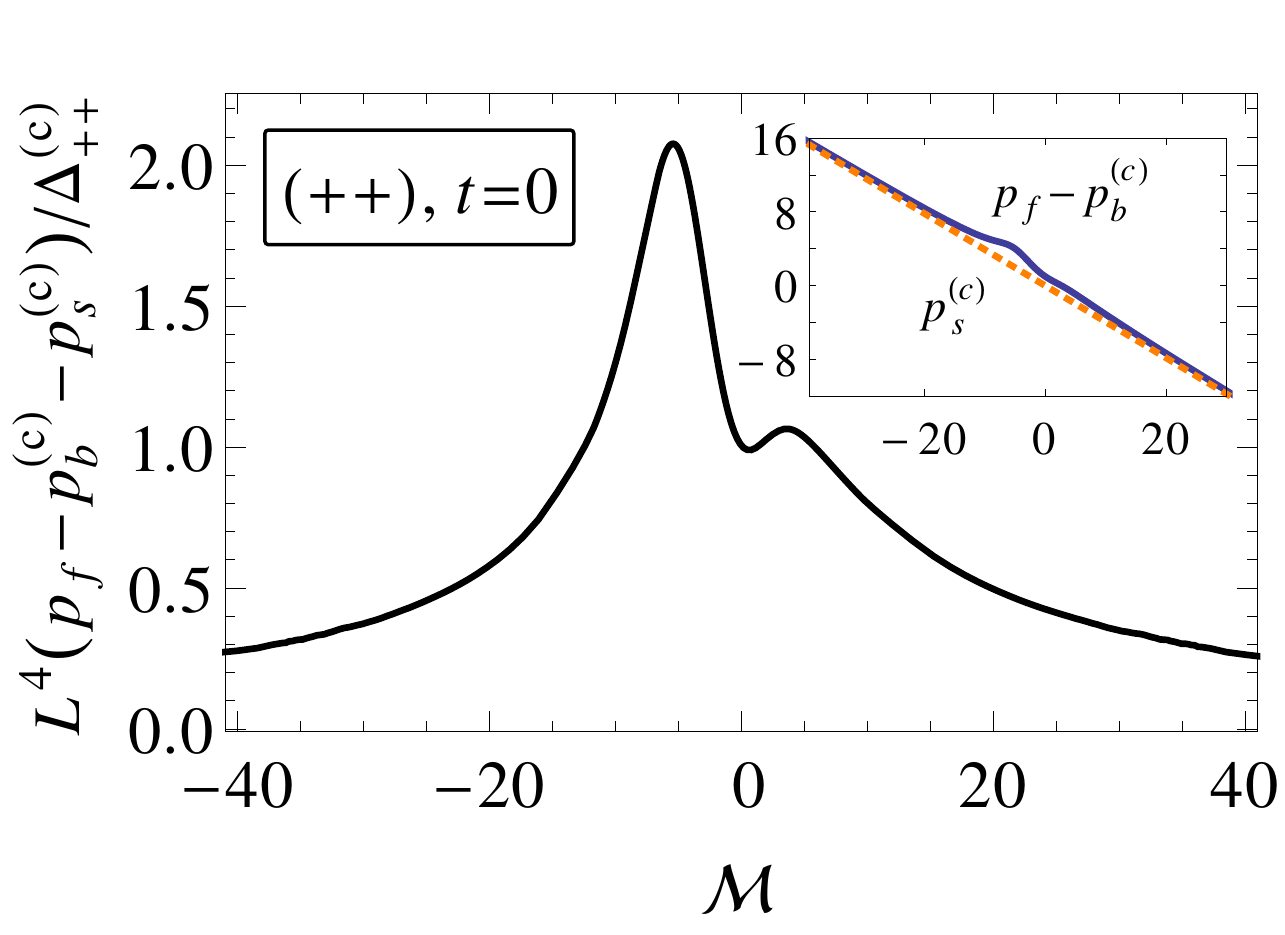}}
	\caption{(a) Pressure $p_f$ [see Eq.\ \eqref{eq_pf_landau}] in films of fixed thicknesses $L$ and $10L$, respectively, with $(++)$ boundary conditions, as well as the grand canonical ($p_b\gc$) and canonical ($p_b\can$) bulk pressures, shown as functions of the imposed mass density $\mden=\mass/L$ of the film at the bulk critical point $t=0$. The bulk pressure $p_b\gc$ is a function of the chemical potential $\mu$ of the film and, therefore, depends implicitly also on $L$. Since $p_b\gc(10L)$ practically coincides with $p_f(10L)$, only $p_b\gc(L)$ is plotted here. (b) Canonical \CCF $\hat\Kcal\can$ as defined by Eq.~\eqref{eq_pCasCan_surfsub}, obtained by subtracting from $p_f$ both the bulk \emph{and} the ``surface'' pressures [with $p_s\can$ given by Eq.~\eqref{eq_pS_can}], as a function of the scaled mass density $\Mass$ [Eq.~\eqref{eq_Mass}]. The inset compares the dependence on $\Mass$ of the surface pressure $p_s\can$ [Eq.~\eqref{eq_pS_can}, dashed line] with the canonical \CCF $\Kcal\can=p_f-p_b\can$ defined by Eq.~\eqref{eq_pCas_pdiff} (the data are normalized as in the main panel).
	Accordingly, the curve in the main panel is the difference between the solid blue and the dashed red curve in the inset. The pressures $p$ and the \CCF are all divided by $k_B T$ and are normalized by the value of the critical Casimir amplitude $\Delta_{++}\can$ [Eq.~\eqref{eq_cas_ampl_can}]. Since both pressure (if expressed as a function of $\mden/\amplPhit$) and $\Delta_{++}\can$ are proportional to $\Delta_0$, their ratio is independent of $g$.
	All data have been obtained from a numerical integration of the ELE [Eqs.~\eqref{eq_ELE0} and \eqref{eq_ELE0_bc}]. For \emph{illustrative} purposes, we have used the parameter values $L/\amplXip=1$ and $H_1=100$. The basic features shown here persist for other parameter values, but are less pronounced for larger $L$ or smaller $H_1$.}
	\label{fig_pCas_mass}
\end{figure*}

In order to gain further insight into the behavior of the \CCF for large values of the mass, in Fig.~\ref{fig_pCas_mass}(a) we compare the pressure $p_f$ of a film with $(++)$ boundary conditions with the canonical and the grand canonical bulk pressure $p_b\can$ and $p_b\gc$, respectively, as a function of $\mden=\mass/L$ at the bulk critical temperature ($t=0$).
To properly elucidate the dependence of the film pressure $p_f$ on the film thickness $L$, we consider here quantities which are dimensionless but not scaled by $L$ [see Eq.\ \eqref{eq_scalvar_mft}].
The grand canonical bulk pressure $p_b\gc$, computed for a chemical potential $\mu$ which ensures the given density $\mden$ in the film, deviates from the film pressure $p_f$ [recall that $p_f\can=p_f\gc \equiv p_f$, see Eq.~\eqref{eq_pf_equal}] only in a limited region (around $\mden/\amplPhit \simeq 3$ for the present setting). 
This marks the region where the grand canonical \CCF $\Kcal\gc = p_f-p_b\gc$ is quantitatively significant. 
For $t=0$, we obtain, from Eqs.~\eqref{eq_pB_gc} and \eqref{eq_bulk_EOS_gc}, $p_b\gc=(3/4)\Delta_0^{1/3}\mu^{4/3}$, where $\mu=\tilde \mu(t=0,\mden,h_1,L)$ is the (numerically determined) chemical potential of the film [compare Eq.\ \eqref{eq_h0} and note that $p_b\gc \propto \Delta_0$ if considered as a function of $\mden/\amplPhit$ \footnote{This follows from the fact that the factor $\Delta_0^{-1}=g/6$ drops out of Eq.\ \eqref{eq_ELE0} when casting MFT in terms of quantities (such as $\mden/\amplPhit$) which involve the same scaling factor $\Delta_0^{-1/2}$ as the scaling variables introduced in Eq.\ \eqref{eq_scalvar_mft}. According to Eq.\ \eqref{eq_pf_landau}, also pressure is proportional to $\Delta_0$.}].
The grand canonical bulk pressure $p_b\gc$ depends on $\mu$ and becomes, upon imposing $\mu=\tilde\mu$, an implicit function of the thermodynamic parameters $t$, $\mden$, $h_1$, and the thickness $L$ of the film [recall also the remark after Eq.\ \eqref{eq_bulk_EOS_can}]. It approaches $p_f$ for sufficiently large values of $\mden$ [see Fig.~\ref{fig_pCas_mass}(a)].
In contrast, in the canonical ensemble, the bulk pressure $p_b\can$ is the one of a uniform system having the same mean mass $\mden$ as the film, which, for $t=0$, is simply given by $p_b\can=3\mden^4/(4\Delta_0)$ [see Eq.~\eqref{eq_pB_can} and note that $p_b\can = (3/4) \Delta_0 (\amplXip)^{-4} (\mden/\amplPhit)^4$] and which deviates significantly from $p_f$ even for large $|\mden|$.
We shall see below that this deviation, which turns out to increase $\propto \mden$, causes the strong dependence of $\Kcal\can$ on the total mass in the film already noted in Fig.~\ref{fig_pCas_MFT}(d).
The canonical film pressure $p_f$, which equals the grand canonical film pressure [see the full purple and the gray dash-dotted curve representing $p_f(L)$ and $p_f(10L)$, respectively, in Fig.~\ref{fig_pCas_mass}(a)] approaches the pressure $p_b\can$ of a uniform bulk system (green dashed curve) only as the thickness $L$ of the film increases (e.g., from $L$ to $10L$), becoming identical to $p_b\can$ for $L\to\infty$.

\begin{figure*}[t!]\centering
	\subfigure[]{\includegraphics[width=0.475\linewidth]{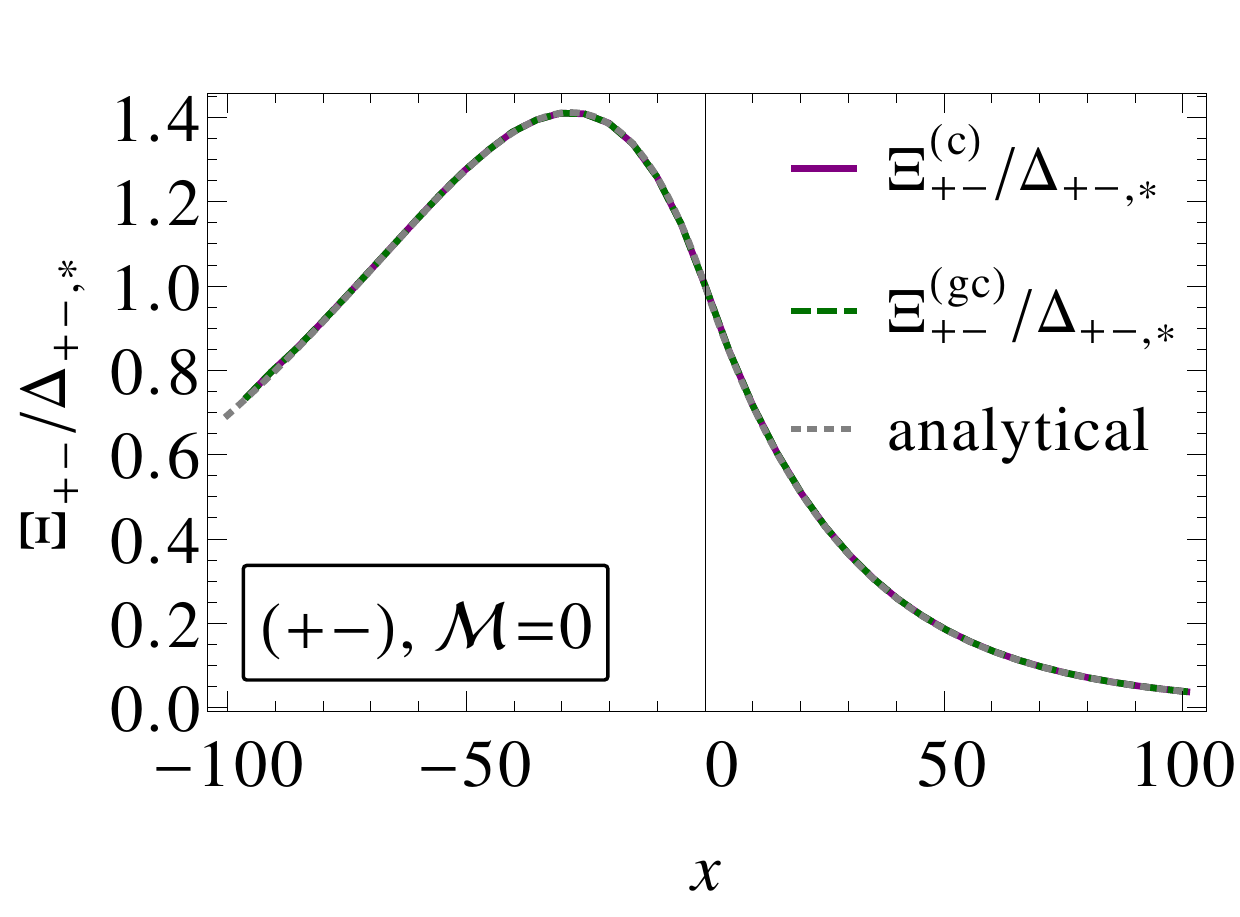}}\qquad
	\subfigure[]{\includegraphics[width=0.46\linewidth]{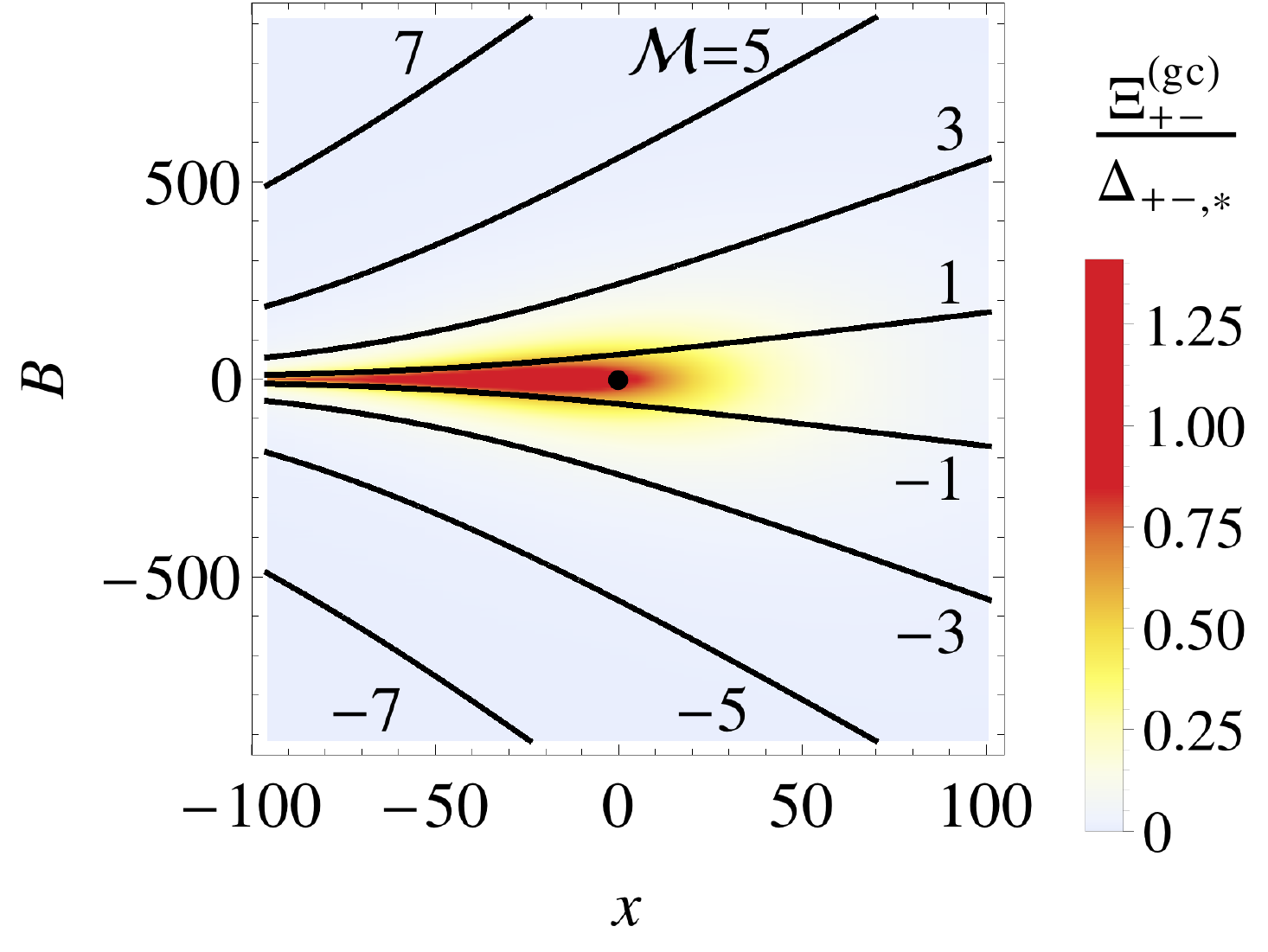}}
	\subfigure[]{\includegraphics[width=0.49\linewidth]{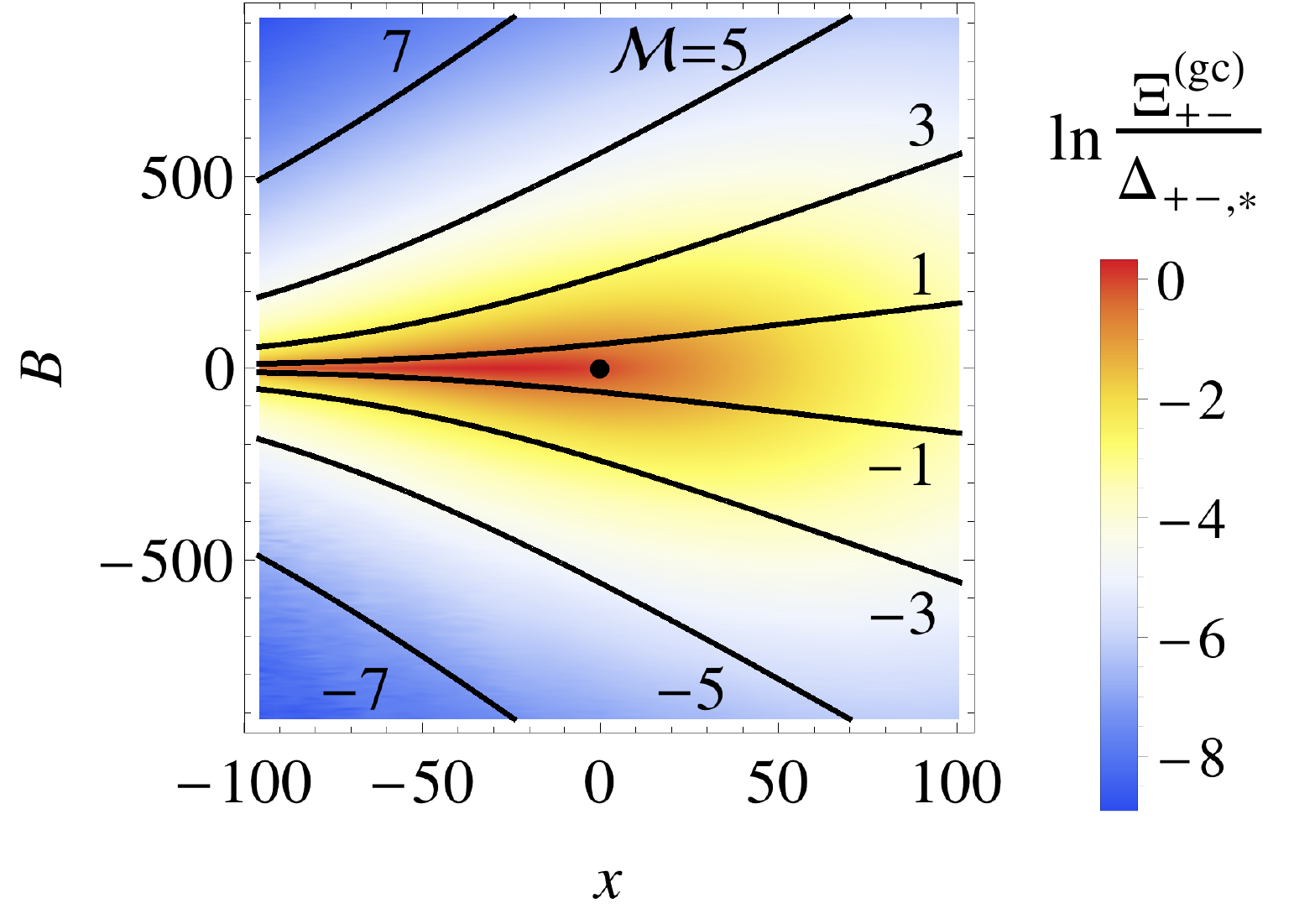}}\qquad
	\subfigure[]{\includegraphics[width=0.47\linewidth]{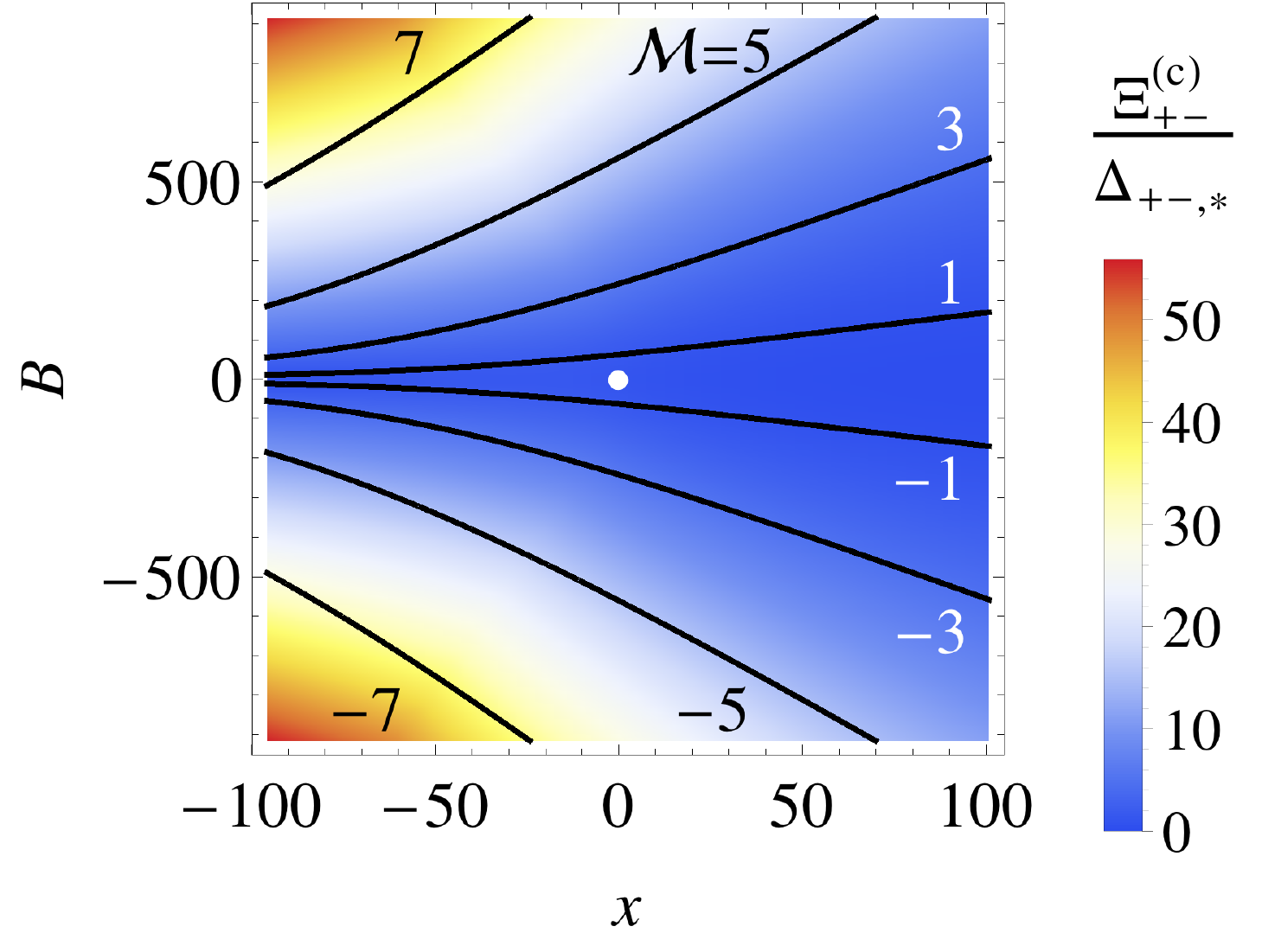}}
	\caption{Critical Casimir force in a film with $(+-)$ boundary conditions in the grand canonical and the canonical ensemble within nonlinear MFT, computed according to Eq.~\eqref{eq_pCas_pdiff} via the stress tensor and the numerically determined OP profiles. (a) Dependence of the normalized scaling functions $\Xi_{+-}\ut{(gc,c)}$ of the \CCF [see Eqs.~\eqref{eq_Casi_force_gc} and \eqref{eq_Casi_force_c}] on the scaled temperature $\tscal=(L/\amplXip)^{1/\nu}t$ for vanishing mass $\Mass=0$. 
	The dotted curve represents the analytical expression of $\Xi\gc_{+-}$ for $H_1=\infty$ \cite{krech_casimir_1997}, which coincides with the numerical data for $\Xi_{+-}\can$ and $\Xi_{+-}\gc$ with $\Mass=0$, indicated by the indistinguishable solid and dashed lines, respectively, across the considered range of $\tscal$. Panels (b), (c), and (d) show the dependences of the \CCF scaling functions $\Xi_{+-}\ut{(gc,c)}$ on the scaled temperature $\tscal$ and on the scaled bulk field $B$. Curves $B=\tilde B(\tscal)$ of constant total mass $\Mass$ [see also Fig.~\ref{fig_phasediag_MFT}(b)] are indicated by the black isolines labeled by the corresponding value of $\Mass$. $\Mass=0$ corresponds to $\tilde B(\tscal)=0$. In order to highlight the variations of the grand canonical \CCF, panel (c) shows the same data as panel (b), but on a logarithmic scale. The scaling functions are normalized by the critical Casimir amplitude $\Delta_{+-,*}$  given by Eq.~\eqref{eq_cas_ampl_pm_lim}. The location of the bulk critical point is indicated by $\bullet$. The strength of the surface field is taken to be $|H_1|\simeq 5100$ for all four panels.}
	\label{fig_pCas_pm_MFT}
\end{figure*}

In order to be applicable to large values of $\mden$, the linear mean field calculations of the previous section have to be extended as discussed in Appendix \ref{app_pert}. This yields the result in Eq.~\eqref{eq1_pcas_can} for the canonical \CCF.  
According to that analysis, the asymptotic behavior of $\Kcal\can$ for large $|\mden|$ is governed by the term
\beq p_s\can\equiv -\frac{2 h_1 \mden}{L} = -2 H_1\Mass \frac{\Delta_0}{L^4} ,
\label{eq_pS_can}\eeq 
which is a surface-like contribution to the pressure, as was already suggested by Eqs.~\eqref{eq_linMF_pCasCan_stressroute} and \eqref{eq_pCasCan_linMF_scalform}.
This is confirmed by the inset of Fig.~\ref{fig_pCas_mass}(b), where the canonical \CCF $\Kcal\can= p_f-p_b\can$ is compared with $p_s\can$ for $t=0$ as a function of the scaled mass density $\Mass$.
Indeed, the dependence of $\Kcal\can$ on $\Mass$ displayed in Fig.~\ref{fig_pCas_MFT}(d) is consistent with the dependence of $p_s\can$ on $\Mass$.
The surface pressure $p_s\can$ has been identified in Sec.~\ref{sec_cas_stress} to stem from a surface contribution to the free energy and is a genuine consequence of the canonical constraint.
In particular, it contributes to the experimentally measurable canonical \CCF for systems in which the bulk fluid exhibits no preferential adsorption (i.e., $H_1=0$) at the outer surfaces bounding the film.
The canonical \CCF $\hat\Kcal\can$ stemming only from the residual finite-size free energy can be obtained by subtracting from $p_f$, in addition to $p_b\can$ as required by Eq.~\eqref{eq_pCas_pdiff}, also the surface pressure $p_s\can$:
\beq \hat \Kcal\can \equiv p_f - p_b\can - p_s\can .
\label{eq_pCasCan_surfsub}\eeq
In the main panel in Fig.~\ref{fig_pCas_mass}(b), $\hat\Kcal\can$ is plotted as a function of the mass $\Mass$ for $t=0$. 
It turns out that the canonical \CCF defined this way indeed tends to zero for large absolute values of $\Mass$.
In Fig.\ \ref{fig_pCas_mass}(b), the decay of $\hat \Kcal\can$ upon increasing $\Mass$ is slower than exponential, but the agreement with the perturbative result in Eq.~\eqref{eq1_pcas_can} is not yet reached. 
Actually, we expect the linear mean field behavior to be attained only for values of $\Mass$ significantly larger than those covered in Fig.~\ref{fig_pCas_mass}(b). However, these larger ones are beyond the reach of our numerical approach.
The structural behavior around $\Mass=0$ exhibited in the main panel of Fig.~\ref{fig_pCas_mass}(b) should be considered with caution, because the expression for $p_s\can$ in Eq.~\eqref{eq_pS_can} is based on linear MFT, which is not expected to be valid around criticality for the strong surface fields considered here.

Figure~\ref{fig_pCas_pm_MFT} illustrates the behavior of the \CCF within MFT for $(+-)$ boundary conditions. 
In contrast to the case of $(++)$ boundary conditions [Fig.~\ref{fig_pCas_MFT}], the \CCF attains a well-defined limit for $H_1\to\infty$ [see Fig.~\ref{fig_casAmpl}(d)] even under the mass constraint. We therefore normalize the scaling functions by $\Delta_{+-,*}$ given in Eq.~\eqref{eq_cas_ampl_pm_lim}.
First, considering in Fig.~\ref{fig_pCas_pm_MFT}(a) the case of vanishing mass, $\Mass=0$, we find that in this case the canonical and grand canonical \CCFs are identical for all values of the scaled temperature $\tscal$ studied.
This is expected because, above the capillary critical temperature (which is located at a large negative value of $\tscal$ not covered by the present data), $\Mass=0$ is realized by $B=0$ in the film as well as in the bulk for \emph{both} ensembles. 
The value of $H_1$ chosen in panel (a) is sufficiently large so that the scaling functions $\Xi_{+-}\gc(\tscal,B,H_1)$ fall on top of the analytical prediction of Ref.~\cite{krech_casimir_1997} derived for $H_1=\infty$.
This finding also provides a welcome, independent check of our numerical calculations.
For $\Mass\neq 0$, the perturbative solution in Eq.~\eqref{eq_pm_B1} as well as the solution of the nonlinear MFT depicted by the solid curves in Fig.~\ref{fig_phasediag_MFT}(b) indicate that the constraint field $\tilde B$ is different from the corresponding one of a homogeneous system with the same mass density [corresponding to $\tilde B\st{hom}$ in Fig.~\ref{fig_phasediag_MFT}(b)]. This implies that the bulk pressures and hence the \CCFs differ for the canonical and grand canonical ensembles.
Indeed, the difference in the \CCFs is clearly exhibited by comparing panel (b) [highlighted on a logarithmic scale in panel (c)] with panel (d) in Fig.~\ref{fig_pCas_pm_MFT}.
Generally, for $(+-)$ boundary conditions, both canonical and grand canonical \CCFs $\Kcal\can$ and $\Kcal\gc$ are repulsive over the whole range of parameters considered.
However, similarly to the case of symmetric boundary fields (Fig.~\ref{fig_pCas_MFT}), $\Kcal\gc$ is largest around the line $B=0$, whereas $\Kcal\can$ increases significantly for larger, nonzero values of $\Mass$.

\begin{figure}[t]\centering
	\includegraphics[width=0.9\linewidth]{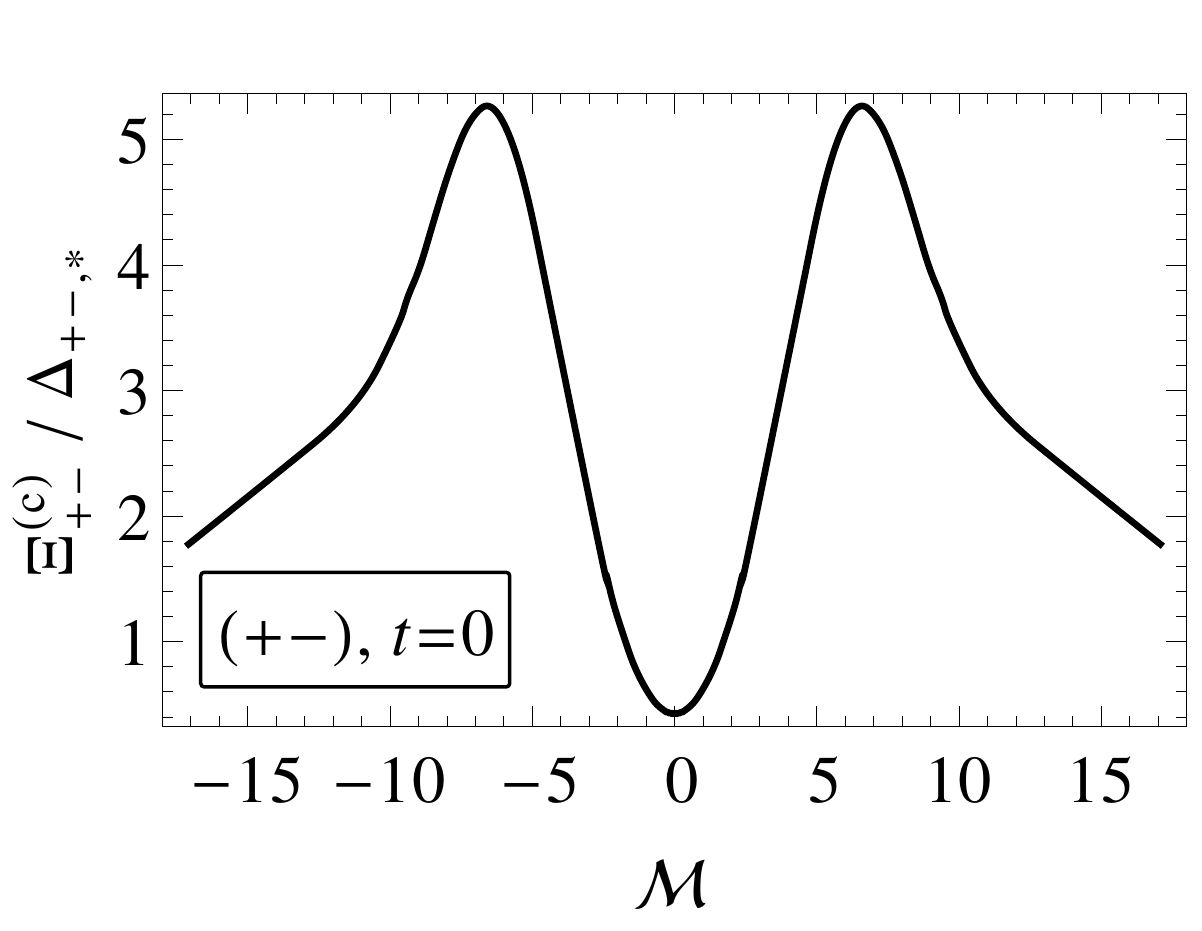}
	\caption{Scaling function $\Xi\can_{+-}$ of the canonical \CCF as function of the mass $\Mass$ for $(+-)$ boundary conditions and at bulk criticality $t=0$. $\Xi_{+-}\can$ is symmetric in $\Mass$ and decays towards zero for $|\Mass|\gg 1$. The data have been obtained from a numerical integration of the ELE [Eqs.~\eqref{eq_ELE} and \eqref{eq_ELE_BC}] for $|H_1|=100$. Due to this rather small value of $|H_1|$, the amplitude of $\Xi\can_{+-}$ at $\Mass=0$ has not yet reached the value $\Delta_{+-,*}$ corresponding to the limit $H_1\to\infty$ [see Figs.~\ref{fig_casAmpl}(d) and \ref{fig_pCas_pm_MFT}(a),(d)] in which $\Xi\can_{+-}(\Mass=0)/\Delta_{+-,*}=1$.}
	\label{fig_pCas_pm_mass}
\end{figure}

The data shown in Fig.~\ref{fig_pCas_pm_MFT}(d) pertain to the strong adsorption regime ($H_1\to\infty$), for which it turns out to be numerically difficult to study large values of $\Mass$.
Therefore, in Fig.~\ref{fig_pCas_pm_mass} we show the scaling function $\Xi_{+-}\can$ as a function of $\Mass$ at bulk criticality but for a smaller value of $|H_1|$.
Similarly to the canonical \CCF shown in Fig.~\ref{fig_pCas_mass}(b), $\Xi_{+-}\can$ attains a maximum at a nonzero value of $|\Mass|$ and decreases for larger $|\Mass|$; the latter feature is in agreement with the predictions of linear MFT discussed in Appendix \ref{app_pert}. However, the range of $\Mass$ covered in Fig.~\ref{fig_pCas_pm_mass} does not allow us to reliably test the detailed prediction given in Appendix \ref{app_pert} concerning the exponential decay of $\Xi_{+-}\can$ as a function of the parameter $\mtscal \equiv \sqrt{\tscal+3\Mass^2}$ for $\mtscal\gg 1$. Upon increasing $|H_1|$, we find that the pronounced maxima of $\Xi\can_{+-}$ move towards larger values of $|\Mass|$, but the shape of the curve remains essentially the same.

\subsection{Monte Carlo simulations of the 3D Ising model}
\label{sec_MC_Casimir}

In order to assess the relevance of thermal fluctuations for the \CCFs discussed above, we have carried out MC simulations of the three-dimensional Ising model [Eq.~\eqref{eq_Ising}] in film geometry.
The basic simulation setup is described in Sec.~\ref{sec_MC_prof}. 
The \CCF at an inverse temperature $\beta=1/(k_B T)$ on a lattice  with transverse area $A=L_{x} L_{y}$ and thickness $L$ is defined in the two ensembles via finite differences following Refs.\ \cite{vasilyev_monte_2007, vasilyev_universal_2009, vasilyev_critical_2013}, i.e.,
\begin{widetext}\begin{subequations}\bal
\Kcal\gc(\beta,\mu,h_1,A,L) &= - \frac{\beta \Delta F\gc(\beta,\mu,h_1,A,L)}{A}+ \beta f_b\gc(\beta,\mu),\label{eq_MC_force_gc}\\
\Kcal\can(\beta,\mden,h_1,A,L) &= - \frac{\beta \Delta F\can(\beta,\mden,h_1,A,L)}{A}+ \beta f_b\can(\beta,\mden),\label{eq_MC_force_can}
\end{align}\label{eq_MC_force}\end{subequations}\end{widetext}
where $\Delta F\ut{(gc,c)}(\beta,\mu|\mden,h_1,A,L)\equiv F\ut{(gc,c)}(\beta,\mu|\mden,h_1,A,L+\frac{1}{2})- F\ut{(gc,c)}(\beta,\mu|\mden,h_1,A,L-\frac{1}{2})$ is the free energy difference and the indicated dependence on $\mu$ or $\mden$ pertains to the grand canonical and canonical cases, respectively. The bulk free energy $f_b$ coincides with the negative bulk pressure.
Moreover, we have reinstated explicitly the transverse area $A$.  
The thickness $L \equiv L_{z}-\frac{1}{2}$, to which the CCF is formally attributed to, is half-integer because it is expressed via the difference of slabs of actual thicknesses $L_{z}$ and $L_{z}-1$.
In general, half-integer values $L=L_{z}-\frac{1}{2}$ are used for the variable in terms of which the \CCF is expressed, while integer values $L_{z}$ are used for the thickness of the system in which computations are performed. 

Before proceeding, here we derive the expressions which follow from the computational scheme described in Eq.~\eqref{eq_MC_force} for the high-temperature limit ($\beta\to 0$) of the free energy and of the \CCF. 
These results will be useful for interpreting our MC results further below, because there we vary the scaling variable $x$ via changing the reduced temperature $t$ in our simulations. (Accordingly, the limit $\tscal \to\infty$ is realized by taking the limit $\beta \to 0$.)
For $\beta\to 0$ and finite bulk and surface fields, the grand canonical partition function is of purely entropic nature and is given by the number of spin configurations, $Z\gc=2^{A L_z}$. This yields a free energy $\beta F\gc=-A L_z \ln 2$ for $\beta\to 0$ and, correspondingly, a bulk free energy density $\beta f_b\gc  = \lim_{A L_z\to\infty}\, \beta F\gc/(A L_z)= -\ln 2$. 
Accordingly, in the limit $\beta\to 0$ the two terms in Eq.~\eqref{eq_MC_force_gc} cancel so that we obtain, as expected, a vanishing \CCF, $\Kcal\gc(\beta\to 0) =0$.
In contrast, in the canonical ensemble with vanishing total magnetization $\mass=0$, an equal number $A L_z/2$ of up and down spins have to be distributed on the lattice, yielding a total number $Z\can = (A L_z)! / [(A L_z/2)!]^2$ of possible configurations. Accordingly, employing the Stirling approximation, the high-temperature limit of the canonical free energy is given by
\beq \beta F\can(\mden=0, A,L_z)\big|_{\beta\to 0} \simeq -A L_z\ln 2 + \onehalf \ln (\pi A L_z/2),
\label{eq_MC_freeE_highT}\eeq 
whereas the corresponding bulk free energy density is $\beta f_b\can = \lim_{A L_z\to\infty}\, \beta F\can/(A L_z)=-\ln 2$, coinciding, as expected, in this limit with $\beta f_b\gc$.
Hence, Eq.~\eqref{eq_MC_force_can} yields the canonical \CCF 
\beq \Kcal\can(\beta\to 0, \mden=0, A, L_z) \simeq - \frac{1}{2A} \ln \frac{L_z}{L_z-1}
\label{eq_MC_cascan_highT}\eeq  
in the high-temperature limit.
The above reasoning can be easily extended to nonzero total magnetizations $\mass$, for which one obtains a value of $\Kcal\can(\beta\to 0)$ which depends on $\mass$.
Thus, in the high-temperature limit and for Ising slabs of finite extent, the canonical \CCF computed according to Eq.~\eqref{eq_MC_force_can} generally attains a finite value, which, however, vanishes if either $A\to\infty$ or $L_z\to\infty$.

In the grand canonical ensemble, $\Delta F\gc$ is computed in our MC simulations via the coupling parameter approach \cite{vasilyev_monte_2007}.
For $(++)$ boundary conditions under the constraint $\mass=0$, the computation of the grand canonical bulk free energy $f_b\gc$ in Eq.~\eqref{eq_MC_force_gc} turns out to be difficult. 
In this case, we consider instead the difference between the \CCFs of a film of thickness $L$ and of a significantly thicker one, which we take here to have a thickness of $2L$:
\beq g\cas(L)\equiv \Kcal\gc(L)-\Kcal\gc(2L).
\label{eq_MC_gcas}\eeq 
In this difference, the contribution of $f_b\gc$ present in Eq.~\eqref{eq_MC_force_gc} drops out, leaving only the free energy differences $\Delta F\gc$ for the two film thicknesses.
It can be shown \cite{vasilyev_monte_2007} that this function $g\cas$ approximates well the true \CCF $\Kcal\gc$. 
(By following the approach used in Ref.\ \cite{vasilyev_monte_2007}, this approximation can be systematically improved.)
In a few cases we have checked this also for the present data by directly computing the bulk free energy density $f_b$ via the energy integration technique (see Ref.~\cite{vasilyev_critical_2013} for details).
For $(+-)$ boundary conditions, we generally use Eq.~\eqref{eq_MC_force_gc} directly.

In the canonical ensemble we make use of the thermodynamic relation $\pd (\beta F)/\pd \beta=E$ between the canonical free energy $F$ and the mean energy $E$ of a system and compute the free energy $F\can$ entering into Eq.~\eqref{eq_MC_force_can} via integration of $E$ over the inverse temperature $\beta$, starting from the known value of $F\can$ in the high-temperature limit:
\begin{multline}
\beta F\can(\beta,\mden,h_1,A,L)
= \beta F\can(\beta, \mden,h_1,A,L)\big|_{\beta\to 0} \\ +\int  \limits_{0}^{\beta} d \beta'\,  E(\beta',\mden,h_1,A,L).
\label{eq_MC_Fcan}
\end{multline}
In the case $\mden=0$, which we focus on here, $\beta F\can|_{\beta\to 0}$ is given by Eq.~\eqref{eq_MC_freeE_highT}.
The energy $E$ of the system is computed by using the Kawasaki method with multispin coding ($5\times 10^{5}$ MC steps, with one step being an attempt of $L_{x}L_{y} L_{z}$ updates) for 100 different values of the inverse temperature $\beta$.
Subsequently, the numerical integration according to Eq.~(\ref{eq_MC_Fcan}) is carried out by using a cubic spline interpolation of these data. 

\begin{figure*}[t]\centering
	\subfigure[]{\includegraphics[width=0.49\linewidth]{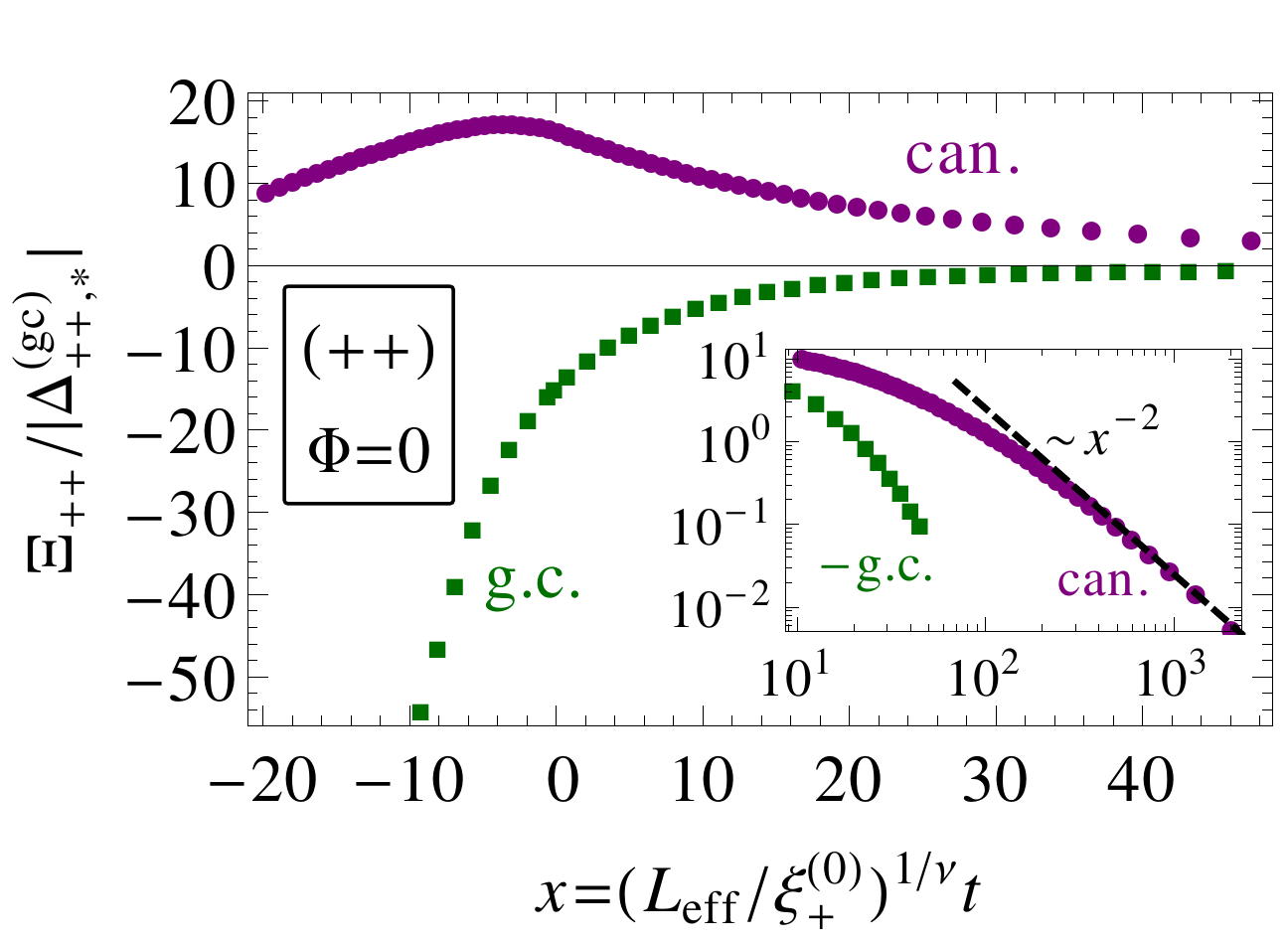}}\qquad
	\subfigure[]{\includegraphics[width=0.467\linewidth]{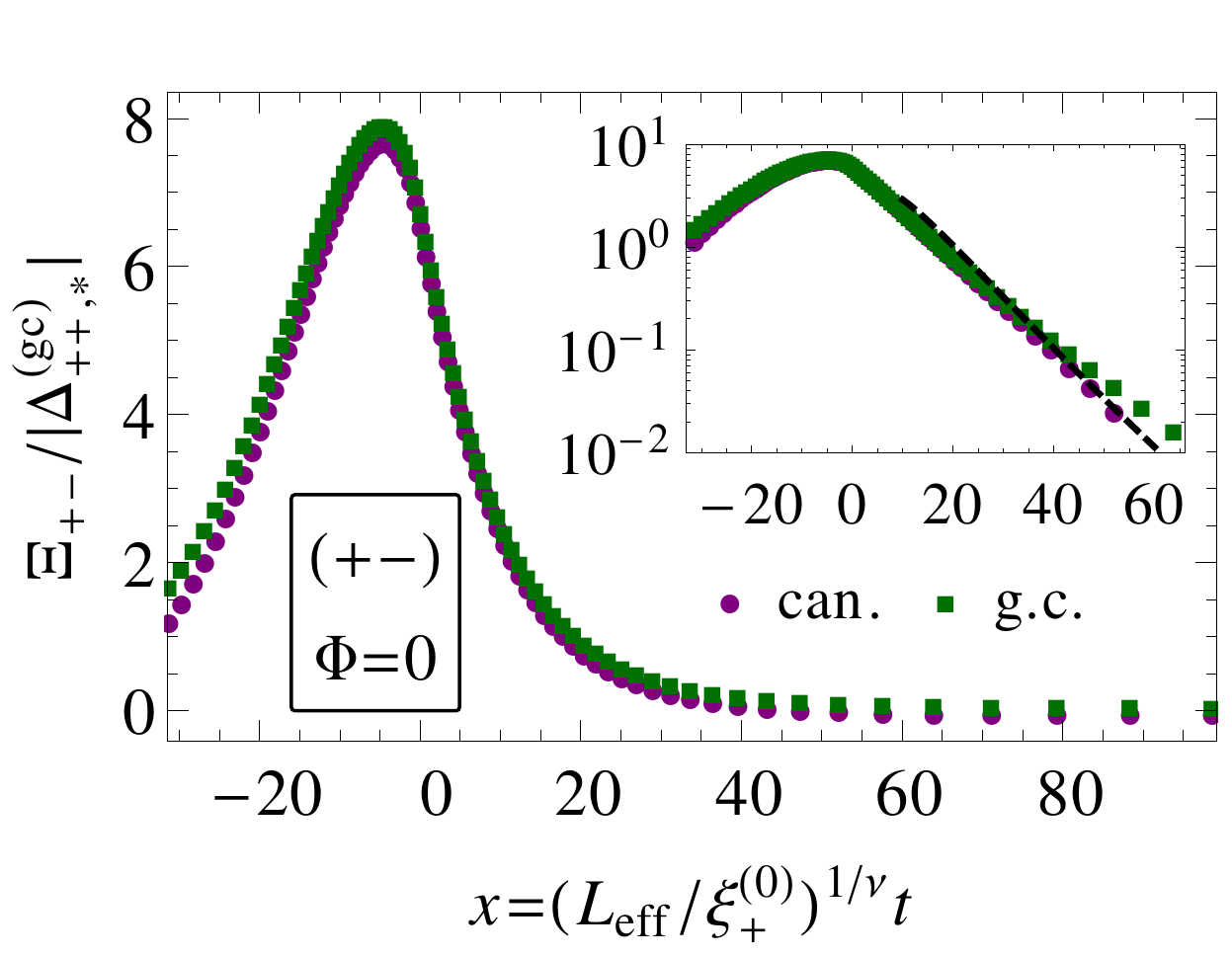}}
	\caption{Scaling functions $\Xi\ut{(gc,c)}$ of the \CCF obtained from MC simulations of the canonical and grand canonical Ising model in a three-dimensional film of size $L_x \times L_y \times L_z = 32\times 32\times 8$ for a vanishing total magnetization $\mass=0$ and for (a) $(++)$ and (b) $(+-)$ boundary conditions. The effective thickness of the film amounts to $L\eff= L_z+2.60=10.60$ and $L\eff=L_z+2.65=10.65$ for $(++)$ and $(+-)$ boundary conditions, respectively, both in the canonical and in the grand canonical ensemble. In order to highlight the approach of the canonical \CCF towards its asymptotic value for large $\tscal$ [Eq.~\eqref{eq_MC_cascan_highT}], the insets show, presented differently than in the main panel, $\left|\Xi\ut{(gc,c)}_{++}(\tscal)-\Xi\ut{(gc,c)}_{++}(\tscal\to\infty)\right|/|\Delta\gc_{++,*}|$ on (a) a double-logarithmic and (b) a semi-logarithmic scale. In the insets, the dashed line indicates a behavior $\propto 1/\tscal^2$ (obtained from a fit) in (a) and $\propto \tscal^{2-\alpha} \exp(-\tscal^\nu)$ [Eq.\ \eqref{eq_Xicas_nlin_asympt_pm}] in (b), where the values $\alpha\simeq 0.11$ and $\nu\simeq 0.63$ for the three-dimensional Ising bulk universality class are used. Error bars are of the order of the symbol size and not shown. Here, the variation of $\tscal$ is realized by changing $t$ at fixed $L\eff$. The scaling functions are normalized by the absolute value of the grand canonical critical Casimir amplitude $\Delta\gc_{++,*}\simeq -0.75$ obtained without constraint for $T=T_c$, $\mu=0$, and $h_1\to\infty$ in the three-dimensional Ising model \cite{vasilyev_universal_2009}. }
	\label{fig_pCas_MC}
\end{figure*}

The scaling form (for a film in $d=3$) of the \CCF in the grand canonical and canonical ensembles is given in Eqs.~\eqref{eq_Casi_force_gc} and \eqref{eq_Casi_force_c}, respectively, which we rewrite here in a slightly modified form in order to account for a number of simulation-specific issues:
\begin{widetext}\begin{subequations}\bal 
\Kcal\gc(\beta, \mu, h_1,A, L) &= \beta^{-1} L\eff^{-3} \Xi\gc\left(\tscal=t \left(\frac{L\eff}{\amplXip}\right)^{1/\nu}, B=\mu \left(\frac{L\eff}{\amplXimu}\right)^{\Delta/\nu},H_1=h_1 \left(\frac{L\eff}{\lenH1}\right)^{\Delta_1/\nu}, \rho\right),\\
\Kcal\can(\beta, \mden, h_1,A, L) &= \beta^{-1} L\eff^{-3} \Xi\can\left(\tscal=t \left(\frac{L\eff}{\amplXip}\right)^{1/\nu}, \Mass=\mden \left(\frac{L\eff}{\lenPhi0}\right)^{\beta/\nu}, H_1=h_1 \left(\frac{L\eff}{\lenH1}\right)^{\Delta_1/\nu}, \rho\right).
\end{align}\label{eq_MC_cas_scalf}\end{subequations}\end{widetext}
Here, $\Xi\ut{(gc,c)}$ are scaling functions, $t=T/T_c-1$ is the reduced temperature, and $\rho\equiv L/\sqrt{A}$ is the aspect ratio of the simulation box.
The values of the critical exponents are reported in Table \ref{tab_crit_exp} and the length scales are stated in Sec.\ \ref{sec_MC_prof}.
As discussed there, $\Delta_1\equiv\Delta_1\ut{ord}$ is the appropriate surface critical exponent and $\lenH1\equiv \lenOrdH1 \simeq 0.21$ is the associated length scale for the Ising model [Eq.\ \eqref{eq_Ising}] employed here.
We use finite values $h_1=\pm 1$ of the surface fields in our MC simulations.
Corrections to scaling, \emph{inter alia}, due to using these finite values of the boundary fields, are accounted for by introducing an effective film thickness $L\eff = L+\delta L$, with $\delta L=2.60$ and $\delta L=2.65$ for $(++)$ and $(+-)$ boundary conditions, respectively \cite{hasenbusch_thermodynamic_2011, vasilyev_critical_2011, vasilyev_critical_2013} \footnote{We remark that these scaling corrections have been established previously in the grand canonical ensemble at a fixed bulk field \cite{vasilyev_critical_2011, vasilyev_critical_2013}. Here, we assume that they remain valid in the canonical case as well.}.
For reasons of simplicity, we focus here only on the case of zero total magnetization, $\mass=0$.
In the grand canonical ensemble, this is realized via having $\mu \ne 0$ and requires, for each value of the temperature $\beta^{-1}$, the computation of the corresponding value of $\mu$ in accordance with the prescription given in Sec.~\ref{sec_MC_prof}.

Figure~\ref{fig_pCas_MC} shows the numerical results for the \CCF scaling functions [Eq.~\eqref{eq_MC_cas_scalf}] obtained from MC simulations in the setting described above with a system size of $L_x\times L_y\times L_z=32\times32\times 8$ (in units of the lattice spacing), for vanishing total magnetization $\mass=0$, and for (a) $(++)$ and (b) $(+-)$ boundary conditions.
These scaling functions are computed based on Eqs.~\eqref{eq_MC_force}, \eqref{eq_MC_Fcan}, and \eqref{eq_MC_cas_scalf}, with the exception of $\Xi_{++}\gc$, which we obtained from the approximation $g\cas\simeq \Kcal$ [Eq.~\eqref{eq_MC_gcas}], as described above.
The scaling functions reported in Fig.\ \ref{fig_pCas_MC} are normalized by the absolute value of the grand canonical critical Casimir amplitude $\Delta\gc_{++,*}\equiv \Xi\gc(\tscal=0,B=0,H_1\gg 1,\rho\ll 1)\simeq -0.75$ for the three-dimensional Ising model \cite{vasilyev_universal_2009}.

Confirming the basic feature of MFT presented in Fig.~\ref{fig_pCas_MFT}, the \CCF inferred from the MC simulations for $(++)$ boundary conditions is repulsive for the canonical and attractive for the grand canonical ensemble [Fig.~\ref{fig_pCas_MC}(a)].
The results for the latter case are consistent with previous MC studies of \CCFs in the presence of a bulk magnetic field \cite{vasilyev_critical_2013}.
From our data we extract the critical Casimir amplitude 
\begin{multline} \Delta_{++,*}\can\equiv \Xi\can(\tscal=0,\Mass=0,H_1\gg 1,\rho=1/4) \\ \simeq -16.9 \Delta\gc_{++,*} \simeq 12.7
\label{eq_MC_casAmpl_c}\end{multline}
for the three-dimensional Ising model in the canonical ensemble.
Within the considered range of $\tscal$ in the supercritical region, we cannot unambiguously determine the precise decay behavior of $\Xi\gc$ upon increasing $\tscal$, although a simple exponential decay appears to describe the data rather well \footnote{Note that the asymptotic behavior in Eq.\ \eqref{eq_Xicas_nlin_asympt_pp} corresponds to $\mu=0$ rather than to the case $\mass=0$ considered in Fig.\ \ref{fig_pCas_MC}.}. 
In the canonical ensemble, in contrast, the \CCF decays significantly slower, as illustrated in the inset of Fig.~\ref{fig_pCas_MC}(a).
However, instead of the decay $\propto 1/\tscal$ predicted by MFT [Eq.~\eqref{eq_Xicas_can_asympt} and Fig.~\ref{fig_pCas_MFT}(a)], we infer from the present data that the high temperature limit [Eq.~\eqref{eq_MC_cascan_highT}] is approached as $\Xi\can(\tscal)-\Xi\can(\tscal\to\infty)\simeq 2.5\times 10^5 |\Delta_{++,*}\gc|/\tscal^n$, with $n\simeq 2$, which, together with the numerical prefactor, has been obtained from a fit.
We remark that, upon increasing the range of $x$ included in the fit, the obtained effective exponent $n$ decreases slightly.
A discrepancy between this value for $n$ obtained for the Ising model and the mean field prediction $n=1$ should not be surprising. 
In fact, analogous differences occur also in the dependence of $\Xi\gc$ on $\tscal$ as described in Eq.\ \eqref{eq_Xicas_nlin_asympt}. 
However, the equivalent expressions in the canonical case are presently not available and deserve further studies.

\begin{figure}[t]\centering
	\includegraphics[width=0.93\linewidth]{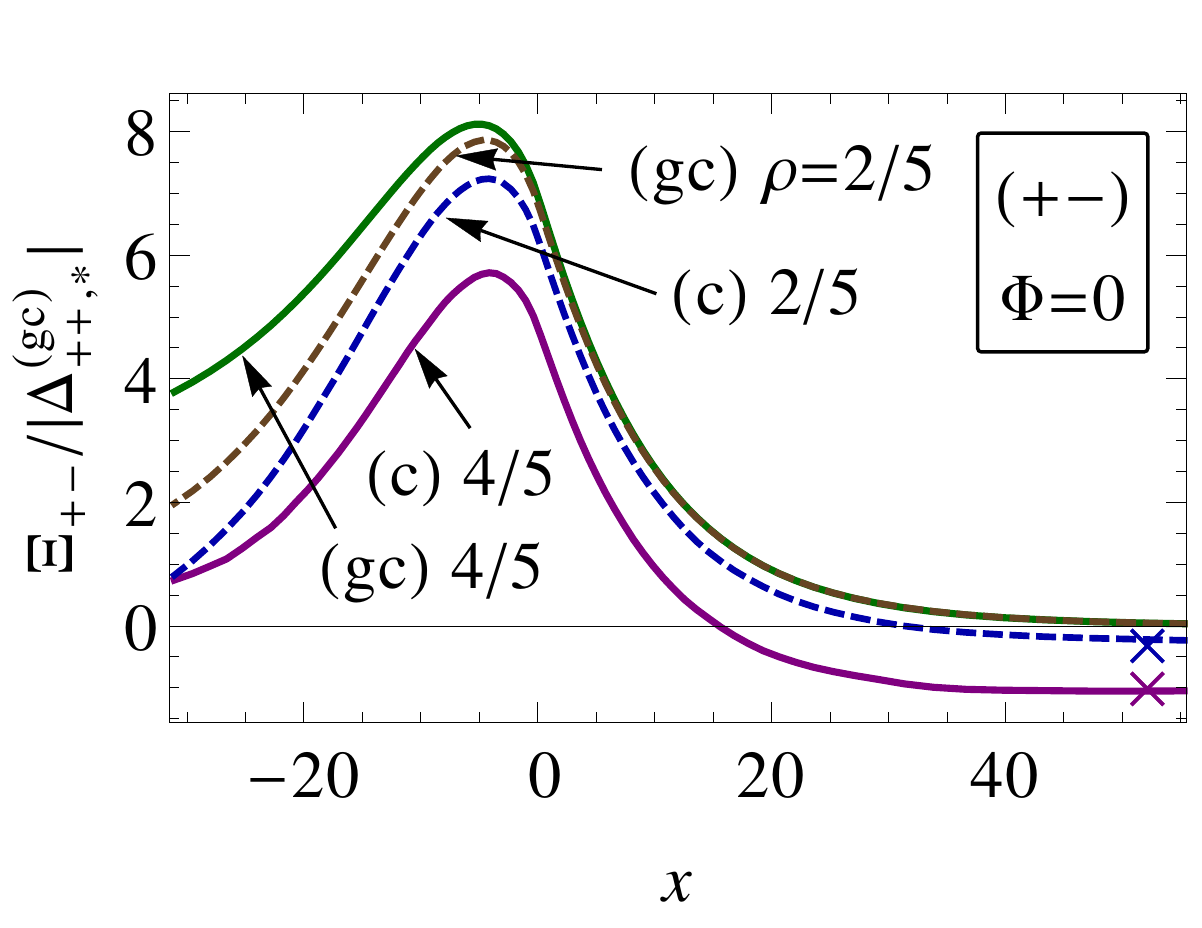}
	\caption{Dependence of the \CCF on the aspect ratio $\rho=L_z/\sqrt{A}$ as obtained from MC simulations of the canonical and grand canonical Ising model in a three-dimensional film for a vanishing total magnetization $\mass=0$ and $(+-)$ boundary conditions. The film thickness is fixed at $L_z = 8$, while the lateral area is chosen as $A=L_x\times L_y=10\times 10$ and $20\times 20$, corresponding to the aspect ratios $\rho$ of $4/5$ (solid curves) and $2/5$ (dashed curves), as indicated by the labels. As in Fig.~\ref{fig_pCas_MC}, the scaling functions $\Xi_{+-}\ut{(gc,c)}$ are plotted as functions of the scaled temperature $\tscal$ and are normalized by the grand canonical critical Casimir amplitude $|\Delta_{++,*}\gc|$ of the Ising model. The analytically predicted high-temperature limits of the canonical \CCF [Eq.~\eqref{eq_MC_cascan_highT}] are indicated by the crosses $\times$.}
	\label{fig_pCas_MC_size}
\end{figure}

As shown in Fig.~\ref{fig_pCas_MC}(b), for $(+-)$ boundary conditions and $\mass=0$ (which is realized by $\mu=0$ in the grand canonical ensemble) and for the considered system size of $32\times32\times 8$, the canonical and grand canonical \CCFs are almost indistinguishable. 
While these data pertain to an aspect ratio of $\rho=L_z/\sqrt{A}=1/4$, we expect the \CCFs to be identical in the limit $\rho\to 0$ (see below).
Accordingly, we extract the critical Casimir amplitudes 
\beq \Delta_{+-,*}\can \simeq \Delta_{+-,*}\gc \simeq -6.7 \Delta_{++,*}\gc\, \simeq\, 5.0,
\eeq
and remark that $\Delta_{+-,*}\gc$ agrees well with previously reported results \cite{vasilyev_critical_2013}.
A fit of the numerical data with the predicted decay behavior of $\Xi_{+-}\gc$ for large $\tscal$ as given in Eq.\ \eqref{eq_Xicas_nlin_asympt_pm}, and the use of appropriate values for the Ising critical exponents, yield reasonable agreement [see inset of Fig.~\ref{fig_pCas_MC}(b)].
We remark, however, that also a simple exponential decay $\propto \exp(-\gamma\tscal)$ with $\gamma\simeq 0.1$ describes the data rather well, within the considered supercritical range of $\tscal$.

Since for $(+-)$ boundary conditions and $\mass=0$ the mean-field expressions of the \CCF are identical in the two ensembles, this situation provides a particularly suitable case to study the effect of fluctuations on the \CCF.
We expect that the more severe restriction of the fluctuation spectrum in the canonical ensemble becomes more significant upon reducing the lateral system size. 
For a fixed thickness $L_z=8$, $(+-)$ boundary conditions, and zero total magnetization $\mass=0$ the dependence of the \CCF on the aspect ratio $\rho=L/\sqrt{A}$ is shown in Fig.~\ref{fig_pCas_MC_size}.
We find that, for each $\rho$, the scaling functions in the two ensembles indeed increasingly deviate upon increasing $\rho$, i.e., upon decreasing the transverse area $A$ (i.e., the dashed lines are closer to each other than the full lines).
Partially, these differences can be attributed directly to the nonzero high-temperature limit of the canonical \CCF predicted by Eq.~\eqref{eq_MC_cascan_highT}, which is indicated by the cross $\times$ in Fig.~\ref{fig_pCas_MC_size}.
However, shifting all scaling functions vertically such that they approach zero for $x\to\infty$ still does not cause the curves to fall upon each other for smaller $x$. 
Near $T_c$, we attribute these remaining differences, which decrease upon decreasing $\rho$, to the presence of critical fluctuations and to the difference in the fluctuation spectra. Since by definition the bulk pressure is independent of the aspect ratio, we conclude that in general, beyond MFT, film pressures can be different in the two ensembles, leading to a violation of Eq.~\eqref{eq_pf_equal}.
Below $T_c$, additional effects due to phase separation might come into play. These aspects deserve further and more detailed analyses.

\section{Summary and Outlook}
Theoretically and experimentally, critical phenomena have been mainly investigated for the grand canonical ensemble in which the system can exchange particles with a reservoir at the same fixed chemical potential $\mu$.
Here, we have studied critical adsorption and \CCFs occurring in a film of thickness $L$ in the canonical ensemble, i.e., under the constraint of having a fixed value $\mass$ of the integrated order parameter, here referred to as the ``mass'';
this conserved quantity gives rise to an additional scaling variable $\Mass\sim \mass L^{\beta/\nu-1}$, where $\beta$ and $\nu$ are standard bulk critical exponents.
Such a situation is encountered naturally in experiments as well as in a variety of numerical methods, such as in molecular dynamics or in the lattice Boltzmann method, which involve a mass conservation law.
We have focused here on mean field theory (MFT) in the presence of symmetry-breaking boundary conditions [$(++)$ and $(+-)$] described by a surface field $h_1$, but with unbroken translational symmetry in the directions parallel to the confining walls. 
This setup is suitable for describing critical adsorption of binary fluids \cite{ gambassi_critical_2009}.
For large lateral system sizes, it is expected that MFT captures well the dominant contributions to the canonical and grand canonical partition functions.
Within this setup, MFT of a canonical system can be completely described in terms of a grand canonical ensemble with a chemical potential $\mu=\tilde\mu$ chosen such that the imposed value of $\mass$ is realized.
The qualitative features emerging from our mean field analysis are confirmed by Monte Carlo simulations of the three-dimensional Ising model.
In the following we summarize our main results:
\benum
\item 
Within MFT, the chemical potential $\mu$ arises naturally as the Lagrange multiplier required for the constrained minimization of the canonical equilibrium free energy (see Sec.~\ref{sec_ads_model}).
As a consequence of this relationship, the constraint-induced chemical potential $\tilde\mu$ acquires a dependence on various system parameters, such as the system size, temperature, ``mass'' $\mass$, and the adsorption strength $h_1$ at the walls.
The ensuing dependences of $\mu=\tilde\mu$ on these parameters have been studied analytically and numerically in Secs.\ \ref{sec_ads_pert} and \ref{sec_ads_num}, and are summarized in the diagram in Fig.~\ref{fig_phasediag_MFT}.
These findings have important repercussions on the behavior of the \CCF in the presence of a mass constraint.

\item As a crucial consequence of the mass constraint we have demonstrated that, within MFT and in the case of symmetric boundary conditions [$(++)$], the limit of infinitely strong surface absorption (corresponding to a surface field $h_1\to\infty$) cannot be taken. 
The reason is the divergence $\propto 1/\hat z$ of the mean field OP profile as the distance $\hat z$ from the wall decreases [Fig.~\ref{fig_prof_H1range}(a,b)], which leads to a macroscopically large amount of adsorbed mass within the film. 
As a consequence, not only the parts of the profile at the wall and near the center of the film [Figs.~\ref{fig_prof_H1range}(c,d)], but also the constraint-induced bulk field (Fig.~\ref{fig_bulkfield_constr}) and the mean field amplitude of the canonical \CCF [Fig.\ \ref{fig_casAmpl}(a)] diverge upon increasing $h_1$.
Critical fluctuations have the profound effect of reducing the degree of this singularity ($\propto \hat z^{-0.52}$) such that the integrated OP profile remains finite even for $h_1\to\infty$, thereby eliminating the above mentioned divergence associated with MFT in the canonical ensemble.
If the boundary conditions are perfectly anti-symmetric [$(+-)$], the potentially divergent contributions to the excess adsorption from the region near the two walls cancel out so that, in this case, the divergence is absent already within MFT.
We remark that the limit $h_1\to\infty$ is of particular interest, because it corresponds to the renormalization-group fixed-point of the so-called normal surface universality class. It has turned out that this theoretical concept describes quite accurately even actual experimental results obtained under conditions of strong adsorption preference (see, e.g., Refs.\ \cite{hertlein_direct_2008, gambassi_critical_2009, gambassi_casimir_2009, gambassi_critical_2011, trondle_trapping_2011, paladugu_nonadditivity_2016}).

\item As revealed by MFT, the mass constraint leads to a significantly different behavior of the \CCF compared to the unconstrained case.
In particular, for $\mass=0$ and $(++)$ boundary conditions with finite $h_1$ [Fig.~\ref{fig_pCas_MFT}(a)], the canonical \CCF is \emph{repulsive} and its scaling function $\Xi\can$ decays, within MFT, algebraically $\propto 1/\tscal$ for large values of the scaling variable $\tscal=(L/\amplXip)^{1/\nu}t$, where $\xi_\pm(t=T/T_c-1\to 0^\pm)=\amplXipm |t|^{-\nu}$ is the bulk correlation length.
In contrast, under the same conditions, but in the grand canonical ensemble, the \CCF is \emph{attractive} and its scaling function $\Xi\gc$ decays $\propto \exp(-\sqtscal)$ upon increasing $\tscal$.
Instead, for $(+-)$ boundary conditions and with the constraint $\mass=0$, the canonical and the grand canonical Casimir forces are identical within MFT [Fig.~\ref{fig_pCas_pm_MFT}(a)].

\item The qualitative features of MFT are confirmed by Monte Carlo simulations of the three-dimensional Ising model [Fig.~\ref{fig_pCas_MC}(a)]: in the case $\mass=0$ with $(++)$ boundary conditions, the canonical \CCF acquires a repulsive character and decays rather slowly for large $\tscal$, while for $(+-)$ boundary conditions (and sufficiently large lateral system sizes), canonical and grand canonical \CCFs are practically indistinguishable.
The asymptotic decay behavior of the \CCF for large $\tscal$ differs from the mean-field prediction.
We have further demonstrated that decreasing the lateral system size for $(+-)$ boundary conditions enhances the difference between the \CCFs in the two ensembles [Fig.~\ref{fig_pCas_MC_size}], which is due to the effect of critical fluctuations. 

\item We have shown that in the canonical and in the grand canonical ensemble the functional forms of the stress tensor, as obtained from the mean field free energy functional, are \emph{identical} (see Sec.~\ref{sec_Casimir_mft} and Appendix \ref{app_stressten}). 
This not only provides an alternative approach to study \CCFs in the canonical case, but also implies that, within MFT and under the same thermodynamic conditions, the \emph{film} pressures in the canonical and in the grand canonical ensemble are identical [see Eq.~\eqref{eq_pf_equal}].
As a crucial consequence of this identity it follows that the difference between the corresponding \CCFs must be due to the different \emph{bulk} pressures that are subtracted [Eq.~\eqref{eq_pCas_pdiff}]: in the grand canonical ensemble, the appropriate bulk pressure is the one of a homogeneous system with the same chemical potential $\mu$ as the film. 
In contrast, in the canonical ensemble, while depending in principle on the specific experimental setup, the most natural choice for the bulk system is one which has the same mean mass density $\mden=\mass/L$ as the film (with $\mass$ taken as per transverse area of the film). 
This choice is indeed in line with standard finite-size scaling arguments invoked to extract the \CCF from the residual free energy.
The ensuing situation can be most easily understood for $(++)$ boundary conditions and a constraint of zero total mass, i.e., $\mass=0$. In the \emph{film}, the constraint is realized by introducing into the free energy functional [Eq.~\eqref{eq_Landau_func_gc}] a chemical potential $\mu=\tilde\mu$ which depends, \emph{inter alia}, on the film thickness, adsorption strength, and temperature and which, in general, is nonzero.
In contrast, in the \emph{bulk}, the analogous constraint of a vanishing OP density $\mden=0$ requires a \emph{vanishing} bulk chemical potential, $\mu_b=0$ in the corresponding free energy functional [Eq.~\eqref{eq_Landau_func_gc_blk}]. 
Due to the identical forms of the mean field stress tensors in the two ensembles, the two distinct chemical potentials give rise to different bulk pressures.
For comparison, for $(+-)$ boundary conditions, the situation $\mass=0$ is realized with $\mu=\mu_b=0$, leading in this case to identical \CCFs in the two ensembles.
We expect that, once the effects of fluctuations are taken into account, film pressures turn out to be different in the canonical and the grand canonical ensemble, thereby providing a further contribution to the difference between the corresponding CCFs (see Fig.~\ref{fig_pCas_MC_size} and the related discussion).

\item 
Within MFT we have found that, once a nonzero total mass $\mass$ is imposed, canonical and grand canonical \CCFs generally differ both under $(++)$ and $(+-)$ boundary conditions.
In particular, the canonical \CCF shows a strong variation with the mass $\mass$ of the film [Figs.~\ref{fig_pCas_MFT}(b-d) and \ref{fig_pCas_pm_MFT}(b-d)].
For $(++)$ boundary conditions, it turns out that this dependence can be explicitly attributed to a term which carries the character of a surface contribution to the film free energy [see Eq.~\eqref{eq_linMF_pCasCan_stressroute}], the presence of which is a genuine consequence of the mass constraint in the canonical ensemble.
Once this term is subtracted, the canonical \CCF decays towards zero for large $\mass$ [Fig.~\ref{fig_pCas_mass}(b)].
For $(+-)$ boundary conditions, the \CCF displays maxima for nonzero values of $\mass$ and decreases for large values of $|\mass|$ (Fig.~\ref{fig_pCas_pm_mass}).
\eenum 

An experimental test of our predictions requires the realization of a system with a constant value of the integrated OP.
Among the simplest possible examples is a binary liquid mixture, for which one prohibits particle exchange between a suitably constructed compartment and its environment.
Importantly, experimental measurements of the \CCF in the canonical ensemble require film and bulk to consist of the same fluid with identical mean values of the OP, i.e., of the concentration.
Only in this case the bulk-like contribution to the film pressure is balanced precisely by the corresponding one of the same surrounding binary liquid mixture so that the canonical \CCF as analyzed in the present study is revealed.
In the grand canonical setup, the required cancellation is guaranteed by construction, because film and bulk fluid are thermodynamically coupled via the same chemical potential.
As we have shown here, the canonical \CCF exhibits novel features compared to the grand canonical one, such as its significantly slower decay upon increasing the film thickness and a change of its character of being attractive or repulsive.
Hence, prohibiting mass exchange between film and environment opens up a further, and hitherto unexplored, route to tune the \CCF. 
Being able to control \CCFs is highly desirable for micro- and nanoscale mechanical devices in order to prevent stiction due to the omnipresent quantum-mechanical Casimir forces which are typically attractive \cite{casimir_attraction_1948, bordag_new_2001}.

Besides our Monte Carlo simulations, we have focused on the already rich MFT of a laterally homogeneous film with transverse symmetry-breaking boundary conditions, covering the so-called normal surface universality class \cite{binder_critical_1983,diehl_field-theoretical_1986}.
We have seen that, within MFT and as long as the surface adsorption strength $h_1$ is finite, a canonical system can be mapped exactly onto a grand canonical one with an appropriately chosen value of the chemical potential. 
Hence, at least within MFT, the ensemble difference for the \CCF is extrinsic in the sense that it is caused by the difference in the corresponding bulk pressures.
\emph{Intrinsic} differences between the two ensembles arise due to the different fluctuation spectra, which lend themselves to future studies.
It will be furthermore interesting to investigate critical adsorption and \CCFs beyond MFT within the canonical ensemble for varying surface fields and nonzero total magnetizations.
This can be accomplished, for instance, by suitably extending the present Monte Carlo simulations of the Ising model.
In addition, further surface universality classes in the canonical ensemble may be considered and the effect of surface enhancements under non-symmetry breaking boundary conditions may be studied.
Finally, the effects of possible lateral inhomogeneities in finite films due to phase separation below the capillary critical point await a detailed investigation.


\acknowledgments{We thank T.\ F.\ Mohry for providing access to a numerical code. 
Useful discussions with D.\ Dantchev, E.\ Eisenriegler, R.\ Evans, F.\ H\"{o}fling, A.\ Maciolek, and M.\ Tr\"{o}ndle are acknowledged.}

\appendix

\section{Scaling behavior and mapping relation}
\label{app_mapping}
In Sec.~\ref{sec_scal_CA} we invoke general scaling hypotheses. Here, we check them for the actual MFT under investigation, in particular concerning the scaling behavior of the OP profiles $\phi(z)$ across the film for finite surface fields. We find that the profiles obtained for such systems can be scaled onto a single universal profile corresponding to the case $|h_1| \to \infty$, but for a thicker film, extending the findings in Ref.\ \cite{mohry_crossover_2010} to the presence of bulk fields.
To this end we consider the function $\hat\phi(z) \equiv b \phi(b z)$ where $b$ is an arbitrary rescaling factor and $\phi(z)$ solves Eq.~\eqref{eq_ELE0} with the boundary condition in Eq.~\eqref{eq_ELE0_bc}. This leads to
\beq \begin{split}
\pd_z^2 \hat\phi(z) &= b^3 \phi''(b z) = b^3 \left(\tau\phi(bz) + \frac{g}{6}\phi^3(bz) - \mu\right)\\
&= b^2 \tau \hat\phi + \frac{g}{6} \hat \phi - b^3 \mu
\end{split}\eeq 
and
\beq \frac{d}{dz} \hat \phi\left(z \right)\Big|_{z=\pm (L/2)/b} = b^2 \phi'\left(\pm \frac{L}{2}\right) = \pm b^2 h_1^\pm.
\label{eq_prof_rescaling_bc}\eeq 
Accordingly, we conclude that, if $\phi(z)$ solves the ELE for the parameters $\tau$, $g$, $\mu$, and $h_1^\pm$ in a film with thickness $L$, then $b \phi(b z)$ solves the ELE for the parameters $b^2 \tau$, $g$, $b^3 \mu$, and $b^2 h_1^\pm$, respectively, in a film of thickness $L/b$. 
This can be expressed in terms of the homogeneity relation 
\beq \phi(z,\tau,\mu,h_1,L) = b^{-1} \phi(z/b, b^3 \mu, b^2 h_1, L/b),
\label{eq_scal_phi_hb}\eeq 
which was anticipated in Eq.~\eqref{eq_gen_hom_phi} for the general case, i.e., beyond MFT.
The mass constraint in Eq.~\eqref{eq_Mass0} transforms into $\mass = \int_{-(L/2)/b}^{(L/2)/b} dz\, \hat \phi(z)$ and thus remains unchanged.

Instead of using the boundary condition \eqref{eq_prof_rescaling_bc} for the rescaled field $h_1$, one may note \cite{mohry_crossover_2010} alternatively, that $\hat \phi$ satisfies, for $b<1$, a boundary condition at $z=\pm L/2$:
\beq \frac{d}{dz} \hat \phi\left(z \right)\Big|_{z=\pm L/2} = b^2 \phi'\left(\pm \frac{b L}{2}\right) \equiv \pm \hat h_1^\pm,
\label{eq_prof_mapping}\eeq 
where the last equation defines the effective surface field $\hat h_1^\pm$. 
Equation \eqref{eq_prof_mapping} provides an implicit equation for the rescaling factor $b$ as a function of $L$ and $\hat h_1^\pm$. 
It  expresses the trivial fact that, if a profile $\phi(z)$ fulfills the ELE in the domain $(-L/2,+L/2)$, a part of the profile around $z=0$ fulfills the same ELE in the subdomain $(-bL/2,+bL/2)$, $b<1$, and the boundary conditions are provided essentially by the slope of $\phi$ at the shifted boundaries $\pm bL/2$.
In contrast to Eq.~\eqref{eq_scal_phi_hb}, this transformation of the profile does not respect mass conservation.

For brevity, in the following we focus on $(++)$ boundary conditions, i.e., $H_1^\pm= H_1$. (The case of $(+-)$ boundary conditions can be analyzed analogously.)
The considerations above imply a relationship between the OP profile $ m_*$ corresponding to the case $H_1\to\infty$ in a film of thickness $L^*$ and the profile for finite $H_1$ in a film of size $L<L^*$. 
If, for certain values of $\tau$ and $\mu$, the quantities
\beq  m_*(\zeta),\quad \tscal, \quad B_{*}\eeq 
satisfy the ELE in Eq.\ \eqref{eq_ELE} in a film of thickness $L^*$ with surface field $H_1^*=\infty$, then the rescaled quantities
\beq b  m_*(b \zeta),\quad b^2\tscal, \quad b^{3}B_{*} \eeq  
satisfy the ELE in a film of thickness $L=b L^*$, with a finite surface field $H_1$ and with the rescaling factor $b<1$ determined by
\beq H_1 = -b^2  m_*'\left( -\frac{b}{2}\right) = b^2  m_*'\left( \frac{b}{2}\right) .
\label{eq_h1_map}\eeq 
Since, for positive $H_1^*$, the slope of $ m_*$ varies between $-\infty$ and $0$ ($0$ and $\infty$) in the interval $-1/2\leq \zeta \leq 0$ ($0\leq \zeta\leq 1/2$), for any positive $H_1$ Eq.~\eqref{eq_h1_map} renders a solution $b(H_1)$. The case $H_1<0$ can analogously be mapped to a universal profile pertaining to $H_1^*=-\infty$.
We thus conclude that any profile $ m(\zeta,\tscal,B)$ solving the Euler-Lagrange Eq.~\eqref{eq_ELE} for finite $H_1$ and given $B$ turns out to be a part of the universal profile $ m_*$: 
\beq  m(\zeta,\tscal, B) = b  m_*(b\zeta, b^{-2}\tscal, b^{-3}B),
\eeq 
with the rescaling factor $b(H_1)$ determined by Eq.~\eqref{eq_h1_map}.
This fact can be used in order to derive a short-distance expansion for the profile close to a wall, alternatively to the explicit construction in Sec.~\ref{sec_sde}.

\section{Perturbation theory for order parameter profiles with large mean values}
\label{app_pert}

In this appendix, we briefly present a perturbative solution of the Ginzburg-Landau model within mean field theory, which is applicable for large values of the prescribed mass $\mass$.
In contrast to Sec.~\ref{sec_ads_pert}---where we effectively consider perturbations around $\mass=0$---here we construct a perturbative solution around a non-vanishing, spatially constant, mean value 
\beq \mden =\frac{1}{L} \int_{-L/2}^{L/2} dz \, \phi(z) = \frac{\mass}{L}
\label{eq1_mass}\eeq
of the OP profile $\phi(z)$. 
To this end we write
\beq \phi(z) = \mden + \phi_0(z),
\label{eq1_m_split}\eeq
with $\int_{-L/2}^{L/2}dz\, \phi_0(z)=0$.
With this decomposition, we naturally account for the fact that, sufficiently far from the critical point, the OP profile $\phi(z)$ varies significantly only close to the boundaries, while it is practically constant in the central region of the film.

First, we focus on films with equal surface fields.
Upon inserting the decomposition \eqref{eq1_m_split} into the free energy functional of Eq.~\eqref{eq_Landau_func_gc} and neglecting terms of higher than second order in $\phi_0$, we have, for $h_1^+=h_1^-\equiv h_1$,
\begin{widetext}\begin{equation} \Fcal_f\gc \simeq \int_{L/2}^{L/2} dz \left[ \onehalf (\phi_0')^2 + \onehalf \tau (\mden+\phi_0)^2 + \frac{1}{4!} g (\mden^4 + 4\mden^3 \phi_0 + 6\mden^2 \phi_0^2) - \mu(\mden+\phi_0)\right]  - h_1\left[  \phi_0(-L/2) +  \phi_0(L/2)\right] -2h_1\mden.
\label{eq1_freeEn}\end{equation}\end{widetext}
Crucially, for the purpose of deriving the ELE for $\phi_0(z)$ from the condition $\delta \Fcal_f\gc/\delta\phi_0(z)=0$, we take $\mden$ to be constant and not participating in the variation with respect to $\phi_0(z)$. 
Once the solution $\phi_0(z)$, which depends on $\mden$, is obtained, Eq.~\eqref{eq1_mass} then renders an implicit, self-consistent equation for $\mden$.
The corresponding ELE, which by construction is linear in $\phi_0(z)$, reads
\beq \phi_0'' = \tau\mden + \frac{1}{6}g\mden^3+ (\tau + \frac{1}{2}g\mden^2) \phi_0 - \mu,
\label{eq1_ELE}\eeq 
with the boundary conditions
\beq \phi_0'\big|_{z=-L/2} = -\phi_0'\big|_{z=L/2} = -h_1.
\eeq 
This equation is solved by
\beq \phi_0(z) = \frac{\mu}{\mtau}-\mden + \frac{g}{3\mtau}\mden^3 + \frac{h_1}{\sqrt{\mtau}}\frac{ \cosh(z\sqrt{\mtau})}{\sinh(L\sqrt{\mtau}/2)},
\label{eq1_phi0}\eeq
where 
\beq \mtau \equiv \tau + \onehalf g\mden^2
\label{eq1_mod_tau}\eeq
is a temperature-like parameter.
For $\mden=0$, the expression in Eq.~\eqref{eq1_m_split} reduces to the one given in Eq.~\eqref{eq_m0_sol}.
Furthermore, all considerations in Sec.~\ref{sec_ads_pert} regarding the divergence of the profile for certain values of $\tau\leq 0$ apply also to Eq.~\eqref{eq1_phi0}  after replacing $\tau$ with $\mtau$.
Due to Eqs.~\eqref{eq1_phi0} and \eqref{eq1_mod_tau}, the consistency condition \eqref{eq1_mass} turns into
\beq \mu = \tau \mden + \frac{1}{6}g\mden^3 -\frac{2h_1}{L} = \mden \mtau -\frac{1}{3}g\mden^3 -\frac{2h_1}{L}.
\label{eq1_B_rel_mass}\eeq 
In the canonical ensemble, Eq.~\eqref{eq1_B_rel_mass} directly yields the bulk field $\tilde \mu=\mu$ as a function of $\mden$ and renders the constrained solution
\beq \tilde\phi_0(z) = -\frac{2h_1}{L\mtau} + \frac{h_1}{\sqrt{\mtau}}\frac{ \cosh(z\sqrt{\mtau})}{\sinh(L\sqrt{\mtau}/2)},
\label{eq1_phi0_constr}\eeq
once the expression for $\mu$ is inserted into Eq.~\eqref{eq1_phi0}.
In the grand canonical ensemble, instead, the bulk field $\mu$ is given and Eq.~\eqref{eq1_B_rel_mass} has to be inverted for $\mden=\mden(\mu)$.

\begin{figure}[t]\centering
	\includegraphics[width=0.9\linewidth]{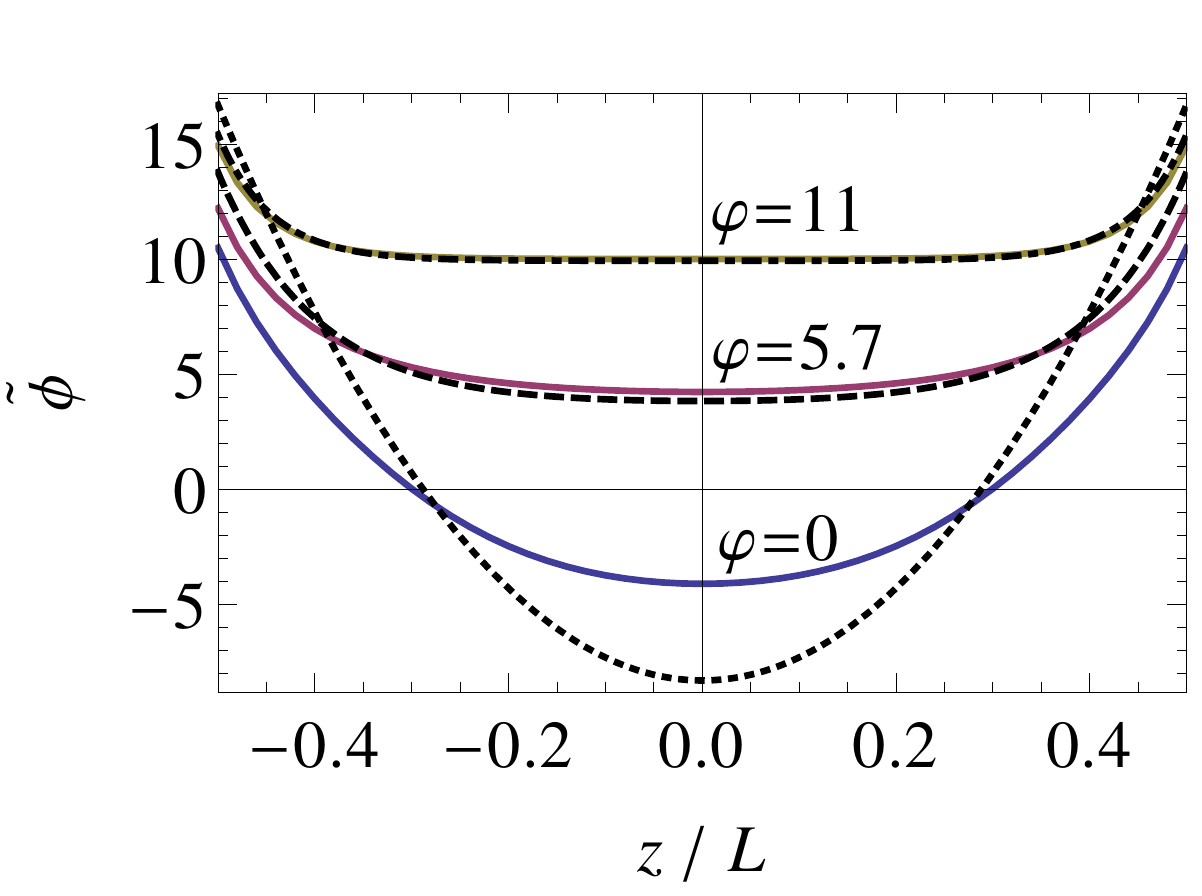}
	\caption{Constrained order parameter profiles in a critical film ($\tau=0$) with $(++)$ boundary conditions and various values of the mean mass $\mden$ [Eq.~\eqref{eq1_mass}]. The colored solid lines represent the profiles obtained from a numerical solution of the ELE in Eq.\ \eqref{eq_ELE0} with the bulk field being determined such that a given value of the mean mass $\mden$ [Eq.~\eqref{eq1_mass}] is recovered. Broken black lines represent the analytical solution $\tilde\phi_0$ in Eq.~\eqref{eq1_phi0_constr} of the linearized ELE in Eq.\ \eqref{eq1_ELE} for the same values of $\mden$. The profiles are computed here for fixed values of $L$ and $h_1$, corresponding to a value of the scaling variable $H_1=100$. As expected from our analysis in Appendix \ref{app_pert}, the accuracy of the analytical solution $\tilde \phi_0$ increases upon increasing $\mden$.}
	\label{fig_OPprof_var_M}
\end{figure}

The expression in Eq.~\eqref{eq1_phi0_constr} can be considered to be an accurate approximation of the solution of the full ELE [Eq.~\eqref{eq_ELE0}] if the terms $(g/2)\mden\phi_0^2 + (g/6)\phi_0^3$ discarded in the expansion of $(\mden+\phi_0)^3$ in Eq.~\eqref{eq1_ELE} are small compared to $(g/6)\mden^3 + (g/2)\mden^2\phi_0$, which are kept in Eq.~\eqref{eq1_ELE}. 
From the dependence of $\phi_0$ on $\mden$ and the fact that $\phi_0$ vanishes as $\mden\to\pm \infty$ [see Eqs.~\eqref{eq1_phi0} and \eqref{eq1_mod_tau}], this condition is easily seen to be fulfilled for sufficiently large $|\mden|$. 
The detailed condition depends on $h_1$ and $\tau$ and is not stated here.
Figure \ref{fig_OPprof_var_M} compares the solution of the linearized ELE in Eq.~\eqref{eq1_phi0_constr} (broken lines) with the numerical solution (solid lines) of the complete ELE in Eq.\ \eqref{eq_ELE0} for $\tau=0$, a large value of $h_1$ and various values of $\mden$. The profile $\phi(z)$ depends significantly on $z$ only close to criticality ($\mden=0$), while the spatial variation diminishes upon increasing $\mden$ and the analytical solution (broken lines) of the linearized ELE becomes more accurate.
Agreement between the exact and the analytical solution improves also upon reducing the strength of $h_1$. Note that, for $\mden=0$, $\tilde\phi_0$ in Eq.~\eqref{eq1_phi0_constr} reduces to the expression given in Eq.~\eqref{eq_phi0_constr}, the accuracy of which has been analyzed in Fig.~\ref{fig_linMFT_valid}.

In order to be consistent with the level of approximation of the free energy in Eq.~\eqref{eq1_freeEn}, we expand also the transverse component of the stress tensor in Eq.~\eqref{eq_stressten_gc} up to quadratic order in $\phi_0$:
\begin{multline} T_{zz}\gc =  \onehalf (\phi_0')^2 - \onehalf \tau (\mden+\phi_0)^2 \\ - \frac{1}{4!} g \left( \mden^4 + 4\mden^3 \phi_0 + 6\mden^2 \phi_0^2\right) +\mu(\mden+\phi_0) .
\label{eq1_stress}\end{multline}
Accordingly, the grand canonical film pressure $p_f\gc= T_{zz}\gc$ follows by inserting the expression \eqref{eq1_phi0} for the OP profile into $T_{zz}\gc$, resulting in
\beq p_f\gc = \frac{\mu^2}{2\mtau} + \frac{g}{6\mtau}\mden^3\left(2\mu - \frac{3}{4}\mtau\mden + \frac{1}{3} g\mden^3\right) + \frac{h_1^2}{1-\cosh (L\sqrt{\mtau})},
\eeq 
where $\mden$ has to be understood as a function of $\mu$ according to Eq.~\eqref{eq1_B_rel_mass}.
In the canonical ensemble, instead, we obtain the pressure $p_f\can$ after inserting Eq.~\eqref{eq1_B_rel_mass} into the previous expression: 
\beq p_f\can =  \onehalf \mtau \mden^2 - \frac{1}{8}g\mden^4 -\frac{2 h_1\mden}{L} + \frac{2h_1^2}{L^2\mtau} + \frac{h_1^2}{1-\cosh (L\sqrt{\mtau})}.
\eeq 

In order to obtain the bulk pressure $p_b\gc$ in the grand canonical ensemble, we decompose the bulk OP as $\phi_b = \mden+\phi_{b,0}$ and expand, for reasons of consistency, the bulk equation of state [i.e., Eq.~\eqref{eq1_ELE} with $\phi_0\to \phi_{b,0}=\const.$] analogously up to first order in $\phi_{b,0}$, obtaining $\tau(\mden+\phi_{b,0}) + (g/6)\mden^3 + (g/2)\mden^2 \phi_{b,0}=\mu$ instead of Eq.~\eqref{eq1_B_rel_mass}. Solving for $\phi_{b,0}$ and inserting the result into the stress tensor in Eq.~\eqref{eq1_stress}, one finds
\beq p_b\gc=\frac{\mu^2}{2\mtau} + \frac{g}{6\mtau}\mden^3\left(2\mu - \frac{3}{4}\mtau\mden + \frac{1}{3} g\mden^3\right).
\eeq 
The bulk pressure $p_b\can$ in the canonical ensemble follows, instead, from Eqs.~\eqref{eq1_stress} and \eqref{eq1_B_rel_mass} (with $\phi_0=0$ and $h_1=0$) immediately as
\beq p_b\can = \onehalf \tau\mden^2 + \frac{1}{8}g\mden^4.
\label{eq1_pB_can}\eeq 
After subtracting the bulk pressures $p_b\cgc$ from the corresponding ones $p_f\cgc$ in the film, we obtain the \CCF per area and $k_B T$ [see Eqs.\ \eqref{eq_pCas_dFdL}, \eqref{eq_scalvar}, and \eqref{eq_scalvar_mft}]
\beq \Kcal\gc = \frac{h_1^2}{1-\cosh(L\sqrt{\mtau})} = \frac{\Delta_0}{L^{4}}\left[  \frac{H_1^2}{1-\cosh(\sqrt{\mtscal})} \right]
\label{eq1_pcas_gc}\eeq
in the grand canonical ensemble and 
\beq\begin{split} \Kcal\can &= -\frac{2 h_1 \mden}{L} + \frac{2h_1^2}{L^2\mtau} + \frac{h_1^2}{1-\cosh(L\sqrt{\mtau})} \\ &=  \frac{\Delta_0}{L^{4}}\left[ - 2 H_1 \Mass + \frac{2H_1^2}{\mtscal} + \frac{H_1^2}{1-\cosh(\sqrt{\mtscal})} \right] \end{split}
\label{eq1_pcas_can}\eeq 
in the canonical ensemble, where $\mtscal=L^2\mtau$.
The terms in the square brackets [including the prefactor $\Delta_0$ defined in Eq.~\eqref{eq_Delta0}] represent the corresponding scaling functions $\Xi\gc$ [Eq.~\eqref{eq_Casi_force_gc}] and $\Xi\can$ [Eq.\ \eqref{eq_Casi_force_c}], respectively, which are obtained after introducing the dimensionless quantities in Eq.~\eqref{eq_scalvar} and after defining 
\beq \mtscal \equiv \tscal+3\Mass^2
\eeq 
in analogy with Eq.~\eqref{eq1_mod_tau}.

The expressions in Eqs.\ \eqref{eq1_pcas_gc} and \eqref{eq1_pcas_can} incorporate the leading dependence of the \CCF on the mass (density) $\mden=\mass/L$ [Eq.~\eqref{eq_mden}] of the film and become identical to the expressions of linear MFT reported in Eqs.~\eqref{eq_linMF_pCasGC_stressroute} and \eqref{eq_linMF_pCasCan_stressroute}, provided therein $\tau$ is replaced by the shifted temperature parameter $\mtau$ [Eq.~\eqref{eq1_mod_tau}], which reduces to $\tau$ for $\mden=0$.
For large $\mden$, as considered here, $\mtau$ is the appropriate quantity to enter into Eqs.\ \eqref{eq1_pcas_gc} and \eqref{eq1_pcas_can}.
We thus conclude that, sufficiently far from criticality ($|\Mass|\gg 1$), the grand canonical Casimir force $\Kcal\gc$ decays exponentially upon increasing $\mden$, whereas the canonical one $\Kcal\can$ scales linearly with $\mden$ due to the presence of the ``surface pressure'' term $-2h_1\mden/L$. 
If this term is subtracted from $\Kcal\can$, the behavior for large $\mden$ is governed by the second term in Eq.~\eqref{eq1_pcas_can}, which yields the algebraic dependence $\Kcal\can(|\mden|\gg \sqrt{|\tau|})+2 h_1\mden/L \propto 1/\mden^2$.

Repeating the above procedure for $(+-)$ boundary conditions yields for the scaling functions of the \CCFs the same expression as the one reported in Eq.~\eqref{eq_pm_Xicas_linMF}, again with $\tau$ replaced by $\mtau$.
Thus, in this case the \CCF decays exponentially as a function of $\mden$ for \emph{both} ensembles.

\section{Derivation of the stress tensor}
\label{app_stressten}
In the following, we derive the stress tensor $T_{ij}\can$ stemming from a generic free energy functional in the presence of a constraint on the integral of the order parameter (OP), i.e., in the canonical ensemble.
Within MFT, we find that this stress tensor exhibits the same expression as in the more common unconstrained (i.e., grand canonical) case (see, e.g., Refs.~\cite{krech_casimir_1994, okamoto_attractive_2013}), with the bulk field playing the role of a Lagrange multiplier.
The derivation consists of evaluating the free energy difference between \emph{two equilibrium configurations} of a fluid confined by arbitrarily shaped boundaries. 
For instance, in the case of a film bounded by two walls, the two configurations can differ by an infinitesimal displacement of the walls.
We consider an arbitrarily shaped volume $V$ and therefore, differently from the main text, do not normalize here thermodynamically extensive quantities by a transverse area.
The initial configuration of the fluid within the volume $V$ in $d$ spatial dimensions is described by coordinates $\rv$, is characterized by an OP field $\phi(\rv)$, and has the free energy
\beq \Fcal = \int_V d^dr\,\Lcal(\phi(\rv), \nabla \phi(\rv)) + \int_S dS\, \Lcal_S(\phi(\rv)),
\label{eq2_stress_Finit}\eeq 
where $\Lcal$ denotes the bulk free energy density and $\Lcal_S$ is the explicit contribution to its counterpart at the surface. The second integral in Eq.~\eqref{eq2_stress_Finit} runs over the boundary $S$ of the volume $V$.
In the final configuration, e.g.,  after the displacement of the walls of a film, the system with the final volume $V'$ is described by coordinates 
\beq \rv' = \rv+\uv(\rv),
\label{eq2_stress_coordtr}\eeq 
where $\uv(\rv)$ characterizes the displacement of the local OP.
In general, upon a change of the configuration, the system attains a new thermodynamic equilibrium state, implying a change of the OP in addition to the one due to the displacement from position $\rv$ to $\rv'$.
Denoting $\phi'(\rv')$ (note that here the symbol $'$ does not indicate a differentiation) as the OP in the final configuration, this change is expressed as
\beq \phi'(\rv'(\rv)) = \phi(\rv) + \delta\phi(\rv),
\label{eq2_stress_OPtr}\eeq 
which essentially serves as the definition of $\delta\phi(\rv)$.
In both the initial and final configurations, the OP is required to obey the constraint of fixed total mass:
\beq \mass = \int_V d^dr\, \phi(\rv) = \int_{V'}d^d r'\, \phi'(\rv').\label{eq2_OPconstr}\eeq 

In order to evaluate the free energy 
\beq \Fcal' = \int_{V'} d^d r'\, \Lcal(\phi'(\rv'), \nabla' \phi'(\rv')) + \int_{S'} dS'\, \Lcal_S(\phi'(\rv'))
\eeq 
in the final configuration, we use the fact that, for small displacements,
\beq d^d r' = d^d r \left[1+\nabla\cdot\uv + O(u^2)\right],
\label{eq2_Jacobian}\eeq 
where the expression in square brackets stems from the Jacobian determinant associated with the coordinate transform described by Eq.~\eqref{eq2_stress_coordtr}.
Furthermore, for any given function $\tilde f(\rv)$ and $\rv=\rv(\rv')$ one can define $f(\rv') \equiv \tilde f(\rv(\rv'))$, from which it follows that
\beq \frac{\pd f(\rv')}{\pd r_i'} = \frac{\pd \tilde f(\rv(\rv'))}{\pd r_i'} = \frac{\pd \tilde f}{\pd r_j}\frac{\pd r_j}{\pd r_i'} .
\label{eq2_stress_OPderivtr}\eeq 
In addition, from the inversion of Eq.~\eqref{eq2_stress_coordtr}, i.e., $\rv(\rv') = \rv'-\uv(\rv(\rv'))$, it follows that
\beqn 
\frac{\pd r_j}{\pd r_i'} = \delta_{ij}-\frac{\pd u_j}{\pd r_k}\frac{\pd r_k}{\pd r_i'} \simeq \delta_{ij}-\frac{\pd u_j}{\pd r_k}\delta_{ki}+O(u^2) = \delta_{ij} - E_{ij},
\eeqn 
where
\beq E_{ij} \equiv \frac{\pd u_j }{ \pd r_i}
\label{eq_strain_ten}\eeq 
is the strain tensor associated with the transformation field $\uv$.
Accordingly, to leading order in $u$, Eq.~\eqref{eq2_stress_OPderivtr} turns into 
\beq 
\frac{\pd f(\rv')}{\pd r_i'} 
= \left\{\left[\frac{\pd}{\pd r_i} - E_{ij} \frac{\pd}{\pd r_j}\right] \tilde f(\rv)\right\}\Bigg|_{\rv=\rv(\rv')}.
\label{eq2_dfdr_strain}\eeq
As a result, the difference $\Delta\Fcal$ between the free energies of the two infinitesimally different configurations is given by
\begin{widetext}
\beq\begin{split}
\Delta\Fcal\equiv \Fcal' - \Fcal &= \int_V d^d r\,(1+\nabla\cdot\uv) \Lcal\left[\phi+\delta\phi, \pd_i(\phi+\delta\phi) - E_{ij}\pd_j(\phi+\delta\phi)\right] +\int_S dS\, \Lcal_S(\phi+\delta\phi) - \Fcal\\
&= \int_V d^d r\, (1+\nabla\cdot \uv) \left\{ \Lcal[\phi,\pd_i\phi] + \frac{\pd \Lcal}{\pd\phi}\delta\phi + \frac{\pd\Lcal}{\pd(\pd_i\phi)}\pd_i \delta\phi - \frac{\pd\Lcal}{\pd(\pd_i\phi)}E_{ij} \pd_j\phi + O(u\delta\phi)\right\} \\ 
&\qquad + \int dS \left\{\Lcal_S(\phi) + \frac{\pd\Lcal_S}{\pd\phi}\delta\phi\right\} - \Fcal\\
&= \int_V d^d r \left[(\nabla\cdot\uv) \Lcal - E_{ij} \frac{\pd \Lcal}{\pd(\pd_i\phi)} \pd_j\phi\right] 
+ \int_V d^d r \left[\frac{\pd\Lcal}{\pd\phi} - \pd_i \frac{\pd\Lcal}{\pd(\pd_i\phi)}\right]\delta\phi 
 + \int_S dS\left[\frac{\pd \Lcal_S}{\pd \phi} - n_i \frac{\pd\Lcal}{\pd(\pd_i\phi)}\right]\delta\phi ,
\end{split}
\label{eq2_stress_variation}\eeq
where $\bv{n}$ is the unit vector normal to the surface $S$ and pointing towards the interior of the volume $V$.
Note that $V'$ and $S'$ are mapped onto $V$ and $S$ under the transformation in Eq.~\eqref{eq2_stress_coordtr}.

In order to evaluate this expression further, we make use of the fact that $\phi(\rv)$ is an equilibrium configuration of the OP, i.e., that it minimizes $\Fcal$ under the constraint of fixed overall mass $\mass$.
Accordingly, we minimize
\beq \bar\Fcal \equiv \Fcal - \mu\left[\int_V d^d r\, \phi(\rv)-\mass \right],
\eeq 
where $\mu$ is a Lagrange multiplier which is eventually determined such that the constraint is obeyed.
The variation $\delta\bar \Fcal$ of $\bar\Fcal$ with respect to a variation $\phi\to \phi+\delta\phi$ within the volume $V$ and at its boundary $S$ yields
\begin{equation} \delta\bar\Fcal = \int_V d^d r \left[\frac{\pd\Lcal}{\pd\phi} - \pd_i \frac{\pd\Lcal}{\pd(\pd_i\phi)} -\mu\right]\delta\phi 
 + \int_S dS\left[\frac{\pd \Lcal_S}{\pd \phi} -n_i \frac{\pd\Lcal}{\pd(\pd_i\phi)}\right]\delta\phi .
\label{eq_Fconstr_var}\end{equation}
This variation vanishes for all possible choices of $\delta\phi$ only if the quantities in both square brackets vanish. This yields the Euler-Lagrange equations in the bulk,
\beq \frac{\pd\Lcal}{\pd\phi} - \pd_i \frac{\pd\Lcal}{\pd(\pd_i\phi)}  = \mu,
\label{eq2_stress_ELE_bulk}\eeq 
and the boundary conditions
\beq \frac{\pd \Lcal_S}{\pd \phi}\Big|_S = n_i \frac{\pd\Lcal}{\pd(\pd_i\phi)}\Big|_S.
\label{eq2_stress_ELE_wall}\eeq 

Using these results, Eq.~\eqref{eq2_stress_variation} turns into ($E_{ii}=\nabla\cdot \uv$)
\beq \Delta \Fcal = \int_V d^d r\, E_{ij}\left[\Lcal\delta_{ij} - \frac{\pd \Lcal}{\pd(\pd_i\phi)} \pd_j\phi\right] + \mu \int_V d\rv\, \delta\phi(\rv).
\label{eq2_stress_deltaF0}\eeq 
We now make use of the fact that $\delta\phi(\rv)$ must be such that the OP constraint is obeyed in both configurations and compute its integral as
\beq \begin{split} 
\int_V d^d r\, \delta\phi(\rv) &= \int_V d^d r\, \phi'(\rv'(\rv)) - \underbrace{\int_V d^d r\, \phi(\rv)}_{\mass} = \underbrace{\int_{V'} d^d r'\, \phi'(\rv')}_{\mass} - \int_{V'} d^d r'\, [\nabla'\cdot\uv(\rv(\rv'))]\, \phi'(\rv') -\mass\\
&= -\int_{V'} d^d r' \left[\nabla'\cdot \uv(\rv(\rv'))\right] \phi'(\rv')= -\int_V d^d r \,(1+\nabla\cdot\uv)\left[(\pd_i - E_{ij} \pd_j) u_i\right]\left[\phi(\rv)+\delta\phi(\rv)\right] \\
&= -\int_V d^d r \,(\nabla\cdot\uv)\,\phi(\rv) + O(u\,\delta\phi)+O(u^2),
\end{split}\eeq \end{widetext}
where we have used Eq.~\eqref{eq2_OPconstr} and the inverse of Eq.~\eqref{eq2_Jacobian}, $d^d r = d^d r'(1-\nabla'\cdot\uv)+ O(u^2)$.
Accordingly, up to $O(u)$ and $O(\delta\phi)$, the change of the free energy in Eq.~\eqref{eq2_stress_deltaF0} can be written as
\beq \Delta \Fcal = -\int_V d^d r E_{ij} \bar T_{ij}[\phi\eq(\rv)],
\label{eq2_stress_deltaF}\eeq 
where $\bar T_{ij}$ is the stress tensor valid for the canonical ensemble, i.e.,
\beq T_{ij}\can[\phi] = \bar T_{ij}[\phi] \equiv \frac{\pd\bar\Lcal}{\pd(\pd_i\phi)} \pd_j\phi -\bar \Lcal\delta_{ij}  ,\quad \text{with}\quad \bar \Lcal= \Lcal-\mu\phi.
\label{eq2_stress_ten}\eeq 
It is important to keep in mind that the stress tensor in \eqref{eq2_stress_deltaF} is evaluated for the \emph{equilibrium} solution $\phi$ as determined by Eqs.~\eqref{eq2_stress_ELE_bulk} and \eqref{eq2_stress_ELE_wall} and with the Lagrange multiplier $\mu$ taken such that the OP constraint [Eq.~\eqref{eq2_OPconstr}] is obeyed.
We note that the formal expression of this stress tensor coincides with the one which can be derived in the grand canonical ensemble where $\mu$ is a fixed external bulk field.
The divergence of the stress tensor is given by
\beq \pd_i \bar T_{ij} = \left[ \pd_i\left(\frac{\pd\bar\Lcal}{\pd(\pd_i\phi)}\right) - \frac{\pd\bar \Lcal}{\pd\phi}\right]\pd_j\phi = -\frac{\delta\bar\Fcal}{\delta\phi}\pd_j\phi 
\label{eq2_stress_div}\eeq 
which, according to Eq.~\eqref{eq_Fconstr_var}, vanishes in equilibrium.

We now apply the general expressions derived above to the special case of a fluid film confined between two planar and parallel surfaces and consider an infinitesimal displacement of the latter.
The film is assumed to be homogeneous along the lateral directions, such that now the thermodynamic limit of infinite transverse area $A$ can be taken from the outset. 
Accordingly, we return to the convention used in the main text and consider all thermodynamically extensive quantities to be divided by $A$.
The coordinate in the direction perpendicular to the two boundaries is denoted by $z$. 
In the initial configuration, characterized by a certain film thickness $L$, total mass (per transverse area) $\mass$, and surface field $h_1$, the equilibrium state is realized with a certain bulk field $\mu$ and the OP profile $\phi(z)$ solves the corresponding Euler-Lagrange equation [Eq.~\eqref{eq2_stress_ELE_bulk} with the boundary condition given in Eq.~\eqref{eq2_stress_ELE_wall}]. In the final configuration, having a wall separation $L'=L+\Delta L$, the Euler-Lagrange equation is solved with a new value $\mu'$ of the Lagrange multiplier and a new field $\phi'(z')$.
In order to compute the corresponding free energy change $\Delta \Fcal$ (per transverse area) according to Eq.~\eqref{eq2_stress_deltaF}, we note that $\uv = (z/L)\Delta L\,\bv{e}_z$ (where $\bv{e}_z$ is the unit vector in $z$-direction and with the left and right wall initially located at $z=0$ and $z=L$, respectively), $E_{ij} = (\Delta L/L)\delta_{iz}\delta_{jz}$, and thus
\beq \Delta \Fcal = -\frac{\Delta L}{L}\int_0^L dz\, \bar T_{zz} \equiv -\,\Delta L\,\bra \bar T_{zz}\ket = -\, \Delta L \, \bar T_{zz}.
\label{eq2_stress_deltaF_film}\eeq 
The last equality in Eq.\ \eqref{eq2_stress_deltaF_film} is a consequence of the stress tensor being spatially constant in equilibrium due to Eq.~\eqref{eq2_stress_div} and due to the fact that $\bar T_{ij}$ does not depend on $x$ and $y$, which in turn is the case because the system is translationally invariant in the lateral directions ($\pd_x \bar T_{xj}+ \pd_y \bar T_{yj} + \pd_z \bar T_{zj} = 0$ together with $\pd_x \bar T_{xj}=0=\pd_y \bar T_{yj}$ imply $\pd_z \bar T_{zz}=0$).

The bulk pressure $p_b$ is typcially defined as the isotropic part of the stress tensor \cite{onuki_phase_2002}; within MFT and for a spatially homogeneous system ($\phi(\rv)=\phi$), the canonical bulk pressure follows from Eq.~\eqref{eq2_stress_ten} as
\beq p_b = \frac{1}{3} \sum_i \bar T_{ii} = \mu \phi -\Lcal[\phi]\,.
\eeq 
This result is consistent with the well-known thermodynamic expression for the pressure obtained from the canonical free energy $F(T,V,N)$ of a $N$-particle system at temperature $T$ within a volume $V$.
Expressing the homogeneous function $F$ in terms of the free energy density $f$ as
\beq F(T,V,N) = V  f(T,\phi) \,,
\label{f_density}
\eeq
with $\phi =N/V$, the pressure results indeed as
\beq 
p_b = -\frac{\pd F}{\pd V} = -f - V\frac{\pd f}{\pd \phi}\frac{\pd \phi}{\pd V} = -f + V\frac{N}{V^2}\mu = \mu \phi - f\,.
\label{eq2_pbulk_thermo}\eeq
The chemical potential is given by
\beq 
\mu = \frac{\pd F}{\pd N} = \frac{\pd F}{\pd \phi}\frac{\pd \phi}{\pd N} = V\frac{\pd f}{\pd \phi}\frac{1}{V} = \frac{\pd f}{\pd \phi}\,.
\label{eq2_mub_thermo}\eeq
Equation \eqref{eq2_mub_thermo} corresponds to Eq.\ \eqref{eq2_stress_ELE_bulk} in the special case $\phi=\const$\ by noting that $\Lcal(\phi=\const)=f$ and $\pd_i \delta\Lcal/\delta(\pd_i \phi)=0$. 

%


\end{document}